\documentclass[9pt]{article}

\usepackage{lipsum} 
\usepackage[version=4]{mhchem}
\usepackage{siunitx}
\usepackage{amssymb}
\usepackage{amsmath}
\usepackage{textcomp,marvosym}
\usepackage{nameref, hyperref}

\usepackage{amsfonts}
\usepackage[dvipdfmx]{graphicx}
\usepackage{epstopdf}
\usepackage{tikz}
\usepackage{mathrsfs}
\usepackage{ulem}
\usepackage{tikz-cd}
\usepackage{fancybox}
\usepackage{empheq}
\usepackage{amscd}
\usepackage[all]{xy}
\usepackage{mathrsfs}
\usepackage{stfloats}
\usepackage{braket}
\usepackage{here}
\usepackage{float}

\newcommand{\veebot}{\mathrel{\vee\!\!\!\!\bot}}
\newcommand{\lambdabar}{{\mkern0.75mu\mathchar '26\mkern -9.75mu\lambda}}
\def\bSig\mathbf{\Sigma}

\DeclareSIUnit\Molar{M}

\title{Biological hierarchies emerged from natural characteristics of number theory}

\author{Shun Adachi \thanks{Independent Researcher; Corresponding: f.peregrinusns@mbox.kyoto-inet.or.jp; present address: 146-2-5-202 Nakagawara, Makishima-cho, Uji, Kyoto, Japan}}

\begin{document}

\maketitle

\begin{abstract}
Ecologists have long debated whether biological communities are fundamentally continuous or composed of discrete units. Continuum views emphasize smooth compositional change and ambiguous boundaries, whereas classification-based approaches rely on discrete community types for analysis and management. We show how biological grouping—particularly species formation—can emerge from interactions among populations governed by number-theoretic structure.  In our framework, a species is identified with a $p$-Sylow subgroup of a community occupying a single niche; this identification is supported by a topological analysis.  We call the resulting framework the patch with zeta dominance (PzDom) model.  We then examine the system's topological properties in detail and demonstrate that both hierarchical organization and temporal ordering are induced by a one-dimensional probability space endowed with an appropriate topology.  To clarify the appearance of induced fractal structure and its relation to renormalization, we develop a theoretical account based on a new observation: the scaling parameters that play the role of magnetization analogs coincide exactly with the imaginary parts of the nontrivial zeros of the Riemann zeta function.  In the PzDom model, all required computations reduce to the time-dependent density of individuals. The PzDom framework reconciles these perspectives by showing that continuous community variation, represented by small $s$, can give rise to discrete species-level structure when number-theoretic constraints stabilize specific configurations. Thus, continuity and discreteness emerge as different dynamical phases of the same system, offering a unified explanation for long-standing debates in community ecology and the **species problem**.
\end{abstract}

\rm 

\section{Introduction}

Understanding whether biological communities should be treated as continuous entities or as discrete, classifiable units has been a long-standing issue in ecology and taxonomy. A persistent viewpoint holds that community change is fundamentally continuous and that imposing categorical boundaries is artificial. This perspective originates from the classical Clements—Gleason debate, in which Clements conceptualized communities as cohesive  ``superorganisms'', whereas Gleason argued that communities arise from continuous species-specific responses to environmental gradients (e.g. \cite{Resasco2024}). The latter view underlies the enduring position that community classification lacks a natural basis.

Several factors contribute to the persistence of this non-classification perspective. Natural communities exhibit gradual transitions driven by environmental variation, succession, competition, and mutualistic interactions, making sharp boundaries difficult to justify biologically. Successional theory further emphasizes continuity, particularly within plant ecology. In addition, philosophical discussions surrounding the species problem highlight the inherent ambiguity of biological boundaries, reinforcing skepticism toward rigid classification schemes.

Despite this, classification-based approaches remain equally influential. Community typologies such as the Braun—Blanquet system, cluster-analytic methods in community ecology, diversity metrics, and comparative analyses of community structure all rely on discrete categories. For research, conservation, and management, operational boundaries are often indispensable.

The theoretical framework developed in the **PzDom model** in this manuscript and the associated parameter small $s$ provides a mathematical bridge between these historically opposing views. The model demonstrates that continuous ecological variation can give rise to discrete structures through phase-transition-like behavior. When \(\Re(s) < 2\), communities behave as continuous systems and species boundaries become indistinct, offering a mathematical explanation for the persistence of non-classification perspectives. Conversely, when \(\Re(s) > 2\), stable, discrete structures analogous to $p$-Sylow-subgroup-type groupings emerge, making classification biologically meaningful.

Thus, the PzDom framework suggests that continuity and discreteness are not mutually exclusive properties of ecological systems. Instead, the appropriateness of classification depends on the system's state, quantified by small $s$. This provides a unified mathematical explanation for long-standing debates in ecology, taxonomy, and the philosophy of biological classification. \\

In other lines of logic, living organisms differ from nonliving matter in three principal traits: reproduction, metabolism, and compartmentalization. These traits are essential for maintaining individual identity and for achieving high reproductive fitness. However, these definitions omit the social dimensions of living systems. Biological societies are not mere aggregates of independent individuals; they include social interactions that can increase collective fitness. Consequently, grouping based on relational interactions is an important characteristic of organisms and should be considered alongside the classical traits. The problem we address here is how to define “grouping” appropriately in the context of reproduction.\\

Group selection is a proposed evolutionary mechanism in which natural selection acts at the level of groups rather than solely at the level of individuals or genes. In the long run, however, naive group selection is unlikely to persist: the cost of supporting unrelated individuals without direct benefit tends to eliminate such traits over evolutionary time. For this reason, biologists have proposed alternative mechanisms, notably kin selection (inclusive fitness theory) and multilevel selection. Kin selection operates among related individuals, so cooperation can increase fitness in proportion to relatedness; this mechanism is well documented in many taxa, for example eusocial insects. Mechanisms such as a “green beard” gene have been proposed to mark cooperators. Multilevel selection, by contrast, allows selection to act simultaneously at multiple hierarchical levels rather than only at the individual level; hierarchies in biological systems can be interpreted as the outcome of selection acting at different levels, although this theory remains controversial except in some human contexts \cite{Sober1998}. For humans, social choice theory \cite{Suzumura1983} and indirect reciprocity \cite{Nowak2005} also account for apparently self-sacrificing behaviors without invoking group selection.

We now summarize how the notion of group can be interpreted within biological hierarchies. Biological organization comprises multiple nested levels: molecule, cell, tissue, organ, organ system, organism, population, community, ecosystem, and biosphere. In principle, these hierarchical levels may arise from the properties of thermodynamically open systems and from set-theoretic aspects of dissipative structures, but a detailed, general theory remains elusive. To investigate the origin of these hierarchies, we focus on one particularly elusive concept: species. Numerous species concepts exist, and none applies uniformly across all biological contexts. The biological species concept—“groups of actually or potentially interbreeding natural populations that are reproductively isolated from other such groups”—is experimentally tractable and widely used, but it fails for asexual organisms and for ring species. The morphological species concept—“a group of organisms in which individuals conform to certain fixed properties”—applies only when diagnostic traits are discrete; cryptic species provide counterexamples. The evolutionary species concept—“an entity composed of organisms that maintains its identity through time and space and has its own independent evolutionary fate”—is useful but inherently vague, since the degree of divergence required to delimit species is not well defined and often overlaps with population-level variation. Although a common species definition would facilitate comparative studies of community dynamics, no single definition is universally satisfactory for the problems we consider.

To investigate the species concept, we begin with the population. A population, which is ranked immediately below species in the taxonomic hierarchy, is often defined in ecological and demographic terms. Krebs \cite{Krebs1972} defines a population as ``a group of organisms of the same species occupying a particular space at a particular time." This definition is qualitative, and researchers adopt population definitions appropriate to their context. By contrast, the concept of species has a long history of refinement. John Ray offered a biological definition in Historia Plantarum: ``no surer criterion for determining species has occurred to me than the distinguishing features that perpetuate themselves in propagation from seed. Thus, no matter what variations occur in the individuals or the species, if they spring from the seed of one and the same plant, they are accidental variations and not such as to distinguish a species... Animals likewise that differ specifically preserve their distinct species permanently; one species never springs from the seed of another nor vice versa" (translated by Silk; \cite{Ray1686}). Although Ray viewed species as static, his distinction between intrapecific variation and interspecific difference remains important. In this paper we extend that idea by proposing a definition based on discontinuities in the spectrum of the Selberg zeta function and on the zeros of the Riemann zeta function. In Systema Naturae, Carl von Linn\'e \cite{vonLinne1735} presented systematic definitions for taxa at different hierarchical levels (species, genus, order, class), later extended to family, phylum, kingdom, and domain, and he established binomial nomenclature as a standard for scientific names. Although Linn\'e treated species as static, his system enabled a systematic, qualitative estimate of relatedness among organisms. The static view of creationism was later challenged by Lamarck in Philosophie Zoologique \cite{deLamarck1809}, where Lamarckian inheritance was proposed; this idea was largely ignored until recent work on transgenerational epigenetic inheritance \cite{Boskovic2018} prompted reevaluation. The theory of evolution by natural and sexual selection was developed by Wallace \cite{Wallace1858} and Darwin \cite{Darwin1859}, yet the mechanisms of speciation remained unclear. In Versuche \"{u}ber Pflanzenhybriden, Mendel \cite{Mendel1866} proposed that genetic factors influence hybridization and evolution; his experiments involved crosses among varieties with different alleles rather than hybridization between distinct species. The modern species concept combines evolutionary and reproductive criteria: Mayr \cite{Mayr1942} defined species as ``groups of actually or potentially interbreeding natural populations, which are reproductively isolated from other such groups." The distinction between population and species has often hinged on whether the concept is ecological/physical or evolutionary/genetic. Complex cases require combined definitions, and empirical data may be ambiguous with respect to population versus species status \cite{Coyne2004, Whitham2006, Hausdorf2025}. Ring species illustrate this ambiguity: neighboring populations may interbreed continuously while terminal populations are reproductively isolated, so reproductive isolation alone does not always delimit species. Conversely, reports of sympatric speciation show that speciation with discontinuity can occur without geographic isolation (e.g., \cite{Barluenga2006}). Some authors treat the biological species as a conceptual unity produced by gene exchange via bisexual reproduction and migration; under this view, the degree of unification should correlate with bisexual reproduction. In practice this correlation is weak: uniparental taxa often resemble their biparental relatives in similar environments, and isolated populations of biparental species can retain distinct identity for long periods despite limited gene flow \cite{Rasnitsyn2007}. We addressed part of this issue for Japanese Dictyostelia by decoupling short-term ecological time scales from long-term evolutionary time scales, which differ by a factor on the order of $10^{17}$ \cite{Adachi2015}. Thus, populations can be effectively neutral on short ecological time scales while an assemblage of populations is not neutral on evolutionary time scales; accordingly, the characteristic time scale of species-level gene flow is assumed to differ fundamentally from that of population-level migration.

As we saw in the first paragraph, species cannot be disentangled from the social interactions of their constituents. The modern study of social evolution began with Hamilton's ``The genetical evolution of social behavior I" and ``II" \cite{Hamilton1964a,Hamilton1964b}, followed by Maynard Smith's ``Group selection and kin selection" \cite{Smith1964} and Price's ``Selection and covariance" \cite{Price1970}. These works established that genetic relatedness is important for the maintenance of cooperative phenotypes. For cooperation among nonrelatives that can coevolve, Trivers's theory of reciprocal altruism explains how reciprocity can sustain cooperation \cite{Trivers1971}. Multilevel selection theories were later advanced in works such as ``Reintroducing group selection to the human behavioral sciences" \cite{Wilson1994} and Unto Others: The Evolution and Psychology of Unselfish Behavior \cite{Sober1998}; these frameworks can account for cooperative selection among genes that are distantly related within a cell or within a reproductive unit. Although some elementary mathematical arguments have been offered in support of group selection \cite{Sober1998}, modern mathematical analysis shows such arguments are often incomplete because they neglect set-theoretic structure and assume unrealistically long periods of environmental constancy. Biological hierarchies therefore remain a central concept for analyzing evolutionary processes. Information theory can help analyze dynamics across scales, but genetic closeness at the species level is not always easily distinguished from population-level relatedness. 

Our goal is to provide a mathematically grounded mechanism, using exponential- and logarithm-like functions, that yields a clear definition of species and, more broadly, of interspecific interactions. These concepts should be defined qualitatively but supported by quantitative calculations. Previously, we showed that populations can be interpreted within Hubbell's unified neutral theory \cite{Hubbell1997,Hubbell2001} (or within MaxEnt frameworks; e.g., \cite{Phillips2006}), while temporal species can be interpreted as a more adaptive concept \cite{Adachi2015}. This distinction suggests a route to resolving the species-definition problem: population and species levels have distinct characteristics that can be evaluated by the distribution of individual counts under population-level or species-level criteria. We develop this logic further by combining an adaptation concept with number theory. First, we construct a metric that discriminates the boundary between population dynamics, which are associated with chaotic behavior, and species dynamics, which are associated with directional adaptation or maladaptation. This approach is inspired by the theory of fractal zeta functions \cite{Lapidus2017} and is supported by recent work showing strong links between developmental noise and macroevolutionary divergence in fractal-like traits \cite{Saito2025}. At the same time, we adopt a formalism that resembles quantum mechanics so that quantization (the discrete nature of species) and wave-like descriptions (functions that represent species-species interactions) appear as complementary aspects of the same framework. This analogy clarifies how ``interaction" and ``independence" can be formalized mathematically; we address the differences between microscopic quantum mechanics and the present theory where relevant. 

We tested the model using data from Dictyostelia, soil mesofauna, and historical world economic indicators. The Dictyostelia community appears to be near equilibrium with a quantized character: the observed individual-fitness indicator $D$, both among observed values and expected values, are approximately equal to the expected values. Soil mesofauna show ordering or reformation, with $D_{observed} >> D_{expected}$. By contrast, regional world-economics data exhibit chaotic behavior, with $D_{observed} << D_{expected}$, despite overall expansion in scale. These empirical patterns support the model's qualitative distinctions.

\section{Results}
\subsection{A compact universal form: Use the Price equation on a logarithmic abundance (or fitness) scale to obtain a single, general relation for population and species dynamics}
First, we introduce exponential- and logarithm-like functions to force convergence of the indicators to discrete values. The neutral logarithmic distribution of ranked biological populations, for example a Dictyostelia metacommunity \cite{Adachi2015}, can be written as follows: 

\begin{equation}N_k = a - b\ln k,\end{equation}

where $N$ is the population density or the mean population density of a species across patches, and  $k$ is the rank index of the population. The parameters $a \approx N_1$ and the rate of decrease $b$
 are estimated from the data after sorting populations by abundance rank. We also applied this approximation to adaptive species to quantify their deviation from neutral populations \cite{Adachi2015}. Note that this approximation is valid only for communities that occupy the same niche and does not apply to coevolving communities in nonoverlapping niches. 
 
Based on the theory of diffusion equations for Markov processes, as applied in population genetics \cite{Kimura1964}, we assume that the relative abundance of a population or species is proportional to the $N$th power of $D$ ($= \bar{w}$ in \cite{Kimura1964}) multiplied by the relative patch quality $P$ ($= C$ in \cite{Kimura1964}). In other words, $PD^{N_k} = C\bar{w}^{N_k} = \phi = N_k/\Sigma N$; see also \cite{Kimura1964}. In this formulation, $D^{N_k}$  denotes the relative fitness of the population; it varies in time and depends on genetic and environmental background as well as on interactions among individuals. The factor $P$ represents a relative environmental variable that depends on the background of the occupying species and may differ within the same environment when a different species is dominant.

To better understand the principles deduced from Kimura's theory, we introduce the Price equation \cite{Price1970}: 

\begin{equation} w_k\Delta z = \mathrm{Cov}(w_k, z_k) + E(w_k \Delta z_k). \end{equation} 

Remember $N_1 = kN_k$ when the distribution is completely harmonic (neutral). Note that $z = \ln(k\cdot N_k)/\ln k = 1 + \ln N_k/\ln k$, where $k \neq 1$ and $z = +\infty$ when $k = 1$; we use this in place of gene frequency in Price's original formulation. Furthermore, $w_k$ plays the role of a selection coefficient rather than absolute fitness. The relative distance between the logarithms of the norms $N$ and the rank $k$ will be discussed below in the context of Selberg zeta analysis \cite{Juhl2001}; here $\ln k$ is the relative entropy from a uniform distribution as $a$ (in other words, it is a Kullback--Leibler divergence $D_1(P|Q) = \sum_{i = 1}^n p_i\ln\frac{p_i}{q_i}$ with $n = 1$, $p_i = 1$, $q_i = 1/k$, the interaction probability from the first ranked population/species). Thus we can quantify the deviation from a logarithmic distribution; both $\ln N_k$ and $\ln k$ serve as topological entropies.

Next, we assume that for a given patch the expected individual population or species abundance equals the averaged maximum fitness, namely $D$ raised to the power $E(N)$, where $E(N)$ is the mean $N$ across all populations or the sum of mean $N$ over all patches and species ($|D_k|^{E(N)}$). This is a virtual assumption for a worldline (the path of an object in a particular space) because a population appears to be in equilibrium when it follows a logarithmic distribution \cite{Hubbell1997,Hubbell2001,Adachi2015} and when species dominate \cite{Adachi2015}. We will show below that the scale-invariant parameter small $s$ indicates adaptation in species within neutral populations. Under the assumptions in this paragraph, $E[w]$ equals $|D_k|^{E(N)}$ and $E[z]$ is approximately $\ln N_1/\ln k$. If we set $w = |D_k|^{E(N)} - \Delta z$ (where $\Delta z$ denotes a deviation from neutrality), we obtain 

\begin{equation}w_k = \frac{\ln\frac{N_1}{N_k}}{\ln k} - 1 + |D_k|^{E(N)} (k \neq 1).\end{equation}

Recall that $N_1 = kN_k$ when the distribution is harmonic (normal). In this case, $\ln\frac{N_1}{N_k} = \ln k$, and hence $\frac{\ln\frac{N_1}{N_k}}{\ln k}$ equals $1$. This value quantifies the deviation from the harmonic (neutral) distribution. We treat the special case $k = 1$ in a later subsection (see Eqs.~(12) and (13)). For notational simplicity, denote $1 + \mathrm{Cov}/\Delta z: \frac{\ln\frac{N_1}{N_k}}{\ln k}$ by $\Re(s)$ and $E(w_k): |D_k|^{E(N)}$ by $\Im(s)$. Now consider the Dirichlet series $\sum_{j = 1}^{\infty}b_{Dj}l_{Dj}^s$. An upper bound for the box dimension (fractal dimension) of the set $A$ (a subset of $\mathbb{R}^2$) is: 

\begin{equation}\overline{\dim}_BA = \underset{n \to \infty}{\lim\sup}\frac{1}{\ln l_{Dn}^{-1}}\ln(\sum_{j = 1}^nb_{Dj})\end{equation} \cite{Cahen1894}.

If we regard $\underset{n \to \infty}{\lim}\sum_{j = 1}^nb_{Dj} = \frac{N_1}{N_k}$ and $\underset{n \to \infty}{\lim}l_{Dn} = \frac{1}{k}$, then a Dirichlet series satisfying these limits characterizes the model under consideration; sufficiently large dimensions permit approximations arbitrarily close to these limits. When $s = 1$, the Dirichlet series reduces to the set $\frac{N_1}{kN_k}$, which measures the deviation of $N$ from a logarithmic distribution and provides a natural datum point at 1. The discreteness implied by this construction is examined in later sections, in particular in Figure 1 and Tables 1--3.

\subsection{Introducing $\Re(s)$ permits a practical classification of neutrality regimes}
Zipf's law is commonly used to analyze discrete probability distributions that follow a power law.  In particular, if the distribution of $N$ can be approximated by a logarithmic relation with parameter $k$, then Eqn.~(1) applies.  Zipf's law is closely related to the Riemann zeta function $ \zeta(s)=\sum_{n=1}^{\infty}\frac{1}{n^{s}} $, as shown below: 

\begin{equation}|P_k||D_k|^{N_k} = f_s(k) = \frac{1}{k^{\Re(s)}|\zeta(s)|} = \frac{N_k}{\Sigma N}.\end{equation}

This normalizes the $k$th abundance by $\Sigma N$. We take absolute values of $|\zeta|$ and $|P_k|$ to approximate both the $\Re(s)>1$ ($\zeta>0$, $N_k<N_1/k$) and $\Re(s)<1$ ($\zeta<0$, $N_k>N_1/k$) cases. Note that the model is presented from the viewpoint of the first-ranked population; cooperation or competition is manifested in the dynamics and dominance of that population. To distinguish population-level dynamics from species-level dynamics, linearization of the model above leads to the following form: 

\begin{equation}\frac{\Delta N_k}{\Sigma N} = -\frac{\Delta\zeta}{k^{\Re(s)}|\zeta|^2}.\end{equation}

Therefore, $\Delta N_k>0$ implies $\Delta\zeta<0$, $\Delta N_k<0$ implies $\Delta\zeta>0$, and $\Delta N_k=0$ implies $\Delta\zeta=0$. Each local extremum of $\zeta$ thus corresponds to a pole for the population or species, and a large (resp.\ small) value of $|\zeta|$ corresponds to a small (resp.\ large) fluctuation. Only those points of $\zeta$ that are close to zero represent growth bursts or collapses of the population or species. According to the Riemann hypothesis, at these points the following relations hold: $[\Re(s)=1/2]$ and $[\Re(s)=-2l_s,\ \Im(s)=0]$, where $l_s$ is a natural number independent of the population or species rank; negative values of $s$ will be characterized later.

Taking the logarithm of Eqn.~(5), we obtain

\begin{equation}N_k = \frac{1}{\ln D_k}\ln\frac{1}{P_k\zeta(s)} - \frac{s}{\ln D_k}\ln k,\end{equation}
\begin{equation}-\ln D_k\cdot N_k = \ln P_k\zeta(s) + s\ln k.\end{equation}
Therefore,
\begin{equation}a = \frac{-\ln P_k\zeta(s)}{\ln D_k},\end{equation}
\begin{equation}b = \frac{s}{\ln D_k},\end{equation}
\begin{equation}\Re(s) = \frac{\ln\frac{N_1}{N_k}}{\ln k} (k \neq 1),\end{equation}
\begin{equation}\zeta(s) = \frac{\Sigma N}{N_1} \geq 1\ (k = 1\mathrm{\ in\ species}),\end{equation}
\begin{equation}\Re(s) = 1\ (k = 1\mathrm{\ in\ population/observer}),\end{equation}
\begin{equation}D_k = e^{\frac{s}{b}},\end{equation}
\begin{equation}P_k = \frac{1}{D_k^a\zeta(s)}.\end{equation}

$\Re(s)$ is scale invariant when $k$ is fixed for a given system.  Note that $s$ and $\zeta(s)$ can therefore be estimated from the empirical distribution of $N$.  When $k=1$ for a population, we set $s=1$ because the distribution is nearly harmonic.  When $k=1$ for a species, $s$ should be computed via an inverse of $\zeta$, since in that case the distribution is no longer harmonic.  For convergence we require $N\sim 0$, $s\sim 1$, $\zeta\sim\pm\infty$, and $P\sim 0$.  We also assume $s=+\infty$ and $\zeta=1$ when a single population or species is observed.

From the data of \cite{Adachi2015} we computed $s$ using both population and species relative abundances in Dictyostelia and found significantly different results (see Figure~1 and Table~1).  Comparable analyses for soil mesofauna and for world economic data are reported in Figure~1 and Tables~2--3.  Population values for Dictyostelia and for soil mesofauna lie in the interval $[0,2]$, whereas species values frequently exceed $2$.  This pattern indicates that populations behave neutrally for $0<\Re(s)<2$, while species are more likely to dominate when $\Re(s)>2$; we discuss this implication in more detail below.

In Table~1, 6 of 54 Dictyostelia $s$ values exceed $2$ (highlighted in red); these cases were not observed within a population of 162 samples ($(p\sim 4\times 10^{-62})$ for the $\chi^2$-test).  Similar results hold for soil mesofauna: Collembola show $16/88$ versus $0/72$ with $(p\sim 6\times 10^{-13})$ ($\chi^2$-test), and Sarcoptiformes show $11/56$ versus $0/72$ with $(p\sim 2\times 10^{-7})$.  For human populations, $\Re(s)$ takes values both above and below $2$, but this variation does not appear to be organized by species as described later.  For GDP, $\Re(s<2)$ after 1870, and the fractal dimension is not apparent in that system. Henceforth $s$ denotes the small-$s$ parameter of this model.

\begin{figure}[H]
\includegraphics*[width=8cm]{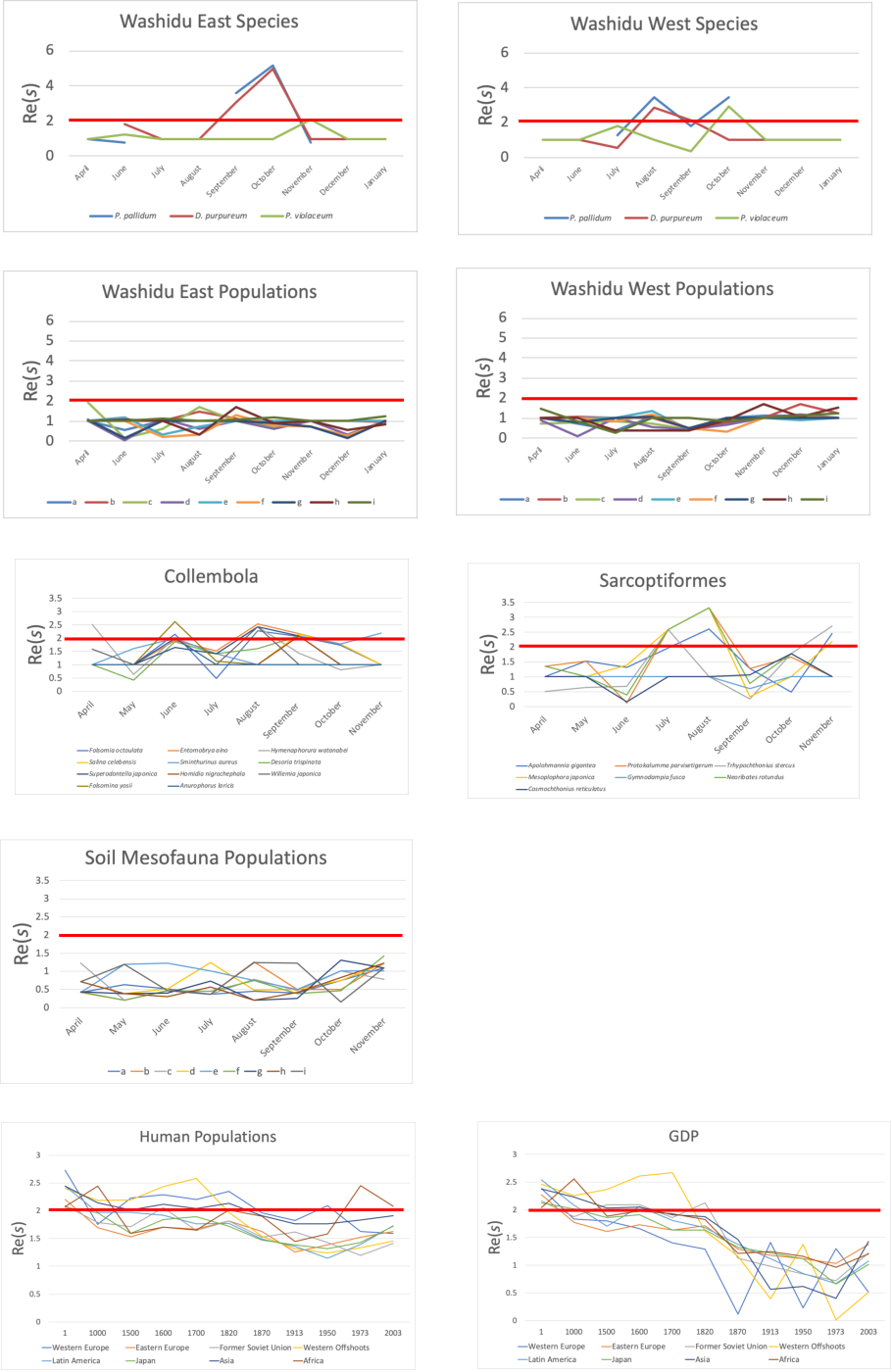}
\vskip2.5cm
\caption{Dynamics of $\Re(s)$. Dynamics of $\Re(s)$ over time for Dictyostelia species and populations in two Washidu quadrats, soil mesofauna species \& populations in Ohdaigahara, and global human populations \& GDP.  \textit{P. pallidum}: \textit{Polysphondylium pallidum}; \textit{D. purpureum}: \textit{Dictyostelium purpureum}; \textit{P. violaceum}: \textit{Polysphondylium violaceum}.  The top two panels present data for three Dictyostelia species; the next two panels present data for nine point quadrats of Dictyostelia.  The following two panels present data for Collembola and Sarcoptiformes in Ohdaigahara.  The next panel presents data for nine point quadrats of soil mesofauna in Ohdaigahara.  The final two panels show global human population and GDP.  Red lines indicate $\Re(s)=2$, which mark the critical threshold for the development of fractal structure.}
\label{fig:1}
\end{figure}

\begin{table}[H]
\caption{$\Re(s)$ and $N$ values for Dictyostelia.}
\label{tab:1}
\centering
\resizebox{\textwidth}{!}{%
\selectfont{\begin{tabular}{lrrrrrr}
\cline{1-7}
\multicolumn{1}{|l|}{$\Re(s)$} & \multicolumn{1}{l|}{\textit{P. pallidum} (WE)} & \multicolumn{1}{l|}{\textit{D. purpureum} (WE)} & \multicolumn{1}{l|}{\textit{P. violaceum} (WE)} & \multicolumn{1}{l|}{\textit{P. pallidum} (WW)} & \multicolumn{1}{l|}{\textit{D. purpureum} (WW)} & \multicolumn{1}{l|}{\textit{P. violaceum} (WW)} \\ \hline
\multicolumn{1}{|l|}{April} & \multicolumn{1}{l|}{1} & \multicolumn{1}{l|}{-} & \multicolumn{1}{l|}{1} & \multicolumn{1}{l|}{1} & \multicolumn{1}{l|}{-} & \multicolumn{1}{l|}{1} \\ \hline
\multicolumn{1}{|l|}{June} & \multicolumn{1}{l|}{0.7693} & \multicolumn{1}{l|}{1.8305} & \multicolumn{1}{l|}{1.258} & \multicolumn{1}{l|}{-} & \multicolumn{1}{l|}{1} & \multicolumn{1}{l|}{1} \\ \hline
\multicolumn{1}{|l|}{July} & \multicolumn{1}{l|}{-} & \multicolumn{1}{l|}{1} & \multicolumn{1}{l|}{1} & \multicolumn{1}{l|}{1.2619} & \multicolumn{1}{l|}{0.5752} & \multicolumn{1}{l|}{1.7742} \\ \hline
\multicolumn{1}{|l|}{August} & \multicolumn{1}{l|}{-} & \multicolumn{1}{l|}{1} & \multicolumn{1}{l|}{1} & \multicolumn{1}{l|}{3.4223} & \multicolumn{1}{l|}{\color[HTML]{CB0000}2.8795} & \multicolumn{1}{l|}{1} \\ \hline
\multicolumn{1}{|l|}{September} & \multicolumn{1}{l|}{3.5762} & \multicolumn{1}{l|}{{\color[HTML]{CB0000} 3.0777}} & \multicolumn{1}{l|}{1} & \multicolumn{1}{l|}{1.7897} & \multicolumn{1}{l|}{\color[HTML]{CB0000}2.1411} & \multicolumn{1}{l|}{0.3186} \\ \hline
\multicolumn{1}{|l|}{October} & \multicolumn{1}{l|}{5.1648} & \multicolumn{1}{l|}{{\color[HTML]{CB0000} 4.9423}} & \multicolumn{1}{l|}{1} & \multicolumn{1}{l|}{3.4417} & \multicolumn{1}{l|}{1} & \multicolumn{1}{l|}{{\color[HTML]{CB0000} 2.9047}} \\ \hline
\multicolumn{1}{|l|}{November} & \multicolumn{1}{l|}{0.7481} & \multicolumn{1}{l|}{1} & \multicolumn{1}{l|}{{\color[HTML]{CB0000} 2.056}} & \multicolumn{1}{l|}{-} & \multicolumn{1}{l|}{1} & \multicolumn{1}{l|}{1} \\ \hline
\multicolumn{1}{|l|}{December} & \multicolumn{1}{l|}{-} & \multicolumn{1}{l|}{1} & \multicolumn{1}{l|}{1} & \multicolumn{1}{l|}{-} & \multicolumn{1}{l|}{1} & \multicolumn{1}{l|}{1} \\ \hline
\multicolumn{1}{|l|}{January} & \multicolumn{1}{l|}{-} & \multicolumn{1}{l|}{1} & \multicolumn{1}{l|}{1} & \multicolumn{1}{l|}{-} & \multicolumn{1}{l|}{1} & \multicolumn{1}{l|}{1} \\ \hline
 &  &  &  &  &  &  \\ \hline
\multicolumn{1}{|l|}{$N$} & \multicolumn{1}{l|}{\textit{P. pallidum} (WE)} & \multicolumn{1}{l|}{\textit{D. purpureum} (WE)} & \multicolumn{1}{l|}{\textit{P. violaceum} (WE)} & \multicolumn{1}{l|}{\textit{P. pallidum} (WW)} & \multicolumn{1}{l|}{\textit{D. purpureum} (WW)} & \multicolumn{1}{l|}{\textit{P. violaceum} (WW)} \\ \hline
\multicolumn{1}{|l|}{April} & \multicolumn{1}{l|}{0} & \multicolumn{1}{l|}{76} & \multicolumn{1}{l|}{0} & \multicolumn{1}{l|}{0} & \multicolumn{1}{l|}{83} & \multicolumn{1}{l|}{0} \\ \hline
\multicolumn{1}{|l|}{June} & \multicolumn{1}{l|}{123} & \multicolumn{1}{l|}{209} & \multicolumn{1}{l|}{52} & \multicolumn{1}{l|}{147} & \multicolumn{1}{l|}{0} & \multicolumn{1}{l|}{0} \\ \hline
\multicolumn{1}{|l|}{July} & \multicolumn{1}{l|}{1282} & \multicolumn{1}{l|}{0} & \multicolumn{1}{l|}{0} & \multicolumn{1}{l|}{80} & \multicolumn{1}{l|}{215} & \multicolumn{1}{l|}{320} \\ \hline
\multicolumn{1}{|l|}{August} & \multicolumn{1}{l|}{1561} & \multicolumn{1}{l|}{0} & \multicolumn{1}{l|}{0} & \multicolumn{1}{l|}{1330} & \multicolumn{1}{l|}{\color[HTML]{CB0000}181} & \multicolumn{1}{l|}{0} \\ \hline
\multicolumn{1}{|l|}{September} & \multicolumn{1}{l|}{901} & \multicolumn{1}{l|}{{\color[HTML]{CB0000} 107}} & \multicolumn{1}{l|}{0} & \multicolumn{1}{l|}{809} & \multicolumn{1}{l|}{\color[HTML]{CB0000}77} & \multicolumn{1}{l|}{649} \\ \hline
\multicolumn{1}{|l|}{October} & \multicolumn{1}{l|}{1069} & \multicolumn{1}{l|}{{\color[HTML]{CB0000} 35}} & \multicolumn{1}{l|}{0} & \multicolumn{1}{l|}{799} & \multicolumn{1}{l|}{0} & \multicolumn{1}{l|}{{\color[HTML]{CB0000} 107}} \\ \hline
\multicolumn{1}{|l|}{November} & \multicolumn{1}{l|}{60} & \multicolumn{1}{l|}{0} & \multicolumn{1}{l|}{{\color[HTML]{CB0000} 101}} & \multicolumn{1}{l|}{336} & \multicolumn{1}{l|}{0} & \multicolumn{1}{l|}{0} \\ \hline
\multicolumn{1}{|l|}{December} & \multicolumn{1}{l|}{190} & \multicolumn{1}{l|}{0} & \multicolumn{1}{l|}{0} & \multicolumn{1}{l|}{711} & \multicolumn{1}{l|}{0} & \multicolumn{1}{l|}{0} \\ \hline
\multicolumn{1}{|l|}{January} & \multicolumn{1}{l|}{29} & \multicolumn{1}{l|}{0} & \multicolumn{1}{l|}{0} & \multicolumn{1}{l|}{99} & \multicolumn{1}{l|}{0} & \multicolumn{1}{l|}{0} \\ \hline
 &  &  &  &  &  &  \\
\end{tabular}}
}
\centering (continues)
\end{table}

\begin{table}
\centering (continued)\\
\resizebox{\textwidth}{!}{%
\selectfont{\begin{tabular}{lrrrrrrrrr}
\cline{1-10}
\multicolumn{1}{|l|}{$\Re(s)$ WE} & \multicolumn{1}{l|}{a} & \multicolumn{1}{l|}{b} & \multicolumn{1}{l|}{c} & \multicolumn{1}{l|}{d} & \multicolumn{1}{l|}{e} & \multicolumn{1}{l|}{f} & \multicolumn{1}{l|}{g} & \multicolumn{1}{l|}{h} & \multicolumn{1}{l|}{i} \\ \hline
\multicolumn{1}{|l|}{April} & \multicolumn{1}{l|}{1.0336} & \multicolumn{1}{l|}{1} & \multicolumn{1}{l|}{1.9442} & \multicolumn{1}{l|}{1} & \multicolumn{1}{l|}{1} & \multicolumn{1}{l|}{1} & \multicolumn{1}{l|}{1.0821} & \multicolumn{1}{l|}{1} & \multicolumn{1}{l|}{1} \\ \hline
\multicolumn{1}{|l|}{June} & \multicolumn{1}{l|}{0.5328} & \multicolumn{1}{l|}{1} & \multicolumn{1}{l|}{0.1545} & \multicolumn{1}{l|}{0.0332} & \multicolumn{1}{l|}{1.1928} & \multicolumn{1}{l|}{1} & \multicolumn{1}{l|}{0.1374} & \multicolumn{1}{l|}{1.076} & \multicolumn{1}{l|}{1.0071} \\ \hline
\multicolumn{1}{|l|}{July} & \multicolumn{1}{l|}{1} & \multicolumn{1}{l|}{1} & \multicolumn{1}{l|}{0.6131} & \multicolumn{1}{l|}{1.1497} & \multicolumn{1}{l|}{0.3117} & \multicolumn{1}{l|}{0.2016} & \multicolumn{1}{l|}{1} & \multicolumn{1}{l|}{1} & \multicolumn{1}{l|}{1.148} \\ \hline
\multicolumn{1}{|l|}{August} & \multicolumn{1}{l|}{1} & \multicolumn{1}{l|}{1.4925} & \multicolumn{1}{l|}{1.7167} & \multicolumn{1}{l|}{0.6348} & \multicolumn{1}{l|}{0.7075} & \multicolumn{1}{l|}{0.3523} & \multicolumn{1}{l|}{1} & \multicolumn{1}{l|}{0.3502} & \multicolumn{1}{l|}{1} \\ \hline
\multicolumn{1}{|l|}{September} & \multicolumn{1}{l|}{1} & \multicolumn{1}{l|}{1.1361} & \multicolumn{1}{l|}{1} & \multicolumn{1}{l|}{1} & \multicolumn{1}{l|}{1} & \multicolumn{1}{l|}{1.3035} & \multicolumn{1}{l|}{1.0325} & \multicolumn{1}{l|}{1.7248} & \multicolumn{1}{l|}{1.085} \\ \hline
\multicolumn{1}{|l|}{October} & \multicolumn{1}{l|}{1} & \multicolumn{1}{l|}{0.6746} & \multicolumn{1}{l|}{0.6937} & \multicolumn{1}{l|}{0.6092} & \multicolumn{1}{l|}{1} & \multicolumn{1}{l|}{0.7836} & \multicolumn{1}{l|}{0.9259} & \multicolumn{1}{l|}{0.886} & \multicolumn{1}{l|}{1.1746} \\ \hline
\multicolumn{1}{|l|}{November} & \multicolumn{1}{l|}{1} & \multicolumn{1}{l|}{1} & \multicolumn{1}{l|}{1} & \multicolumn{1}{l|}{1} & \multicolumn{1}{l|}{1} & \multicolumn{1}{l|}{0.7481} & \multicolumn{1}{l|}{0.472} & \multicolumn{1}{l|}{1} & \multicolumn{1}{l|}{1} \\ \hline
\multicolumn{1}{|l|}{December} & \multicolumn{1}{l|}{1} & \multicolumn{1}{l|}{1} & \multicolumn{1}{l|}{1} & \multicolumn{1}{l|}{0.3429} & \multicolumn{1}{l|}{1} & \multicolumn{1}{l|}{0.2455} & \multicolumn{1}{l|}{0.1712} & \multicolumn{1}{l|}{0.5647} & \multicolumn{1}{l|}{1} \\ \hline
\multicolumn{1}{|l|}{January} & \multicolumn{1}{l|}{1} & \multicolumn{1}{l|}{0.9516} & \multicolumn{1}{l|}{1} & \multicolumn{1}{l|}{1} & \multicolumn{1}{l|}{1} & \multicolumn{1}{l|}{1} & \multicolumn{1}{l|}{1} & \multicolumn{1}{l|}{0.8666} & \multicolumn{1}{l|}{1.215} \\ \hline
 &  &  &  &  &  &  &  &  &  \\ \hline
\multicolumn{1}{|l|}{$\Re(s)$ WW} & \multicolumn{1}{l|}{a} & \multicolumn{1}{l|}{b} & \multicolumn{1}{l|}{c} & \multicolumn{1}{l|}{d} & \multicolumn{1}{l|}{e} & \multicolumn{1}{l|}{f} & \multicolumn{1}{l|}{g} & \multicolumn{1}{l|}{h} & \multicolumn{1}{l|}{i} \\ \hline
\multicolumn{1}{|l|}{April} & \multicolumn{1}{l|}{1} & \multicolumn{1}{l|}{1} & \multicolumn{1}{l|}{0.7125} & \multicolumn{1}{l|}{0.8782} & \multicolumn{1}{l|}{1} & \multicolumn{1}{l|}{1} & \multicolumn{1}{l|}{1} & \multicolumn{1}{l|}{1} & \multicolumn{1}{l|}{1.473} \\ \hline
\multicolumn{1}{|l|}{June} & \multicolumn{1}{l|}{0.708} & \multicolumn{1}{l|}{1.0735} & \multicolumn{1}{l|}{0.7614} & \multicolumn{1}{l|}{0.1056} & \multicolumn{1}{l|}{1} & \multicolumn{1}{l|}{1} & \multicolumn{1}{l|}{0.7883} & \multicolumn{1}{l|}{1} & \multicolumn{1}{l|}{0.8612} \\ \hline
\multicolumn{1}{|l|}{July} & \multicolumn{1}{l|}{0.3888} & \multicolumn{1}{l|}{1} & \multicolumn{1}{l|}{0.8635} & \multicolumn{1}{l|}{1} & \multicolumn{1}{l|}{1} & \multicolumn{1}{l|}{0.8614} & \multicolumn{1}{l|}{1} & \multicolumn{1}{l|}{0.3629} & \multicolumn{1}{l|}{0.263} \\ \hline
\multicolumn{1}{|l|}{August} & \multicolumn{1}{l|}{1.0524} & \multicolumn{1}{l|}{1} & \multicolumn{1}{l|}{0.756} & \multicolumn{1}{l|}{0.5644} & \multicolumn{1}{l|}{1.3367} & \multicolumn{1}{l|}{1.1911} & \multicolumn{1}{l|}{1.0473} & \multicolumn{1}{l|}{0.3985} & \multicolumn{1}{l|}{1} \\ \hline
\multicolumn{1}{|l|}{September} & \multicolumn{1}{l|}{0.4918} & \multicolumn{1}{l|}{0.4236} & \multicolumn{1}{l|}{0.4243} & \multicolumn{1}{l|}{0.4427} & \multicolumn{1}{l|}{0.4535} & \multicolumn{1}{l|}{0.4969} & \multicolumn{1}{l|}{0.5051} & \multicolumn{1}{l|}{0.3985} & \multicolumn{1}{l|}{1} \\ \hline
\multicolumn{1}{|l|}{October} & \multicolumn{1}{l|}{1} & \multicolumn{1}{l|}{0.8073} & \multicolumn{1}{l|}{0.8982} & \multicolumn{1}{l|}{0.6913} & \multicolumn{1}{l|}{1} & \multicolumn{1}{l|}{0.3219} & \multicolumn{1}{l|}{1} & \multicolumn{1}{l|}{0.9284} & \multicolumn{1}{l|}{0.8523} \\ \hline
\multicolumn{1}{|l|}{November} & \multicolumn{1}{l|}{1.1334} & \multicolumn{1}{l|}{1} & \multicolumn{1}{l|}{1} & \multicolumn{1}{l|}{1} & \multicolumn{1}{l|}{1} & \multicolumn{1}{l|}{1} & \multicolumn{1}{l|}{1} & \multicolumn{1}{l|}{1.7225} & \multicolumn{1}{l|}{1} \\ \hline
\multicolumn{1}{|l|}{December} & \multicolumn{1}{l|}{1.1214} & \multicolumn{1}{l|}{1.7164} & \multicolumn{1}{l|}{1} & \multicolumn{1}{l|}{1.1833} & \multicolumn{1}{l|}{0.9208} & \multicolumn{1}{l|}{1.0718} & \multicolumn{1}{l|}{1} & \multicolumn{1}{l|}{1.0594} & \multicolumn{1}{l|}{1.1375} \\ \hline
\multicolumn{1}{|l|}{January} & \multicolumn{1}{l|}{1} & \multicolumn{1}{l|}{1.2501} & \multicolumn{1}{l|}{1} & \multicolumn{1}{l|}{1} & \multicolumn{1}{l|}{1} & \multicolumn{1}{l|}{1} & \multicolumn{1}{l|}{1} & \multicolumn{1}{l|}{1.5151} & \multicolumn{1}{l|}{1.2228} \\ \hline
 &  &  &  &  &  &  &  &  &  \\ \hline
\multicolumn{1}{|l|}{$N$ WE} & \multicolumn{1}{l|}{a} & \multicolumn{1}{l|}{b} & \multicolumn{1}{l|}{c} & \multicolumn{1}{l|}{d} & \multicolumn{1}{l|}{e} & \multicolumn{1}{l|}{f} & \multicolumn{1}{l|}{g} & \multicolumn{1}{l|}{h} & \multicolumn{1}{l|}{i} \\ \hline
\multicolumn{1}{|l|}{April} & \multicolumn{1}{l|}{680} & \multicolumn{1}{l|}{0} & \multicolumn{1}{l|}{94} & \multicolumn{1}{l|}{0} & \multicolumn{1}{l|}{0} & \multicolumn{1}{l|}{1392} & \multicolumn{1}{l|}{424} & \multicolumn{1}{l|}{0} & \multicolumn{1}{l|}{0} \\ \hline
\multicolumn{1}{|l|}{June} & \multicolumn{1}{l|}{1120} & \multicolumn{1}{l|}{0} & \multicolumn{1}{l|}{2131} & \multicolumn{1}{l|}{2580} & \multicolumn{1}{l|}{221} & \multicolumn{1}{l|}{2640} & \multicolumn{1}{l|}{2270} & \multicolumn{1}{l|}{384} & \multicolumn{1}{l|}{372} \\ \hline
\multicolumn{1}{|l|}{July} & \multicolumn{1}{l|}{0} & \multicolumn{1}{l|}{0} & \multicolumn{1}{l|}{1573} & \multicolumn{1}{l|}{469} & \multicolumn{1}{l|}{2613} & \multicolumn{1}{l|}{3200} & \multicolumn{1}{l|}{3680} & \multicolumn{1}{l|}{0} & \multicolumn{1}{l|}{580} \\ \hline
\multicolumn{1}{|l|}{August} & \multicolumn{1}{l|}{0} & \multicolumn{1}{l|}{331} & \multicolumn{1}{l|}{170} & \multicolumn{1}{l|}{1728} & \multicolumn{1}{l|}{1800} & \multicolumn{1}{l|}{3760} & \multicolumn{1}{l|}{0} & \multicolumn{1}{l|}{3267} & \multicolumn{1}{l|}{4800} \\ \hline
\multicolumn{1}{|l|}{September} & \multicolumn{1}{l|}{0} & \multicolumn{1}{l|}{1240} & \multicolumn{1}{l|}{0} & \multicolumn{1}{l|}{4320} & \multicolumn{1}{l|}{0} & \multicolumn{1}{l|}{418} & \multicolumn{1}{l|}{820} & \multicolumn{1}{l|}{1307} & \multicolumn{1}{l|}{960} \\ \hline
\multicolumn{1}{|l|}{October} & \multicolumn{1}{l|}{0} & \multicolumn{1}{l|}{1413} & \multicolumn{1}{l|}{1680} & \multicolumn{1}{l|}{2360} & \multicolumn{1}{l|}{3600} & \multicolumn{1}{l|}{1020} & \multicolumn{1}{l|}{594} & \multicolumn{1}{l|}{736} & \multicolumn{1}{l|}{313} \\ \hline
\multicolumn{1}{|l|}{November} & \multicolumn{1}{l|}{0} & \multicolumn{1}{l|}{0} & \multicolumn{1}{l|}{0} & \multicolumn{1}{l|}{907} & \multicolumn{1}{l|}{0} & \multicolumn{1}{l|}{540} & \multicolumn{1}{l|}{540} & \multicolumn{1}{l|}{0} & \multicolumn{1}{l|}{0} \\ \hline
\multicolumn{1}{|l|}{December} & \multicolumn{1}{l|}{0} & \multicolumn{1}{l|}{0} & \multicolumn{1}{l|}{0} & \multicolumn{1}{l|}{580} & \multicolumn{1}{l|}{0} & \multicolumn{1}{l|}{787} & \multicolumn{1}{l|}{773} & \multicolumn{1}{l|}{376} & \multicolumn{1}{l|}{933} \\ \hline
\multicolumn{1}{|l|}{January} & \multicolumn{1}{l|}{0} & \multicolumn{1}{l|}{457} & \multicolumn{1}{l|}{0} & \multicolumn{1}{l|}{0} & \multicolumn{1}{l|}{1300} & \multicolumn{1}{l|}{0} & \multicolumn{1}{l|}{0} & \multicolumn{1}{l|}{391} & \multicolumn{1}{l|}{560} \\ \hline
 &  &  &  &  &  &  &  &  &  \\ \hline
\multicolumn{1}{|l|}{$N$ WW} & \multicolumn{1}{l|}{a} & \multicolumn{1}{l|}{b} & \multicolumn{1}{l|}{c} & \multicolumn{1}{l|}{d} & \multicolumn{1}{l|}{e} & \multicolumn{1}{l|}{f} & \multicolumn{1}{l|}{g} & \multicolumn{1}{l|}{h} & \multicolumn{1}{l|}{i} \\ \hline
\multicolumn{1}{|l|}{April} & \multicolumn{1}{l|}{840} & \multicolumn{1}{l|}{0} & \multicolumn{1}{l|}{384} & \multicolumn{1}{l|}{457} & \multicolumn{1}{l|}{0} & \multicolumn{1}{l|}{0} & \multicolumn{1}{l|}{0} & \multicolumn{1}{l|}{0} & \multicolumn{1}{l|}{109} \\ \hline
\multicolumn{1}{|l|}{June} & \multicolumn{1}{l|}{1088} & \multicolumn{1}{l|}{421} & \multicolumn{1}{l|}{869} & \multicolumn{1}{l|}{3160} & \multicolumn{1}{l|}{0} & \multicolumn{1}{l|}{0} & \multicolumn{1}{l|}{1140} & \multicolumn{1}{l|}{3400} & \multicolumn{1}{l|}{1320} \\ \hline
\multicolumn{1}{|l|}{July} & \multicolumn{1}{l|}{1680} & \multicolumn{1}{l|}{0} & \multicolumn{1}{l|}{613} & \multicolumn{1}{l|}{0} & \multicolumn{1}{l|}{0} & \multicolumn{1}{l|}{720} & \multicolumn{1}{l|}{2880} & \multicolumn{1}{l|}{1933} & \multicolumn{1}{l|}{2400} \\ \hline
\multicolumn{1}{|l|}{August} & \multicolumn{1}{l|}{704} & \multicolumn{1}{l|}{0} & \multicolumn{1}{l|}{1627} & \multicolumn{1}{l|}{2496} & \multicolumn{1}{l|}{288} & \multicolumn{1}{l|}{457} & \multicolumn{1}{l|}{860} & \multicolumn{1}{l|}{3520} & \multicolumn{1}{l|}{4640} \\ \hline
\multicolumn{1}{|l|}{September} & \multicolumn{1}{l|}{1760} & \multicolumn{1}{l|}{1760} & \multicolumn{1}{l|}{1627} & \multicolumn{1}{l|}{1386} & \multicolumn{1}{l|}{1440} & \multicolumn{1}{l|}{2016} & \multicolumn{1}{l|}{1147} & \multicolumn{1}{l|}{2640} & \multicolumn{1}{l|}{3480} \\ \hline
\multicolumn{1}{|l|}{October} & \multicolumn{1}{l|}{3520} & \multicolumn{1}{l|}{960} & \multicolumn{1}{l|}{613} & \multicolumn{1}{l|}{1350} & \multicolumn{1}{l|}{0} & \multicolumn{1}{l|}{2816} & \multicolumn{1}{l|}{0} & \multicolumn{1}{l|}{667} & \multicolumn{1}{l|}{1380} \\ \hline
\multicolumn{1}{|l|}{November} & \multicolumn{1}{l|}{760} & \multicolumn{1}{l|}{0} & \multicolumn{1}{l|}{0} & \multicolumn{1}{l|}{0} & \multicolumn{1}{l|}{0} & \multicolumn{1}{l|}{0} & \multicolumn{1}{l|}{2640} & \multicolumn{1}{l|}{800} & \multicolumn{1}{l|}{0} \\ \hline
\multicolumn{1}{|l|}{December} & \multicolumn{1}{l|}{590} & \multicolumn{1}{l|}{124} & \multicolumn{1}{l|}{0} & \multicolumn{1}{l|}{440} & \multicolumn{1}{l|}{1600} & \multicolumn{1}{l|}{784} & \multicolumn{1}{l|}{4400} & \multicolumn{1}{l|}{1013} & \multicolumn{1}{l|}{2000} \\ \hline
\multicolumn{1}{|l|}{January} & \multicolumn{1}{l|}{1307} & \multicolumn{1}{l|}{331} & \multicolumn{1}{l|}{0} & \multicolumn{1}{l|}{0} & \multicolumn{1}{l|}{0} & \multicolumn{1}{l|}{0} & \multicolumn{1}{l|}{0} & \multicolumn{1}{l|}{160} & \multicolumn{1}{l|}{560} \\ \hline
\end{tabular}
}
}

\tiny{\begin{raggedright}
\noindent WE: the Washidu East quadrat; WW: the Washidu West quadrat (see \cite{Adachi2015}). Scientific names of Dictyostelia species: \textit{P. pallidum} = \textit{Polysphondylium pallidum}; \textit{D. purpureum} = \textit{Dictyostelium purpureum}; \textit{P. violaceum} = \textit{Polysphondylium violaceum}. For the calculation of $\Re(s)$, see the main text.  $N$ denotes the number of cells per 1~g of soil for Dictyostelia; species labels for Dictyostelia indicate the corresponding values. Indices (a)--(i) label the point quadrats. Red marks indicate $\Re(s)$ values that are approximately integral and greater than or equal to 2. The data relating identical $N$ values are integrated to $k$ of the first rank among them.\\
\end{raggedright}}
\end{table}

\begin{table}[H]
\caption{$\Re(s)$ and $N$ values for soil mesofauna.}
\label{tab:2}
\centering

\resizebox{\columnwidth}{!}{%
\begin{tabular}{|l|l|l|l|l|l|l|l|l|l|l|l|}
\hline
Collembola $\Re(s)$ &
  \textit{Folsomia octoulata} &
  \textit{Entomobrya aino} &
  \textit{Hymenaphorura watanabei} &
  \textit{Salina celebensis} &
  \textit{Sminthurinus aureus} &
  \textit{Desoria trispinata} &
  \textit{Superodontella japonica} &
  \textit{Homidia nigrochephala} &
  \textit{Willemia japonica} &
  \textit{Folsomina yosii} &
  \textit{Anurophorus laricis} \\ \hline
April     & 1      & 1      & 2.5214 & 1      & 1      & 1      & 1      & 1      & 1.5850 & 1      & 1      \\ \hline
May       & 1      & 1      & 0.6309 & 1      & 1.6135 & 0.4150 & 1      & 1      & 1      & 1      & 1      \\ \hline
June      & 2.1402 & 1.9116 & 1.9874 & 1      & 1.9874 & 1.8716 & 1.6464 & 1      & 1      & 2.6147 & 1.9874 \\ \hline
July      & 0.4854 & 1.5106 & 1.4037 & 1      & 1.4037 & 1.4037 & 1.4037 & 1      & 1      & 1.1403 & 1      \\ \hline
August    & 2.2709 & 2.5361 & 2.4290 & 1      & 1      & 1.6001 & 2.4290 & 1      & 2.4290 & 1      & 1      \\ \hline
September & 2.0671 & 2.1575 & 1.4341 & 2.1365 & 2.0922 & 2.0922 & 2.0867 & 2.0867 & 1      & 1      & 1      \\ \hline
October   & 1.7286 & 1.7712 & 0.8074 & 1.7712 & 1.7712 & 1      & 1      & 1      & 1      & 1      & 1      \\ \hline
November  & 1      & 1      & 1      & 1      & 2.1853 & 1      & 1      & 1      & 1      & 1      & 1      \\ \hline
\end{tabular}%
}

\resizebox{\columnwidth}{!}{%
\begin{tabular}{|l|l|l|l|l|l|l|l|l|l|l|l|}
\hline
Collembola $N$ &
  \textit{Folsomia octoulata} &
  \textit{Entomobrya aino} &
  \textit{Hymenaphorura watanabei} &
  \textit{Salina celebensis} &
  \textit{Sminthurinus aureus} &
  \textit{Desoria trispinata} &
  \textit{Superodontella japonica} &
  \textit{Homidia nigrochephala} &
  \textit{Willemia japonica} &
  \textit{Folsomina yosii} &
  \textit{Anurophorus laricis} \\ \hline
April     & 0           & 0           & 0.3333 & 0           & 0           & 0           & 0           & 0           & 0.1111 & 0           & 0           \\ \hline
May       & 0           & 0           & 0.2222 & 0           & 0.4444 & 0.3333 & 0           & 0           & 0           & 0           & 0           \\ \hline
June      & 5.4444 & 0.6667 & 0.2222 & 0           & 0.2222 & 0.1111 & 0.5556 & 0           & 0           & 0.8889 & 0.2222 \\ \hline
July      & 0.5556 & 0.7778 & 0.1111 & 0           & 0.1111 & 0.1111 & 0.1111 & 0           & 0           & 0.2222 & 0           \\ \hline
August    & 3.2222 & 0.5556 & 0.1111 & 0           & 0           & 0.5556 & 0.1111 & 0           & 0.1111 & 0           & 0           \\ \hline
September & 6.4444 & 1.4444 & 1.3333 & 0.3333 & 0.2222 & 0.2222 & 0.1111 & 0.1111 & 0           & 0           & 0           \\ \hline
October   & 0.7778 & 0.1111 & 0.4444 & 0.1111 & 0.1111 & 0           & 0           & 0           & 0           & 0           & 0           \\ \hline
November  & 0           & 0.1111 & 0           & 0           & 0.2222 & 0           & 0           & 0           & 0           & 0           & 0           \\ \hline
\end{tabular}%
}

\resizebox{\columnwidth}{!}{%
\begin{tabular}{|l|l|l|l|l|l|l|l|}
\hline
Sarcoptiformes $\Re(s)$ &
  \textit{Apolohmannia gigantea} &
  \textit{Protokalumma parvisetigerum} &
  \textit{Trhypochthonius stercus }&
  \textit{Mesoplophora japonica} &
  \textit{Gymnodampia fusca} &
  \textit{Neoribates rotundus} &
  \textit{Cosmochthonius reticulatus} \\ \hline
April     & 1           & 1.3450 & 0.5         & 1.3450 & 1          & 1.3450 & 1           \\ \hline
May       & 1.5303 & 1.5303 & 0.6309 & 1           & 1          & 1           & 1           \\ \hline
June      & 1.3174 & 0.1255 & 0.6826 & 1.3869 & 1          & 0.3888 & 0.1660 \\ \hline
July      & 1.9774 & 2.5850 & 2.5850 & 2.5850 & 1          & 2.5850 & 1           \\ \hline
August    & 2.6166 & 3.3219 & 1           & 3.3219 & 1          & 3.3219 & 1           \\ \hline
September & 1.2835 & 1.2835 & 0.2619 & 0.3390 & 0.6094& 0.7737 & 1.0686 \\ \hline
October   & 0.4854 & 1.6579  & 1.7712& 1           & 1          & 1.7712 & 1.7712 \\ \hline
November  & 2.4594 & 1           & 2.7053 & 2.1827 & 1          & 1           & 1 \\ \hline          
\end{tabular}%
}

\resizebox{\columnwidth}{!}{%
\begin{tabular}{|l|l|l|l|l|l|l|l|}
\hline
Sarcoptiformes $N$        & 
  \textit{Apolohmannia gigantea} &
  \textit{Protokalumma parvisetigerum} &
  \textit{Trhypochthonius stercus }&
  \textit{Mesoplophora japonica} &
  \textit{Gymnodampia fusca} &
  \textit{Neoribates rotundus} &
  \textit{Cosmochthonius reticulatus}      \\ \hline
April    & 0           & 0.2222 & 0.1111 & 0.2222 & 0         & 0.22222 & 0           \\ \hline
May      & 0.4444 & 0.44444 & 0.2222 & 0           & 0         & 0           & 0           \\ \hline
June      & 1.3333 & 1.2222 & 0.4444 & 0.1111 & 0           & 0.7778 & 1.1111 \\ \hline
July     & 0.6667 & 0.1111 & 0.1111 & 0.1111 & 0         & 0.1111 & 0           \\ \hline
August   & 1.1111 & 0.1111 & 0           & 0.1111 & 0         & 0.1111 & 0           \\ \hline
September & 0.8889 & 0.8889 & 0.6667 & 0.5556 & 0.3333 & 0.2222 & 0.1111 \\ \hline
October  & 0.5556 & 0.7778 & 0.1111 & 0           & 0         & 0.1111 & 0.1111 \\ \hline
November & 0.2222 & 0           & 1.2222 & 0.1111 & 0         & 0           & 0           \\ \hline
\end{tabular}%
}

\resizebox{\columnwidth}{!}{%
\begin{tabular}{|l|l|l|l|l|l|l|l|l|l|}
\hline
$\Re(s)$    & a          & b          & c          & d          & e       & f          & g          & h          & i          \\ \hline
April     & 0.4150 & 0.4150  & 1.2269 & 0.7124 & 0.4150  & 0.4307 & 0.4307 & 0.7124 & 0.7124 \\ \hline
May      & 0.6309 & 0.2075 & 0.2075 & 0.3869 & 1.1849 & 0.2075 & 0.3869 & 0.3869 & 1.1849    \\ \hline
June      & 0.5101  & 0.4709 & 0.4650 & 0.5101 & 1.2249  & 0.4717 & 0.3959 & 0.2990 & 0.4709 \\ \hline
July     & 0.3691 & 0.3652& 0.3691 & 1.2456  & 1       & 0.4526 & 0.7233 & 0.5646 & 0.3626\\ \hline
August    & 0.4534 & 1.2583  & 0.7740 & 0.4735 & 0.7377 & 0.7377 & 0.2065 & 0.2031 & 1.2325 \\ \hline
September & 0.3930 & 0.5040& 0.4614  & 0.5040 & 0.5040 & 0.3834 & 0.2451 & 0.4150  & 1.2191   \\ \hline
October  & 0.7481 & 0.5        & 1          & 0.7481 & 1       & 0.4650 & 1.3125    & 0.8271 & 0.1520 \\ \hline
November & 1.0860& 1.2224 & 0.7784  & 1.2224 & 1       & 1.4178   & 1.0860 & 1.2224 & 1.0860 \\ \hline
\end{tabular}%
}

\resizebox{\columnwidth}{!}{%
\begin{tabular}{|l|l|l|l|l|l|l|l|l|l|}
\hline
$N$         & a  & b  & c  & d  & e  & f  & g  & h  & i  \\ \hline
April     & 3  & 3  & 4  & 1  & 3  & 2  & 2  & 1  & 1  \\ \hline
May       & 1  & 3  & 3  & 2  & 4  & 3  & 2  & 2  & 4  \\ \hline
June      & 11 & 10 & 9  & 11 & 25 & 13 & 19 & 18 & 10 \\ \hline
July      & 6  & 5  & 6  & 9  & 0  & 4  & 2  & 3  & 7  \\ \hline
August    & 8  & 15 & 3  & 7  & 4  & 4  & 13 & 12 & 1  \\ \hline
September & 17 & 12 & 14 & 12 & 12 & 21 & 27 & 18 & 32 \\ \hline
October   & 3  & 5  & 0  & 3  & 0  & 6  & 10 & 2  & 9  \\ \hline
November  & 1  & 3  & 2  & 3  & 0  & 7  & 1  & 3  & 1  \\ \hline
\end{tabular}%
}
\tiny{\begin{raggedright}
\noindent  For calculation of $\Re(s)$, see the main text. $N$ is the number of individuals per 100~cc of litter for soil mesofauna. Indices (a)--(i) label the point quadrats. The data relating identical $N$ values are integrated to $k$ of the first rank among them.\\
\end{raggedright}}
\end{table}

\begin{table}[H]
\caption{$\Re(s)$ and raw data for the world economics.}
\label{tab:3}
\centering

\resizebox{\columnwidth}{!}{%
\begin{tabular}{|l|l|l|l|l|l|l|l|l|}
\hline
$\Re(s)$ for populations & Western Europe & Eastern Europe & Former Soviet Union & Western Offshoots & Latin America & Japan & Asia & Africa \\ \hline
1    & 2.7231 & 2.2059 & 2.0915 & 2.4021 & 2.4422 & 2.0606 & 2.4412 & 2.0709 \\ \hline
1000 & 1.7516 & 1.6926 & 1.7889 & 2.1830 & 1.9705 & 1.9575 & 2.1457 & 2.4386 \\ \hline
1500 & 2.2270 & 1.5364 & 1.7163 & 2.1943 & 1.9695 & 1.5951 & 2.0121& 1.5935 \\ \hline
1600 & 2.2867 & 1.7055& 2.0601 & 2.4301 & 1.9191 & 1.8443 & 2.1205 & 1.7049 \\ \hline
1700 & 2.2008& 1.6697 & 1.6444& 2.5805 & 1.7660 & 1.8970 & 2.0380 & 1.6506 \\ \hline
1820 & 2.3523 & 1.8174 & 1.8164 & 1.9729 & 1.7724 & 1.7230 & 2.1364 & 2.0152 \\ \hline
1870 & 1.9625 & 1.6238 & 1.5215& 1.5424 & 1.4879 & 1.4691 & 1.9080& 1.9016 \\ \hline
1913 & 1.8266 & 1.2613 & 1.6197  & 1.3156 & 1.3601  & 1.3877& 1.7664 & 1.4461 \\ \hline
1950 & 2.0908 & 1.3855  & 1.4274 & 1.2403 & 1.1484 & 1.3181 & 1.7607 & 1.5831 \\ \hline
1973 & 1.6254 & 1.5233 & 1.1989 & 1.3320& 1.3986 & 1.4330 & 1.8364 & 2.4553 \\ \hline
2003 & 1.5961 & 1.6308 & 1.4115 & 1.4561 & 1.7263 & 1.7188& 1.9077 & 2.0794  \\ \hline
\end{tabular}%
}

\resizebox{\columnwidth}{!}{%
\begin{tabular}{|l|l|l|l|l|l|l|l|l|}
\hline
population & Western Europe & Eastern Europe & Former Soviet Union & Western Offshoots & Latin America & Japan & Asia & Africa \\ \hline
1    & 25050  & 4750   & 3900   & 1120   & 5600   & 3000   & 165400  & 17000  \\ \hline
1000 & 25560  & 6500   & 7100   & 1870   & 11400  & 7500   & 175100  & 32300  \\ \hline
1500 & 57332  & 13500  & 16950  & 2800   & 17500  & 15400  & 268400  & 46610  \\ \hline
1600 & 73778  & 16950  & 20700  & 2300   & 8600   & 18500  & 360000  & 55320  \\ \hline
1700 & 81460  & 18800  & 26550  & 1750   & 12050  & 27000  & 374500  & 61080  \\ \hline
1820 & 133040 & 36457  & 54765  & 11231  & 21591  & 31000  & 679375  & 74236  \\ \hline
1870 & 187504 & 53557  & 88672  & 46088  & 40399  & 34437  & 730796  & 90466  \\ \hline
1913 & 260975 & 79530  & 156192 & 111401 & 80935  & 51672  & 925689  & 124697 \\ \hline
1950 & 304941 & 87637  & 179571 & 176457 & 165938 & 83805  & 1298972 & 228181 \\ \hline
1973 & 358825 & 110418 & 249712 & 250841 & 307873 & 108707 & 2139915 & 390202 \\ \hline
2003 & 394604 & 121434 & 287601 & 346233 & 541359 & 127214 & 3606753 & 853422 \\ \hline
\end{tabular}%
}

\resizebox{\columnwidth}{!}{%
\begin{tabular}{|l|l|l|l|l|l|l|l|l|}
\hline
$\Re(s)$ for GDP & Western Europe & Eastern Europe & Former Soviet Union & Western Offshoots & Latin America & Japan       & Asia        & Africa      \\ \hline
1    & 2.3878 & 2.2702 & 2.1654 & 2.4658 & 2.5378 & 2.1287 & 2.3739     & 2.0402 \\ \hline
1000 & 1.8312 & 1.7716 & 1.8747 & 2.2570 & 2.0815 & 2.0153 & 2.2272    & 2.5617  \\ \hline
1500 & 1.7978 & 1.6100 & 2.0914 & 2.3666 & 1.7013 & 1.8598 & 2.0304     & 1.8843 \\ \hline
1600 & 1.6611 & 1.7336 & 2.0913 & 2.6057  & 2.0606 & 1.9082 & 2.0527    & 1.9835 \\ \hline
1700 & 1.3997 & 1.6377 & 1.8629  & 2.6690 & 1.8086& 1.6363 & 1.9142     & 1.9277 \\ \hline
1820 & 1.2932 & 1.7121 & 2.1313 & 1.6197  & 1.6793 & 1.6401 & 1.8805      & 1.8236 \\ \hline
1870 & 0.1233  & 1.2904 & 1.1293 & 1.1634 & 1.3797 & 1.3261 & 1.4617      & 1.2168 \\ \hline
1913 & 1.4127    & 1.1812  & 0.9786 & 0.3976 & 1.1222  & 1.2181 & 0.5667 & 1.2484 \\ \hline
1950  & 0.2283     & 1.1199    & 0.8402         & 1.3761        & 0.8516   & 1.1150 & 0.6169 & 1.1641 \\ \hline
1973 & 1.3019   & 1.0312 & 0.7185 & 0.0136 & 0.6718 & 0.6657   & 0.4063 & 0.9657 \\ \hline
2003 & 0.5163  & 1.3797 & 1.2217 & 0.5132 & 1.0726 & 1.0163 & 1.4264     & 1.2074 \\ \hline
\end{tabular}%
}

\resizebox{\columnwidth}{!}{%
\begin{tabular}{|l|l|l|l|l|l|l|l|l|}
\hline
GDP & Western Europe & Eastern Europe & Former Soviet Union & Western Offshoots & Latin America & Japan & Asia & Africa \\ \hline
1    & 14433   & 1956   & 1560    & 448     & 2240    & 1200    & 75535    & 8030    \\ \hline
1000 & 10925   & 2600   & 2840    & 748     & 4560    & 3188    & 81683    & 13835   \\ \hline
1500 & 44183   & 6696   & 8458    & 1120    & 7288    & 7700    & 153617   & 19383   \\ \hline
1600 & 65602   & 9289   & 11426   & 920     & 3763    & 9620    & 207469   & 23473   \\ \hline
1700 & 81213   & 11393  & 16196   & 833     & 6346    & 15390   & 214281   & 25776   \\ \hline
1820 & 159851  & 24906  & 37678   & 13499   & 14921   & 20739   & 391738   & 31266   \\ \hline
1870 & 367466  & 50163  & 83646   & 111493  & 27311   & 25393   & 400245   & 45234   \\ \hline
1913 & 902210  & 134793 & 232351  & 582941  & 120796  & 71653   & 609135   & 79486   \\ \hline
1950 & 1396078 & 185023 & 510243  & 1635490 & 415328  & 160966  & 830428   & 203131  \\ \hline
1973 & 4096764 & 550756 & 1513070 & 4058289 & 1389460 & 1242932 & 2621624  & 549993  \\ \hline
2003 & 7857394 & 786408 & 1552231 & 9708029 & 3132145 & 2699261 & 13855834 & 1322087 \\ \hline
\end{tabular}%
}

\tiny{\begin{raggedright}
\noindent  For the calculation of $\Re(s)$ see the main text.  For each region, one unit of population corresponds to $10^{3}$ individuals, and one unit of GDP corresponds to $10^{6}$ (1990) international dollars.  Raw data are taken from \cite{Maddison2007}.\\
\end{raggedright}}
\end{table}

If $D_k\sim 1$two limiting scenarios arise.

\begin{enumerate}
\item[(i)] For the coefficient $a$ to converge we require
\[
P_k \approx \frac{1}{\zeta(s)} \approx \sum_{n=1}^{\infty}\frac{\mu(n)}{n^{s}},
\]
where $\mu$ denotes the M\"obius function.  In this regime, convergence of the coefficient $b$ typically requires $s\sim 0$.
\item[(ii)] If $P_k\ll 1$, convergence of $N$ requires $s\sim 1$.  In that case one finds
\[
kP_k \approx \frac{1}{\zeta(s)}.
\]
\end{enumerate}

When case (i) holds the patches exhibit **true neutrality**.  Both (i) and (ii) admit a simple interpretation in terms of a zero-sum Markov process (cf.\ Hubbell, 2001): in either limit $f_s(k)\sim P_k$, so populations appear neutral with respect to the diversity index $D$.  If both populations and environment satisfy
\[
P_k \approx \frac{1}{\zeta(s)} \approx \sum_{n=1}^{\infty}\frac{\mu(n)}{n^{s}},
\]
we call this situation **M\"obius neutrality**.  If $D$ shows apparent neutrality with $s\sim 1$, we call it **harmonic neutrality**.  Thus the parameter $s$ serves as a compact indicator of the system's characteristic status.

Focusing on harmonic neutrality, values $\Re(s)>2$ indicate adaptation that exceeds fluctuations expected under individual-level harmonic neutrality.  The M\"obius function $\mu(n)$ has the combinatorial sign structure $\mu(n)=+1$ for integers with an even number of distinct prime factors (bosonic sign), $\mu(n)=-1$ for an odd number (fermionic sign), and $\mu(n)=0$ when $n$ is divisible by a squared prime (observational null).  This sign pattern naturally arises when multiplicative (quantized) interactions occur between components.

A classical identity of Madhava (circa 1400 CE) is relevant:
\[
\sum_{\substack{n=1\\\ n\ \text{odd}}}^{\infty}\frac{(-1)^{(n-1)/2}}{n}=\frac{\pi}{4}.
\]
Interpreted probabilistically for large $n$, this shows that expected interactions among many fermionic contributions produce a factor $\pi/4$ when $s=1$.  Pairwise interaction (multiplication by $2$) then yields $\pi/2$, which motivates in the next subsection a rotation of the real axis by angle $\pi/2$ and the induction of a hierarchical structure in an orthogonal (imaginary) dimension.

There is a way of describing quantization of $w$. First, think of the tube zeta function \begin{equation} \tilde{\zeta}A(s):=\int_{0}^{\delta} t_t^{s-N-1}|A_{t_t}|dt_t. \end{equation} Think of $r\to+\infty$ and then \begin{equation} \int_{1/r}^{\delta} t_t^{\dim_B A - N - 1}|A_{t_t}dt_t \sim \mathrm{res}\big(\tilde{\zeta}_A,\dim_B A\big)\ln r \end{equation} \cite{Lapidus2017}. We can set an average Minkowski content as \begin{equation} \tilde{\mathscr{M}}^{\dim_B A}(A):=\lim_{r\to+\infty}\frac{1}{\ln r}\int_{1/r}^{\delta} t_t^{\dim_B A - N - 1}|A_{t_t}|dt_t. \end{equation} From Lemma 2.4.7 of \cite{Lapidus2017},\begin{equation}\tilde{\mathscr{M}}^{*s}(A) = \Bigg\{\begin{array}{l}0\ (\mathrm{for}\ \Re(s) > \overline{D}_{av})\\ +\infty\ (\mathrm{for}\ 0 \leq \Re(s)  < \overline{D}_{av})\end{array},\end{equation}
\begin{equation}\tilde{\mathscr{M}}_*^{s}(A) = \Bigg\{\begin{array}{l}0\ (\mathrm{for}\ \Re(s) > \underline{D}_{av})\\ +\infty\ (\mathrm{for}\ 0 \leq \Re(s)  < \underline{D}_{av})\end{array},\end{equation} where $D_{\mathrm{av}}$ is an average Minkowski dimension. When $\dim_B A$ exists, $D_{\mathrm{av}}=\dim_B A$ (Proposition 2.4.9 of \cite{Lapidus2017}) and as $\Re(s)=\dim_B A\to D_{\mathrm{av}}$, $\tilde{\mathscr{M}}^{s}(A)$ will converge to \begin{equation} \lim_{r\to+\infty}\frac{1}{\ln r}\int_{1/r}^{\delta} t_t^{\dim_B A - N - 1}|A_{t_t}|dt_t = F(t_t), \end{equation} independent of $\delta$ and $r$. Thus quantization related with $w$ occurs.\\

There is a way of describing quantization of $s$. First, think of the tube zeta function of the second kind $ \tilde{\zeta}_{2A}(s):=\int_{0}^{\delta} t_t^{\,s-N}\,|A_{t_t}|\,d_tt.$ Think of $r\to+\infty$ and then

\begin{equation}\int_{1/r}^{\delta}t_t^{\dim_BA - N}|A_{t_t}|dt_t \sim \mathrm{res}(\tilde{\zeta}_{2A}, \dim_BA)\delta.\end{equation}

We can set an average Minkowski content as

\begin{equation}\tilde{\mathscr{M}}_2^{\dim_BA}(A) := \underset{r \to +\infty}{\lim}\frac{1}{\delta}\int_{1/r}^{\delta}t_t^{\dim_BA - N}|A_{t_t}|dt_t.\end{equation}

Then,

\begin{equation}\tilde{\mathscr{M}}_2^{*s}(A) = \Bigg\{\begin{array}{l}0\ (\mathrm{for}\ \Re(s) > \overline{D}_{av})\\ +\infty\ (\mathrm{for}\ 0 \leq \Re(s)  < \overline{D}_{av})\end{array},\end{equation}
\begin{equation}\tilde{\mathscr{M}}_{2*}^{s}(A) = \Bigg\{\begin{array}{l}0\ (\mathrm{for}\ \Re(s) > \underline{D}_{av})\\ +\infty\ (\mathrm{for}\ 0 \leq \Re(s)  < \underline{D}_{av})\end{array},\end{equation}

Here $D_{\mathrm{av}}$ is an average Minkowski dimension. When $\dim_B A$ exists, $D_{\mathrm{av}}=\dim_B A$, and as $\Re(s)=\dim_B A\to D_{\mathrm{av}}$, $\widetilde{\mathscr{M}}_{2}^{s}(A)$ will converge to

\begin{equation}
\lim_{r\to+\infty}\frac{1}{\ln r}\int_{1/r}^{\delta} t_t^{\,\dim_B A - N}|A_{t_t}|dt_t = F(t_t),
\end{equation}

independent of $\delta$ and $r$. Thus quantization related with $s$ also occurs. In these ways, quantization is an expected outcome of Minkowski components for both $w$ and $s$. This appears to hold for Dictyostelia but not for soil mesofauna; we will discuss this discrepancy later.

Also note that here we assume continuity of functions or the existence of $ \dim_B A $, which is likely to hold in natural systems. This is a mere description, and not a mathematical proof.

\subsection{Introducing $\Im(s)$ clarifies adaptation and maladaptation in species}
Next, we need to consider $R = T$ theory, Weil's explicit formula, and some algebraic number theory to define $\Im(s)$ precisely. $R = T$ theory is based on an ordinary representation of a Galois deformation ring. If we consider the mapping to $T$ that is shown below, they become isomorphic and fulfill the conditions for a theta function for zeta analysis. Let ($s \in \mathbb{C}, D \in \mathbb{C}$), where $\mathbb{C}$ is complex. First, we introduce a small $s$ that fulfills the requirements from a higher-dimensional theta function. Assuming $\mathbb{H}$, $\mathbb{C}$, $\mathbb{R}$, $\mathbb{R^*_+}$, $\mathbb{R_\pm}$ as a higher-dimensional analogue of the upper half-plane,  complex planes constituted by $\prod \mathrm{C} = \frac{1}{k^s\zeta}$, $[\prod \mathrm{C}]^+ = \{z \in \mathrm{\mathbb{C}} | z = \bar{z}\}/T_H$ (Hecke ring, $R = T$ theorem), $|\Im(s)|$ dual with $\mathbb{|R|}$ \cite{Weil1952}, $\ln|\Im(s)|$, $s$ could be set on $\mathbb{H}$ and $\ln|\Im(s)| = \ln(|D|^{E(N)}) = E(N) \ln|D|$ is a part of $\Re(s) = b\ln|D|$.  Thus, $\mathbb{H}\subseteq\mathbb{C}\supseteq\mathbb{R}\supseteq\mathbb{R_\pm}\supseteq\mathbb{R^*_+}$, and the functions described here constitute a theta function. The series converges absolutely and uniformly on every compact subset of $\mathbb{R} \times \mathbb{R} \times \mathbb{H}$ \cite{Neukirch1999}, and this describes a (3 + 1)-dimensional system. This is based on the $R = T$ theorem and Weil's explicit formula (correspondence of zeta zero points, Hecke operator, and Hecke ring); for a more detailed discussion, see \cite{Weil1952} \cite{Taylor1995} \cite{Wiles1995} \cite{Kisin2009}. Since $b$ is real, $\Re(s) = b\ln|D|$ and $\Im(s) = b\arg D$. Thus, $\Re(s)$ is related to the absolute value of an individual's fitness, and $\Im(s)$ is the time scale for oscillations of $D$ and is the argument multiplied by the $b$ scale. Therefore, 

\begin{equation}\frac{\partial\Re(s)}{\partial t} = b\frac{1}{|D|}\frac{\partial|D|}{\partial t},\end{equation}
\begin{equation}\frac{\partial\Im(s)}{\partial t} = b\frac{\partial\arg D}{\partial t}.\end{equation}

When $0<\Re(s)<1$ and $|D|' > 0$, one typically finds $\Re(s)\sim 1$, which corresponds to harmonic neutrality. This behavior was frequently observed in the Dictyostelia data. When $0<\Re(s)<1$ and $|D|'<0$, one typically finds $\Re(s)\sim 0$, corresponding to M\"obius neutrality. For $1<\Re(s)<2$, the population or species may diverge when $\Im(s)=T$, i.e., when $\Im(s)$ equals the imaginary part of a nontrivial zero of $\zeta$. Equivalently, divergence occurs when $\arg D = \frac{T}{b}$. 

Note that $D = e^{s/b} = e^{\Re(s)/b}\bigg(\cos\frac{\Im(s)}{b} + i\sin\frac{\Im(s)}{b}\bigg)$, so quantization of $\Im(s)$ can be expressed as $\Im(s)\sim \pm m\frac{\pi}{2}b$, with $m\in\mathbb{N}$. Here ``quantization'' refers to constraining a continuous degree of freedom to discrete values; the factor $\pi/2$ can be computed from the Madhava series discussed earlier. Under the assumption that the distribution of population or species numbers is in equilibrium and depends on pairwise interactions as described above, we have $|D| = e^{\Re(s)/b}$. 

By the Riemann--von Mangoldt formula \cite{vonMangoldt1905}, the number of nontrivial zeros up to height $T$ is \begin{equation} N(T)=\frac{T}{2\pi}\log\frac{T}{2\pi}-\frac{T}{2\pi}+O(\log T), \end{equation} which implies heuristically that $T\sim 2\pi e^{2\pi N(T)/T + 1}$. Using Stirling's approximation $\ln n!\approx n\ln n - n$, one may interpret $N(T)\approx \ln n!$ and view the number of species as the sum of relative entropies. Since $T=\pm m(\pi/2)b$, for the population or species as a whole we obtain $|D|^{E(N)} = e^{m\pi\Re(s)E(N)/(2T)}$. Because the $|D|^{E(N)}$ axis and the $T/\arg D$ axis are orthogonal and the latter has scale $2\pi$ times that of the former, the approximation $|D|^{E(N)}\approx |T|$ (see Table~4) provides a good fit for large adaptive growth bursts or collapses at the population or species level. Solving for $m$ gives \begin{equation} m=\frac{1}{\Re(s)E(N)}\Big(4N(T)+\frac{2T}{\pi}(\ln 2\pi + 1)\Big). \end{equation}

If one fixes a unit of space for measuring population density, then $\Im(s)=e^{\Re(s)E(N)/b}$ is scale invariant for species when $\Re(s)$ is scale invariant with respect to system size. Here $b$ denotes the order of the ratio of the sum of population densities of a given species to the number of patches, and $E(N)$ denotes the ratio of the sum of population densities to the number of patches. For a given population, if $b$ is of the order of the population density in a typical patch, then $b$ is also scale invariant with respect to sampling size, provided a sufficiently large number of samples are taken.

We interpret nontrivial zeros of $\zeta$ as \textit{prime states} (states associated with prime numbers $p$); these zeros signal imminent growth bursts or collapses of a species. Recall the functional equation for the Riemann zeta function \cite{Riemann1859}: \begin{equation} \zeta(s)=2^s\pi^{s-1}\sin\frac{\pi s}{2}\Gamma(1-s)\zeta(1-s) =2^s\pi^s\sin\frac{\pi s}{2}\frac{1}{\Gamma(s)\sin\pi s}\zeta(1-s). \end{equation} To avoid discontinuities at zeros of $\zeta$, $\Re(s)$ must be $1/2$ or an integer; consequently, zeros of $\zeta$ constrain both $\Re(s)$ and $\Im(s)$ to specific values. Note that the set $T$ consists of the imaginary parts of the nontrivial zeros of $\zeta$, and these values are not integers in the quantization described above.

This model is consistent with several empirical observations for Dictyostelia species (values shown in red in Table~1): ($\Re(s), \Im(s) \approx T, m$) = (3.078, 14.99, 0.01003), (4.942, 38.74, 0.01723), (2.056, 275.5, 2.994), (2.8795, 13.80, 0.009451), (2.1411, 115.9, 0.05094), (2.9047, 13.93, 0.004941). Thus the model offers a plausible explanation for observed quantization phenomena in some Dictyostelia cases at the level of the $O(\log T)$ term. For broader populations or for macroeconomic data, the observed $\Im(s)$ values do not generally coincide with zeta zeros, indicating the system is not at a prime-state zero point.

Consider $\mathrm{res}\big(\tilde{\zeta}_A,\dim_B A\big)=\mathrm{res}\big(\tilde{\zeta}_{2A},\dim_B A\big)$. As $\delta\to r$, we have \begin{equation} \frac{\tilde{\zeta}_{2A}}{\tilde{\zeta}_A}\sim\frac{r}{\ln r}\sim\pi(r), \end{equation} where $\pi(r)$ denotes the prime counting function for sufficiently large $r$. Thus $r$ can be interpreted as the number of possible quantization levels: as $r$ grows, the distribution approaches the characteristics of prime numbers and quantization by primes is effectively realized.

Next, consider the absolute zeta function for the multiplicative group: \begin{equation} \zeta_{\mathbb{G}_m/F_1}(s)=\frac{s}{s-1}=\frac{s}{w}, \end{equation} when $\mathbb{G}_m\cong GL(1)$. The tube zeta functions acting on the denominator $\tilde{\zeta}_A$ and the numerator $\tilde{\zeta}_{2A}$ transform the absolute zeta function into the prime counting function. Consequently, the number of primes can be recovered from the $F_1$-theoretic construction.

\begin{table}[h]
\caption{$T$, $\Im(s)$, $p$, and $|N(p)|$ values.}
\label{tab:4}
\centering
\resizebox{\textwidth}{!}{%
\begin{tabular}{|l|r|r|r|r|r|r|l|r|r|r|r|r|r|}
\hline
$T$         & WE \textit{P. pallidum} & WE \textit{D. purpureum} & WE \textit{P. violaceum} & WW \textit{P. pallidum} & WW \textit{D. purpureum} & WW \textit{P. violaceum} & $\Im(s)$     & WE \textit{P. pallidum} & WE \textit{D. purpureum} & WE \textit{P. violaceum} & WW \textit{P. pallidum} & WW \textit{D. purpureum} & WW \textit{P. violaceum} \\ \hline
April       &                &                 &                 &                &                 &                 &April       &                &                 &                 &                &                 &                 \\ \hline
June      &                & 147.4228        & 30.4249         &                &                 &                 & June      & 8.1822         & 148.6187        & 31.1005         &                &                 &                 \\ \hline
July      &                &                 &                 & 40.9187        &                 & 174.7542        & July      &                &                 &                 & 39.3062        & 5.3315          & 174.5203        \\ \hline
August    &                &                 &                 & 21.0220        & 14.1347         &                 & August    &                &                 &                 & 22.6267        & 13.7962         & 2.4878          \\ \hline
September & 21.0220        & 14.1347         &                 & 52.9703        & 116.2267        &                 & September & 23.2403        & 14.9897         & 2.4101          & 53.1153        & 115.8800        & 2.0281          \\ \hline
October   & 48.0052        & 37.5862         &                 & 21.0220        &                 & 14.1347         & October   & 45.6795        & 38.7450         & 2.0958          & 22.6675        & 2.4764          & 13.9291         \\ \hline
November  &                & 14.1347         & 275.5875        &                &                 &                 & November  & 7.7262         & 15.3777         & 275.5449        &                &                 &                 \\ \hline
December  &                &                 &                 &                &                 &                 & December  &                &                 &                 &                &                 &                 \\ \hline
January   &                &                 &                 &                &                 &                 & January   &                &                 &                 &                &                 &                 \\ \hline
\end{tabular}
}

\centering

\resizebox{\columnwidth}{!}{%
\begin{tabular}{|l|l|l|l|l|l|l|l|l|l|l|l|l|l|}
\hline
$p$ &
  WE \textit{P. pallidum} &
  WE \textit{D. purpureum} &
  WE \textit{P. violaceum} &
  WW \textit{P. pallidum} &
  WW \textit{D. purpureum} &
  WW \textit{P. violaceum} &
  $|N(p)|$ for $a_1$ &
  WE \textit{P. pallidum} &
  WE \textit{D. purpureum} &
  WE \textit{P. violaceum} &
  WW \textit{P. pallidum} &
  WW \textit{D. purpureum} &
  WW \textit{P. violaceum} \\ \hline
April    &    &     &     &    &   &     & April    &                         &             &           &             &                         &             \\ \hline
June     &    & 239 & 7   &    &   &     & June     & {\color[HTML]{FF0000} } & 0.9833  & 1.0138 &             &                         &             \\ \hline
July     &    &     &     & 17 &   & 317 & July     &                         &             &           & 0.9952 & {\color[HTML]{FF0000} } & 1.0160 \\ \hline
August   &    &     &     & 3  & 2 &     & August   &                         &             &           & 0.6482 & 0.2425             &             \\ \hline
September &
  3 &
  2 &
   &
  31 &
  157 &
   &
  September &
  0.7866 &
  0.2889 &
   &
  0.9546 &
  1.0028 &
  {\color[HTML]{FF0000} } \\ \hline
October  & 23 & 13  &     & 3  &   & 2   & October  & 1.0874              & 1.0998 &           & 0.6567 &                         & 0.2595 \\ \hline
November &    & 2   & 677 &    &   &     & November & {\color[HTML]{FF0000} } & 0.8161 & 1.0041         &             &                         &             \\ \hline
December &    &     &     &    &   &     & December &                         &             &           &             &                         &             \\ \hline
January  &    &     &     &    &   &     & January  &                         &             &           &             &                         &             \\ \hline
\end{tabular}%
}

\resizebox{\columnwidth}{!}{%
\begin{tabular}{|l|l|l|l|l|l|l|l|l|l|l|l|l|l|l|l|l|l|l|l|l|l|l|l|}
\hline
Collembola $T$ &
  \textit{F. octoulata} &
  \textit{E. aino} &
  \textit{H. watanabei} &
  \textit{S. celebensis} &
  \textit{S. aureus} &
  \textit{D. trispinata} &
  \textit{S. japonica} &
  \textit{H. nigrochephala} &
  \textit{W. japonica} &
  \textit{F. yosii} &
  \textit{A. laricis} &
  Collembola $\Im(s)$ &
  \textit{F. octoulata} &
  \textit{E. aino} &
  \textit{H. watanabei} &
  \textit{S. celebensis} &
  \textit{S. aureus} &
  \textit{D. trispinata} &
  \textit{S. japonica} &
  \textit{H. nigrochephala} &
  \textit{W. japonica} &
  \textit{F. yosii} &
  \textit{A. laricis} \\ \hline
April &
   &
   &
  32.9351 &
   &
   &
   &
   &
   &
   &
   &
   &
  April &
  3.9931 &
  3.9931 &
  32.8194 &
  3.9931 &
  3.9931 &
  3.9931 &
  3.9931 &
  3.9931 &
  8.9753 &
  3.9931 &
  3.9931 \\ \hline
May &
  156.1129 &
  156.1129 &
  25.0109 &
  156.1129 &
  3460.0461 &
   &
  156.1129 &
  156.1129 &
  156.1129 &
  156.1129 &
  156.1129 &
  May &
  156.1013 &
  156.1013 &
  24.2039 &
  156.1013 &
  3459.8183 &
  8.1348 &
  156.1013 &
  156.1013 &
  156.1013 &
  156.1013 &
  156.1013 \\ \hline
June &
  3738.5881 &
  1552.7361 &
  2078.1555 &
  48.0052 &
  2078.1555 &
  1331.8276 &
  560.2408 &
  48.0052 &
  48.0052 &
  23164.6618 &
  2078.1555 &
  June &
  3738.8644 &
  1552.5027 &
  2078.3449 &
  46.7021 &
  2078.3449 &
  1331.3103 &
  560.2473 &
  46.7021 &
  46.7021 &
  23164.8534 &
  2078.3449 \\ \hline
July &
  14.1347 &
  3088.0834 &
  1748.5640 &
  204.1897 &
  1748.5640 &
  1748.5640 &
  1748.5640 &
  204.1897 &
  204.1897 &
  430.3287 &
  204.1897 &
  July &
  13.2243 &
  3088.0979 &
  1748.2481 &
  204.2100 &
  1748.2481 &
  1748.2481 &
  1748.2481 &
  204.2100 &
  204.2100 &
  430.7380 &
  204.2100 \\ \hline
August &
  653.6496 &
  1391.8532 &
  1025.2657 &
  14.1347 &
  14.1347 &
  95.8706 &
  1025.2657 &
  14.1347 &
  1025.2657 &
  14.1347 &
  14.1347 &
  August &
  653.1512 &
  1391.9755 &
  1025.4619 &
  17.3611 &
  17.3611 &
  96.2522 &
  1025.4619 &
  17.3611 &
  1025.4619 &
  17.3611 &
  17.3611 \\ \hline
September &
  2365.1915 &
  3321.6636 &
  219.0676 &
  3069.7829 &
  2598.5988 &
  2598.5988 &
  2545.8780 &
  2545.8780 &
  43.3271 &
  43.3271 &
  43.3271 &
  September &
  2365.0919 &
  3322.2806 &
  219.1262 &
  3069.7912 &
  2599.0799 &
  2599.0799 &
  2545.3001 &
  2545.3001 &
  42.8699 &
  42.8699 &
  42.8699 \\ \hline
October &
  393.4277 &
  456.3284 &
  14.1347 &
  456.3284 &
  456.3284 &
  32.9351 &
  32.9351 &
  32.9351 &
  32.9351 &
  32.9351 &
  32.9351 &
  October &
  393.6927 &
  456.1483 &
  16.2950 &
  456.1483 &
  456.1483 &
  31.7150 &
  31.7150 &
  31.7150 &
  31.7150 &
  31.7150 &
  31.7150 \\ \hline
November &
   &
   &
   &
   &
  94.6513 &
   &
   &
   &
   &
   &
   &
  November &
  8.0312 &
  8.0312 &
  8.0312 &
  8.0312 &
  94.8862 &
  8.0312 &
  8.0312 &
  8.0312 &
  8.0312 &
  8.0312 &
  8.0312 \\ \hline
\end{tabular}%
}

\resizebox{\columnwidth}{!}{%
\begin{tabular}{|l|l|l|l|l|l|l|l|l|l|l|l|l|l|l|l|l|l|l|l|l|l|l|l|}
\hline
$p$ &
  \textit{F. octoulata} &
  \textit{E. aino} &
 \textit{ H. watanabei} &
  \textit{S. celebensis} &
  \textit{S. aureus} &
  \textit{D. trispinata} &
  \textit{S. japonica} &
  \textit{H. nigrochephala} &
  \textit{W. japonica} &
  \textit{F. yosii} &
  \textit{A. laricis} &
  $|N(p)|$ for $a_1$ &
  \textit{F. octoulata} &
   \textit{E. aino} &
   \textit{H. watanabei} &
   \textit{S. celebensis} &
   \textit{S. aureus} &
   \textit{D. trispinata} &
   \textit{S. japonica} &
   \textit{H. nigrochephala} &
   \textit{W. japonica} &
    \textit{F. yosii} &
  \textit{A. laricis} \\ \hline
April &
   &
   &
  11 &
   &
   &
   &
   &
   &
   &
   &
   &
  April &
   &
   &
  0.9389 &
   &
   &
   &
   &
   &
   &
   &
   \\ \hline
May &
  263 &
  263 &
  5 &
  263 &
  26687 &
   &
  263 &
  263 &
  263 &
  263 &
  263 &
  May &
  1.0153 &
  1.0153 &
  0.8443 &
  1.0153 &
   &
   &
  1.0153 &
  1.0153 &
  1.0153 &
  1.0153 &
  1.0153 \\ \hline
June &
  29567 &
  8971 &
  13399 &
  23 &
  13399 &
  7237 &
  2081 &
  23 &
  23 &
  307259 &
  13399 &
  June &
   &
   &
   &
  1.0166 &
   &
   &
   &
  1.0166 &
  1.0166 &
   &
   \\ \hline
July &
  2 &
  22901 &
  10567 &
  421 &
  10567 &
  10567 &
  10567 &
  421 &
  421 &
  1399 &
  421 &
  July &
  0.6824 &
   &
   &
  1.0089 &
   &
   &
   &
  1.0089 &
  1.0089 &
  0.9949 &
  1.0089 \\ \hline
August &
  2617 &
  7691 &
  4999 &
  2 &
  2 &
  107 &
  4999 &
  2 &
  4999 &
  2 &
  2 &
  August &
   &
   &
   &
  0.7918 &
  0.7918 &
  0.9802 &
   &
  0.7918 &
   &
  0.7918 &
  0.7918 \\ \hline
September &
  15937 &
  25261 &
  463 &
  22717 &
  18143 &
  18143 &
  17659 &
  17659 &
  19 &
  19 &
  19 &
  September &
   &
   &
  0.9891 &
   &
   &
   &
   &
   &
  0.9557 &
  0.9557 &
  0.9557 \\ \hline
October &
  1213 &
  1511 &
  2 &
  1511 &
  1511 &
  5 &
  5 &
  5 &
  5 &
  5 &
  5 &
  October &
  1.0058 &
  1.0059 &
  0.8221 &
  1.0059 &
  1.0059 &
  1.0153 &
  1.0153 &
  1.0153 &
  1.0153 &
  1.0153 &
  1.0153 \\ \hline
November &
   &
   &
   &
   &
  103 &
   &
   &
   &
   &
   &
   &
  November &
   &
   &
   &
   &
  1.0364 &
   &
   &
   &
   &
   &
   \\ \hline
\end{tabular}%
}

\resizebox{\columnwidth}{!}{%
\begin{tabular}{|l|l|l|l|l|l|l|l|l|l|l|l|l|l|l|l|}
\hline
Sarcoptiformes $T$ &
  \textit{A. gigantea} &
  \textit{P. parvisetigerum} &
  \textit{T. stercus} &
  \textit{M. japonica} &
  \textit{G. fusca} &
  \textit{N. rotundus} &
  \textit{C. reticulatus} &
  Sarcoptiformes $\Im(s)$ &
  \textit{A. gigantea} &
  \textit{P. parvisetigerum} &
  \textit{T. stercus} &
  \textit{M. japonica} &
  \textit{G. fusca} &
  \textit{N. rotundus} &
  \textit{C. reticulatus} \\ \hline
April &
  426327.526 &
  37342757.27 &
  653.6496 &
  37342757.27 &
  426327.5258 &
  37342757.27 &
  426327.526 &
  April &
  426327.4376 &
  37342757.37 &
  652.9375 &
  37342757.37 &
  426327.4376 &
  37342757.37 &
  426327.4376 \\ \hline
May &
  12665.1014 &
  12665.1014 &
  49.7738 &
  478.9422 &
  478.9422 &
  478.9422 &
  478.9422 &
  May &
  12664.5182 &
  12664.5182 &
  49.1378 &
  479.5458 &
  479.5458 &
  479.5458 &
  479.5458 \\ \hline
June &
  25989.9928 &
   &
  193.0797 &
  44401.4393 &
  2244.3707 &
  21.0220 &
   &
  June &
  25989.7757 &
  2.6343 &
  193.8381 &
  44401.3320 &
  2244.07656 &
  20.0861 &
  3.5985 \\ \hline
July &
  794.4838 &
  6185.4601 &
  6185.4601 &
  6185.4601 &
   &
  6185.4601 &
   &
  July &
  794.7808 &
  6185.9266 &
  6185.9266 &
  6185.9266 &
  29.2897 &
  6185.9266 &
  29.2897 \\ \hline
August &
  173.4115 &
  696.6261 &
   &
  696.6261 &
   &
  696.6261 &
   &
  August &
  173.5014 &
  696.5693 &
  7.1749 &
  696.5693 &
  7.1749 &
  696.5693 &
  7.1749 \\ \hline
September &
  64420.6672 &
  64420.6672 &
   &
  21.0220 &
  192.0267 &
  792.4277 &
  10091.7991 &
  September &
  64420.8957 &
  64420.8957 &
  9.5752 &
  18.6345 &
  192.0530 &
  792.4324&
  10091.8166 \\ \hline
October &
   &
  366.2127 &
  549.4976 &
  32.9351 &
  32.9351 &
  549.4976&
  549.4976 &
  October &
  5.6335 &
  366.6181 &
  548.8626&
  35.2073 &
  35.2073 &
  548.8626 &
  548.8626 \\ \hline
November &
  37.5862 &
   &
  52.9703 &
  25.0109 &
   &
   &
   &
  November &
  37.1904 &
  4.3505 &
  53.3860 &
  24.7572 &
  4.3505 &
  4.3505 &
  4.3505 \\ \hline
\end{tabular}%
}

\resizebox{\columnwidth}{!}{%
\begin{tabular}{|l|l|l|l|l|l|l|l|l|l|l|l|l|l|l|l|}
\hline
$p$ &
  \textit{A. gigantea} &
  \textit{P. parvisetigerum} &
  \textit{T. stercus} &
 \textit{M. japonica} &
  \textit{G. fusca} &
  \textit{N. rotundus} &
  \textit{C. reticulatus} &
  $|N(p)|$ for $a_1$ &
 \textit{ A. gigantea} &
 \textit{ P. parvisetigerum} &
  \textit{T. stercus} &
  \textit{M. japonica} &
  \textit{G. fusca} &
  \textit{N. rotundus} &
  \textit{C. reticulatus} \\ \hline
April &
  10361909 &
  1755205931 &
  2617 &
  1755205931 &
  10361909 &
  1755205931 &
  10361909 &
  April &
   &
   &
   &
   &
   &
   &
   \\ \hline
May &
  143719 &
  143719 &
  29 &
  1619 &
  1619 &
  1619 &
  1619 &
  May &
   &
   &
  1.0219 &
   &
   &
   &
   \\ \hline
June &
  354247 &
   &
  383 &
  686863 &
  14887 &
  3 &
   &
  June &
   &
   &
  1.0163 &
   &
   &
  1.0767 &
   \\ \hline
July &
  3469 &
  57203 &
  57203 &
  57203 &
   &
  57203 &
   &
  July &
   &
   &
   &
   &
   &
   &
   \\ \hline
August &
  313 &
  2843 &
   &
  2843 &
   &
  2843 &
   &
  August &
  0.9966 &
   &
   &
   &
   &
   &
   \\ \hline
September &
  1082161 &
  1082161 &
   &
  3 &
  379 &
  3463 &
  107719 &
  September &
   &
   &
   &
  0.9391 &
  1.0157 &
   &
   \\ \hline
October &
   &
  1069 &
  2011 &
  11 &
  11 &
  2011 &
  2011 &
  October &
   &
  1.0072 &
   &
  1.0708 &
  1.0708 &
   &
   \\ \hline
November &
  13 &
   &
  31 &
  5 &
   &
   &
   &
  November &
  0.9949 &
   &
  0.9973 &
  0.7500 &
   &
   &
   \\ \hline
\end{tabular}%
}

\tiny{\begin{raggedright}
\noindent WE denotes the Washidu East quadrat and WW denotes the Washidu West quadrat; see \cite{Adachi2015}. For scientific names consult Tables~1 and~2. The set $T$ consists of the theoretical imaginary parts of the nontrivial zeros of the Riemann $\zeta$ function associated with primes $p$, and we take $\Im(s)=|D|^{E(N)}$. The $T, \Im(s)$ for populations and for world economic data are not displayed because the values of $\Im(s)$ are extremely small (they never reach the value corresponding to $p=2$), so $T$ and $\Im(s)$ do not align in those cases. Procedures for computing $p$ and $|N(p)|$ are given in the main text. Blank entries in the tables indicate values that are undefined or that overflow the numerical range.\\
\end{raggedright}}
\end{table}

\subsection{Group theoretical insight to a species concept}
When we examine a particular hierarchical level, that level must exhibit a form of homeostasis to be recognized as an identifiable unit. Maintaining identity in natural systems requires adaptation to the environment; accordingly, the biological notion of adaptation is a fundamental concept for evolvable hierarchical systems. We begin at a level where adaptation is clearly applicable, for example the individual level or, in this work, the metapopulation level. By metapopulation we mean a collection of local populations linked by dispersal; this concept captures population dynamics tied to physical space (a union of habitat patches) and serves as a dynamical unit that, in our analogy, obeys a heat-bath type rule integrating stochastic behavior.

If one can select an appropriate set of indicator values and morphisms, the actual hierarchical structure in nature that is compatible with adaptation becomes detectable. The existence of such morphisms and indicators is supported empirically by the discrete clustering often observed in genetic distances on phylogenetic trees or networks. If no suitable morphisms or indicators exist, hierarchical identity collapses and phylogenetic clustering disappears. Below we propose a concrete example of such morphisms and indicators based on $p$-Sylow subgroups.

A clear definition of ``species'' requires a labeling scheme. Prime numbers are convenient labels because a prime cannot be factored into smaller positive integers; if a label admits nontrivial divisors, the corresponding identity may split into two or more distinct identities and thus become unstable. Interpreting these numbers as counts of interaction types that govern long-term dynamics, one expects that stable long-term interaction patterns are associated with prime counts. Consequently, a $p$-Sylow subgroup from group theory---which is associated to a particular prime $p$ and whose structure reflects $p$-power symmetries---is a natural candidate for a morphism and indicator: it can label a species by a prime $p$, and its group-theoretic properties encode the number and stability of interactions that define a species' equilibrium in a niche. In the sequel we outline why $p$-Sylow subgroups are suitable for this role.

Model interactions among populations or patches by a group $G$. Let the elements of $G$ represent elementary interactions (for example, directed effects from population $X$ to population $Y$), and let the group operation represent composition or synthesis of interactions. The group axioms admit natural biological readings: \begin{itemize} \item \textbf{Closure.} Composing two interactions yields another interaction in the same set. \item \textbf{Associativity.} The order of binary composition does not affect the final combined interaction. \item \textbf{Identity element $1_G$.} There exists a neutral interaction (a self-interaction) that leaves other interactions unchanged. \item \textbf{Inverse elements.} For each interaction there exists a reverse interaction that restores the neutral state. \end{itemize}

A group $G$ is \emph{nilpotent} if its lower central series $ G = G_0 \supseteq G_1 \supseteq \cdots \supseteq G_j =  \{1_G\} $ terminates at the trivial subgroup after finitely many steps, and for each $i$ the quotient $G_{i-1}/G_i$ lies in the center of $G/G_i$. Biologically, nilpotency models a system that, after finitely many interaction steps (or time steps), converges to a unique trivial subgroup representing an equilibrium identity. Adaptation processes that drive a population or community toward a stable state are naturally modeled by such convergence.

Two algebraic facts motivate this choice: \begin{itemize} \item Every finite $p$-group is nilpotent; hence a subgroup encoding a species' interactions that is a $p$-group automatically satisfies the nilpotency condition that models convergence to equilibrium. \item If a group is not nilpotent, it cannot be a $p$-Sylow subgroup; thus nilpotency is a necessary structural property for the $p$-Sylow interpretation. \end{itemize}

The identity element $1_G$ corresponds to an equilibrium status in a niche: multiplying by $1_G$ leaves the niche unchanged. In a ring-theoretic analogy, the additive identity $0_G$ represents absence of contribution; adding a different niche in equilibrium does not alter the remaining niche. These analogies justify modeling community states by algebraic structures (groups or rings) that reflect stability and invariance properties observed in biological systems.

We propose modeling a species identity at a chosen hierarchical level by a nilpotent subgroup, and in particular by a $p$-Sylow subgroup when a prime-based labeling is appropriate. This choice links discrete empirical clustering (phylogenetic or genetic) to algebraic structure: the prime $p$ labels the species and the subgroup structure encodes the number and stability of interactions that maintain the species' identity in its niche. 

We next examine the Sylow theorems \cite{Sylow1872} in the present biological context.  A finite group $G$ of order $ |G|=p_1^{\ell_1}p_2^{\ell_2}\cdots p_g^{\ell_g} $ is nilpotent if and only if, for each $i=1,\dots,g$, every subgroup $N_i$ of order $p_i^{\ell_i}$ a $p_i$-Sylow subgroup is normal in $G$.  A subgroup $H$ is normal precisely when every left coset equals the corresponding right coset, i.e. $a_GH=Ha_G$ for all $a_G\in G$.  Each $N_i$ therefore has order $p_i^{\ell_i}$, and the Sylow theorems imply that the number of such subgroups is congruent to $1$ modulo $p_i$ and divides $|G|/p_i^{\ell_i}$; in particular, these subgroups are mutually conjugate.  Conjugacy of a subgroup $H$ means that for any $a_G\in G$ the subgroup $a_GHa_G^{-1}$ is again a subgroup of the same order and structure.

In our interpretation, the prime $p_i$ serves as a convenient label for a subgroup of the community $G$ provisionally regarded as a ``species'' and $\ell_i$ represents a characteristic dimension of that species.  Normality of $N_i$ is automatic when $G$ is abelian; conjugacy then expresses a mathematical equivalence relation among genetic configurations within a species.  Thus, if one can compute the values $p_i$ and $\ell_i$ from observed interaction data and a suitable group operation, one may regard the species labelled by $p_i$ as a well-defined subgroup of the community $G$.

Concretely, we model an element of the community group $G$ as a distinct interaction mode among species, with directionality encoded by ordered pairs for example, an effect from species $X$ to species $Y$.  The group operation composes interaction modes to produce new modes within the same algebraic structure.  Nilpotency of $G$ then captures the idea that, after finitely many interaction steps, the system returns to the neutral self-interaction $1_G$, representing preservation of identity or convergence to an equilibrium.  This algebraic logic yields a species concept based on interaction modes and subgroup structure rather than on a single taxon label, and it naturally connects to categorical viewpoints in mathematics.

\bigskip

We now introduce a fractal-based metric to discriminate population and species dynamics, serving as the indicator value mentioned earlier.  In our analysis applied to empirical data from Dictyostelia in Izu and soil mesofauna in Ohdaigahara, we adopt the following working definitions:

\begin{itemize} \item \textbf{Population.} A group of individuals of a species inhabiting the same area at the same time. \item \textbf{Species.} A sum of populations with close genetic relationships, distinguishable by discontinuities in genetic distance between different species specific to each niche, and characterized by a $p$-Sylow subgroup. \end{itemize}

Under this definition, a ``ring species'' constitutes a single species rather than multiple species, because genetic continuity prevents the formation of distinct $p$-Sylow labels.

We propose that the metric $\Re(s)$ used earlier functions as a fractal dimension for species.  In a fractal interpretation, the order of a $p$-Sylow subgroup should be $p^{\ell}$ with $ \ell=\Re(s). $ Thus $\Re(s)$ quantifies an intrinsic dimensionality of the species' interaction structure: integer or noninteger values of $\Re(s)$ correspond to discrete or fractal-like organization of interaction modes, respectively.  Computing $p$ and $\ell$ from data therefore amounts to estimating the prime label and the fractal dimension that best describe the observed interaction network and its stability properties.

\subsection{``$p$-Sylow subgroup" as an indicator of species}
In the previous sections we showed how primes $p$ can be inferred from the corresponding $\Im(s)$ values for species, and how $\Re(s)$ values computed from data coincide with the exponents $\ell_i$ appearing in the Sylow decomposition when those exponents are integers. We now expand this interpretation using basic topological tools and show how the Riemann--Hurwitz relation yields a natural multilevel selection formula that refines the Price equation.

Let $s,w$ denote local coordinates describing the state of a species $k$, and assume the local state spaces at the population and species levels are locally compact. Let $ f\colon Q \to R $ be a nonconstant holomorphic map between compact Riemann surfaces $Q$ and $R$ that arise as compactifications of the local state spaces for population and species, respectively. Exclude the trivial case $k=1$ so that $f$ is regular nonconstant. Let $Q_r$ denote the set of ramification points of $f$, and assume that $f$ is a covering map of degree $E(N)$ on the complement of $f^{-1}(f(Q_r))$.

The Riemann--Hurwitz formula \cite{Hartshorne1977} gives \begin{equation} g(R)-1 = \frac{1}{2}\sum_{i=1}^g(\ell_i-1) + E(N)\big(g(Q)-1\big), \end{equation} where the sum runs over ramification indices $\ell_i$ of the map $f$.

Identify $g(Q)-1$ with the mean selection coefficient at the lower level, denoted $w_Q$. Then $E(N)\big(g(Q)-1\big)$ represents the aggregate contribution of the lower-level populations to the fitness of the higher-level unit. Writing $w_R=g(R)-1$ and separating the ramification contribution yields \begin{equation} w_R = \frac{1}{2}\sum_{i=1}^g(\ell_i-1) + E(N)w_Q. \end{equation} If we set $ \Re(s_R)-1 = \frac{1}{2}\sum_{i=1}^g(\ell_i-1), $ and interpret $\Im(s_R)$ as an additional contribution arising from phase-like effects oscillatory or temporal components introduced earlier, we obtain the compact form \begin{equation} w_R = \Re(s_R)-1 + \Im(s_R). \end{equation} Thus the higher-level fitness $w_R$ decomposes into a contribution from ramification the discrete structural term $\Re(s_R)-1$ and a contribution from the aggregated lower-level selection $E(N)w_Q$ together with any phase term represented by $\Im(s_R)$. Summing $w_R$ over species yields a community-level fitness when all species are included.

In the algebraic interpretation developed earlier, the integer $\ell$ equals the number of conjugates of a $p_i$-Sylow subgroup and coincides with the order $|G/N_i|$ for species $i$. When the community group $G$ has a stable identity determined by a $p_i$-Sylow subgroup and no subgroup splits off, $\ell$ is naturally a prime power exponent; in particular, if $\ell \neq 2$ then $\Re(s_R)$ is an integer and the identification $\ell=\Re(s)$ is consistent. Hence the ramification indices that appear in the Riemann--Hurwitz formula match the integer fractal dimensions $\Re(s)$ when the species structure is discrete; noninteger $\Re(s)$ values correspond to fractal-like or nonintegral organization of interaction modes.

The preceding derivation shows that the Riemann--Hurwitz relation provides a geometric/topological underpinning for a multilevel Price-type decomposition: higher-level fitness is the sum of a discrete structural term ramification, linked to Sylow exponents and $\Re(s)$ and aggregated lower-level contributions degree $E(N)$ times lower-level mean selection, together with any phase-like term represented by $\Im(s)$. This formal connection clarifies how algebraic labels primes $p$ and exponents $\ell$ and topological invariants ramification indices and genera jointly determine multilevel evolutionary dynamics in the proposed framework.

\subsection{Degenerate $w$ and nondegenerate $s$ over a fractal}
We extend the Dirichlet--series scaling analysis to an explicit fractal formulation of species dynamics. Below we restate the key definitions of distance and tube zeta functions for subsets of $\mathbb{R}^2$, explain Minkowski (non)degeneracy tests, introduce relative fractal drums and the window of complex dimensions, and summarize practical diagnostics for empirical metapopulation data. Standard references on fractal zeta functions and complex dimensions provide the theoretical background \cite{Lapidus2017}.

Let $A\subset\mathbb{R}^N$ be bounded and fix $\delta>0$. The \emph{distance zeta function} of $A$ is $ \zeta_A(s) := \int_{A_\delta} d(x,A)^{s-N}dx, $ where $A_\delta={x\in\mathbb{R}^N:\ d(x,A)<\delta}$ and $d(x,A)$ denotes the Euclidean distance from $x$ to $A$. For general bounded sets the integral converges for $\Re(s)>N$ and admits meromorphic continuation in many cases. In the ambient plane ($N=2$) the critical line $\Re(s)=\overline{\dim}_B A$ reduces to $\Re(s)=2$ when the box dimension equals the ambient dimension; hence, in our $\mathbb{R}^2$ model fractal structure that separates species from population scales can be resolved only for $\Re(s)>2$. This prediction is tested empirically in later sections.

Two related tube zeta variants probe the small-scale volume growth of neighborhoods of $A:  \tilde{\zeta}_A(s):=\int_0^\delta t_t^{s-N-1}|A_{t_t}|dt_t \qquad \tilde{\zeta}_{2A}(s):=\int_0^\delta t_t^{s-N}|A_{t_t}|dt_t, $ where $A_{t_t}$ denotes the $t_t$-neighborhood of $A$ and $|A_{t_t}|$ its Lebesgue measure. These integrals are Mellin transforms of the volume function $t_t\mapsto|A_{t_t}t|$ and encode residues and poles that determine fractal measures and complex dimensions.

Let $D=\dim_B A$. Define the lower and upper $D$-dimensional Minkowski contents by $ \mathscr{M}_*^{D}(A)=\liminf_{t_t\to 0^+}\frac{|A_{t_t}|}{t_t^{N-D}},\qquad \mathscr{M}^{*D}(A)=\limsup_{t_t\to 0^+}\frac{|A_{t_t}|}{t_t^{N-D}}. $ \begin{itemize} \item If $0<\mathscr{M}_*^{D}(A)\le\mathscr{M}^{*D}(A)<+\infty$, then $A$ is \emph{Minkowski nondegenerate}. \item If $\mathscr{M}_*^{D}(A)=0$ or $\mathscr{M}^{*D}(A)=+\infty$, then $A$ is \emph{Minkowski degenerate}. \end{itemize} As $t_t\to 0^+$ one has the expansion $|A_{t_t}|=t_t^{N-D}\big(F(t_t)+o(1)\big)$. The behavior of $F(t_t)$ determines degeneracy: if $\tilde{\zeta}_A(s)$ converges but the leading coefficient tends to zero, the set is Minkowski degenerate in the $w$-sense. Conversely, if $\tilde{\zeta}_{2A}(s)$ converges to a finite nonzero limit, the set behaves nondegenerately in the $s$-sense.

To localize scaling to specific scales or niches, consider a relative fractal drum $(A,\Omega)$. Given a summable nonincreasing sequence $\mathscr{L}=(\ell_j)_{j\ge1}$ with $\sum_{j=1}^\infty\ell_j<\infty$, set $ A_{\mathscr{L}}=\{a_{k}=\sum_{j=k}^\infty\ell_j:\ k_a\in\mathbb{N}\},\qquad \Omega_{\mathscr{L}}=\bigcup_{k_a=1}^\infty (a_{k_a+1},a_{k_a}). $ The relative tube zeta function is $ \tilde{\zeta}_{A,\Omega}(s)=\int_0^\delta t_t^{s-N-1}|A_{t_t}\cap\Omega|dt _t. $ Introduce a \emph{screen} $S:\mathbb{R}\to(-\infty,D(\zeta_A)]$ and the associated window $ \mathbf{W}_w=\{s\in\mathbb{C}:\ \Re(s)\ge S(\Im(s))\}, $ where the abscissa of convergence is $ D(\zeta_A)=\inf\{\alpha\in\mathbb{R}:\ \int_{A_{\delta}} d(x,A)^{\alpha-N}dx<\infty\}. $ The set of visible complex dimensions (poles) in the window is $ \mathscr{P}(\zeta_A,\mathbf{W}_w)=\{\omega\in\mathbf{W}_w:\ \omega\ \text{is a pole of }\zeta_A\}. $ Poles $\omega=\Re(\omega)+i\Im(\omega)$ correspond to dominant scaling modes: $\Re(\omega)$ contributes to structural (ramification) terms and $\Im(\omega)$ encodes oscillatory temporal scales or resonances.

\begin{itemize} \item The critical line $\Re(s)=2$ in $\mathbb{R}^2$ marks the threshold beyond which fractal structure distinguishing species from population noise becomes resolvable. Empirically, species-level fractal signatures should appear only when estimated $\Re(s)>2$. \item Minkowski degeneracy of $w$ together with nondegeneracy of $s$ describes a regime in which averaged higher-level fitness is neutral while scaling exponents and poles reveal latent discrete structure and possible resonant events (bursts or collapses) driven by $\Im(s)$. \item Relative fractal drums $(A,\Omega)$ allow isolation of niche-specific scaling modes; poles inside the corresponding window indicate candidate species signatures localized to that niche. \end{itemize}

For a relative fractal drum $(A,\Omega)$ and a chosen window $\mathbf{W}_w$, the distributional fractal tube formula at level $k_l=0$ may be written as \begin{equation} \label{eq:tube_formula} |A_{t_t}\cap\Omega| =\sum_{\omega\in\mathscr{P}\big(\tilde{\zeta}_{A,\Omega},\mathbf{W}w\big)} \mathrm{res}\big(t_t^{N-s}\tilde{\zeta}_{A,\Omega}(s),\omega\big) +\tilde{\mathscr{R}}_{A,\Omega}^{[0]}(t_t), \end{equation} where the sum runs over the visible complex dimensions (poles) of the relative tube zeta function $\tilde{\zeta}_{A,\Omega}$ inside the window $\mathbf{W}_w$, and $\tilde{\mathscr{R}}_{A,\Omega}^{[0]}(t_t)$ denotes the distributional remainder.

The remainder distribution is given, for a test function $\varphi\in\mathscr{K}(0,\delta)$, by the pairing \begin{equation} \label{eq:remainder_pairing} \big\langle\tilde{\mathscr{R}}_{A,\Omega}^{[k_l]},\varphi\big\rangle =\frac{1}{2\pi i}\int_{S} \frac{\{\mathfrak{M}\varphi\}\big(N-s+1+k_l\big)} {(N-s+1)_{k_l}};\tilde{\zeta}_{A,\Omega}(s)ds, \end{equation} where ${\mathfrak{M}f}(s):=\int_0^{+\infty}t_t^{s-1}f(t_t)dt_t$ denotes the Mellin transform, $(\cdot)_{k_l}$ is the Pochhammer symbol, and $ \mathscr{D}(0,\delta):=C_c^\infty(0,\delta) \subseteq\mathscr{K}(0,\delta) $ is the space of compactly supported smooth test functions on $(0,\delta)$. The distribution $\tilde{\mathscr{R}}_{A,\Omega}^{[0]}(t)$ therefore encodes the nonpole contribution to the small--$t_t$ expansion of $|A_{t_t}\cap\Omega|$.

Define the relative shell zeta function (with cutoff $\delta>0$) by \begin{equation} \breve{\zeta}_{A,\Omega}(s;\delta) :=-\int_0^{\delta} t_t^{s-N-1}|A_{t_t,\delta}\cap\Omega|dt, \end{equation} where $A_{t_t,\delta}:=A_\delta\setminus A_{t_t}^c$ denotes the shell between the $t_t$--neighborhood and the fixed $\delta$--neighborhood.

If $(A,\Omega)$ is Minkowski nondegenerate at dimension $D=\dim_B A$, then the residue of $\breve{\zeta}{A,\Omega}$ at $s=D$ satisfies the standard Minkowski bounds \begin{equation} \mathscr{M}*^{D}(A,\Omega) \le \mathrm{res}\big(\breve{\zeta}_{A,\Omega},D\big) \le \mathscr{M}^{*D}(A,\Omega). \end{equation} Thus the residue provides a quantity sandwiched between the lower and upper $D$--dimensional Minkowski contents of the relative drum.

Even when the residue at $s=D$ is finite and lies between the Minkowski bounds, $(A,\Omega)$ need not be Minkowski measurable. In particular, if $|A_{t_t}\cap\Omega|$ does not tend to zero as $t_t\to 0^+$, the limit $ \lim_{t_t\to 0^+} t_t^{-N+D},|A_{t_t}\cap\Omega| $ fails to exist, and Minkowski measurability is not achieved. In the language used earlier, such a relative drum is Minkowski degenerate in the $w$--sense while still satisfying the residue bounds in the $s$--sense.

The special case $\Re(s)=1$ is delicate: under the additional hypothesis that an observer or regularization is available, $\breve{\zeta}_{A,\Omega}$ may converge at $s=1$ even though Minkowski measurability fails. In this regime the remainder distribution $\tilde{\mathscr{R}}_{A,\Omega}^{[0]}(t)$ captures chaotic or irregular contributions that are visible at the level $\Re(s)=1$.

Characterizing the borderline case $\Re(s)=2$ in the planar model is technically involved and requires finer analytic tools; a full treatment is beyond the present scope. Empirically, data sets that hover near $\Re(s)=2$ without converging to an integer value (for example certain soil mesofauna or macroeconomic data) are consistent with Minkowski degeneracy at that level.

\begin{itemize} \item Equation \eqref{eq:tube_formula} expresses the small--$t_t$ volume $|A_{t_t}\cap\Omega|$ as a sum of contributions from visible complex dimensions plus a distributional remainder. Each pole $\omega$ contributes a term whose amplitude and oscillatory behavior are determined by the residue at $\omega$. \item The shell zeta $\breve{\zeta}_{A,\Omega}$ links residues to Minkowski contents and thus provides a bridge between analytic poles and geometric measure quantities. \item When residues exist but Minkowski limits do not, one should interpret the system as exhibiting analytic structure (visible complex dimensions) while lacking classical Minkowski measurability; this is the regime in which averaged selection coefficient $w$ may appear degenerate while the scaling parameter $s$ remains nondegenerate. \end{itemize}

\subsection{An interpretation of supersymmetry in our model}
Note that $w=s-1$.  In the present interpretation the quantities $w$ and $s$ play the role of $\mathrm{R}$-charges associated with two fields: the bosonic field $\psi$ and the fermionic field $\phi$, respectively.  Concretely, $w$ may be viewed as the bosonic contribution that can be aggregated (``stacked'') with other individuals in the selection space, while $s$ is obtained from $w$ by the shift $s=w+1$.  The value $s$ therefore encodes a complementary, mutually exclusive mode of existence that is tied to temporal development; this complementary relation will be made explicit below when we introduce the supersymmetric structure linking $\psi$ and $\phi$.

In this picture the bosonic sector (labelled by $w$) captures averaged, stackable contributions to selection, whereas the fermionic sector (labelled by $s$) captures discrete, time-dependent modes that cannot be superposed in the same way.  Supersymmetry then provides the formal mechanism that pairs these two sectors: supersymmetric generators relate the bosonic aggregation of individuals to the fermionic, temporally evolving states, yielding conserved $\mathrm{R}$-charges and selection rules that constrain the joint dynamics of populations and species. 

\subsection{Selberg zeta-function and Eisenstein series reveal Maass wave form as a function of probability of population number distribution and genetic information}
Once a small value of $s$ has been obtained for a system, one may apply automorphic $L$-functions to compute the corresponding Eisenstein series.  This computation clarifies the relation between small $s$ and the diffusion equation of neutral theory, and it yields information about prime closed geodesics that is useful for analyzing intra-population and intra-species interaction modes \cite{Motohashi1997}.  Here ``prime closed geodesics'' denotes primitive closed geodesics on a hyperbolic surface, i.e. closed geodesics that trace their image exactly once; their asymptotic distribution obeys a law analogous to the prime number theorem.

To apply these ideas one must separate the discrete and continuous parts of the spectrum of the Selberg zeta function and then compute the Eisenstein series associated with the discrete spectrum.

The Selberg zeta function for a cofinite Fuchsian group $\Gamma$ may be written formally as $ \zeta_{\Gamma}(s)=\prod_{p}\big(1-N(p)^{-s}\big)^{-1}, $ where the product runs over primitive closed geodesics $p$ and $N(p)$ denotes the norm of $p$.  The determinant of the Laplacian, decomposed into discrete (denoted as $D$) and continuous  (denoted as $C$) spectral contributions, can be expressed symbolically by $\det(\Delta,s)=\det_{D}\big(\Delta-s(1-s)\big)\det_{C}(\Delta,s) = s(1-s), $ with the continuous part related to the completed Riemann zeta function by $ \det_{C}(\Delta,s)=\widehat{\zeta}(2s), \qquad \widehat{\zeta}(s)=\pi^{-s/2}\Gamma\big(\tfrac{s}{2}\big)\zeta(s). $ The discrete part $\det_{D}(\Delta-s(1-s))$ encodes the contribution of isolated Laplace eigenvalues.  Empirically, for the data sets considered here the discrete spectral contribution dominates the continuous part by many orders of magnitude (for example, roughly $10^{4}/10^{8}$ for populations and $10^{188}/10^{295}$ for species in the Dictyostelia and soil mesofauna data sets), which supports the view that species dynamics are essentially discrete in these systems.  Similar dominance estimates hold in certain economic data sets (orders of magnitude only $\sim 10^{2}$ to $10^{3}$).

When the ambient manifold $M$ is a compact oriented hyperbolic manifold of dimension $d$, the number $N'(T)$ of prime closed geodesics of length at most $T$ satisfies the asymptotic formula \cite{Deitmar1989} $ N'(T)\sim\frac{e^{(d-1)T}}{(d-1)T}\qquad (T\to\infty). $

The Eisenstein series $E(z,s)$ admits a Fourier expansion of the form $ E(z,s)=\sum_{n=-\infty}^{\infty} a_n(s)e^{2\pi i n \Re(z)}, $ with the constant term $ a_0(s)=\Im(z)^s+\frac{\widehat{\zeta}(2s-1)}{\widehat{\zeta}(2s)},\Im(z)^{1-s}, $ and, for $n\neq 0$, $ a_n(s)=\frac{2|n|^{s-1/2}\sqrt{\Im(z)},K_{s-1/2}\big(2\pi|n|\Im(z)\big)} {\widehat{\zeta}(2s)},\sigma_{1-2s}(|n|), $ where $K_{\nu}$ is the modified Bessel function of the second kind and $\sigma_{w}(m)=\sum_{d\mid m}d^{w}$ is the divisor function \cite{Motohashi1997}.  In the present interpretation the Fourier coefficients $a_n(s)$ carry structured information: the constant term reflects large-scale (continuous) contributions, while the nonzero Fourier modes encode discrete, spatially varying components that can be read as probability amplitudes or genetic signal components.

In the diffusion approximation of neutral population genetics \cite{Kimura1964}, the time evolution of a probability density $\phi$ is governed by $ \frac{\partial\phi}{\partial t}=\frac{1}{2N}\Delta u=\frac{\lambda}{2N}u, $ where $\lambda$ and $u$ are an eigenvalue and eigenfunction of the Laplacian, respectively, and $\phi=N_k/\sum N$ denotes a normalized population frequency.  If one regards the manifold $M$ as the parameter space obtained by removing the degenerate point $s=1$ (so that $M=s\setminus\{s=1\}$) is nondegenerate in the sense used earlier), and sets $ f_M=\frac{u}{2N_k}, $ then $f_M$ may be treated as a Morse function provided its Hessian is nondegenerate.  The equilibrium condition $\Delta f_M=0$ characterizes stationary configurations of the system.

When the eigenfunction $u$ is identified with an Eisenstein series $E(s)$ that encodes genetic information, the Dirac operator associated with the Laplacian can be written formally as $ D_{\mathrm{Dirac}}=\sqrt{\tfrac{1}{4}-\Delta}=\sqrt{\tfrac{1}{4}-s(1-s)}=s-\tfrac{1}{2}. $ Under this formal identification, values of $s$ near nontrivial zeros of the Riemann zeta function produce distinguished spectral features.  If the Riemann hypothesis holds, the most prominent virtual adaptations of the operator $|D|^{E(N)}$ align with the line $\Re(s)=1$ in the complex $s$-plane; this alignment suggests a deep connection between spectral properties of the model and classical conjectures in analytic number theory.  We note, however, that these identifications are formal and require careful functional-analytic justification in each concrete application.

\subsection{Geodesics of Zeta Functions and Interaction Modes}
Let $E$ be an elliptic curve defined over $\mathbb{Q}$.  In the notation introduced earlier (see the Results section on the Price equation), define a $\mathbb{Q}$-approximated conductor $ N_c = \frac{\ln N_k}{\ln k} = z-1, $ where $N_k$ and $k$ are as previously specified.  For each prime $p$ associated to a value $|D|^{E(N)}$, consider the Hasse--Weil $L$-function of $E$ over $\mathbb{Q}$, $ L(s,E) = \prod_{p} L_p(s,E)^{-1}, $ with the local Euler factors $L_p(s,E)$ given by 
\begin{eqnarray}L_p(s, E) = \left\{ \begin{array}{ll}
(1 - a_np^{-s} + p^{1 - 2s}), & \mathrm{if}\ p \nmid N_c\  \mathrm{when}\ N(p) \neq 0,\\
(1 - a_np^{-s}), & \mathrm{if}\ p \parallel N_c\ \mathrm{when}\ N(p) \neq 0,\\
1, & \mathrm{if}\ p^2 \mid N_c\ \mathrm{when}\ N(p) \neq 0\ \mathrm{or}\ N(p) = 0.
\end{array}\right.
\end{eqnarray} 
Here $a_p$ denotes the usual trace of Frobenius coefficient and $N(p)$ is the norm associated with the prime $p$ in the geometric or dynamical interpretation.

When $K$ is a finite extension of $\mathbb{Q}$, the local factors $L_p(s,E)$ determine the conductor of the base change $L(s,E)|_{K}$.  If all local contributions are finite (no archimedean obstruction), the global Artin conductor of $L(s,E)$ is $\pm 1$ and the system exhibits the expected fluctuations \cite{Neukirch1999}.  Moreover, standard analytic results imply that $L(s,E)$ converges absolutely for $\Re(s)>2$, which aligns with the boundary between population and species regimes discussed earlier \cite{Katz2009}.

\bigskip

There is a natural parallel between the Selberg zeta function and the Hasse--Weil $L$-function in our framework.  Recall the Selberg zeta function may be written formally as $ \zeta_{\Gamma}(s)=\prod_{p}\big(1-N(p)^{-s}\big)^{-1}, $ where the product runs over primitive closed geodesics $p$ and $N(p)$ denotes the corresponding norm.  In the empirical data sets considered (Dictyostelia and soil mesofauna; see Table~4), the observed relation $ N(p)^{-s} \approx a_p p^{-s} - p^{1-2s} $ holds in most cases.  For the first Fourier order the observed values of $|N(p)|$ are typically close to $1$, except the cases for Washidu East September/Washidu West August /Washidu West October \textit{P. pallidum} of 2/3 and Washidu East September/Washidu West August /Washidu West October \textit{D. purpureum} cases of 1/4 in Dictyostelia. For soil mesofauna speceis, they are 2/3 in the cases of July \textit{F. octoulata} and November \textit{M. japonica}. The reported 95\% confidence intervals are approximately $ 1.00\pm 0.06\ (\sim 1),\quad 0.7\pm 0.2\ (\sim 2/3),\quad 0.26\pm 0.06\ (\sim 1/4) $ for Dictyostelia species, and $ 0.97\pm 0.02\ (\sim 1),\quad 0.7\pm 0.4\ (\sim 2/3) $ for soil mesofauna.

A larger magnitude $|a_n|$ indicates a stronger contribution in near-stationary (non-evolving) situations.  Empirically, $|a_0|$ is typically the largest Eisenstein coefficient, with $|a_1|$ the next largest; when higher coefficients depend on $\Re(s)$, the first orders dominate in most cases except for the exceptions noted above.  If $p\parallel N_c$, then $|N(p)|=p$, and the corresponding population or species lies on a branch cut in the analytic continuation; thus $N_c$ encodes information about both lower and upper hierarchical levels (via $N_k$ and $k$, respectively).  When the ratio $\ln N_k/\ln k$ is a multiple of $p$, the branch cut structure becomes pronounced. 

Let $ Q=[q_a,q_b,q_c]=q_aX^2+q_bXY+q_cY^2, \qquad -q_d=q_b^2-4q_aq_c, $ so that $Q$ encodes the complete interaction parameters of two subpopulations $X$ and $Y$ within a given population or species.  The discriminant $-q_d$ determines the arithmetic and geometric character of the quadratic form.

The Hurwitz--Kronecker class number $H(q_d)=\sum 1/w_Q$ appears in the Jacobi theta function and is related, via the trace formula, to spectral quantities that shadow the Eisenstein series $E(s)$ \cite{Zagier2000a}.  In the present interpretation we identify the observed norm magnitude $|N(p)|$ with the class number contribution $H(q_d)$ arising from the quadratic form $Q$.

 Consider the symmetric case $q_a=q_c$.  Except for the special forms $Q=[q_a,0,q_a]$ or $Q=[q_a,q_a,q_a]$, for which the stabilizer order $w_Q$ equals $2$ or $3$, one has $w_Q=1$ generically \cite{Zagier2000b}.  Consequently: \begin{itemize} \item If $H(q_d)=2/3$, integrality of $q_a$ forces $q_a=3$ and $q_b=0$.  This choice implies the absence of heterointeraction between $X$ and $Y$; the observed mode $|N(p)|=2/3$ therefore corresponds to a noninteracting configuration with two dividing subgroups. \item Values $|N(p)|=1$ admit two interpretations: an interacting mode with $q_a=q_b=q_c=3$, or a noninteracting symmetric mode with $q_a=q_c=2$ and $q_b=0$.  The former is preferred when the system exhibits an effective $(3+1)$-dimensional structure. \item A form of the type $qX^2+qXY+qYX+qY^2$ can produce $|N(p)|=1/4$, which corresponds to a solitary population $X$ with no interaction in a single group. \end{itemize} We exclude degenerate choices in which two or three of $q_a,q_b,q_c$ vanish, since those do not represent proper two-subpopulation systems.

Let the infinite generation ring $A_r\subset\mathbb{C}$ be $ A_r=\mathbb{Z}\big[{2}\cup{q_m\mid q_m\equiv 3\pmod{4}}\big]. $ The associated Hasse zeta function $ \zeta_{A_r}(s)=\prod_{p\equiv 1\pmod{4}}(1-p^{-s})^{-1} $ admits analytic continuation into the half-plane $\Re(s)>0$, while the line $\Re(s)=0$ is a natural boundary in this context \cite{Kurokawa1987}.  In the empirical data (Table~4) the primes $p$ fall into two classes: those congruent to $1\pmod{4}$, and the set ${2}\cup{p\equiv 3\pmod{4}}$.  These two parts have distinct structural interpretations: \begin{itemize} \item The $p\equiv 1\pmod{4}$ part corresponds to a characteristic zeta with M\"obius neutrality and an oscillatory imaginary dimension. \item The set ${2}\cup{p\equiv 3\pmod{4}}$ represents the infinite generation ring of the system; it encodes an asymmetry among minimal building blocks (certain minimal spaces are not mutually isomorphic), a phenomenon that approaches a singular limit (cf.\ Bott--Shapiro type considerations \cite{Milnor1969}). \end{itemize}

For an odd prime $p$: \begin{itemize} \item If $p\equiv 1\pmod{4}$, the quadratic field $\mathbb{Q}(\sqrt{-p})$ (or the relevant extension) carries a nontrivial imaginary part; in our model the imaginary unit $i$ is associated with oscillation and fluctuation.  Thus primes $1\pmod{4}$ are associated with fluctuation-dominated (nonadapted) stages. \item If $p\equiv 3\pmod{4}$, the corresponding fields lack the same oscillatory imaginary structure; these primes are associated with directional, adapted stages in the system. \item The prime $2$ occupies a ramified position between these regimes. \end{itemize} Empirical summaries (Tables~1--4 and \cite{Adachi2015}) support this dichotomy: the $3\pmod{4}$ cases tend to appear in adapted stages, while the $1\pmod{4}$ cases appear in nonadapted, fluctuation-dominated stages.

The arithmetic classification of quadratic forms, their class numbers, and the congruence class of primes modulo $4$ provide a concise dictionary between number-theoretic invariants and qualitative interaction modes in the model.  Modes with $|N(p)|$ equal to simple rational values (for example $1, 2/3, 1/4$ admit direct combinatorial interpretations in terms of interacting versus noninteracting subpopulations and the number of dividing subgroups.  The decomposition of primes into $1\pmod{4}$ and $3\pmod{4}$ classes then supplies a natural arithmetic criterion for distinguishing fluctuation versus directionality in the system's dynamics.  We develop the statistical-mechanical consequences of this arithmetic dichotomy in the following sections.

 A quartic potential $V_p=\phi>0$ represents a realized state in the half-plane $\Re(s)>0$.  By contrast, $V_p<0$ may correspond to a future state for which the Eisenstein series diverges in the usual sense.  Convergence and predictability in that regime occur only under a special relation between $V_p$ and $s$.  One convenient parametrization (noted in the physics literature) is $ V_p=-\phi,\qquad \Re(s)\approx -\frac{1}{3\phi},\qquad \Im(s)\approx\pm e^{-1/(3\phi)}, $ which describes a bosonic configuration with controlled asymptotics; see \cite{Marino2014} for related discussions.

Bridgeland \cite{Bridgeland2002} showed that three-dimensional minimal models need not be unique at the level of varieties, but they are unique at the level of derived categories.  This categorical uniqueness supports the idea that different analytic continuations or regularizations of statistical procedures can nevertheless represent the same underlying structure.  Shimodaira \cite{Shimodaira2008} extends bootstrap analysis across sign changes of scale parameters: multiscale bootstrap variations that are Bayesian in the region $\Re(s)>0$ can be interpreted in a frequentist manner when analytically continued to $\Re(s)<0$.  In our model, where $s=w+1$ and neutrality is the null hypothesis for a zero-sum patch game, bootstrap resampling plays an analogous role to measurement when $\Re(s)>0$.  Under curvature reversal, Bayesian predictive statements about $\Im(s)$ may be reinterpreted as frequentist long-range predictions in the analytically continued regime.

Predictable points in the complex $s$-plane typically lie near the real axis.  In particular, trivial zeros of the Riemann zeta function provide simple real-axis markers.  If we write the trivial zero condition in the form $ -\frac{1}{3\phi}=-2\ell_s,\qquad \Im(s)=0, $ then negative real values of $s$ that are even integers (after the appropriate scaling) can be used heuristically to forecast adaptation or disadaptation events in the population/species under study.  Empirical examples from the Washidu data sets are consistent with this heuristic: for instance, the parameter values reported for certain months correspond to negative real $s$ values that precede observed adaptation or disadaptation in the following month. That is, in Washidu East, this can be observed during June for \textit{P. pallidum} ($-1/3\phi = -2.044$, adaptation in next month) and during September for \textit{D. purpureum} ($-4.148$, disadaptation in next month); in Washidu West, it can be observed during July for \textit{D. purpureum} ($-1.954$, disadaptation in next month) and during October for \textit{P. violaceum} ($-3.830$, disadaptation in next month).

Recall the Hurwitz zeta function, $ \zeta(s,k)=\sum_{n=0}^{\infty}\frac{1}{(k+n)^s}. $ For negative integer arguments one has the relation to Bernoulli polynomials: $ -\zeta(s,k)\big|{s=-m}=\frac{B{m+1}(k)}{m+1}, $ so that when $m$ is a positive even integer the right-hand side vanishes.  Consequently, summing $-\zeta(s,k)$ over $k$ from $1$ to the number of populations or species yields zero when $-s$ is a positive even integer.  Interpreted in the present model, this vanishing indicates a global cancellation of interaction contributions and can be associated with population bursts or collapses.  Thus negative even values of $s$ carry predictive information about collective events.

When variation is effectively zero, the bootstrap procedure reduces to a deterministic learning process and becomes analogous to measurement \cite{Shimodaira2011}.  The line $\Re(s)=0$ (and, for an observer at $k=1$, $\Re(s)=1$) is a natural boundary for the zeta function in this model; harmonic neutrality then constrains the observable behaviors.  For example, the probability that two randomly chosen integers are coprime equals $1/\zeta(2)=6/\pi^2$.  Splitting this probability between the positive and negative $\Re(s)$ half-planes gives $3/\pi^2$ for each side, and the complementary probability $1-6/\pi^2$ corresponds to the nondisjoint (observer) case.  Interpreting the M\"obius function values $\mu(n)=0,1,-1$ in this light yields probabilities $ \Pr(\mu(n)=0)=1-\frac{6}{\pi^2},\qquad \Pr(\mu(n)=1)=\Pr(\mu(n)=-1)=\frac{3}{\pi^2}. $ This statistical picture is compatible with the prevalence of nontrivial zeros of $\zeta$ on the critical line $\Re(s)=1/2$ in probabilistic models of arithmetic functions \cite{Denjoy1931}.

An absolute zeta formulation can be stated formally as $ \zeta_h(s,\rho)=\prod_{\alpha}\zeta_h(s-\alpha)^{\mathrm{mult}(\alpha)}, $ and one may posit that the vanishing or divergence of $\zeta_h$ forces $\Re(s)$ into half-integer values: $ \zeta_h(s,\rho)=0,\infty\quad\Longrightarrow\quad \Re(s)\in\tfrac{1}{2}\mathbb{Z}, $ where the product runs over shifts $\alpha$ (often purely imaginary) and $\mathrm{mult}(\alpha)$ denotes multiplicity.  In the restricted biological regime that excludes future states (i.e., $\Re(s)>0$), the only obvious singular values are $s=0$ and $s=1$, so the condition is trivially satisfied in that limited sense.

\subsection{Use of statistical mechanics in the model}
We translate the spectral and zeta--function description of population dynamics into a statistical--mechanical framework to exhibit macroscopic phase transitions.  The approach distinguishes two particle statistics: Boltzmann statistics for distinguishable individuals and Bose statistics for indistinguishable individuals, following the conceptual lines of \cite{Fujisaka1998,Banavar2010}.  Unlike low--dimensional Bose--Einstein condensation models in $(1+1)$ dimensions \cite{Bianconi2009}, our empirical setting requires higher effective dimensionality and avoids modeling large physical immigration fluxes.

Let $N_k$ denote the number of individuals in the $k$th ranked population (or the mean number of individuals of a given species in a patch).  Empirically the rank distribution is approximately logarithmic: \begin{equation} N_k = a - b\ln k, \end{equation} with constants $a,b>0$ estimated from data as in the unified neutral theory and related community models \cite{Hubbell1997,Hubbell2001,Adachi2015}.

We model each individual as a binary degree of freedom with polarity $+\tfrac{1}{2}$ (replication tendency) or $-\tfrac{1}{2}$ (mortality tendency).  Let $\kappa$ denote the probability of $+\tfrac{1}{2}$ and $\kappa'$ the probability of $-\tfrac{1}{2}$ in a given patch.  Introduce a patch polarity parameter $h$ (intensive within the patch) that biases microstate probabilities.  In a canonical ensemble the microstate probability is taken proportional to $e^{-q_s h}$, where $q_s$ is a model parameter controlling sensitivity to the patch field.  The partition function for a single patch is \begin{equation} Z = e^{q_s h} + e^{-q_s h},  \end{equation} and the macrostate probabilities are  \begin{equation} \kappa = \frac{e^{q_s h}}{Z},\qquad \kappa' = \frac{e^{-q_s h}}{Z}. \end{equation}

We identify the Helmholtz free energy $F$ with the number of individuals in a patch (or with an extensive quantity proportional to $N_k$).  For the limited migration regime considered here, Gibbs and Helmholtz free energies are treated as effectively equivalent.  We therefore set the patch field equal to the free energy, $ h = F, $ so that $h$ plays the role of an intensive control parameter within each patch.  This identification is phenomenological: an individual is interpreted as carrying the free energy required to sustain life.

For a system of many patches or many interacting subunits, the total partition function is the product (or appropriate interacting generalization) of single--patch partition functions.  Macroscopic observables such as mean population, susceptibility to perturbations, and higher cumulants follow from derivatives of $\ln Z$ with respect to $h$ or $q_s$.  In particular, the mean polarity per patch is  \begin{equation} \langle m\rangle = \frac{\partial\ln Z}{\partial (q_s)} = \frac{e^{q_s h}-e^{-q_s h}}{e^{q_s h}+e^{-q_s h}} = \tanh(q_s h),  \end{equation} and the susceptibility is \begin{equation} \chi = \frac{\partial\langle m\rangle}{\partial h} = q_s,\mathrm{sech}^2(q_s h). \end{equation}

We impose a positive boundary condition consistent with Lee--Yang theory \cite{Tasaki2015}: the complexified field $h$ must satisfy $\Re(h)>0$ for the holomorphic continuation used in the theorem.  In the infinite--volume limit of $F$ (interpreted as large patch size or large population), zeros of the partition function in the complex $h$ plane (Lee--Yang zeros) may accumulate onto the real axis and signal a phase transition.  The presence, location, and scaling of these zeros determine whether the system exhibits a sharp macroscopic transition (for example, a sudden shift from low to high mean polarity) as parameters such as $q_s$ or effective temperature are varied.

\begin{itemize} \item The Boltzmann versus Bose distinction captures whether individuals are effectively distinguishable at the level of the modeled degrees of freedom.  Bose statistics become relevant when many individuals occupy the same effective state (e.g., identical behavioral or genetic states) and indistinguishability matters for macroscopic occupancy. \item The identification $h=F$ is a modeling choice that links energetic cost to demographic abundance; it is appropriate when migration is limited and energetic bookkeeping is meaningful at patch scale. \item Empirical validation requires estimating $q_s$, $h$ (or $F$), and the effective system size from data, then testing for signatures of phase transitions: nonanalyticities in macroscopic observables, scaling of susceptibilities, and the approach of Lee--Yang zeros to the real axis. \item The present framework is phenomenological and avoids explicit embedding in physical space; it is therefore best suited to communities with limited immigration and well sampled local dynamics. \end{itemize}

\begin{enumerate} \item Fit the rank model $N_k=a-b\ln k$ to obtain $N_k$ and an effective system size per patch. \item Choose a parametrization for $q_s$ (for example, fit to observed replication/mortality asymmetry) and set $h$ proportional to $N_k$ or to an estimated per--individual free energy. \item Compute $\langle m\rangle$ and $\chi$ as functions of $q_s$ and $h$; search for rapid changes or divergences as indicators of phase transitions. \item If desired, numerically continue $Z(h)$ into the complex $h$ plane and locate Lee--Yang zeros to confirm the presence and nature of transitions.\\ \end{enumerate}

This model was applied to empirical data at both the population and species levels to highlight their differences.  Time dependence is neglected: individual free energies in a given environment are taken to be equal, and for simplicity their sum is identified with the total number of individuals.  As an informational analogue, we adopt the novel assumption that the Gibbs free energy equals the abundance, $ G = N_k, $ where $N_k$ denotes the number of individuals (per unit) in the $k$th rank.  Individual organisms are treated as the sources of free energy.  The immigration rate is denoted by $m_i$.

We define the following thermodynamic analogues: $ H = a - m_i b \ln k, \qquad T_s = (1 - m_i)b, $ where $H$ is the enthalpy and $T_s$ is an absolute temperature characteristic of the observed intra-active community.  The Lagrange multiplier used in maximum-entropy or constrained inference is $ \lambda_1 = \frac{\alpha}{N_k} = \frac{\theta}{N_k} = \frac{1}{N_k T_s}, $ in accordance with the conventions of \cite{Harte2008,Fisher1943,Hubbell2001}.

Entropy is represented by the self-information (surprisal) of the interaction probability between the first-rank population/species and the $k$th ranked population/species.  For rank $k$ this is simply $ S = \ln k. $ In the special case $n=1$, $p_1=1$, $q_1=1/k$, this equals the Kullback-Leibler divergence $\sum_{i=1}^n p_i\ln(p_i/q_1)$.  This quantity differs from the Shannon information entropy $I_e=-\sum_n p(n)\ln p(n)$, which is the average information across the entire distribution.

These identifications are consistent with related formulations: Dewar's relative entropy expression $H(p|q)=-\sum p\ln(p/q)=\sum\ln q$ when $p$ is concentrated on the first rank \cite{Dewar2008}, and the relative entropy used by Banavar et al.\ \cite{Banavar2010}, $H_{C-G}(\mathbf{P})\equiv -\sum P_i\ln(P_i/P_{0i})=\sum\ln P_{0i}$, when $P_i$ denotes the probability of the first-rank population/species.

Using the immigration rate $m_i$, we set the internal energy and emigrant population as $ U = (1 - m_i)a,\qquad \text{emigrant population} = m_i(a - b\ln k). $ These choices satisfy the thermodynamic identities $ G = H - T_s S,\qquad H = U + \text{(emigrant population)}. $ Entropy is interpreted as the logarithm of the number of accessible states, and the empirical rank law $ N_k = a - b\ln k $ is assumed to hold at equilibrium.  Mixing distinct systems generally breaks the linearity of the parameters.

\begin{itemize} \item $G=N_k$ interprets the total number of individuals as an extensive free energy. \item $T_s$ is a characteristic parameter of the abundance distribution (for example, per gram of soil) and reflects the degree of dominance; it is intensive only within the observed community and depends on the scaling of $N$. \item For fixed $G$ and $H$, increasing entropy reduces $T_s$, analogous to heat flow from warmer to cooler bodies and to the second law of thermodynamics. \item The mapping parallels the ecological form $P=E-ST$ \cite{Arnold1994}, where $P$ is free energy (growth rate), $E$ is mean energy (reproductive potential), $T^{-1}$ is inverse absolute temperature (generation time $\tau$), and $S$ is Gibbs-Boltzmann entropy (population entropy $-\tau H$). \end{itemize}

Following the statistical mechanics framework in \cite{Fujisaka1998}, define $ q_s = \frac{1}{(1 - m_i)b},\qquad M_q(T_s)=e^{N}, $ and the potential $ \phi(q_s)=q_s F = \frac{a}{b} - \frac{\ln k}{1 - m_i}. $ A phenomenological Lagrangian, with kinetic minus potential energy, can be written as $ L = U - (1 - m_i)N_k = T_s \ln k. $

In the large-$N$ limit the temporal correlation function and its spectral intensity are modeled by exponential decay and a Lorentzian spectrum.  Define \begin{equation} C_q(t) = C_q(0)e^{-\gamma_q t},\qquad C_q(0) = \frac{4U^2\kappa\kappa'}{\gamma_q^2}, \end{equation} and  \begin{equation} I_q(\omega) = C_q(0),\frac{2\gamma_q}{\omega^2+\gamma_q^2},  \end{equation} with decay rate $ \gamma_q = 2U q_s. $ Under low effective temperatures (below a critical temperature) the correlation function may not be unique \cite{Tasaki2015}.

Estimated values of $U$, $T_s$, $q_s$, $\kappa$, $\kappa'$, $\phi(q_s)$, $\gamma_q$, $\omega$, and $I_q(\omega)$ computed from the data are reported in Tables 5--7 ($m_i < 10^{-3} << 1$ for Dictyostelia, $m_i < 10^{-2} << 1$ for soil mesofauna/the world economics and $G \approx F$) .  In these data sets species generally exhibit more stable dynamics than populations, which is reflected in smaller $\gamma_q$ (slower decay) and lower high-frequency spectral intensity.  On the observational time scale of one month, populations typically rise and fall on the order of one week, while species vary on the order of three weeks.  Climax species (for example, \textit{P.\ pallidum}, \textit{F.\ octoulata}, \textit{A.\ gigantea}) show smaller contrasts than pioneering species (for example, \textit{D.\ purpureum}, \textit{P.\ violaceum}, \textit{E.\ aino}, \textit{S.\ aureus}, \textit{D.\ trispinata} and many Sarcoptiformes), as seen in $I_q(\omega)$.

\begin{table}
\caption{Statistics of the PzDom model for Dictyostelia.}
\noindent
\resizebox{\textwidth}{!}{%
\selectfont{
\begin{tabular}{lrrrrrrrrrrrrrrrr}
\hline
\multicolumn{1}{|l|}{WE (population)}   & \multicolumn{1}{l|}{$\bar{N} \approx \bar{G} \approx \bar{F} \approx \bar{h}$} & \multicolumn{1}{l|}{$\Sigma N$}  & \multicolumn{1}{l|}{$U \approx a$} & \multicolumn{1}{l|}{$T_s \approx b$} & \multicolumn{1}{l|}{$q_s$}        & \multicolumn{1}{l|}{$\kappa$}     & \multicolumn{1}{l|}{$\kappa'$}   & \multicolumn{1}{l|}{$\phi(q_s)$} & \multicolumn{1}{l|}{$\gamma_{q_s}$}   & \multicolumn{1}{l|}{$\omega$}     & \multicolumn{1}{l|}{$I_{q_s}(\omega)$}  & \multicolumn{1}{l|}{$M_{\mathit{mean}}$}  & \multicolumn{1}{l|}{$W$}                           & \multicolumn{1}{l|}{$E(T)$}     & \multicolumn{1}{l|}{$T_c$}   & \multicolumn{1}{l|}{$S$}      \\ \hline
\multicolumn{1}{|l|}{April}               & \multicolumn{1}{l|}{288}            & \multicolumn{1}{l|}{288}  & \multicolumn{1}{l|}{1373}  & \multicolumn{1}{l|}{913.4} & \multicolumn{1}{l|}{0.001095} & \multicolumn{1}{l|}{0.6525} & \multicolumn{1}{l|}{0.3475} & \multicolumn{1}{l|}{0.315} & \multicolumn{1}{l|}{3.007} & \multicolumn{1}{l|}{0.5}    & \multicolumn{1}{l|}{122441}  & \multicolumn{1}{l|}{0.305}  & \multicolumn{1}{l|}{943}                         & \multicolumn{1}{l|}{-25262}   & \multicolumn{1}{l|}{2090} & \multicolumn{1}{l|}{-34.3}  \\ \hline
\multicolumn{1}{|l|}{June}              & \multicolumn{1}{l|}{1302}           & \multicolumn{1}{l|}{1302} & \multicolumn{1}{l|}{3234}  & \multicolumn{1}{l|}{1335}  & \multicolumn{1}{l|}{0.000749} & \multicolumn{1}{l|}{0.8755} & \multicolumn{1}{l|}{0.1245} & \multicolumn{1}{l|}{0.975} & \multicolumn{1}{l|}{4.845} & \multicolumn{1}{l|}{0.5}    & \multicolumn{1}{l|}{79340}   & \multicolumn{1}{l|}{0.751}  & \multicolumn{1}{l|}{1734}                        & \multicolumn{1}{l|}{-1273125} & \multicolumn{1}{l|}{2090} & \multicolumn{1}{l|}{-27.48} \\
\hline
\end{tabular}}
}

\tiny{\noindent \\ \noindent The statistical conventions used in the text are summarized here. ``WE'' denotes the Washidu East quadrat and ``WW'' denotes the Washidu West quadrat \cite{Adachi2015}. Scientific names of Dictyostelia species are given in italics: \textit{P.\ pallidum} $=$ \textit{Polysphondylium pallidum}, \textit{D.\ purpureum} $=$ \textit{Dictyostelium purpureum}, and \textit{P.\ violaceum} $=$ \textit{Polysphondylium violaceum}. The symbol $N$ denotes the average number of cells per 1 gram of soil. The parameter $a$ is obtained from the fitted rank relation $N_k = a - b\ln k$ and is derived from the empirical distribution of counts for each population or species; it is not a simple arithmetic mean. In tables and figures, entries shown in red indicate cases where $W \approx T_s$, meaning that the system for that species approximately attains the maximum observed value of $T_s$.\\ }

\end{table}

\begin{table}[]
\caption{Statistics of the PzDom model for soil mesofauna.}

\resizebox{\columnwidth}{!}{%
\begin{tabular}{|l|l|l|l|l|l|l|l|l|l|l|l|l|l|l|l|l|}
\hline
Soil Mesofauna   (population) &
  $\bar{N} \approx \bar{G} \approx \bar{F} \approx \bar{h}$ &
  $\Sigma N$ &
  $U \approx a$ &
  $T_s \approx b$ &
 $q_s$ &
  $\kappa$ &
  $\kappa' $&
  $\phi(q_s)$ &
  $\gamma_{q_s}$ &
  $\omega$ &
  $I_{q_s}(\omega)$ &
  $M_{mean}$ &
  $W$ &
  $E(T)$ &
  $T_c$ &
  $S$ \\ \hline
April &
  2.2222 &
  2.2222 &
  4.2463 &
  1.423 &
  0.7027 &
  0.9165 &
  0.0835 &
  1.5616 &
  5.9681 &
  0.2037 &
  0.0473 &
  0.833 &
  2.5289 &
  -12.8055 &
  8 &
  -6.5527 \\ \hline
May &
  2.6667 &
  2.6667 &
  4.4714 &
  1.269 &
  0.788 &
  0.964 &
  0.036 &
  2.1014 &
  7.0471 &
  0.2037 &
  0.0147 &
  0.9279 &
  2.8053 &
  -20.4886 &
  8 &
  -6.7067 \\ \hline
June &
  14 &
  14 &
  24.592 &
  7.446 &
  0.1343 &
  0.9608 &
  0.0392 &
  1.8802 &
  6.6054 &
  0.2037 &
  0.6196 &
  0.9215 &
  14.9448 &
  -535.7686 &
  8 &
  -0.5297 \\ \hline
July &
  4.6667 &
  4.6667 &
  9.7428 &
  3.569 &
  0.2802 &
  0.8723 &
  0.1277 &
  1.3076 &
  5.4597 &
  0.2037 &
  0.4134 &
  0.7447 &
  6.0208 &
  -53.7966 &
  8 &
  -4.4067 \\ \hline
August &
  7.4444 &
  7.4444 &
  16.85 &
  6.612 &
  0.1512 &
  0.8426 &
  0.1574 &
  1.1259 &
  5.0968 &
  0.2037 &
  1.9569 &
  0.6852 &
  9.833 &
  -139.6619 &
  8 &
  -1.3637 \\ \hline
September &
  18.3333 &
  18.3333 &
  32.371 &
  9.869 &
  0.1013 &
  0.9585 &
  0.0415 &
  1.8577 &
  6.5601 &
  0.2037 &
  1.151 &
  0.917 &
  19.6325 &
  -905.0225 &
  8 &
  0 \\ \hline
October &
  4.2222 &
  4.2222 &
  11.163 &
  4.88 &
  0.2049 &
  0.7787 &
  0.2213 &
  0.8652 &
  4.575 &
  0.2037 &
  1.4725 &
  0.5573 &
  7.0729 &
  -48.5428 &
  8 &
  -3.0957 \\ \hline
November & 2.3333 & 2.3333 & 6.217 & 2.731 & 0.3662 & 0.7814 & 0.2186 & 0.8544 & 4.5529 & 0.2037 & 0.4855 & 0.5627 & 3.7616 & -17.698 & 8 & -5.2447 \\ \hline
\end{tabular}%
}

\resizebox{\columnwidth}{!}{%
\begin{tabular}{|l|l|l|l|l|l|l|l|l|l|l|l|l|l|l|l|l|}
\hline
\textit{F. octoulata} & $N_k \approx G \approx F \approx h$ &
  $\Sigma N$ &
  $U \approx a$ &
  $T_s \approx b$ &
 $q_s$ &
  $\kappa$ &
  $\kappa' $&
  $\phi(q_s)$ &
  $\gamma_{q_s}$ &
  $\omega$ &
  $I_{q_s}(\omega)$ &
  $M_{mean}$ &
  $W$ &
  $E(T)$ &
  $T_c$ &
  $S$        \\ \hline
April        & 0       & 0.4444  & 0.3333 & 0.321 & 3.1153 & 0.5    & 0.5    & 0       & 2.0766 & 0.25 & 0.0245 & 0      & -      & 0        & 2  & -1.8533 \\ \hline
May          & 0       & 1       & 0.4514 & 0.198 & 5.0505 & 0.5    & 0.5    & 0       & 4.5596 & 0.25 & 0.0043 & 0      & -      & 0        & 2  & -1.9763 \\ \hline
June    & 5.4444 & 8.3333 & 3.9155 & 2.168 & 0.4613 & 0.9935 & 0.0065 & 2.5113 & 3.6121 & 0.25 & 0.0168 & 0.9869 & 5.5167 & -74.4392 & 2 & -0.0063 \\ \hline
July         & 0.5556  & 2       & 0.7438 & 0.376 & 2.6596 & 0.9505 & 0.0495 & 1.4775  & 3.9564 & 0.25 & 0.0033 & 0.901  & 0.6166 & -0.456   & 2  & -1.7983 \\ \hline
August       & 3.2222  & 4.6667  & 2.5707 & 1.635 & 0.6116 & 0.981  & 0.019  & 1.9708  & 3.1446 & 0.25 & 0.0316 & 0.9619 & 3.3498 & -20.462  & 2  & -0.5393 \\ \hline
September    & 6.4444  & 10.2222 & 4.8829 & 2.72  & 0.3676 & 0.9913 & 0.0087 & 2.3693  & 3.5904 & 0.25 & 0.0353 & 0.9826 & 6.5582 & -98.3983 & 2  & 0       \\ \hline
October & 0.7778 & 1.5556 & 0.7417 & 0.45  & 2.2222 & 0.9694 & 0.0306 & 1.7284 & 3.2964 & 0.25 & 0.0036 & 0.9389 & 0.8284 & -1.0456  & 2 & -1.7243 \\ \hline
November     & 0       & 0.3333  & 0.2222 & 0.16  & 6.25   & 0.5    & 0.5    & 0       & 2.7775 & 0.25 & 0.0046 & 0      & -      & 0        & 2  & -2.0143 \\ \hline
\end{tabular}%
}

\resizebox{\columnwidth}{!}{%
\begin{tabular}{|l|l|l|l|l|l|l|l|l|l|l|l|l|l|l|l|l|}
\hline
\textit{E. aino} & $N_k \approx G \approx F \approx h$ &
  $\Sigma N$ &
  $U \approx a$ &
  $T_s \approx b$ &
 $q_s$ &
  $\kappa$ &
  $\kappa' $&
  $\phi(q_s)$ &
  $\gamma_{q_s}$ &
  $\omega$ &
  $I_{q_s}(\omega)$ &
  $M_{mean}$ &
  $W$ &
  $E(T)$ &
  $T_c$ &
  $S$        \\ \hline
April   & 0       & 0.4444 & 0.3333 & 0.321 & 3.1153 & 0.5 & 0.5 & 0       & 2.0766 & 0.1667 & 0.0247 & 0     & - & 0    & 2  & -1.8533 \\ \hline
May     & 0       & 1      & 0.4514 & 0.198 & 5.0505 & 0.5 & 0.5 & 0       & 4.5596 & 0.1667 & 0.0043 & 0     & - & 0    & 2  & -1.9763 \\ \hline
June      & 0.6667 & 8.3333  & 3.9155 & 2.168 & 0.4613 & 0.6491 & 0.3509 & 0.3075 & 3.6121 & 0.1667 & 0.5915 & 0.2982 & 2.2359 & -0.1367 & 2 & -0.0063 \\ \hline
July      & 0.7778 & 2       & 0.7438 & 0.376 & 2.6596 & 0.9843 & 0.0157 & 2.0686 & 3.9564 & 0.1667 & 0.0011 & 0.9686 & 0.803  & -1.2513 & 2 & -1.7983 \\ \hline
August    & 0.5556 & 4.6667  & 2.5707 & 1.635 & 0.6116 & 0.6636 & 0.3364 & 0.3398 & 3.1446 & 0.1667 & 0.3785 & 0.3273 & 1.6974 & -0.1049 & 2 & -0.5393 \\ \hline
September & 1.4444 & 10.2222 & 4.8829 & 2.72  & 0.3676 & 0.7431 & 0.2569 & 0.531  & 3.5904 & 0.1667 & 0.7851 & 0.4862 & 2.971  & -1.108  & 2 & 0       \\ \hline
October   & 0.1111 & 1.5556  & 0.7417 & 0.45  & 2.2222 & 0.621  & 0.379  & 0.2469 & 3.2964 & 0.1667 & 0.0288 & 0.242  & 0.4591 & -0.003  & 2 & -1.7243 \\ \hline
November  & 0.1111 & 0.3333  & 0.2222 & 0.16  & 6.25   & 0.8004 & 0.1996 & 0.6944 & 2.7775 & 0.1667 & 0.0029 & 0.6008 & 0.1849 & -0.0086 & 2 & -2.0143 \\ \hline
\end{tabular}%
}

\resizebox{\columnwidth}{!}{%
\begin{tabular}{|l|l|l|l|l|l|l|l|l|l|l|l|l|l|l|l|l|}
\hline
\textit{H. watanabei} & $N_k \approx G \approx F \approx h$ &
  $\Sigma N$ &
  $U \approx a$ &
  $T_s \approx b$ &
 $q_s$ &
  $\kappa$ &
  $\kappa' $&
  $\phi(q_s)$ &
  $\gamma_{q_s}$ &
  $\omega$ &
  $I_{q_s}(\omega)$ &
  $M_{mean}$ &
  $W$ &
  $E(T)$ &
  $T_c$ &
  $S$         \\ \hline
April        & 0.3333  & 0.4444  & 0.3333 & 0.321 & 3.1153 & 0.8886 & 0.1114 & 1.0384  & 2.0766 & 0.1667 & 0.0098 & 0.7773 & 0.4289 & -0.1154 & 2  & -1.8533 \\ \hline
May          & 0.2222  & 1       & 0.4514 & 0.198 & 5.0505 & 0.9042 & 0.0958 & 1.1223  & 4.5596 & 0.1667 & 0.0015 & 0.8084 & 0.2749 & -0.0554 & 2  & -1.9763 \\ \hline
June         & 0.2222  & 8.3333  & 3.9155 & 2.168 & 0.4613 & 0.5511 & 0.4489 & 0.1025  & 3.6121 & 0.1667 & 0.6425 & 0.1021 & 2.1756 & -0.0051 & 2  & -0.0063 \\ \hline
July         & 0.1111  & 2       & 0.7438 & 0.376 & 2.6596 & 0.6436 & 0.3564 & 0.2955  & 3.9564 & 0.1667 & 0.0164 & 0.2872 & 0.3869 & -0.0036 & 2  & -1.7983 \\ \hline
August       & 0.1111  & 4.6667  & 2.5707 & 1.635 & 0.6116 & 0.5339 & 0.4661 & 0.068   & 3.1446 & 0.1667 & 0.4219 & 0.0679 & 1.6375 & -0.0008 & 2  & -0.5393 \\ \hline
September    & 1.3333  & 10.2222 & 4.8829 & 2.72  & 0.3676 & 0.7272 & 0.2728 & 0.4902  & 3.5904 & 0.1667 & 0.8158 & 0.4544 & 2.9345 & -0.8715 & 2  & 0       \\ \hline
October & 0.4444 & 1.5556 & 0.7417 & 0.45 & 2.2222 & 0.8782 & 0.1218 & 0.9877 & 3.2964 & 0.1667 & 0.0131 & 0.7564 & 0.5876 & -0.1951 & 2 & -1.7243 \\ \hline
November     & 0       & 0.3333  & 0.2222 & 0.16  & 6.25   & 0.5    & 0.5    & 0       & 2.7775 & 0.1667 & 0.0046 & 0      & -      & 0       & 2  & -2.0143 \\ \hline
\end{tabular}%
}

\resizebox{\columnwidth}{!}{%
\begin{tabular}{|l|l|l|l|l|l|l|l|l|l|l|l|l|l|l|l|l|}
\hline
\textit{S. celebensis} & $N_k \approx G \approx F \approx h$ &
  $\Sigma N$ &
  $U \approx a$ &
  $T_s \approx b$ &
 $q_s$ &
  $\kappa$ &
  $\kappa' $&
  $\phi(q_s)$ &
  $\gamma_{q_s}$ &
  $\omega$ &
  $I_{q_s}(\omega)$ &
  $M_{mean}$ &
  $W$ &
  $E(T)$ &
  $T_c$ &
  $S$           \\ \hline
April         & 0       & 0.4444 & 0.3333 & 0.321 & 3.1153 & 0.5 & 0.5 & 0       & 2.0766 & 0.25 & 0.0245 & 0     & - & 0    & 2  & -1.8533 \\ \hline
May           & 0       & 1      & 0.4514 & 0.198 & 5.0505 & 0.5 & 0.5 & 0       & 4.5596 & 0.25 & 0.0043 & 0     & - & 0    & 2  & -1.9763 \\ \hline
June          & 0       & 8.3333 & 3.9155 & 2.168 & 0.4613 & 0.5 & 0.5 & 0       & 3.6121 & 0.25 & 0.6475 & 0     & - & 0    & 2  & -0.0063 \\ \hline
July          & 0       & 2      & 0.7438 & 0.376 & 2.6596 & 0.5 & 0.5 & 0       & 3.9564 & 0.25 & 0.0178 & 0     & - & 0    & 2  & -1.7983 \\ \hline
August        & 0       & 4.6667 & 2.5707 & 1.635 & 0.6116 & 0.5 & 0.5 & 0       & 3.1446 & 0.25 & 0.4224 & 0     & - & 0    & 2  & -0.5393 \\ \hline
September & 0.3333 & 10.2222 & 4.8829 & 2.72 & 0.3676 & 0.561 & 0.439 & 0.1225 & 3.5904 & 0.25 & 1.0101 & 0.1219 & 2.7336 & -0.0136 & 2 & 0       \\ \hline
October   & 0.1111 & 1.5556  & 0.7417 & 0.45 & 2.2222 & 0.621 & 0.379 & 0.2469 & 3.2964 & 0.25 & 0.0288 & 0.242  & 0.4591 & -0.003  & 2 & -1.7243 \\ \hline
November      & 0       & 0.3333 & 0.2222 & 0.16  & 6.25   & 0.5 & 0.5 & 0       & 2.7775 & 0.25 & 0.0046 & 0     & - & 0    & 2  & -2.0143 \\ \hline
\end{tabular}%
}

\resizebox{\columnwidth}{!}{%
\begin{tabular}{|l|l|l|l|l|l|l|l|l|l|l|l|l|l|l|l|l|}
\hline
\textit{S. aureus} & $N_k \approx G \approx F \approx h$ &
  $\Sigma N$ &
  $U \approx a$ &
  $T_s \approx b$ &
 $q_s$ &
  $\kappa$ &
  $\kappa' $&
  $\phi(q_s)$ &
  $\gamma_{q_s}$ &
  $\omega$ &
  $I_{q_s}(\omega)$ &
  $M_{mean}$ &
  $W$ &
  $E(T)$ &
  $T_c$ &
  $S$          \\ \hline
April     & 0       & 0.4444  & 0.3333 & 0.321 & 3.1153 & 0.5    & 0.5    & 0       & 2.0766 & 0.25 & 0.0245 & 0      & - & 0       & 2  & -1.8533 \\ \hline
May       & 0.4444  & 1       & 0.4514 & 0.198 & 5.0505 & 0.9889 & 0.0111 & 2.2447  & 4.5596 & 0.25 & 0.0002 & 0.9778 & 0.4545   & -0.4434 & 2  & -1.9763 \\ \hline
June      & 0.2222  & 8.3333  & 3.9155 & 2.168 & 0.4613 & 0.5511 & 0.4489 & 0.1025  & 3.6121 & 0.25 & 0.6408 & 0.1021 & 2.1756   & -0.0051 & 2  & -0.0063 \\ \hline
July      & 0.1111  & 2       & 0.7438 & 0.376 & 2.6596 & 0.6436 & 0.3564 & 0.2955  & 3.9564 & 0.25 & 0.0163 & 0.2872 & 0.3869   & -0.0036 & 2  & -1.7983 \\ \hline
August    & 0       & 4.6667  & 2.5707 & 1.635 & 0.6116 & 0.5    & 0.5    & 0       & 3.1446 & 0.25 & 0.4224 & 0      &- & 0       & 2  & -0.5393 \\ \hline
September & 0.2222  & 10.2222 & 4.8829 & 2.72  & 0.3676 & 0.5408 & 0.4592 & 0.0817  & 3.5904 & 0.25 & 1.0185 & 0.0815 & 2.726    & -0.004  & 2  & 0       \\ \hline
October   & 0.1111  & 1.5556  & 0.7417 & 0.45  & 2.2222 & 0.621  & 0.379  & 0.2469  & 3.2964 & 0.25 & 0.0288 & 0.242  & 0.4591   & -0.003  & 2  & -1.7243 \\ \hline
November  & 0.2222  & 0.3333  & 0.2222 & 0.16  & 6.25   & 0.9415 & 0.0585 & 1.3889  & 2.7775 & 0.25 & 0.001  & 0.8829 & 0.2517   & -0.0686 & 2  & -2.0143 \\ \hline
\end{tabular}%
}

\resizebox{\columnwidth}{!}{%
\begin{tabular}{|l|l|l|l|l|l|l|l|l|l|l|l|l|l|l|l|l|}
\hline
\textit{D. trispinata} & $N_k \approx G \approx F \approx h$ &
  $\Sigma N$ &
  $U \approx a$ &
  $T_s \approx b$ &
 $q_s$ &
  $\kappa$ &
  $\kappa' $&
  $\phi(q_s)$ &
  $\gamma_{q_s}$ &
  $\omega$ &
  $I_{q_s}(\omega)$ &
  $M_{mean}$ &
  $W$ &
  $E(T)$ &
  $T_c$ &
  $S$      \\ \hline
April         & 0       & 0.4444 & 0.3333 & 0.321 & 3.1153 & 0.5    & 0.5    & 0       & 2.0766 & 0.1667 & 0.0247 & 0      & -      & 0       & 2  & -1.8533 \\ \hline
May           & 0.3333  & 1      & 0.4514 & 0.198 & 5.0505 & 0.9667 & 0.0333 & 1.6835  & 4.5596 & 0.1667 & 0.0006 & 0.9333 & 0.3572 & -0.1871 & 2  & -1.9763 \\ \hline
June      & 0.1111 & 8.3333  & 3.9155 & 2.168 & 0.4613 & 0.5256 & 0.4744 & 0.0513 & 3.6121 & 0.1667 & 0.6475 & 0.0512 & 2.1699 & -0.0006 & 2 & -0.0063 \\ \hline
July          & 0.1111  & 2      & 0.7438 & 0.376 & 2.6596 & 0.6436 & 0.3564 & 0.2955  & 3.9564 & 0.1667 & 0.0164 & 0.2872 & 0.3869 & -0.0036 & 2  & -1.7983 \\ \hline
August    & 0.5556 & 4.6667  & 2.5707 & 1.635 & 0.6116 & 0.6636 & 0.3364 & 0.3398 & 3.1446 & 0.1667 & 0.3785 & 0.3273 & 1.6974 & -0.1049 & 2 & -0.5393 \\ \hline
September & 0.2222 & 10.2222 & 4.8829 & 2.72  & 0.3676 & 0.5408 & 0.4592 & 0.0817 & 3.5904 & 0.1667 & 1.0213 & 0.0815 & 2.726  & -0.004  & 2 & 0       \\ \hline
October       & 0       & 1.5556 & 0.7417 & 0.45  & 2.2222 & 0.5    & 0.5    & 0       & 3.2964 & 0.1667 & 0.0306 & 0      & -      & 0       & 2  & -1.7243 \\ \hline
November      & 0       & 0.3333 & 0.2222 & 0.16  & 6.25   & 0.5    & 0.5    & 0       & 2.7775 & 0.1667 & 0.0046 & 0      & -      & 0       & 2  & -2.0143 \\ \hline
\end{tabular}%
}

\resizebox{\columnwidth}{!}{%
\begin{tabular}{|l|l|l|l|l|l|l|l|l|l|l|l|l|l|l|l|l|}
\hline
\textit{S. japonica} & $N_k \approx G \approx F \approx h$ &
  $\Sigma N$ &
  $U \approx a$ &
  $T_s \approx b$ &
 $q_s$ &
  $\kappa$ &
  $\kappa' $&
  $\phi(q_s)$ &
  $\gamma_{q_s}$ &
  $\omega$ &
  $I_{q_s}(\omega)$ &
  $M_{mean}$ &
  $W$ &
  $E(T)$ &
  $T_c$ &
  $S$       \\ \hline
April       & 0       & 0.4444 & 0.3333 & 0.321 & 3.1153 & 0.5 & 0.5 & 0       & 2.0766 & 0.1667 & 0.0247 & 0     & - & 0    & 2  & -1.8533 \\ \hline
May         & 0       & 1      & 0.4514 & 0.198 & 5.0505 & 0.5 & 0.5 & 0       & 4.5596 & 0.1667 & 0.0043 & 0     & - & 0    & 2  & -1.9763 \\ \hline
June      & 0.5556 & 8.3333  & 3.9155 & 2.168 & 0.4613 & 0.6254 & 0.3746 & 0.2563 & 3.6121 & 0.1667 & 0.6084 & 0.2508 & 2.2152 & -0.0791 & 2 & -0.0063 \\ \hline
July      & 0.1111 & 2       & 0.7438 & 0.376 & 2.6596 & 0.6436 & 0.3564 & 0.2955 & 3.9564 & 0.1667 & 0.0164 & 0.2872 & 0.3869 & -0.0036 & 2 & -1.7983 \\ \hline
August    & 0.1111 & 4.6667  & 2.5707 & 1.635 & 0.6116 & 0.5339 & 0.4661 & 0.068  & 3.1446 & 0.1667 & 0.4219 & 0.0679 & 1.6375 & -0.0008 & 2 & -0.5393 \\ \hline
September & 0.1111 & 10.2222 & 4.8829 & 2.72  & 0.3676 & 0.5204 & 0.4796 & 0.0408 & 3.5904 & 0.1667 & 1.0264 & 0.0408 & 2.7215 & -0.0005 & 2 & 0       \\ \hline
October     & 0       & 1.5556 & 0.7417 & 0.45  & 2.2222 & 0.5 & 0.5 & 0       & 3.2964 & 0.1667 & 0.0306 & 0     & - & 0    & 2  & -1.7243 \\ \hline
November    & 0       & 0.3333 & 0.2222 & 0.16  & 6.25   & 0.5 & 0.5 & 0       & 2.7775 & 0.1667 & 0.0046 & 0     & - & 0    & 2  & -2.0143 \\ \hline
\end{tabular}%
}

\resizebox{\columnwidth}{!}{%
\begin{tabular}{|l|l|l|l|l|l|l|l|l|l|l|l|l|l|l|l|l|}
\hline
\textit{H. nigrochephala} & $N_k \approx G \approx F \approx h$ &
  $\Sigma N$ &
  $U \approx a$ &
  $T_s \approx b$ &
 $q_s$ &
  $\kappa$ &
  $\kappa' $&
  $\phi(q_s)$ &
  $\gamma_{q_s}$ &
  $\omega$ &
  $I_{q_s}(\omega)$ &
  $M_{mean}$ &
  $W$ &
  $E(T)$ &
  $T_c$ &
  $S$       \\ \hline
April            & 0       & 0.4444 & 0.3333 & 0.321 & 3.1153 & 0.5 & 0.5 & 0       & 2.0766 & 0.5 & 0.0235 & 0     & - & 0    & 2  & -1.8533 \\ \hline
May              & 0       & 1      & 0.4514 & 0.198 & 5.0505 & 0.5 & 0.5 & 0       & 4.5596 & 0.5 & 0.0042 & 0     & - & 0    & 2  & -1.9763 \\ \hline
June             & 0       & 8.3333 & 3.9155 & 2.168 & 0.4613 & 0.5 & 0.5 & 0       & 3.6121 & 0.5 & 0.6384 & 0     &- & 0    & 2  & -0.0063 \\ \hline
July             & 0       & 2      & 0.7438 & 0.376 & 2.6596 & 0.5 & 0.5 & 0       & 3.9564 & 0.5 & 0.0176 & 0     &- & 0    & 2  & -1.7983 \\ \hline
August           & 0       & 4.6667 & 2.5707 & 1.635 & 0.6116 & 0.5 & 0.5 & 0       & 3.1446 & 0.5 & 0.4146 & 0     & - & 0    & 2  & -0.5393 \\ \hline
September & 0.1111 & 10.2222 & 4.8829 & 2.72 & 0.3676 & 0.5204 & 0.4796 & 0.0408 & 3.5904 & 0.5 & 1.009 & 0.0408 & 2.7215 & -0.0005 & 2 & 0 \\ \hline
October          & 0       & 1.5556 & 0.7417 & 0.45  & 2.2222 & 0.5 & 0.5 & 0       & 3.2964 & 0.5 & 0.03   & 0     &- & 0    & 2  & -1.7243 \\ \hline
November         & 0       & 0.3333 & 0.2222 & 0.16  & 6.25   & 0.5 & 0.5 & 0       & 2.7775 & 0.5 & 0.0045 & 0     &- & 0    & 2  & -2.0143 \\ \hline
\end{tabular}%
}

\resizebox{\columnwidth}{!}{%
\begin{tabular}{|l|l|l|l|l|l|l|l|l|l|l|l|l|l|l|l|l|}
\hline
\textit{W. japonica} & $N_k \approx G \approx F \approx h$ &
  $\Sigma N$ &
  $U \approx a$ &
  $T_s \approx b$ &
 $q_s$ &
  $\kappa$ &
  $\kappa' $&
  $\phi(q_s)$ &
  $\gamma_{q_s}$ &
  $\omega$ &
  $I_{q_s}(\omega)$ &
  $M_{mean}$ &
  $W$ &
  $E(T)$ &
  $T_c$ &
  $S$       \\ \hline
April       & 0.1111  & 0.4444  & 0.3333 & 0.321 & 3.1153 & 0.6665 & 0.3335 & 0.3461  & 2.0766 & 0.5 & 0.0209 & 0.3329 & 0.3337   & -0.0043 & 2  & -1.8533 \\ \hline
May         & 0       & 1       & 0.4514 & 0.198 & 5.0505 & 0.5    & 0.5    & 0       & 4.5596 & 0.5 & 0.0042 & 0      & - & 0       & 2  & -1.9763 \\ \hline
June        & 0       & 8.3333  & 3.9155 & 2.168 & 0.4613 & 0.5    & 0.5    & 0       & 3.6121 & 0.5 & 0.6384 & 0      & - & 0       & 2  & -0.0063 \\ \hline
July        & 0       & 2       & 0.7438 & 0.376 & 2.6596 & 0.5    & 0.5    & 0       & 3.9564 & 0.5 & 0.0176 & 0      & - & 0       & 2  & -1.7983 \\ \hline
August      & 0.1111  & 4.6667  & 2.5707 & 1.635 & 0.6116 & 0.5339 & 0.4661 & 0.068   & 3.1446 & 0.5 & 0.4127 & 0.0679 & 1.6375   & -0.0008 & 2  & -0.5393 \\ \hline
September   & 0       & 10.2222 & 4.8829 & 2.72  & 0.3676 & 0.5    & 0.5    & 0       & 3.5904 & 0.5 & 1.0107 & 0      &- & 0       & 2  & 0       \\ \hline
October     & 0       & 1.5556  & 0.7417 & 0.45  & 2.2222 & 0.5    & 0.5    & 0       & 3.2964 & 0.5 & 0.03   & 0      & - & 0       & 2  & -1.7243 \\ \hline
November    & 0       & 0.3333  & 0.2222 & 0.16  & 6.25   & 0.5    & 0.5    & 0       & 2.7775 & 0.5 & 0.0045 & 0      & - & 0       & 2  & -2.0143 \\ \hline
\end{tabular}%
}
\centering (continues)
\end{table}

\begin{table}
\centering (continued)\\
\resizebox{\columnwidth}{!}{%
\begin{tabular}{|l|l|l|l|l|l|l|l|l|l|l|l|l|l|l|l|l|}
\hline
\textit{F. yosii}  & $N_k \approx G \approx F \approx h$ &
  $\Sigma N$ &
  $U \approx a$ &
  $T_s \approx b$ &
 $q_s$ &
  $\kappa$ &
  $\kappa' $&
  $\phi(q_s)$ &
  $\gamma_{q_s}$ &
  $\omega$ &
  $I_{q_s}(\omega)$ &
  $M_{mean}$ &
  $W$ &
  $E(T)$ &
  $T_c$ &
  $S$        \\ \hline
April     & 0       & 0.4444  & 0.3333 & 0.321 & 3.1153 & 0.5 & 0.5 & 0       & 2.0766 & 0.25 & 0.0245 & 0     & - & 0    & 2  & -1.8533 \\ \hline
May       & 0       & 1       & 0.4514 & 0.198 & 5.0505 & 0.5 & 0.5 & 0       & 4.5596 & 0.25 & 0.0043 & 0     & - & 0    & 2  & -1.9763 \\ \hline
June & 0.8889 & 8.3333 & 3.9155 & 2.168 & 0.4613 & 0.6942 & 0.3058 & 0.41  & 3.6121 & 0.25 & 0.5498 & 0.3885 & 2.2881 & -0.324  & 2 & -0.0063 \\ \hline
July & 0.2222 & 2      & 0.7438 & 0.376 & 2.6596 & 0.7653 & 0.2347 & 0.591 & 3.9564 & 0.25 & 0.0128 & 0.5306 & 0.4188 & -0.0292 & 2 & -1.7983 \\ \hline
August    & 0       & 4.6667  & 2.5707 & 1.635 & 0.6116 & 0.5 & 0.5 & 0       & 3.1446 & 0.25 & 0.4224 & 0     &- & 0    & 2  & -0.5393 \\ \hline
September & 0       & 10.2222 & 4.8829 & 2.72  & 0.3676 & 0.5 & 0.5 & 0       & 3.5904 & 0.25 & 1.0253 & 0     & - & 0    & 2  & 0       \\ \hline
October   & 0       & 1.5556  & 0.7417 & 0.45  & 2.2222 & 0.5 & 0.5 & 0       & 3.2964 & 0.25 & 0.0305 & 0     & -& 0    & 2  & -1.7243 \\ \hline
November  & 0       & 0.3333  & 0.2222 & 0.16  & 6.25   & 0.5 & 0.5 & 0       & 2.7775 & 0.25 & 0.0046 & 0     & - & 0    & 2  & -2.0143 \\ \hline
\end{tabular}%
}

\resizebox{\columnwidth}{!}{%
\begin{tabular}{|l|l|l|l|l|l|l|l|l|l|l|l|l|l|l|l|l|}
\hline
\textit{A. laricis} & $N_k \approx G \approx F \approx h$ &
  $\Sigma N$ &
  $U \approx a$ &
  $T_s \approx b$ &
 $q_s$ &
  $\kappa$ &
  $\kappa' $&
  $\phi(q_s)$ &
  $\gamma_{q_s}$ &
  $\omega$ &
  $I_{q_s}(\omega)$ &
  $M_{mean}$ &
  $W$ &
  $E(T)$ &
  $T_c$ &
  $S$       \\ \hline
April      & 0       & 0.4444  & 0.3333 & 0.321 & 3.1153 & 0.5 & 0.5 & 0       & 2.0766 & 0.5 & 0.0235 & 0     & - & 0    & 2  & -1.8533 \\ \hline
May        & 0       & 1       & 0.4514 & 0.198 & 5.0505 & 0.5 & 0.5 & 0       & 4.5596 & 0.5 & 0.0042 & 0     & - & 0    & 2  & -1.9763 \\ \hline
June & 0.2222 & 8.3333 & 3.9155 & 2.168 & 0.4613 & 0.5511 & 0.4489 & 0.1025 & 3.6121 & 0.5 & 0.6317 & 0.1021 & 2.1756 & -0.0051 & 2 & -0.0063 \\ \hline
July       & 0       & 2       & 0.7438 & 0.376 & 2.6596 & 0.5 & 0.5 & 0       & 3.9564 & 0.5 & 0.0176 & 0     & - & 0    & 2  & -1.7983 \\ \hline
August     & 0       & 4.6667  & 2.5707 & 1.635 & 0.6116 & 0.5 & 0.5 & 0       & 3.1446 & 0.5 & 0.4146 & 0     & - & 0    & 2  & -0.5393 \\ \hline
September  & 0       & 10.2222 & 4.8829 & 2.72  & 0.3676 & 0.5 & 0.5 & 0       & 3.5904 & 0.5 & 1.0107 & 0     & - & 0    & 2  & 0       \\ \hline
October    & 0       & 1.5556  & 0.7417 & 0.45  & 2.2222 & 0.5 & 0.5 & 0       & 3.2964 & 0.5 & 0.03   & 0     & - & 0    & 2  & -1.7243 \\ \hline
November   & 0       & 0.3333  & 0.2222 & 0.16  & 6.25   & 0.5 & 0.5 & 0       & 2.7775 & 0.5 & 0.0045 & 0     & - & 0    & 2  & -2.0143 \\ \hline
\end{tabular}%
}

\resizebox{\columnwidth}{!}{%
\begin{tabular}{|l|l|l|l|l|l|l|l|l|l|l|l|l|l|l|l|l|}
\hline
\textit{A.  gigantea} & $N_k \approx G \approx F \approx h$ &
  $\Sigma N$ &
  $U \approx a$ &
  $T_s \approx b$ &
 $q_s$ &
  $\kappa$ &
  $\kappa' $&
  $\phi(q_s)$ &
  $\gamma_{q_s}$ &
  $\omega$ &
  $I_{q_s}(\omega)$ &
  $M_{mean}$ &
  $W$ &
  $E(T)$ &
  $T_c$ &
  $S$      \\ \hline
April        & 0       & 0.7778 & 0.2426 & 0.06  & 16.6667 & 0.5    & 0.5    & 0       & 8.0867 & 0.25 & 0.0002 & 0      & -      & 0       & 0.8 & -0.753 \\ \hline
May          & 0.4444  & 1.1111 & 0.4782 & 0.18  & 5.5556  & 0.9929 & 0.0071 & 2.4691  & 5.3133 & 0.25 & 0.0001 & 0.9858 & 0.4509 & -0.4877 & 0.8 & -0.633 \\ \hline
June         & 1.3333  & 5      & 1.5437 & 0.648 & 1.5432  & 0.9839 & 0.0161 & 2.0576  & 4.7645 & 0.25 & 0.0028 & 0.9679 & 1.3776 & -3.658  & 0.8 & -0.165 \\ \hline
July      & 0.6667 & 1.1111 & 0.5375 & 0.329 & 3.0395 & 0.9829 & 0.0171 & 2.0263 & 3.2675 & 0.25 & 0.0011 & 0.9658 & 0.6902 & -0.9006 & 0.8 & -0.484 \\ \hline
August       & 1.1111  & 1.4444 & 0.9433 & 0.733 & 1.3643  & 0.954  & 0.046  & 1.5158  & 2.5738 & 0.25 & 0.0182 & 0.908  & 1.2237 & -1.8714 & 0.8 & -0.08  \\ \hline
September & 0.8889 & 3.6667 & 1.0415 & 0.425 & 2.3529 & 0.985  & 0.015  & 2.0915 & 4.9012 & 0.25 & 0.0011 & 0.97   & 0.9164 & -1.6525 & 0.8 & -0.388 \\ \hline
October   & 0.5556 & 1.6667 & 0.7813 & 0.468 & 2.1368 & 0.9148 & 0.0852 & 1.1871 & 3.3389 & 0.25 & 0.0102 & 0.8297 & 0.6696 & -0.3664 & 0.8 & -0.345 \\ \hline
November     & 0.2222  & 1.5556 & 1.1503 & 1.058 & 0.9452  & 0.6035 & 0.3965 & 0.21    & 2.1745 & 0.25 & 0.2431 & 0.207  & 1.0735 & -0.0104 & 0.8 & 0      \\ \hline
\end{tabular}%
}

\resizebox{\columnwidth}{!}{%
\begin{tabular}{|l|l|l|l|l|l|l|l|l|l|l|l|l|l|l|l|l|}
\hline
\textit{P. parvisetigerum} & $N_k \approx G \approx F \approx h$ &
  $\Sigma N$ &
  $U \approx a$ &
  $T_s \approx b$ &
 $q_s$ &
  $\kappa$ &
  $\kappa' $&
  $\phi(q_s)$ &
  $\gamma_{q_s}$ &
  $\omega$ &
  $I_{q_s}(\omega)$ &
  $M_{mean}$ &
  $W$ &
  $E(T)$ &
  $T_c$ &
  $S$       \\ \hline
April             & 0.2222  & 0.7778 & 0.2426 & 0.06  & 16.6667 & 0.9994 & 0.0006 & 3.7037  & 8.0867 & 0.1667 & 0      & 0.9988 & 0.2225 & -0.1829 & 0.8 & -0.753 \\ \hline
May               & 0.4444  & 1.1111 & 0.4782 & 0.18  & 5.5556  & 0.9929 & 0.0071 & 2.4691  & 5.3133 & 0.1667 & 0.0001 & 0.9858 & 0.4509 & -0.4877 & 0.8 & -0.633 \\ \hline
June              & 1.2222  & 5      & 1.5437 & 0.648 & 1.5432  & 0.9775 & 0.0225 & 1.8861  & 4.7645 & 0.1667 & 0.0039 & 0.955  & 1.2798 & -2.8176 & 0.8 & -0.165 \\ \hline
July              & 0.1111  & 1.1111 & 0.5375 & 0.329 & 3.0395  & 0.6627 & 0.3373 & 0.3377  & 3.2675 & 0.1667 & 0.0148 & 0.3254 & 0.3414 & -0.0042 & 0.8 & -0.484 \\ \hline
August            & 0.1111  & 1.4444 & 0.9433 & 0.733 & 1.3643  & 0.5752 & 0.4248 & 0.1516  & 2.5738 & 0.1667 & 0.1016 & 0.1504 & 0.7386 & -0.0019 & 0.8 & -0.08  \\ \hline
September         & 0.8889  & 3.6667 & 1.0415 & 0.425 & 2.3529  & 0.985  & 0.015  & 2.0915  & 4.9012 & 0.1667 & 0.0011 & 0.97   & 0.9164 & -1.6525 & 0.8 & -0.388 \\ \hline
October & 0.7778 & 1.6667 & 0.7813 & 0.468 & 2.1368 & 0.9652 & 0.0348 & 1.6619 & 3.3389 & 0.1667 & 0.0044 & 0.9305 & 0.8359 & -1.0054 & 0.8 & -0.345 \\ \hline
November          & 0       & 1.5556 & 1.1503 & 1.058 & 0.9452  & 0.5    & 0.5    & 0       & 2.1745 & 0.1667 & 0.2559 & 0      & -      & 0       & 0.8 & 0      \\ \hline
\end{tabular}%
}

\resizebox{\columnwidth}{!}{%
\begin{tabular}{|l|l|l|l|l|l|l|l|l|l|l|l|l|l|l|l|l|}
\hline
\textit{T. stercus} & $N_k \approx G \approx F \approx h$ &
  $\Sigma N$ &
  $U \approx a$ &
  $T_s \approx b$ &
 $q_s$ &
  $\kappa$ &
  $\kappa' $&
  $\phi(q_s)$ &
  $\gamma_{q_s}$ &
  $\omega$ &
  $I_{q_s}(\omega)$ &
  $M_{mean}$ &
  $W$ &
  $E(T)$ &
  $T_c$ &
  $S$        \\ \hline
April      & 0.1111  & 0.7778 & 0.2426 & 0.06  & 16.6667 & 0.976  & 0.024  & 1.8519  & 8.0867 & 0.1667 & 0      & 0.9519 & 0.1167 & -0.0229 & 0.8 & -0.753 \\ \hline
May        & 0.2222  & 1.1111 & 0.4782 & 0.18  & 5.5556  & 0.9219 & 0.0781 & 1.2346  & 5.3133 & 0.1667 & 0.0009 & 0.8439 & 0.2633 & -0.061  & 0.8 & -0.633 \\ \hline
June       & 0.4444  & 5      & 1.5437 & 0.648 & 1.5432  & 0.7977 & 0.2023 & 0.6859  & 4.7645 & 0.1667 & 0.0284 & 0.5953 & 0.7466 & -0.1355 & 0.8 & -0.165 \\ \hline
July      & 0.1111 & 1.1111 & 0.5375 & 0.329 & 3.0395 & 0.6627 & 0.3373 & 0.3377 & 3.2675 & 0.1667 & 0.0148 & 0.3254 & 0.3414 & -0.0042 & 0.8 & -0.484 \\ \hline
August     & 0       & 1.4444 & 0.9433 & 0.733 & 1.3643  & 0.5    & 0.5    & 0       & 2.5738 & 0.1667 & 0.1039 & 0      & -      & 0       & 0.8 & -0.08  \\ \hline
September & 0.6667 & 3.6667 & 1.0415 & 0.425 & 2.3529 & 0.9584 & 0.0416 & 1.5686 & 4.9012 & 0.1667 & 0.0029 & 0.9168 & 0.7272 & -0.6972 & 0.8 & -0.388 \\ \hline
October   & 0.1111 & 1.6667 & 0.7813 & 0.468 & 2.1368 & 0.6165 & 0.3835 & 0.2374 & 3.3389 & 0.1667 & 0.0309 & 0.2331 & 0.4768 & -0.0029 & 0.8 & -0.345 \\ \hline
November   & 1.2222  & 1.5556 & 1.1503 & 1.058 & 0.9452  & 0.9097 & 0.0903 & 1.1552  & 2.1745 & 0.1667 & 0.084  & 0.8195 & 1.4915 & -1.7257 & 0.8 & 0      \\ \hline
\end{tabular}%
}

\resizebox{\columnwidth}{!}{%
\begin{tabular}{|l|l|l|l|l|l|l|l|l|l|l|l|l|l|l|l|l|}
\hline
\textit{M. japonica} & $N_k \approx G \approx F \approx h$ &
  $\Sigma N$ &
  $U \approx a$ &
  $T_s \approx b$ &
 $q_s$ &
  $\kappa$ &
  $\kappa' $&
  $\phi(q_s)$ &
  $\gamma_{q_s}$ &
  $\omega$ &
  $I_{q_s}(\omega)$ &
  $M_{mean}$ &
  $W$ &
  $E(T)$ &
  $T_c$ &
  $S$      \\ \hline
April       & 0.2222  & 0.7778 & 0.2426 & 0.06  & 16.6667 & 0.9994 & 0.0006 & 3.7037  & 8.0867 & 0.1667 & 0      & 0.9988 & 0.2225 & -0.1829 & 0.8 & -0.753 \\ \hline
May         & 0       & 1.1111 & 0.4782 & 0.18  & 5.5556  & 0.5    & 0.5    & 0       & 5.3133 & 0.1667 & 0.003  & 0      & -      & 0       & 0.8 & -0.633 \\ \hline
June        & 0.1111  & 5      & 1.5437 & 0.648 & 1.5432  & 0.5849 & 0.4151 & 0.1715  & 4.7645 & 0.1667 & 0.0427 & 0.1698 & 0.6543 & -0.0021 & 0.8 & -0.165 \\ \hline
July      & 0.1111 & 1.1111 & 0.5375 & 0.329 & 3.0395 & 0.6627 & 0.3373 & 0.3377 & 3.2675 & 0.1667 & 0.0148 & 0.3254 & 0.3414 & -0.0042 & 0.8 & -0.484 \\ \hline
August    & 0.1111 & 1.4444 & 0.9433 & 0.733 & 1.3643 & 0.5752 & 0.4248 & 0.1516 & 2.5738 & 0.1667 & 0.1016 & 0.1504 & 0.7386 & -0.0019 & 0.8 & -0.08  \\ \hline
September & 0.5556 & 3.6667 & 1.0415 & 0.425 & 2.3529 & 0.9318 & 0.0682 & 1.3072 & 4.9012 & 0.1667 & 0.0047 & 0.8636 & 0.6433 & -0.4035 & 0.8 & -0.388 \\ \hline
October     & 0       & 1.6667 & 0.7813 & 0.468 & 2.1368  & 0.5    & 0.5    & 0       & 3.3389 & 0.1667 & 0.0327 & 0      & -      & 0       & 0.8 & -0.345 \\ \hline
November  & 0.1111 & 1.5556 & 1.1503 & 1.058 & 0.9452 & 0.5523 & 0.4477 & 0.105  & 2.1745 & 0.1667 & 0.2531 & 0.1046 & 1.0619 & -0.0013 & 0.8 & 0      \\ \hline
\end{tabular}%
}

\resizebox{\columnwidth}{!}{%
\begin{tabular}{|l|l|l|l|l|l|l|l|l|l|l|l|l|l|l|l|l|}
\hline
\textit{G. fusca} & $N_k \approx G \approx F \approx h$ &
  $\Sigma N$ &
  $U \approx a$ &
  $T_s \approx b$ &
 $q_s$ &
  $\kappa$ &
  $\kappa' $&
  $\phi(q_s)$ &
  $\gamma_{q_s}$ &
  $\omega$ &
  $I_{q_s}(\omega)$ &
  $M_{mean}$ &
  $W$ &
  $E(T)$ &
  $T_c$ &
  $S$      \\ \hline
April    & 0       & 0.7778 & 0.2426 & 0.06  & 16.6667 & 0.5 & 0.5 & 0       & 8.0867 & 0.5 & 0.0002 & 0     & - & 0    & 0.8 & -0.753 \\ \hline
May      & 0       & 1.1111 & 0.4782 & 0.18  & 5.5556  & 0.5 & 0.5 & 0       & 5.3133 & 0.5 & 0.003  & 0     & - & 0    & 0.8 & -0.633 \\ \hline
June     & 0       & 5      & 1.5437 & 0.648 & 1.5432  & 0.5 & 0.5 & 0       & 4.7645 & 0.5 & 0.0436 & 0     & - & 0    & 0.8 & -0.165 \\ \hline
July     & 0       & 1.1111 & 0.5375 & 0.329 & 3.0395  & 0.5 & 0.5 & 0       & 3.2675 & 0.5 & 0.0162 & 0     & - & 0    & 0.8 & -0.484 \\ \hline
August   & 0       & 1.4444 & 0.9433 & 0.733 & 1.3643  & 0.5 & 0.5 & 0       & 2.5738 & 0.5 & 0.1006 & 0     & - & 0    & 0.8 & -0.08  \\ \hline
September & 0.3333 & 3.6667 & 1.0415 & 0.425 & 2.3529 & 0.8276 & 0.1724 & 0.7843 & 4.9012 & 0.5 & 0.0104 & 0.6552 & 0.5088 & -0.0871 & 0.8 & -0.388 \\ \hline
October  & 0       & 1.6667 & 0.7813 & 0.468 & 2.1368  & 0.5 & 0.5 & 0       & 3.3389 & 0.5 & 0.0321 & 0     & - & 0    & 0.8 & -0.345 \\ \hline
November & 0       & 1.5556 & 1.1503 & 1.058 & 0.9452  & 0.5 & 0.5 & 0       & 2.1745 & 0.5 & 0.2445 & 0     & - & 0    & 0.8 & 0      \\ \hline
\end{tabular}%
}

\resizebox{\columnwidth}{!}{%
\begin{tabular}{|l|l|l|l|l|l|l|l|l|l|l|l|l|l|l|l|l|}
\hline
\textit{N. rotundus} & $N_k \approx G \approx F \approx h$ &
  $\Sigma N$ &
  $U \approx a$ &
  $T_s \approx b$ &
 $q_s$ &
  $\kappa$ &
  $\kappa' $&
  $\phi(q_s)$ &
  $\gamma_{q_s}$ &
  $\omega$ &
  $I_{q_s}(\omega)$ &
  $M_{mean}$ &
  $W$ &
  $E(T)$ &
  $T_c$ &
  $S$      \\ \hline
April       & 0.2222  & 0.7778 & 0.2426 & 0.06  & 16.6667 & 0.9994 & 0.0006 & 3.7037  & 8.0867 & 0.1667 & 0      & 0.9988 & 0.2225 & -0.1829 & 0.8 & -0.753 \\ \hline
May         & 0       & 1.1111 & 0.4782 & 0.18  & 5.5556  & 0.5    & 0.5    & 0       & 5.3133 & 0.1667 & 0.003  & 0      & -      & 0       & 0.8 & -0.633 \\ \hline
June        & 0.7778  & 5      & 1.5437 & 0.648 & 1.5432  & 0.9169 & 0.0831 & 1.2003  & 4.7645 & 0.1667 & 0.0134 & 0.8337 & 0.9329 & -0.7261 & 0.8 & -0.165 \\ \hline
July        & 0.1111  & 1.1111 & 0.5375 & 0.329 & 3.0395  & 0.6627 & 0.3373 & 0.3377  & 3.2675 & 0.1667 & 0.0148 & 0.3254 & 0.3414 & -0.0042 & 0.8 & -0.484 \\ \hline
August  & 0.1111 & 1.4444 & 0.9433 & 0.733 & 1.3643 & 0.5752 & 0.4248 & 0.1516 & 2.5738 & 0.1667 & 0.1016 & 0.1504 & 0.7386 & -0.0019 & 0.8 & -0.08  \\ \hline
September   & 0.2222  & 3.6667 & 1.0415 & 0.425 & 2.3529  & 0.74   & 0.26   & 0.5229  & 4.9012 & 0.1667 & 0.0142 & 0.4799 & 0.463  & -0.0258 & 0.8 & -0.388 \\ \hline
October & 0.1111 & 1.6667 & 0.7813 & 0.468 & 2.1368 & 0.6165 & 0.3835 & 0.2374 & 3.3389 & 0.1667 & 0.0309 & 0.2331 & 0.4768 & -0.0029 & 0.8 & -0.345 \\ \hline
November    & 0       & 1.5556 & 1.1503 & 1.058 & 0.9452  & 0.5    & 0.5    & 0       & 2.1745 & 0.1667 & 0.2559 & 0      & -      & 0       & 0.8 & 0      \\ \hline
\end{tabular}%
}

\resizebox{\columnwidth}{!}{%
\begin{tabular}{|l|l|l|l|l|l|l|l|l|l|l|l|l|l|l|l|l|}
\hline
\textit{C. reticulatus} & $N_k \approx G \approx F \approx h$ &
  $\Sigma N$ &
  $U \approx a$ &
  $T_s \approx b$ &
 $q_s$ &
  $\kappa$ &
  $\kappa' $&
  $\phi(q_s)$ &
  $\gamma_{q_s}$ &
  $\omega$ &
  $I_{q_s}(\omega)$ &
  $M_{mean}$ &
  $W$ &
  $E(T)$ &
  $T_c$ &
  $S$      \\ \hline
April          & 0       & 0.7778 & 0.2426 & 0.06  & 16.6667 & 0.5 & 0.5 & 0       & 8.0867 & 0.25 & 0.0002 & 0     &- & 0    & 0.8 & -0.753 \\ \hline
May            & 0       & 1.1111 & 0.4782 & 0.18  & 5.5556  & 0.5 & 0.5 & 0       & 5.3133 & 0.25 & 0.003  & 0     & - & 0    & 0.8 & -0.633 \\ \hline
June      & 1.1111 & 5      & 1.5437 & 0.648 & 1.5432 & 0.9686 & 0.0314 & 1.7147 & 4.7645 & 0.25 & 0.0053 & 0.9372 & 1.1855 & -2.1169 & 0.8 & -0.165 \\ \hline
July           & 0       & 1.1111 & 0.5375 & 0.329 & 3.0395  & 0.5 & 0.5 & 0       & 3.2675 & 0.25 & 0.0165 & 0     & - & 0    & 0.8 & -0.484 \\ \hline
August         & 0       & 1.4444 & 0.9433 & 0.733 & 1.3643  & 0.5 & 0.5 & 0       & 2.5738 & 0.25 & 0.1034 & 0     & - & 0    & 0.8 & -0.08  \\ \hline
September & 0.1111 & 3.6667 & 1.0415 & 0.425 & 2.3529 & 0.6278 & 0.3722 & 0.2614 & 4.9012 & 0.25 & 0.0172 & 0.2556 & 0.4346 & -0.0032 & 0.8 & -0.388 \\ \hline
October   & 0.1111 & 1.6667 & 0.7813 & 0.468 & 2.1368 & 0.6165 & 0.3835 & 0.2374 & 3.3389 & 0.25 & 0.0308 & 0.2331 & 0.4768 & -0.0029 & 0.8 & -0.345 \\ \hline
November       & 0       & 1.5556 & 1.1503 & 1.058 & 0.9452  & 0.5 & 0.5 & 0       & 2.1745 & 0.25 & 0.254  & 0     &- & 0    & 0.8 & 0      \\ \hline
\end{tabular}%
}

\raggedright
\tiny{\noindent \\ \noindent The statistical procedures are described in the main text.  For scientific names, see Table~2.  The symbol $N$ denotes the average number of individuals per 100~cc of litter.  The parameter $a$ is not computed as a simple average; instead, $a$ is estimated from the empirical distribution of counts for each population or species via the fitted rank relation $N_k = a - b\ln k$. \\ }

\end{table}

\begin{table}[]
\caption{Statistics of the PzDom model for the world economics.}
\centering Population\\

\resizebox{\columnwidth}{!}{%
\begin{tabular}{|l|l|l|l|l|l|l|l|l|l|l|l|l|l|l|l|l|}
\hline
Western   Europe & $N_k \approx G \approx F \approx h$ &
  $\Sigma N$ &
  $U \approx a$ &
  $T_s \approx b$ &
 $q_s$ &
  $\kappa$ &
  $\kappa' $&
  $\phi(q_s)$ &
  $\gamma_{q_s}$ &
  $\omega$ &
  $I_{q_s}(\omega)$ &
  $M_{mean}$ &
  $W$ &
  $E(T)$ &
  $T_c$ &
  $S$ \\ \hline
1 &
  25050 &
  225800 &
  117300 &
  67200 &
  0.00001488 &
  0.6782 &
  0.3218 &
  0.3728 &
  3.491 &
  0 &
  564600000 &
  0.3564 &
  70290 &
  -233900000 &
  900000 &
  -832800 \\ \hline
1000 &
  25560 &
  267300 &
  127900 &
  71250 &
  0.00001404 &
  0.6721 &
  0.3279 &
  0.3587 &
  3.589 &
  0 &
  623500000 &
  0.3441 &
  74280 &
  -234400000 &
  900000 &
  -828800 \\ \hline
1500 &
  57330 &
  438500 &
  199900 &
  100000 &
  0.00001 &
  0.7589 &
  0.2411 &
  0.5733 &
  3.998 &
  0 &
  915300000 &
  0.5178 &
  110700 &
  -1884000000 &
  900000 &
  -800000 \\ \hline
1600 &
  73780 &
  556100 &
  266600 &
  100000 &
  0.00001 &
  0.8139 &
  0.1861 &
  0.7378 &
  5.332 &
  0 &
  568100000 &
  0.6278 &
  117500 &
  -4016000000 &
  900000 &
  -800000 \\ \hline
1700 &
  81460 &
  603200 &
  280200 &
  200000 &
  0.000005 &
  0.6931 &
  0.3069 &
  0.4073 &
  2.802 &
  0 &
  6073000000 &
  0.3862 &
  210900 &
  -2703000000 &
  900000 &
  -700000 \\ \hline
1820 & 133000 & 1042000 & 494400 & 300000 & 0.000003333 & 0.7083 & 0.2917 & 0.4435 & 3.296 & 0 & 11280000000 & 0.4165 & 319400 & -7849000000 & 900000 & -600000 \\ \hline
1870 &
  187500 &
  1272000 &
  547400 &
  300000 &
  0.000003333 &
  0.7773 &
  0.2227 &
  0.625 &
  3.649 &
  0 &
  8538000000 &
  0.5546 &
  338100 &
  -21970000000 &
  900000 &
  -600000 \\
\hline
\end{tabular}%
}

\resizebox{\columnwidth}{!}{%
\begin{tabular}{|l|l|l|l|l|l|l|l|l|l|l|l|l|l|l|l|l|}
\hline
Eastern Europe & $N_k \approx G \approx F \approx h$ &
  $\Sigma N$ &
  $U \approx a$ &
  $T_s \approx b$ &
 $q_s$ &
  $\kappa$ &
  $\kappa' $&
  $\phi(q_s)$ &
  $\gamma_{q_s}$ &
  $\omega$ &
  $I_{q_s}(\omega)$ &
  $M_{mean}$ &
  $W$ &
  $E(T)$ &
  $T_c$ &
  $S$ \\ \hline
1 &
  4750 &
  225800 &
  117300 &
  67200 &
  0.00001488 &
  0.5353 &
  0.4647 &
  0.07068 &
  3.491 &
  0 &
  643500000 &
  0.07057 &
  67310 &
  -1595000 &
  900000 &
  -832800 \\ \hline
1000 &
  6500 &
  267300 &
  127900 &
  71250 &
  0.00001404 &
  0.5455 &
  0.4545 &
  0.09123 &
  3.589 &
  0 &
  701400000 &
  0.09098 &
  71450 &
  -3854000 &
  900000 &
  -828800 \\ \hline
1500 &
  13500 &
  438500 &
  199900 &
  100000 &
  0.00001 &
  0.5671 &
  0.4329 &
  0.135 &
  3.998 &
  0 &
  1228000000 &
  0.1342 &
  100600 &
  -24600000 &
  900000 &
  -800000 \\ \hline
1600 &
  16950 &
  556100 &
  266600 &
  100000 &
  0.00001 &
  0.5839 &
  0.4161 &
  0.1695 &
  5.332 &
  0 &
  911300000 &
  0.1679 &
  101000 &
  -48700000 &
  900000 &
  -800000 \\ \hline
1700 &
  18800 &
  603200 &
  280200 &
  200000 &
  0.000005 &
  0.5469 &
  0.4531 &
  0.094 &
  2.802 &
  0 &
  7075000000 &
  0.09372 &
  200600 &
  -33220000 &
  900000 &
  -700000 \\ \hline
1820 &
  36460 &
  1042000 &
  494400 &
  300000 &
  0.000003333 &
  0.5605 &
  0.4395 &
  0.1215 &
  3.296 &
  0 &
  13450000000 &
  0.1209 &
  301500 &
  -161500000 &
  900000 &
  -600000 \\ \hline
1870 &
  53560 &
  1272000 &
  547400 &
  300000 &
  0.000003333 &
  0.5883 &
  0.4117 &
  0.1785 &
  3.649 &
  0 &
  11950000000 &
  0.1767 &
  303200 &
  -512100000 &
  900000 &
  -600000 \\
\hline
\end{tabular}%
}

\resizebox{\columnwidth}{!}{%
\begin{tabular}{|l|l|l|l|l|l|l|l|l|l|l|l|l|l|l|l|l|}
\hline
Former Soviet Union & $N_k \approx G \approx F \approx h$ &
  $\Sigma N$ &
  $U \approx a$ &
  $T_s \approx b$ &
 $q_s$ &
  $\kappa$ &
  $\kappa' $&
  $\phi(q_s)$ &
  $\gamma_{q_s}$ &
  $\omega$ &
  $I_{q_s}(\omega)$ &
  $M_{mean}$ &
  $W$ &
  $E(T)$ &
  $T_c$ &
  $S$ \\ \hline
1 &
  3900 &
  225800 &
  117300 &
  67200 &
  0.00001488 &
  0.529 &
  0.471 &
  0.05803 &
  3.491 &
  0 &
  644600000 &
  0.05797 &
  67280 &
  -882700 &
  900000 &
  -832800 \\ \hline
1000 &
  7100 &
  267300 &
  127900 &
  71250 &
  0.00001404 &
  0.5497 &
  0.4503 &
  0.09965 &
  3.589 &
  0 &
  700200000 &
  0.09932 &
  71490 &
  -5023000 &
  900000 &
  -828800 \\ \hline
1500 &
  16950 &
  438500 &
  199900 &
  100000 &
  0.00001 &
  0.5839 &
  0.4161 &
  0.1695 &
  3.998 &
  0 &
  1215000000 &
  0.1679 &
  101000 &
  -48700000 &
  900000 &
  -800000 \\ \hline
1600 &
  20700 &
  556100 &
  266600 &
  100000 &
  0.00001 &
  0.602 &
  0.398 &
  0.207 &
  5.332 &
  0 &
  898600000 &
  0.2041 &
  101400 &
  -88700000 &
  900000 &
  -800000 \\ \hline
1700 &
  26550 &
  603200 &
  280200 &
  200000 &
  0.000005 &
  0.566 &
  0.434 &
  0.1328 &
  2.802 &
  0 &
  7014000000 &
  0.132 &
  201200 &
  -93580000 &
  900000 &
  -700000 \\ \hline
1820 &
  54770 &
  1042000 &
  494400 &
  300000 &
  0.000003333 &
  0.5903 &
  0.4097 &
  0.1826 &
  3.296 &
  0 &
  13210000000 &
  0.1805 &
  303300 &
  -547500000 &
  900000 &
  -600000 \\ \hline
1870 & 88670 & 1272000 & 547400 & 300000 & 0.000003333 & 0.6436 & 0.3564 & 0.2956 & 3.649 & 0 & 11310000000 & 0.2873 & 308700 & -2324000000 & 900000 & -600000 \\
\hline
\end{tabular}%
}

\resizebox{\columnwidth}{!}{%
\begin{tabular}{|l|l|l|l|l|l|l|l|l|l|l|l|l|l|l|l|l|}
\hline
Western Offshoots & $N_k \approx G \approx F \approx h$ &
  $\Sigma N$ &
  $U \approx a$ &
  $T_s \approx b$ &
 $q_s$ &
  $\kappa$ &
  $\kappa' $&
  $\phi(q_s)$ &
  $\gamma_{q_s}$ &
  $\omega$ &
  $I_{q_s}(\omega)$ &
  $M_{mean}$ &
  $W$ &
  $E(T)$ &
  $T_c$ &
  $S$\\ \hline
1 &
  1120 &
  225800 &
  117300 &
  67200 &
  0.00001488 &
  0.5083 &
  0.4917 &
  0.01667 &
  3.491 &
  0 &
  646600000 &
  0.01666 &
  67210 &
  -20910 &
  900000 &
  -832800 \\ \hline
1000 &
  1870 &
  267300 &
  127900 &
  71250 &
  0.00001404 &
  0.5131 &
  0.4869 &
  0.02625 &
  3.589 &
  0 &
  706700000 &
  0.02624 &
  71270 &
  -91780 &
  900000 &
  -828800 \\ \hline
1500 &
  2800 &
  438500 &
  199900 &
  100000 &
  0.00001 &
  0.514 &
  0.486 &
  0.028 &
  3.998 &
  0 &
  1250000000 &
  0.02799 &
  100000 &
  -219500 &
  900000 &
  -800000 \\ \hline
1600 &
  2300 &
  556100 &
  266600 &
  100000 &
  0.00001 &
  0.5115 &
  0.4885 &
  0.023 &
  5.332 &
  0 &
  937200000 &
  0.023 &
  100000 &
  -121700 &
  900000 &
  -800000 \\ \hline
1700 &
  1750 &
  603200 &
  280200 &
  200000 &
  0.000005 &
  0.5044 &
  0.4956 &
  0.00875 &
  2.802 &
  0 &
  7137000000 &
  0.00875 &
  200000 &
  -26800 &
  900000 &
  -700000 \\ \hline
1820 &
  11230 &
  1042000 &
  494400 &
  300000 &
  0.000003333 &
  0.5187 &
  0.4813 &
  0.03744 &
  3.296 &
  0 &
  13630000000 &
  0.03742 &
  300100 &
  -4722000 &
  900000 &
  -600000 \\ \hline
1870 &
  46090 &
  1272000 &
  547400 &
  300000 &
  0.000003333 &
  0.5762 &
  0.4238 &
  0.1536 &
  3.649 &
  0 &
  12040000000 &
  0.1524 &
  302400 &
  -326300000 &
  900000 &
  -600000 \\
\hline
\end{tabular}%
}

\resizebox{\columnwidth}{!}{%
\begin{tabular}{|l|l|l|l|l|l|l|l|l|l|l|l|l|l|l|l|l|}
\hline
Latin America& $N_k \approx G \approx F \approx h$ &
  $\Sigma N$ &
  $U \approx a$ &
  $T_s \approx b$ &
 $q_s$ &
  $\kappa$ &
  $\kappa' $&
  $\phi(q_s)$ &
  $\gamma_{q_s}$ &
  $\omega$ &
  $I_{q_s}(\omega)$ &
  $M_{mean}$ &
  $W$ &
  $E(T)$ &
  $T_c$ &
  $S$ \\ \hline
1 &
  5600 &
  225800 &
  117300 &
  67200 &
  0.00001488 &
  0.5416 &
  0.4584 &
  0.08333 &
  3.491 &
  0 &
  642300000 &
  0.08314 &
  67360 &
  -2613000 &
  900000 &
  -832800 \\ \hline
1000 &
  11400 &
  267300 &
  127900 &
  71250 &
  0.00001404 &
  0.5793 &
  0.4207 &
  0.16 &
  3.589 &
  0 &
  689400000 &
  0.1586 &
  71860 &
  -20790000 &
  900000 &
  -828800 \\ \hline
1500 &
  17500 &
  438500 &
  199900 &
  100000 &
  0.00001 &
  0.5866 &
  0.4134 &
  0.175 &
  3.998 &
  0 &
  1213000000 &
  0.1732 &
  101000 &
  -53590000 &
  900000 &
  -800000 \\ \hline
1600 &
  8600 &
  556100 &
  266600 &
  100000 &
  0.00001 &
  0.5429 &
  0.4571 &
  0.086 &
  5.332 &
  0 &
  930800000 &
  0.08579 &
  100200 &
  -6361000 &
  900000 &
  -800000 \\ \hline
1700 &
  12050 &
  603200 &
  280200 &
  200000 &
  0.000005 &
  0.5301 &
  0.4699 &
  0.06025 &
  2.802 &
  0 &
  7112000000 &
  0.06018 &
  200200 &
  -8748000 &
  900000 &
  -700000 \\ \hline
1820 & 21590 & 1042000 & 494400 & 300000 & 0.000003333 & 0.5359 & 0.4641 & 0.07197 & 3.296 & 0 & 13580000000 & 0.07185 & 300500 & -33550000  & 900000 & -600000 \\ \hline
1870 & 40400 & 1272000 & 547400 & 300000 & 0.000003333 & 0.5669 & 0.4331 & 0.1347  & 3.649 & 0 & 12110000000 & 0.1339  & 301800 & -219800000 & 900000 & -600000 \\
\hline
\end{tabular}%
}

\resizebox{\columnwidth}{!}{%
\begin{tabular}{|l|l|l|l|l|l|l|l|l|l|l|l|l|l|l|l|l|}
\hline
Japan & $N_k \approx G \approx F \approx h$ &
  $\Sigma N$ &
  $U \approx a$ &
  $T_s \approx b$ &
 $q_s$ &
  $\kappa$ &
  $\kappa' $&
  $\phi(q_s)$ &
  $\gamma_{q_s}$ &
  $\omega$ &
  $I_{q_s}(\omega)$ &
  $M_{mean}$ &
  $W$ &
  $E(T)$ &
  $T_c$ &
  $S$       \\ \hline
1     & 3000    & 225800  & 117300  & 67200   & 0.00001488  & 0.5223 & 0.4777 & 0.04464 & 3.491 & 0 & 645500000   & 0.04461 & 67250   & -401800     & 900000 & -832800 \\ \hline
1000  & 7500    & 267300  & 127900  & 71250   & 0.00001404  & 0.5524 & 0.4476 & 0.1053  & 3.589 & 0 & 699400000   & 0.1049  & 71510   & -5921000    & 900000 & -828800 \\ \hline
1500  & 15400   & 438500  & 199900  & 100000  & 0.00001     & 0.5764 & 0.4236 & 0.154   & 3.998 & 0 & 1221000000  & 0.1528  & 100800  & -36520000   & 900000 & -800000 \\ \hline
1600  & 18500   & 556100  & 266600  & 100000  & 0.00001     & 0.5915 & 0.4085 & 0.185   & 5.332 & 0 & 906300000   & 0.1829  & 101100  & -63320000   & 900000 & -800000 \\ \hline
1700  & 27000   & 603200  & 280200  & 200000  & 0.000005    & 0.5671 & 0.4329 & 0.135   & 2.802 & 0 & 7009000000  & 0.1342  & 201200  & -98420000   & 900000 & -700000 \\ \hline
1820  & 31000   & 1042000 & 494400  & 300000  & 0.000003333 & 0.5515 & 0.4485 & 0.1033  & 3.296 & 0 & 13510000000 & 0.103   & 301100  & -99300000   & 900000 & -600000 \\ \hline
1870 & 34440 & 1272000 & 547400 & 300000 & 0.000003333 & 0.5571 & 0.4429 & 0.1148 & 3.649 & 0 & 12170000000 & 0.1143 & 301300 & -136100000 & 900000 & -600000 \\
\hline
\end{tabular}%
}

\resizebox{\columnwidth}{!}{%
\begin{tabular}{|l|l|l|l|l|l|l|l|l|l|l|l|l|l|l|l|l|}
\hline
Asia & $N_k \approx G \approx F \approx h$ &
  $\Sigma N$ &
  $U \approx a$ &
  $T_s \approx b$ &
 $q_s$ &
  $\kappa$ &
  $\kappa' $&
  $\phi(q_s)$ &
  $\gamma_{q_s}$ &
  $\omega$ &
  $I_{q_s}(\omega)$ &
  $M_{mean}$ &
  $W$ &
  $E(T)$ &
  $T_c$ &
  $S$      \\ \hline
1    & 165400  & 225800  & 117300  & 67200   & 0.00001488  & 0.9928 & 0.007228 & 2.461   & 3.491 & 0 & 18560000   & 0.9855 & 167800  & -67330000000 & 900000 & -832800 \\ \hline
1000 & 175100  & 267300  & 127900  & 71250   & 0.00001404  & 0.9927 & 0.007282 & 2.458   & 3.589 & 0 & 20450000   & 0.9854 & 177700  & -75350000000 & 900000 & -828800 \\ \hline
1500 & 268400  & 438500  & 199900  & 100000  & 0.00001     & 0.9954 & 0.004642 & 2.684   & 3.998 & 0 & 23110000   & 0.9907 & 270900  & -1.934E+11   & 900000 & -800000 \\ \hline
1600 & 360000  & 556100  & 266600  & 100000  & 0.00001     & 0.9993 & 0.000746 & 3.6     & 5.332 & 0 & 2796000    & 0.9985 & 360500  & -4.666E+11   & 900000 & -800000 \\ \hline
1700 & 374500  & 603200  & 280200  & 200000  & 0.000005    & 0.9769 & 0.02309  & 1.873   & 2.802 & 0 & 644000000  & 0.9538 & 392600  & -2.626E+11   & 900000 & -700000 \\ \hline
1820 & 679400  & 1042000 & 494400  & 300000  & 0.000003333 & 0.9893 & 0.01067  & 2.265   & 3.296 & 0 & 576700000  & 0.9787 & 694200  & -1.045E+12   & 900000 & -600000 \\ \hline
1870 & 730800  & 1272000 & 547400  & 300000  & 0.000003333 & 0.9924 & 0.0076   & 2.436   & 3.649 & 0 & 372000000  & 0.9848 & 742100  & -1.301E+12   & 900000 & -600000 \\
\hline
\end{tabular}%
}

\resizebox{\columnwidth}{!}{%
\begin{tabular}{|l|l|l|l|l|l|l|l|l|l|l|l|l|l|l|l|l|}
\hline
Africa & $N_k \approx G \approx F \approx h$ &
  $\Sigma N$ &
  $U \approx a$ &
  $T_s \approx b$ &
 $q_s$ &
  $\kappa$ &
  $\kappa' $&
  $\phi(q_s)$ &
  $\gamma_{q_s}$ &
  $\omega$ &
  $I_{q_s}(\omega)$ &
  $M_{mean}$ &
  $W$ &
  $E(T)$ &
  $T_c$ &
  $S$       \\ \hline
1      & 17000   & 225800  & 117300  & 67200   & 0.00001488  & 0.6239 & 0.3761 & 0.253   & 3.491 & 0 & 607100000   & 0.2477 & 68630   & -73110000    & 900000 & -832800 \\ \hline
1000   & 32300   & 267300  & 127900  & 71250   & 0.00001404  & 0.7123 & 0.2877 & 0.4533  & 3.589 & 0 & 579700000   & 0.4246 & 76070   & -473000000   & 900000 & -828800 \\ \hline
1500   & 46610   & 438500  & 199900  & 100000  & 0.00001     & 0.7175 & 0.2825 & 0.4661  & 3.998 & 0 & 1014000000  & 0.435  & 107100  & -1013000000  & 900000 & -800000 \\ \hline
1600   & 55320   & 556100  & 266600  & 100000  & 0.00001     & 0.7515 & 0.2485 & 0.5532  & 5.332 & 0 & 700500000   & 0.5029 & 110000  & -1693000000  & 900000 & -800000 \\ \hline
1700   & 61080   & 603200  & 280200  & 200000  & 0.000005    & 0.6481 & 0.3519 & 0.3054  & 2.802 & 0 & 6511000000  & 0.2962 & 206200  & -1139000000  & 900000 & -700000 \\ \hline
1820   & 74240   & 1042000 & 494400  & 300000  & 0.000003333 & 0.6213 & 0.3787 & 0.2475  & 3.296 & 0 & 12850000000 & 0.2425 & 306100  & -1364000000  & 900000 & -600000 \\ \hline
1870   & 90470   & 1272000 & 547400  & 300000  & 0.000003333 & 0.6464 & 0.3536 & 0.3016  & 3.649 & 0 & 11270000000 & 0.2927 & 309000  & -2468000000  & 900000 & -600000 \\
\hline
\end{tabular}%
}

\centering (continues)
\end{table}

\begin{table}
\centering (continued)\\
GDP

\resizebox{\columnwidth}{!}{%
\begin{tabular}{|l|l|l|l|l|l|l|l|l|l|l|l|l|l|l|l|l|}
\hline
Western Europe & $N_k \approx G \approx F \approx h$ &
  $\Sigma N$ &
  $U \approx a$ &
  $T_s \approx b$ &
 $q_s$ &
  $\kappa$ &
  $\kappa' $&
  $\phi(q_s)$ &
  $\gamma_{q_s}$ &
  $\omega$ &
  $I_{q_s}(\omega)$ &
  $M_{mean}$ &
  $W$ &
  $E(T)$ &
  $T_c$ &
  $S$\\ \hline
1 &
  14430 &
  105400 &
  54730 &
  31350 &
  0.0000319 &
  0.7152 &
  0.2848 &
  0.4604 &
  3.492 &
  0 &
  114700000 &
  0.4304 &
  33530 &
  -95900000 &
  4500000 &
  -4469000 \\ \hline
1000 &
  10930 &
  120400 &
  58960 &
  33150 &
  0.00003016 &
  0.659 &
  0.341 &
  0.3295 &
  3.556 &
  0 &
  138900000 &
  0.3181 &
  34350 &
  -39330000 &
  4500000 &
  -4467000 \\ \hline
1500 &
  44180 &
  248400 &
  117100 &
  64930 &
  0.0000154 &
  0.7959 &
  0.2041 &
  0.6805 &
  3.608 &
  0 &
  379700000 &
  0.5918 &
  74660 &
  -1.328E+09 &
  4500000 &
  -4435000 \\ \hline
1600 &
  65600 &
  331600 &
  160500 &
  89810 &
  0.00001113 &
  0.8117 &
  0.1883 &
  0.7304 &
  3.574 &
  0 &
  690000000 &
  0.6233 &
  105200 &
  -3.144E+09 &
  4500000 &
  -4410000 \\ \hline
1700 &
  81210 &
  371400 &
  171200 &
  94140 &
  0.00001062 &
  0.8488 &
  0.1512 &
  0.8627 &
  3.638 &
  0 &
  625200000 &
  0.6977 &
  116400 &
  -5.69E+09 &
  4500000 &
  -4406000 \\ \hline
1820 &
  159900 &
  694600 &
  313000 &
  200000 &
  0.000005 &
  0.8318 &
  0.1682 &
  0.7993 &
  3.13 &
  0 &
  3576000000 &
  0.6636 &
  240900 &
  -2.042E+10 &
  4500000 &
  -4300000 \\ \hline
1870 &
  367500 &
  1111000 &
  411300 &
  200000 &
  0.000005 &
  0.9753 &
  0.02473 &
  1.837 &
  4.113 &
  0 &
  469100000 &
  0.9505 &
  386600 &
  -2.481E+11 &
  4500000 &
  -4300000 \\
\hline
\end{tabular}%
%
}

\raggedright
\tiny{\noindent \\ \noindent The statistical procedures are described in the main text.  The symbol $N$ denotes either population or gross domestic product (GDP), depending on the context.  For population data, one unit corresponds to one thousand individuals.  For GDP data, one unit corresponds to one million 1990 international dollars.  Numerical values are reported to four significant digits.  For the original raw data, see \cite{Maddison2007}. \\ }

\end{table}

\subsection{Adaptation of species}
When a system is driven by the field $h=F$, the macroscopic polarity $M$ is \begin{equation} M = N_kZ^{-1}\bigl(e^{q_s h}-e^{-q_s h}\bigr) = N_k\tanh\bigl(q_s h\bigr), \end{equation} where $Z=e^{q_s h}+e^{-q_s h}$ is the single-patch partition function.  A spontaneous polarity corresponds to $M\neq 0$.

Define the mean polarity per individual by $ M_{\mathrm{mean}}=\frac{M}{N_k}. $ Empirically, the three dominant species \textit{P.\ pallidum}, \textit{D.\ purpureum}, and \textit{P.\ violaceum} exhibit larger $M_{\mathrm{mean}}$ values and therefore show stronger adaptation than do the averaged individual populations.

In a Weiss mean field the local field is $h=W M_{\mathrm{mean}}$, so that $M_{\mathrm{mean}}$ satisfies the self-consistency equation for $M$.  Here $W$ is an intensive parameter characterizing the intra-patch coupling.  Under the Weiss approximation: \begin{itemize} \item If $T_s>W$, the system is in a disordered (chaotic) regime. \item If $T_s<W$, the system exhibits long-range order. \end{itemize} Thus $W$ acts as an upper bound on $T_s$ for ordering of a given species.  The equality $T_s=W$ is not applicable within the strict mean-field approximation used here.  Values of $M_{\mathrm{mean}}$ and $W$ are reported in Tables~5--7; entries where $T_s\approx W$ indicate that the species has effectively reached the ordering limit set by $W$.

If we set the long-range order parameter per individual to $ p(T_s)=M_{\mathrm{mean}}^2, $ then empirically $p(T_s)\leq 0.01$ appears to be a practical indicator of a stably adapted condition \cite{Tasaki2015}.

Comparing Table~1 with Table~5 shows correspondence between the listed triples $(\Re(s),\Im(s),m)$ and the behavior expected for Bose-Einstein condensation in an integer system, except for the case $(\Re(s),\Im(s),m)=(2.056,275.5,2.994)$.  Although individuals are distinguishable and in principle obey exclusion (fermion-like) behavior, when their number is large they may be approximated as bosons and treated with Bose statistics; under small migration ($m\sim 0$) this approximation can lead to condensation-like macroscopic occupancy.

Near the critical point $T_s\approx T_c$, the mean polarity $M_{\mathrm{mean}}$ is small and the susceptibility follows the Curie-Weiss law: \begin{equation} \chi_T = U'(q_s)\propto \lvert T_s-T_c\rvert^{-1}. \end{equation} In Figure 2, values of $U'(q_s)$ that diverge were used to locate $T_c$ graphically by plotting $q_s$ versus $U(q_s)$ and averaging temperatures near the divergence.  Estimated critical temperatures (95\% confidence intervals) are: $T_c = 2090 \pm 50$ (95\% confidence) for Dictyostelia populations, $1500 \pm 500$ for Dictyostelia species, $8 \pm 4$ for soil mesofauna population, $2 \pm 1$ for Collembola species, $0.8 \pm 0.5$ for Sarcoptiformes species, $10^6 \pm 10^6$ for human population, and $10^6 \pm 10^7$ for GDP. Critical lines for world economics are effectively negligible in the biological range but are included for completeness.  These $T_c$ estimates are consistent with the Weiss field analysis: $T_c$ marks a global phase discontinuity, whereas $W$ marks the local ordering threshold separating ordered and disordered regimes \cite{Tasaki2015}.  For species above their critical temperature, the transition between a dominating-species phase and an increasing-population phase is continuous rather than discontinuous.

The dynamics of $\phi(q_s)$ (Figure~3) visualize the presence and prominence of species: climax species appear as persistently high values, while pioneering species appear as transient peaks.  For world economic data, analogous shifts in prominence are visible as regional changes in GDP and population growth.  Figures~4--6 display domination phases and growth tendencies for each population or species.  In general, when $W>T_s$ there is potential for growth in $N$; this pattern is evident in the economic data where Asia shows strong population growth and regions such as Western Europe and the Western Offshoots show prominence in GDP growth.

\begin{figure}
\includegraphics[width=12cm]{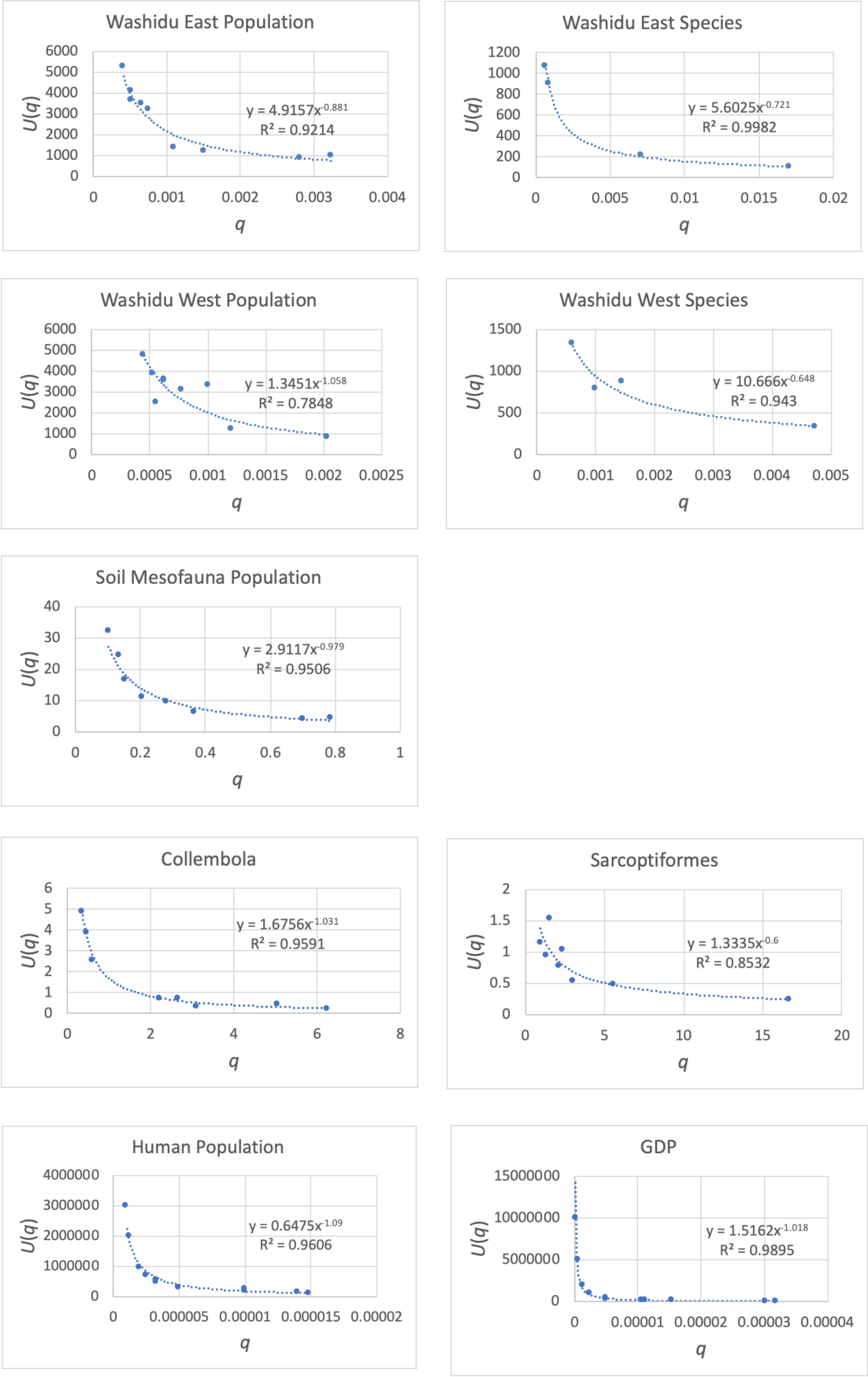}
\caption{$U(q_s)$ fitted with power functions.}
\label{fig:2}
\end{figure}

\begin{figure}
\includegraphics[width=12cm]{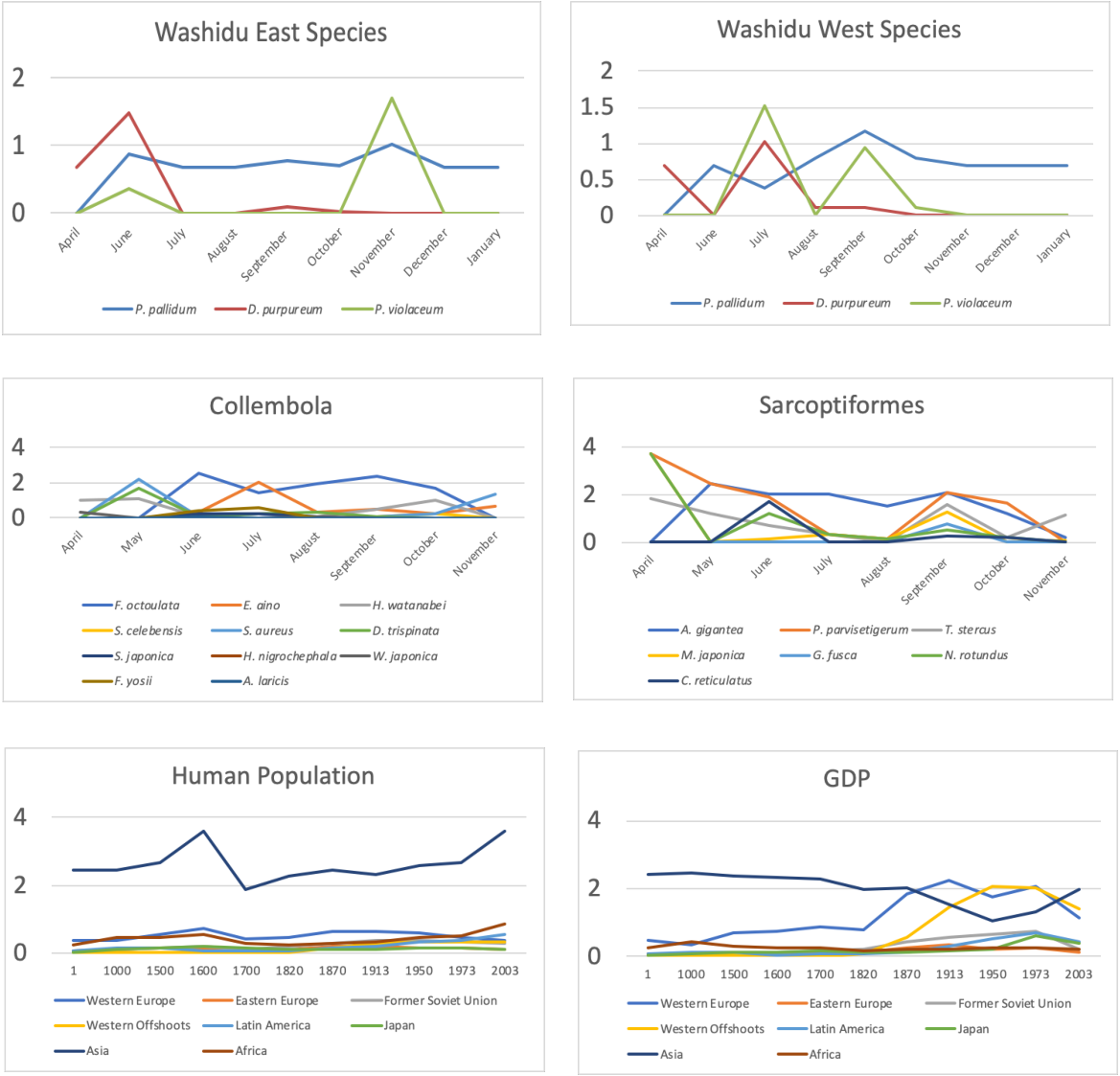}
\vskip5.5cm
\caption{$\phi(q_s)$ dynamics. \textit{P. pallidum} = \textit{Polysphondylium pallidum}; \textit{D. purpureum} = \textit{Dictyostelium purpureum}; \textit{P. violaceum} = \textit{Polysphondylium violaceum}. For scientific names of soil mesofauna, refer Table 2.}
\label{fig:3}
\end{figure}

\begin{figure}
\includegraphics[width=12cm]{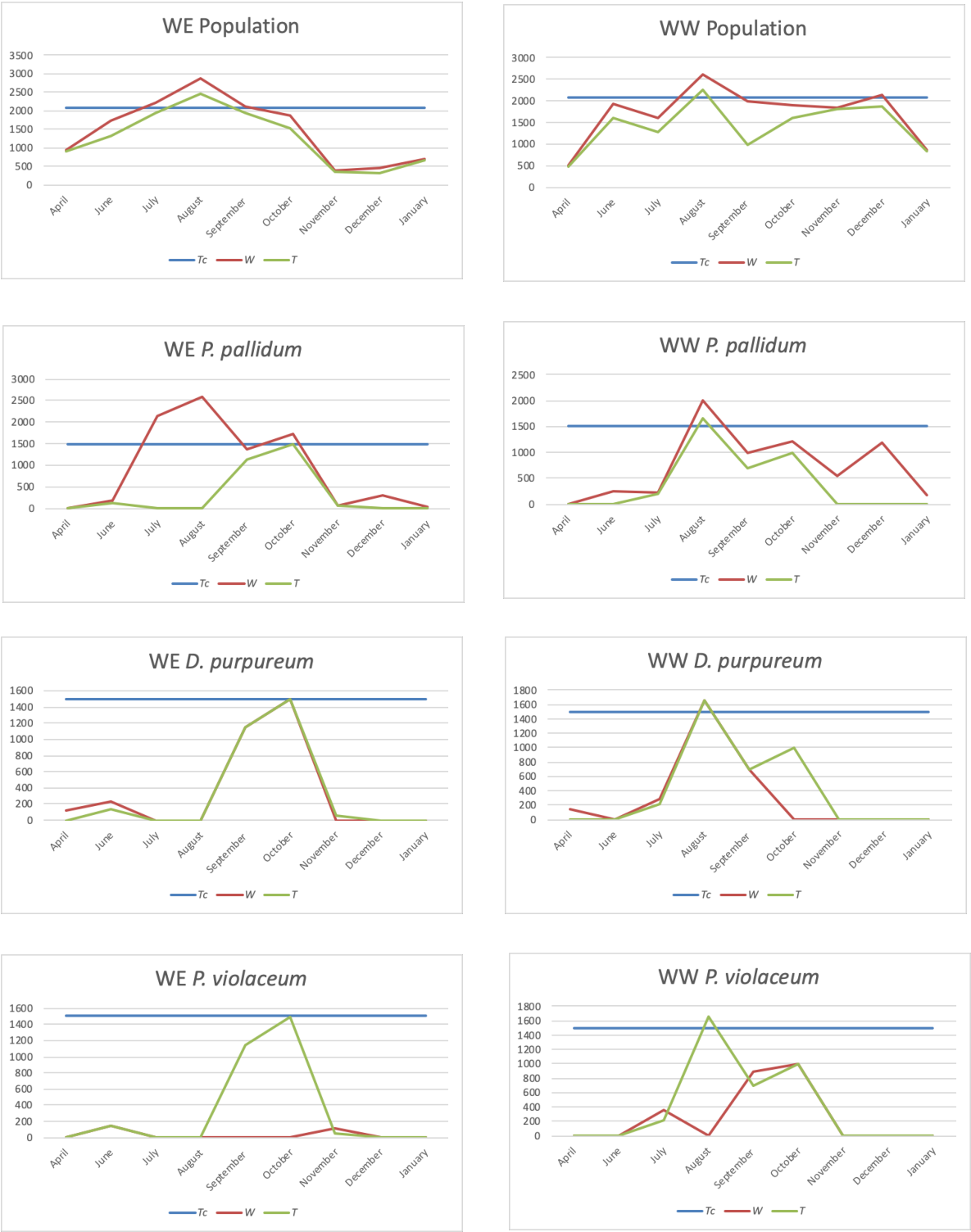}
\vskip2.5cm
\caption{Dynamics of $W$, $T_c$, and $T_s$ for Dictyostelia. WE: Washidu East; WW: Washidu West. \textit{P. pallidum} = \textit{Polysphondylium pallidum}; \textit{D. purpureum} = \textit{Dictyostelium purpureum}; \textit{P. violaceum} = \textit{Polysphondylium violaceum}.}
\label{fig:4}
\end{figure}

\begin{figure}
\includegraphics[width=12cm]{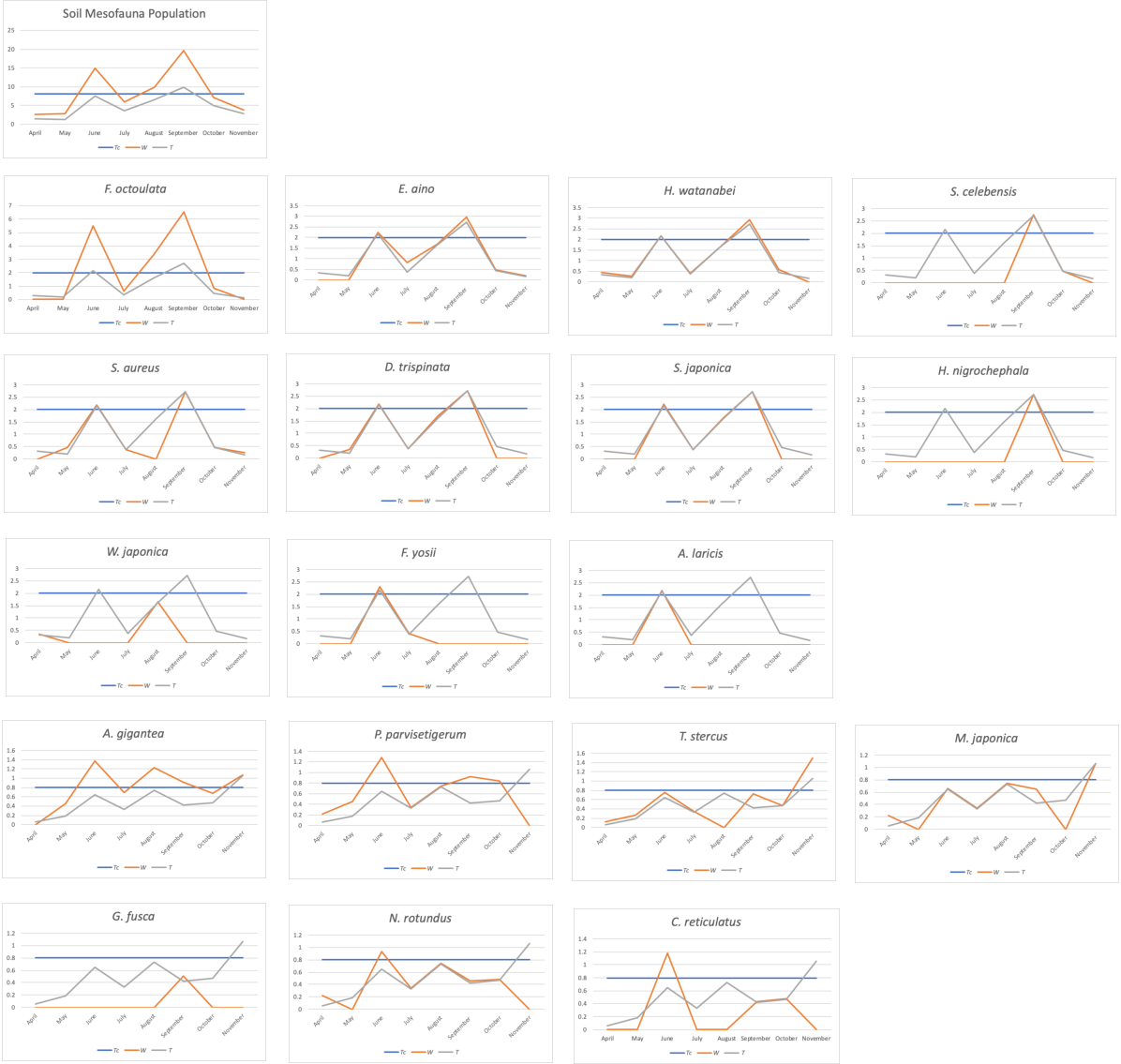}
\vskip5.5cm
\caption{Dynamics of $W$, $T_c$, and $T_s$ for soil mesofauna. For scientific names, refer Table 2.}
\label{fig:5}
\end{figure}

\begin{figure}
\includegraphics[width=12cm]{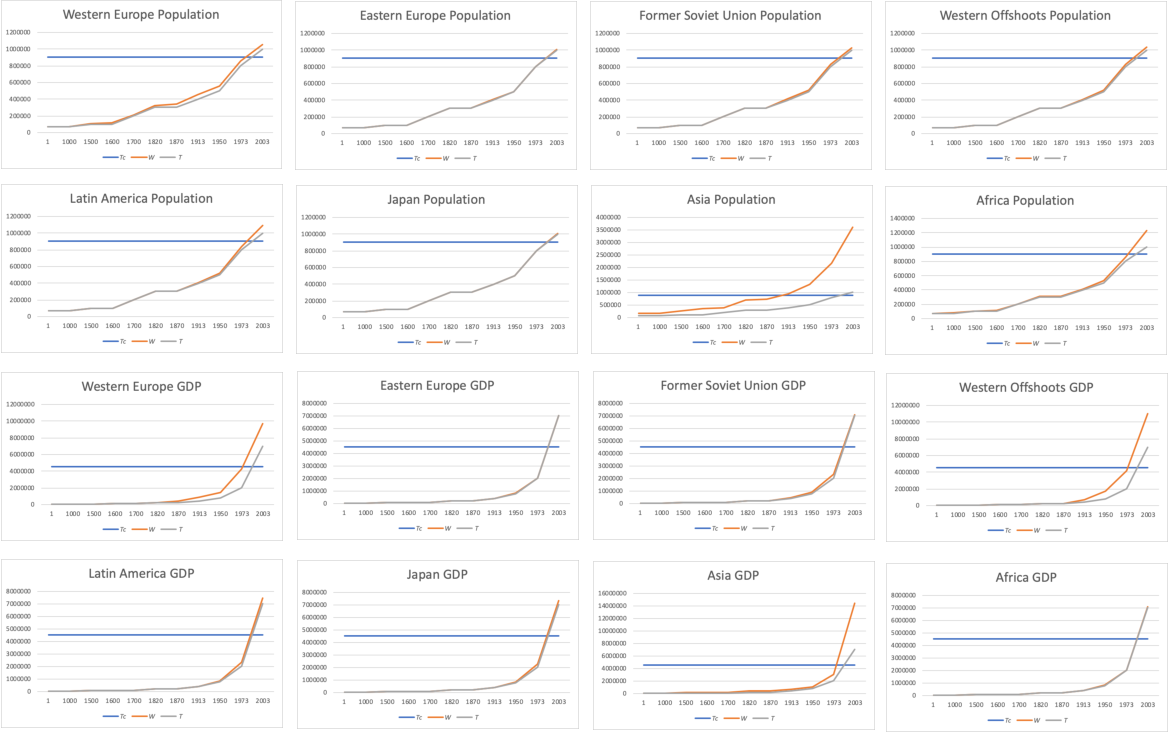}
\vskip9.5cm
\caption{Dynamics of $W$, $T_c$, and $T_s$ for the world economics.}
\label{fig:6}
\end{figure}

When the system is driven by the field $h=F$, the internal energy per patch averaged over patches, denoted $\bar{N}$, is given by \begin{equation} E(T_s) = -\bar{N}h\tanh\bigl(q_s h\bigr) \approx -\bar{N}q_s h^2, \end{equation} where the quadratic approximation holds for small $q_s h$.

The energetic response depends on the relation between the effective temperature $T_s$ and the critical temperature $T_c$: \begin{itemize} \item If $T_s > T_c$, the ordered contribution vanishes and $E=0$. \item If $T_s < T_c$, the internal energy is nonzero and, to leading order near criticality, is proportional to $T_c-T_s$. \end{itemize}

Define the specific heat at constant field (evaluated at $h=0$) by $ C_s = \left.\frac{\partial E}{\partial T_s}\right|_{h=0}. $ The behavior of $C_s$ across the transition is: \begin{itemize} \item $C_s=0$ for $T_s\ge T_c$, \item $C_s$ is finite for $T_s<T_c$, \item $C_s$ exhibits a sudden increase at $T_s=T_c$, marking the energetic signature of the phase change. \end{itemize}

In empirical applications the following points are relevant: \begin{itemize} \item Species typically exhibit higher internal energy than populations, reflecting stronger collective ordering or larger effective coupling at the species scale. \item The critical temperature $T_c$ should be estimated using data specific to each species or population; different taxa may therefore have distinct critical points. \item Multiple dynamical phases are observed in practice, including domination (ordered phase), increasing (growth phase), and chaotic or disordered regimes; classification can be made by comparing $T_s$, $W$, and $T_c$ and by inspecting spectral diagnostics such as $\gamma_q$ and $I_q(\omega)$.\\ \end{itemize}

\subsection{Introducing large \textit{S}, an order parameter}
For the population space defined above and for a system volume $V=1$ (one gram of soil), we introduce an order parameter $S$ to describe ordering near critical points.  Expand the Hamiltonian $f\approx F\approx G$ in powers of $S$ and include an external flow term $h'$ representing immigration or influx from outside: \begin{equation} f(S)=f_0 + A' S^2 + B' S^4 - h' S, \end{equation} where $B'>0$ and $A'$ is taken to have the form $A' = A''(T_s-T_c)$ with $A''$ a constant coefficient.

At equilibrium the stationary condition $\partial f/\partial S=0$ gives the cubic equation \begin{equation} \frac{\partial f}{\partial S}=4B' S^3 + 2A''(T_s-T_c)S - h' = 0, \end{equation} whose real roots determine the possible values of the order parameter $S$.

Key limiting cases and scaling relations follow from this equation: \begin{itemize} \item If $h'=0$ and $T_s>T_c$, the only real solution is $S=0$ (disordered phase). \item If $h'=0$ and $T_s<T_c$, symmetry is broken and nonzero solutions appear with the leading scaling $ S \approx \pm (T_c-T_s)^{1/2}, $ corresponding to the onset of long-range order. \item For small symmetry breaking $h'\neq 0$ at $T_s=T_c$, the order parameter scales as $ S \approx h'^{1/3}. $ \item The isothermal susceptibility near criticality follows the Curie-Weiss form $ \chi_T \approx \lvert T_s-T_c\rvert^{-1}. $ \end{itemize}

Define $C_0=-T_s,\partial^2 f_0/\partial T_s^2$.  The specific heat $C_s=\partial E/\partial T_s\big|_{h=0}$ (or equivalently the second derivative of the free energy with respect to temperature) has the following behavior: \begin{itemize} \item For $T_s\ge T_c$: $C_s = C_0$. \item For $T_s<T_c$: $C_s = \dfrac{T_s A''^{,2}}{2B'}$. \item At $T_s=T_c$ there is a discontinuous jump in $C_s$ corresponding to the phase transition. \end{itemize}

Calculated values of $S$ for the data sets are listed in Tables~5--7.  These results indicate a tendency toward order when a population or species attains dominance: dominant taxa typically exhibit larger $S$ and correspondingly stronger signatures of symmetry breaking and reduced fluctuation.

\subsection{Application of the type IV Painlev\'{e} equation to an $X^2$ system}
After the equilibrium discussion, we consider the time development of the system.  Time will be treated through a variable $t$ (the model uses $t^2$ as the physical time scale in some interpretations).  We model interactions by the quadratic combinations $X^2$, $XY$, and $Y^2$, where $X^2$ and $Y^2$ represent abundances $N_X$ and $N_Y$, and $XY$ represents an interaction term.  The following Lotka--Volterra type system, equivalent to a type IV Painleve system, is assumed: \begin{equation} \frac{dX^2}{dt} = X^2 (XY - Y^2), \frac{dXY}{dt} = -XY(X^2 + Y^2), \frac{dY^2}{dt} = Y^2 (XY - X^2). \end{equation} The sign in the middle equation determines whether the interaction is competitive (as written) or cooperative (change the sign to $+$).  Dividing the first and third equations by $2X$ and $2Y$, respectively, yields evolution equations for $\dot X$ and $\dot Y$.

Define $N_P(t)=1/X(t)$.  Using the Verhulst logistic formulation (see \cite{Lizama2013}), we write \begin{equation} \frac{dN_P}{dt}=N_P\bigl(a_P(t)-b_P(t)N_P\bigr). \end{equation} Under the quadratic interaction mapping one may identify \begin{equation} a_P(t)=-\tfrac{1}{2}XY,\qquad b_P(t)=-\tfrac{1}{2}XY^2, \end{equation} and, with the auxiliary choice $c_P(t)=e^{Y}$, an iterated update consistent with the logistic map can be written as \begin{equation} N_P(t+1)=N_P(t)^{\frac{1}{1+\frac{1}{b}}};c_P(t)^{\frac{1}{b(1+\frac{1}{b})}}, \end{equation} where the discrete time index is related to a continuous phase by $t=b\arg D$ for a suitably chosen branch of $\arg D$.  A gauge factor of the form $h_t(t_t)=(\ln t_t^{-1})^{m_h-1}$ with $t_t=e^{-1/b}$ and $m_h=\Re(s)$ may be introduced to account for scale factors in generation time \cite{Lapidus2017}.

The complex phase $\arg D$ is linked to the inverse temperature $1/b$ in this construction.  Using the absolute zeta function for $\mathbb{G}m$, \begin{equation} \zeta{\mathbb{G}_m/F_1}(s)=\frac{s}{s-1}=\frac{s}{w}, \end{equation} one may interpret $s/w$ as the previous-step value of $s$, giving $\arg w=\arg D=1/b$.  Embedding the dynamics in a group $\mathbb{G}=\mathbb{H}/\lvert\mathbb{H}\rvert\times\mathbb{R}\times\mathbb{R}$ and invoking the uniqueness of the irreducible unitary representation (Stone--von Neumann theorem) yields a scalar action $\rho_G(c_G)=c_G\mathrm{Id}_W$ for central elements $c_G$, which characterizes the linear part of the evolution in representation space \cite{Stone1930, vonNeumann1931, vonNeumann1932, Stone1932}.

To obtain analytic approximations for the discrete factor $D$ we use the Gauss hypergeometric equation with rational parameters $\alpha,\beta,\gamma$: \begin{equation} t(1-t)y''+\bigl(\gamma-(\alpha+\beta+1)t\bigr)y'-\alpha\beta y=0. \end{equation} Form the quotient $w_y(t)=y_1(t)/y_2(t)$ of two linearly independent solutions.  The associated Schwarzian differential equation for $w_y$ is \begin{equation} \left(\frac{w_y''}{w_y'}\right)'-\tfrac{1}{2}\left(\frac{w_y''}{w_y'}\right)^{2} =\tfrac{1}{2}\left(\frac{1-\lambda^2}{t^2}+\frac{1-\mu^2}{(t-1)^2}+\frac{\lambda^2+\mu^2-\nu^2-1}{t(t-1)}\right), \end{equation} with parameter identifications $\lambda^2=(1-\gamma)^2$, $\mu^2=(\gamma-\alpha-\beta)^2$, and $\nu^2=(\alpha-\beta)^2$.

Choosing $\gamma=1$ and matching branch point arguments so that the branch points $t=0,1,\infty$ correspond to $\arg D=0,1/b,\infty$ leads to the condition \begin{equation} \gamma-\alpha-\beta=\pm\frac{1}{b\pi}. \end{equation} From the local form of the quotient one obtains, up to multiplicative constants, \begin{equation} w_y'(t)\propto t^{-1}(t-1)^{\pm\frac{1}{b\pi}-1}y_2(t)^{-2}. \end{equation} In the symmetric choice $y_1=y_2=D$, an approximate closed form is \begin{equation} D(t)\approx\pm\sqrt{\frac{\mathrm{const}}{t(t-1)^{1\mp\frac{1}{b\pi}}}}. \end{equation} Propagating this relation between successive integer time steps (assuming $t$ decreases along the chosen branch because $D>1$) yields the recurrence \begin{equation} D_{t-1}=\sqrt{(t-1)^{\mp\frac{1}{b_t\pi}}\frac{t}{(t-2)^{1\mp\frac{1}{b_{t-1}\pi}}}}D_t, \end{equation} and hence the general estimate \begin{equation} D(t)\approx\{(t-1)^{\mp\frac{1}{b_t\pi}}\frac{t}{(t-2)^{1\mp\frac{1}{b_{t-1}\pi}}}\}^{1/(2t)}. \end{equation}

The Gauss hypergeometric integral representation \begin{equation} {}_2F_1(a,b;c;z)=\frac{\Gamma(c)}{\Gamma(a)\Gamma(c-a)}\int_0^1 t^{a-1}(1-t)^{c-a-1}(1-tz)^{-b}dt \end{equation} provides the local solution behaviour at $t=0,1,\infty$.  The Frobenius solutions at these singular points exhibit the expected divergence or vanishing limits consistent with the biological interpretation: $t\to0,1$ give divergent local amplitudes (transient or pioneering phases), while $t\to\infty$ yields decaying modes (extinction or saturation).

When $b_t\approx b_{t-1}$ for large $t$, the expected $D$ from the hypergeometric/Schwarz analysis can be compared with the observed value $D_{\mathrm{obs}}=e^{\Re(s)/b}$.  Empirically (Table~8), three regimes emerge: \begin{itemize} \item Dictyostelia (integer system): $D_{\mathrm{obs}}\approx D_{\mathrm{exp}}$ — near equilibrium between observed and predicted amplitudes. \item Soil mesofauna: $D_{\mathrm{obs}}\gg D_{\mathrm{exp}}$ — developing order and growth beyond the simple equilibrium prediction. \item World economics: $D_{\mathrm{obs}}\ll D_{\mathrm{exp}}$ — behaviour consistent with a more chaotic, strongly fluctuating regime.\\ \end{itemize}.

\begin{table}[]
\caption{Correspondence between observed and expected $D$ values.}
\centering Dictyostelia\\
\resizebox{\columnwidth}{!}{%
\begin{tabular}{|l|l|l|l|l|l|l|l|l|l|l|l|l|l|l|l|}
\hline
$D$ (observed) &
  WE \textit{P. pallidum} &
  WE \textit{D. purpureum} &
  WE  \textit{P. violaceum} &
  $D$ (expected) &
  WE  \textit{P. pallidum} &
  WE  \textit{D. purpureum} &
  WE  \textit{P. violaceum} &
  $D$ (observed) &
  WW  \textit{P. pallidum} &
  WW  \textit{D. purpureum} &
  WW  \textit{P. violaceum} &
  $D$ (expected) &
  WW  \textit{P. pallidum} &
  WW  \textit{D. purpureum} &
  WW  \textit{P. violaceum} \\ \hline
April       &        &        &        & April     &        &        &        & April     &        &        &        & April     &        &        &        \\ \hline
June      & 1.0055 & 1.0131 & 1.009  & June      & 1.0195 & 1.0004 & 1.0022 & June      &        &        &        & June      &        &        &        \\ \hline
July      &        &        &        & July      &        &        &        & July      & 1.006  & 1.0027 & 1.0084 & July      & 1.0013 & 1.0466 & 1.0002 \\ \hline
August    &        &        &        & August    &        &        &        & August    & 1.0021 & 1.0017 & 1.0006 & August    & 1.0022 & 1.0058 & 1.387  \\ \hline
September & 1.0031 & 1.0027 & 1.0009 & September & 1.0021 & 1.005  & 1.4432 & September & 1.0026 & 1.0031 & 1.0005 & September & 1.0005 & 1.0002 & 2.8602 \\ \hline
October   & 1.0035 & 1.0033 & 1.0007 & October   & 1.0006 & 1.0008 & 2.0852 & October   & 1.0035 & 1.001  & 1.0029 & October   & 1.0022 & 1.3942 & 1.0058 \\ \hline
November  & 1.0128 & 1.0171 & 1.0356 & November  & 1.0251 & 1.0086 & 1.0005 & November  &        &        &        & November  &        &        &        \\ \hline
December  &        &        &        & December  &        &        &        & December  &        &        &        & December  &        &        &        \\ \hline
January   &        &        &        & January   &        &        &        & January   &        &        &        & January   &        &        &        \\ \hline
\end{tabular}%
}

\centering Collembola\\
\resizebox{\columnwidth}{!}{%
\begin{tabular}{|l|l|l|l|l|l|l|l|l|l|l|l|l|l|l|l|l|l|l|l|l|l|l|l|}
\hline
$D$   (observed) &
  \textit{F. octoulata} &
   \textit{E. aino} &
   \textit{H. watanabei} &
   \textit{S. celebensis} &
   \textit{S. aureus} &
   \textit{D. trispinata} &
   \textit{S. japonica} &
   \textit{H. nigrochephala} &
   \textit{W. japonica} &
   \textit{F. yosii} &
   \textit{A. laricis} &
  $D$   (expected) &
   \textit{F. octoulata} &
   \textit{E. aino} &
   \textit{H. watanabei} &
   \textit{S. celebensis} &
   \textit{S. aureus} &
   \textit{D. trispinata} &
   \textit{S. japonica} &
   \textit{H. nigrochephala} &
   \textit{W. japonica} &
  \textit{F. yosii} &
   \textit{A. laricis} \\ \hline
April &
  22.5394 &
  22.5394 &
  2578.0711 &
  22.5394 &
  22.5394 &
  22.5394 &
  22.5394 &
  22.5394 &
  139.4321 &
  22.5394 &
  22.5394 &
  April &
  1.0372 &
  1.0372 &
  1.0005 &
  1.0372 &
  1.0372 &
  1.0372 &
  1.0372 &
  1.0372 &
  1.0067 &
  1.0372 &
  1.0372 \\ \hline
May &
  156.1013 &
  156.1013 &
  24.2039 &
  156.1013 &
  3459.8183 &
  8.1348 &
  156.1013 &
  156.1013 &
  156.1013 &
  156.1013 &
  156.1013 &
  May &
  1 &
  1 &
  1.0003 &
  1 &
  1 &
  1.0024 &
  1 &
  1 &
  1 &
  1 &
  1 \\ \hline
June &
  2.6837 &
  2.4150 &
  2.5011 &
  1.5861 &
  2.5011 &
  2.3709 &
  2.1370 &
  1.5861 &
  1.5861 &
  3.3403 &
  2.5011 &
  June &
  1 &
  1 &
  1 &
  1.0004 &
  1 &
  1 &
  1 &
  1.0004 &
  1.0004 &
  1 &
  1 \\ \hline
July &
  3.6365 &
  55.5707 &
  41.8121 &
  14.2902 &
  41.8121 &
  41.8121 &
  41.8121 &
  14.2902 &
  14.2902 &
  20.7542 &
  14.2902 &
  July &
  1.0035 &
  1 &
  1 &
  1 &
  1 &
  1 &
  1 &
  1 &
  1 &
  1 &
  1 \\ \hline
August &
  4.0107 &
  4.7167 &
  4.4177 &
  1.8434 &
  1.8434 &
  2.6608 &
  4.4177 &
  1.8434 &
  4.4177 &
  1.8434 &
  1.8434 &
  August &
  1 &
  1 &
  1 &
  1.0032 &
  1.0032 &
  1.0001 &
  1 &
  1.0032 &
  1 &
  1.0032 &
  1.0032 \\ \hline
September &
  2.1382 &
  2.2105 &
  1.6943 &
  2.1935 &
  2.1580 &
  2.1580 &
  2.1536 &
  2.1536 &
  1.4443 &
  1.4443 &
  1.4443 &
  September &
  1 &
  1 &
  1 &
  1 &
  1 &
  1 &
  1 &
  1 &
  1.0005 &
  1.0005 &
  1.0005 \\ \hline
October &
  46.5924 &
  51.2183 &
  6.0142 &
  51.2183 &
  51.2183 &
  9.2278 &
  9.2278 &
  9.2278 &
  9.2278 &
  9.2278 &
  9.2278 &
  October &
  1 &
  1 &
  1.0026 &
  1 &
  1 &
  1.0007 &
  1.0007 &
  1.0007 &
  1.0007 &
  1.0007 &
  1.0007 \\ \hline
November &
  518.0128 &
  518.0128 &
  518.0128 &
  518.0128 &
  854296.9028 &
  518.0128 &
  518.0128 &
  518.0128 &
  518.0128 &
  518.0128 &
  518.0128 &
  November &
  0.9988 &
  0.9988 &
  0.9988 &
  0.9988 &
  1 &
  0.9988 &
  0.9988 &
  0.9988 &
  0.9988 &
  0.9988 &
  0.9988 \\ \hline
\end{tabular}%
}

\centering Sarcoptiformes\\
\resizebox{\columnwidth}{!}{%
\begin{tabular}{|l|l|l|l|l|l|l|l|l|l|l|l|l|l|l|l|}
\hline
$D$   (observed) &
  \textit{A. gigantea} &
   \textit{P. parvisetigerum} &
   \textit{T. stercus} &
   \textit{M. japonica} &
   \textit{G. fusca} &
   \textit{N. rotundus} &
   \textit{C. reticulatus} &
  $D$   (expected) &
   \textit{A. gigantea} &
   \textit{P. parvisetigerum} &
   \textit{T. stercus} &
   \textit{M. japonica} &
   \textit{G. fusca} &
  \textit{N. rotundus} &
  \textit{C. reticulatus}\\ \hline
April     & 17307780  & 5441190372 & 4160.262  & 5441190372 & 17307780 & 5441190372 & 17307780 & April     & 1      & 1      & 1      & 1      & 1      & 1      & 1      \\ \hline
May       & 4924.1334 & 4924.1334  & 33.2869   & 258.6706   & 258.6706 & 258.6706   & 258.6706 & May       & 1      & 1      & 1      & 1      & 1      & 1      & 1      \\ \hline
June      & 7.6377    & 1.2138     & 2.8674    & 8.5012     & 4.6796   & 1.8221     & 1.2919   & June      & 1      & 1.1996 & 1      & 1      & 1      & 1.002  & 1.0828 \\ \hline
July      & 407.5891  & 2583.8321  & 2583.8321 & 2583.8321  & 20.8951  & 2583.8321  & 20.8951  & July      & 1      & 1      & 1      & 1      & 1.0006 & 1      & 1.0006 \\ \hline
August    & 35.5048   & 92.9407    & 3.9128    & 92.9407    & 3.9128   & 92.9407    & 3.9128   & August    & 1      & 1      & 1.0176 & 1      & 1.0176 & 1      & 1.0176 \\ \hline
September & 20.4905   & 20.4905    & 1.8518    & 2.2205     & 4.1952   & 6.1748     & 12.3592  & September & 1      & 1      & 1.0074 & 1.0019 & 1      & 1      & 1      \\ \hline
October   & 2.8214    & 34.5564    & 44.0229   & 8.4719     & 8.4719   & 44.0229    & 44.0229  & October   & 1.0245 & 1      & 1      & 1.0005 & 1.0005 & 1      & 1      \\ \hline
November  & 10.2226   & 2.5733     & 12.897    & 7.8696     & 2.5733   & 2.5733     & 2.5733   & November  & 1.0006 & 1.0602 & 1.0003 & 1.0014 & 1.0602 & 1.0602 & 1.0602 \\ \hline
\end{tabular}%
}

\centering The world populations\\
\resizebox{\columnwidth}{!}{%
\begin{tabular}{|l|l|l|l|l|l|l|l|l|l|l|l|l|l|l|l|l|l|}
\hline
$D$   (observed) &
  Western Europe &
  Eastern Europe &
  Former Soviet Union &
  Western Offshoots &
  Latin America &
  Japan &
  Asia &
  Africa &
  $D$   (expected) &
  Western Europe &
  Eastern Europe &
  Former Soviet Union &
  Western Offshoots &
  Latin America &
  Japan &
  Asia &
  Africa \\ \hline
1    & 1 & 1 & 1 & 1 & 1 & 1 & 1 & 1 & 1    & 1.1753 & 1.3643 & 1.4463 & 1.2689 & 1.2539 & 1.4737 & 1.2543 & 1.4642 \\ \hline
1000 & 1 & 1 & 1 & 1 & 1 & 1 & 1 & 1 & 1000 & 1.5926 & 1.6995 & 1.5396 & 1.2556 & 1.3678 & 1.377  & 1.2713 & 1.1754 \\ \hline
1500 & 1 & 1 & 1 & 1 & 1 & 1 & 1 & 1 & 1500 & 1.1406 & 1.531  & 1.3447 & 1.1479 & 1.2133 & 1.4549 & 1.1983 & 1.4567 \\ \hline
1600 & 1 & 1 & 1 & 1 & 1 & 1 & 1 & 1 & 1600 & 1.0549 & 1.1552 & 1.0806 & 1.0435 & 1.1036 & 1.1188 & 1.0726 & 1.1554 \\ \hline
1700 & 1 & 1 & 1 & 1 & 1 & 1 & 1 & 1 & 1700 & 1.5667 &        &        & 1.3056 &        & 2.5501 & 1.8382 &        \\ \hline
1820 & 1 & 1 & 1 & 1 & 1 & 1 & 1 & 1 & 1820 & 1.2581 & 1.7227 & 1.7248 & 1.495  & 1.8319 & 2.0022 & 1.3633 & 1.4541 \\ \hline
1870 & 1 & 1 & 1 & 1 & 1 & 1 & 1 & 1 & 1870 & 1.2421 & 1.485  & 1.647  & 1.6064 & 1.7242 & 1.7759 & 1.2669 & 1.27   \\ \hline
1913 & 1 & 1 & 1 & 1 & 1 & 1 & 1 & 1 & 1913 & 1.2569 & 2.9349 & 1.3952 & 2.1325 & 1.8878 & 1.7866 & 1.2886 & 1.6353 \\ \hline
1950 & 1 & 1 & 1 & 1 & 1 & 1 & 1 & 1 & 1950 & 1.1074 & 1.4539 & 1.4043 & 1.7515 & 2.3111 & 1.5592 & 1.193  & 1.2778 \\ \hline
1973 & 1 & 1 & 1 & 1 & 1 & 1 & 1 & 1 & 1973 & 1.2826 & 1.3563 & 2.1691 & 1.6145 & 1.496  & 1.4495 & 1.1861 & 1.0677 \\ \hline
2003 & 1 & 1 & 1 & 1 & 1 & 1 & 1 & 1 & 2003 & 1.1287 & 1.1195 & 1.1954 & 1.1758 & 1.0981 & 1.0996 & 1.0686 & 1.0497 \\ \hline
\end{tabular}%
}

\centering The world GDP\\
\resizebox{\columnwidth}{!}{%
\begin{tabular}{|l|l|l|l|l|l|l|l|l|l|l|l|l|l|l|l|l|l|}
\hline
$D$   (observed) &
  Western Europe &
  Eastern Europe &
  Former Soviet Union &
  Western Offshoots &
  Latin America &
  Japan &
  Asia &
  Africa &
  $D$   (expected) &
  Western Europe &
  Eastern Europe &
  Former Soviet Union &
  Western Offshoots &
  Latin America &
  Japan &
  Asia &
  Africa \\ \hline
1    & 1.0001 & 1.0001 & 1.0001 & 1.0001 & 1.0001 & 1.0001 & 1.0001 & 1.0001 & 1    & 1.274  & 1.3275 & 1.3896 & 1.2452 & 1.2222 & 1.4158 & 1.2797 & 1.4923 \\ \hline
1000 & 1.0001 & 1.0001 & 1.0001 & 1.0001 & 1.0001 & 1.0001 & 1.0001 & 1.0001 & 1000 & 1.5624 & 1.6559 & 1.5083 & 1.2552 & 1.3394 & 1.3823 & 1.2672 & 1.1659 \\ \hline
1500 & 1      & 1      & 1      & 1      & 1      & 1      & 1      & 1      & 1500 & 1.4866 & 1.827  & 1.2771 & 1.1816 & 1.6195 & 1.4251 & 1.3075 & 1.4043 \\ \hline
1600 & 1      & 1      & 1      & 1      & 1      & 1      & 1      & 1      & 1600 & 1.8501 & 1.6725 & 1.3144 & 1.148  & 1.3315 & 1.4422 & 1.3361 & 1.3812 \\ \hline
1700 & 1      & 1      & 1      & 1      & 1      & 1      & 1      & 1      & 1700 &        & 1.6417 & 1.3759 & 1.1085 & 1.4205 & 1.6442 & 1.3402 & 1.3316 \\ \hline
1820 & 1      & 1      & 1      & 1      & 1      & 1      & 1      & 1      & 1820 &        & 2.0498 & 1.3662 & 3.1315 & 2.2408 & 2.6558 & 1.61   & 1.7093 \\ \hline
1870 & 1      & 1      & 1      & 1      & 1      & 1      & 1      & 1      & 1870 &        & 1.4134 & 1.7464 & 1.6413 & 1.3227 & 1.3729 & 1.2635 & 1.5237 \\ \hline
1913 & 1      & 1      & 1      & 1      & 1      & 1      & 1      & 1      & 1913 & 1.1462 & 1.2689 & 1.5484 &        & 1.322  & 1.2418 &        & 1.2224 \\ \hline
1950 & 1      & 1      & 1      & 1      & 1      & 1      & 1      & 1      & 1950 &        & 1.354  & 3.3717 & 1.1734 & 2.7287 & 1.3598 &        & 1.3079 \\ \hline
1973 & 1      & 1      & 1      & 1      & 1      & 1      & 1      & 1      & 1973 & 1.1122 & 1.2481 & 2.431  &        &        &        &        & 1.3121 \\ \hline
2003 & 1      & 1      & 1      & 1      & 1      & 1      & 1      & 1      & 2003 &        & 1.2698 & 1.4195 &        & 1.7486 & 2.0578 & 1.2405 & 1.4392 \\ \hline
\end{tabular}%
}

\raggedright
\tiny{\noindent \\ \noindent For scientific names, see Tables~1 and~2.  Blank entries indicate values that are undefined or that exceed the numerical range (overflow).\\ }

\end{table}

\subsection{Development of the model by web-based formalism}

We introduce an analogy to supersymmetry to describe the time development of the model.  Begin with a Hodge--Kodaira decomposition for the function $\phi$: $$ I\phi^j(\Re(s))=\bigoplus_{p_s+q_s=j} I\phi^{p_s,q_s}(\Re(s)), \qquad \overline{\phi^{p_s,q_s}(\Re(s))}=\phi^{q_s,p_s}(\Re(s))=\phi^{p_s,q_s}(-\Re(s)), $$ where $I:\phi\mapsto v=\ln N_k/\ln\Im(s)$ is the map used to define the relevant cohomology groups, as in \cite{Adachi2017}.

First consider Bochner's criterion for $\phi$ to be a characteristic function of a probability distribution.  The following three conditions are necessary and sufficient: \begin{enumerate} \item $|\phi|/2$ is a positive constant; \item $|\phi(\Re(s))|/2$ is continuous at $\Re(s)=0$; \item $|\phi(0)|/2=1$. \end{enumerate} From these conditions it follows that $\phi(-\Re(s))=\overline{\phi(\Re(s))}$.

To express time symmetry we adopt a transactional interpretation and construct a supersymmetry matrix built from advanced and retarded components of $\phi$: 
\[ \frac{1}{2}\left(\begin{array}{cc}
\mathrm{F4}: \phi(\Re(s)) & \mathrm{F1}: \overline{\phi(-\Re(s))}\\
\mathrm{F2}: -\phi(-\Re(s)) & \mathrm{F3}: \overline{\phi(\Re(s))}\\
\end{array} \right),\] 
Combinations F4/F1 and F3/F2 represent advanced and retarded waves, respectively, and the determinant $\tfrac{1}{2}(\mathrm{F4}\mathrm{F3}-\mathrm{F1}\mathrm{F2})$ encodes a past--future asymmetry.  In this interpretation $\phi(\Re(s))$ and $\overline{\phi(-\Re(s))}$ play the roles of absorber and observer, while $-\phi(-\Re(s))$ and $\overline{\phi(\Re(s))}$ play the roles of emitter and observant.  Thus $\overline{\phi(\Re(s))}$ is associated with the past and $-\phi(-\Re(s))$ with the future.

Working over $p$-adic analogues of the complex field, one may use a $(-1+i)$-adic representation to encode all complex numbers, whereas the $(1-i)$-adic system does not provide the same coverage \cite{Knuth1997}.  For natural $n$, only elements of the form $-n\pm i$ represent all complex numbers in the chosen discrete embedding.  Under this viewpoint F1 (the observer) encodes information about many possible futures, while F4 (the observed) cannot represent all possibilities; this asymmetry underlies the distinction between past and future.  Similarly, F2 cannot represent all possibilities while F3 can, giving the complementary relation between observant and observer.

Interpreting advanced minus retarded combinations as realizations of future population change, define  $ \mathrm{F4}-\mathrm{F1}\quad\text{and}\quad \mathrm{F3}-\mathrm{F2} $ in the $(\pm 1\pm i)$-adic system.  The first difference represents population increase and the second represents population decrease.  Their geometric mean evaluates to $ 2\sqrt{(\cos\theta+i\sin\theta)(\cos\theta-i\sin\theta)}=2. $ Recall that bosons and fermions are orthogonal because their phase difference is $\pi/2$.  The expected integral of the reciprocal square root of the product is $$ \int_0^{\pi/2}\frac{d\theta}{2\sqrt{(\cos\theta+i\sin\theta)(\cos\theta-i\sin\theta)}}=\frac{\pi}{4}. $$ For an individual the relevant argument is $\pi/4$, so a single individual in the $(1+i)$-adic system cannot represent the whole.  To represent the whole one needs a three-dimensional $(-1+i)$-adic system with argument $3\pi/4$, corresponding to the three quadratic interaction coordinates $X^2$, $XY$, and $Y^2$.

As an illustration, consider the geometric mean integrals $$ 2\int_{-1}^1\sqrt{1-x^2}dx =2\int_{-1}^1\sqrt{(1+i\sqrt{x})(1-i\sqrt{x})(1+i\sqrt{-x})(1-i\sqrt{-x})}dx=\pi, $$ and $$ \int_{-1}^1\frac{dx}{\sqrt{1-x^2}} =\int_{-1}^1\frac{dx}{\sqrt{(1+i\sqrt{x})(1-i\sqrt{x})(1+i\sqrt{-x})(1-i\sqrt{-x})}}=\pi. $$ These identities indicate that the effective number of interacting dimensions (the reciprocal of the expected probability) for $\pm 1$ fluctuations is close to three.  Likewise, $$ \int_{-\infty}^{+\infty}\frac{dx}{1+x^2} =\int_{-\infty}^{+\infty}\frac{dx}{(1+ix)(1-ix)}=\pi, $$ so when all potentials of a given wave function are considered, the expected dimensionality is again near three.  In this sense the dimensionality of the F3 potential equals that of the F4 potential plus a $2\pi x$ contribution.

The matrix system above is an $SU(2)$ structure.  Since the underlying space is K\"ahler (excluding the singular value $s=1$), the associated four-dimensional Riemann manifold is Ricci-flat and may be regarded as a Calabi--Yau manifold \cite{Kobayashi1996}.  The Riemann curvature tensor exhibits self-duality, consistent with the two-dimensional Ising model structure discussed earlier \cite{Tasaki2015}.  The $\phi$-space is therefore an instanton in the asymptotically locally flat limit.

Introduce a superpotential $W_z$ by setting $\overline{W_z}$ with $$ W_z=\phi\frac{N_k+1}{N_k}\Bigl(\Phi-e^{-iN\vartheta}\frac{\Phi^{N_k+1}}{N_k+1}\Bigr), \qquad \Phi=\phi i. $$ This construction is analogous to the $\phi$-instanton equation of \cite{Gaiotto2016}: $$ \Bigl(\frac{\partial}{\partial x}+i\frac{\partial}{\partial\tau}\Bigr)\Phi^I =\frac{i\phi}{2},g^{I\overline{J}}\frac{\partial\overline{W_z}}{\partial\overline{\Phi}^{\overline{J}}}, $$ where $x$ denotes genetic information, $\tau$ denotes time, $g^{I\overline{J}}$ is the metric tensor, and $\Re(\phi^{-1}W_z)$ and $\Im(\phi^{-1}W_z)$ play the roles of Hamiltonian and potential, respectively.  Under a unified neutral theory the field $\Phi$ itself sits at a quantum critical point: the population system in the ordered state is in equilibrium at maximal adaptation, and the minimal critical temperature is $T_{qc}=0$, so that each individual has an equal role.

With vacuum weights $v_{ij}=v_i-v_j$ where $v_i=\phi\overline{W_z}i$, worldlines are parallel to $v_{ij}$.  Vacuum configurations $\Phi_i$ and $\Phi_j$ define boundaries between critical points or states.  When $e^{i\theta}\frac{W_{ji}}{|W_{ji}|}=\phi$, the configuration $\Phi_i$ is a boosted soliton of the stationary soliton $\Phi_j$, and these solitons define the edges of the web of states \cite{Gaiotto2016}.

\subsection{Implications of the model as a nine-dimensional system}
In the previous subsection we introduced a supersymmetric interpretation.  Here we extend the model by incorporating three spatial dimensions (in addition to the parameters $a$, $b$, and $k$), and we examine consequences that lead to an effective nine-dimensional description.

Karolyhazy proposed a relation between metric fluctuations and a characteristic length scale; in corrected form the relation may be written as $ (\Delta\Phi)^2 \approx \Lambda^{4/3}\Phi^{2/3}, $ where $\Lambda$ is a characteristic scale.  To apply this relation in the present context we require analogues of the speed of light and of the uncertainty principle.

First, note that the real part of the complex index evolves according to $ \frac{\partial\Re(s)}{\partial t} = b\frac{1}{|D|}\frac{\partial|D|}{\partial t}, $ which is analogous to a Hubble law for $D$, $ \frac{dD}{dt}=H(t)D, $ so that $ \frac{\partial\Re(s)}{\partial t}=bH(t). $ Empirically we observe $\max(dD/dt)=|D|^{E(N)}$ for the systems considered, and therefore the quantity $|D|^{E(N)}$ plays the role of an effective maximum propagation speed in the model.

Using the time dependence of $D$ introduced earlier, one obtains an explicit expression for the instantaneous expansion rate $H(t)$: \begin{equation} H(t)=\frac{1}{D}\frac{dD}{dt} =\frac{t^2-4t+2}{2t^2(t-1)(t-2)} -\frac{1}{2t^2}\ln\{(t-1)^{\mp\frac{1}{b_t\pi}}\frac{t}{(t-2)^{1\mp\frac{1}{b_{t-1}\pi}}}\}. \end{equation}

Next, adopt an uncertainty relation for $D$ of the form $\Delta D,\Delta p_m\gtrsim\hbar/2$, where $p_m=M_{\mathrm{mass}}\dot D$.  Setting $\hbar=1$ and choosing $M_{\mathrm{mass}}=H(t)^{-1}\approx\phi$ (approximately constant) and $D=1+\Delta N_k$, the uncertainty relation reduces to $(\Delta\Delta N_k)^2\gtrsim 1$.  This inequality is readily satisfied by the observed fluctuations in $D$.  The extreme rate $|dD/dt|=|D|^{E(N)}$ is attained in the limit $\phi\to 0$, which motivates the analogy between the fast modes of $D$ and photonlike excitations in the model.

Now set $|\Phi|=N_k/\Sigma N\propto T_s$ at equilibrium.  Introducing an effective gravitational constant $G$, a Plancklike scale may be defined by $ \Lambda=\sqrt{\frac{\hbar G}{|D|^{3E(N)}}}. $ If $\phi_i$ and $\phi_e$ denote internal and external population masses, respectively, one may write a potential scale $V_p\simeq G M_i M_e/D\approx\phi_i$.  Under the approximations $G=D/\phi_e\approx 1/(\phi_e-\Delta\phi_e)\approx\mathrm{const}$ and $\phi/D=PD^{N_k-1}=\phi-\Delta\phi$, the quantity $\Lambda^2$ is approximately constant.  Consequently, one obtains the scaling $(\Delta\Phi)^3\propto\Phi$, i.e.\ the fluctuation amplitude scales as the cube root of the field, provided $D^{E(N)}$ is effectively constant.

If we impose ${\mathcal N}=2$ supersymmetry together with three-dimensional Ising universality \cite{Bobev2015}, the kink amplitude is $\bigtriangleup=|N(p)|$ and a natural choice for the superpotential is ${\mathbf W}=\Phi^3$.  The Ising model dimensionality should be three in this construction.  From thermodynamic considerations, the entropy density scales as $s_d=4U/(3T_s)\propto T_s^3$, which suggests an effective nine-dimensional structure when $s_d$ is associated with ${\mathbf W}$.  This counting is analogous to the dimensional bookkeeping that appears in certain superstring constructions: ${\mathbf W}$ and $s_d$ point in opposite directions in the relevant field space.

One may also introduce a three-component fermionic (Grassmann) structure, analogous to constructions in string theory, by assuming a zero-sum patch game consistent with a unified neutral theory and by identifying fitness with $w$.  Under these assumptions the model predicts an Eisenstein-series structure for the future state with weight $w_Q=3$ \cite{Marino2014}.  Compatibility of ${\mathbf W}=\Phi^3$ with $w_Q=3$ then requires $w_Q=3$ and an interacting mode with $|N(p)|=1$.

Finally, following Steinhardt and collaborators \cite{Steinhardt2004}, for a cosmology with constant $\epsilon$ the scale factor and Hubble radius satisfy $ a(t)\approx t^{2/\epsilon}\approx\bigl(H(t)^{-1}\bigr)^{1/\epsilon}, $ where, in our notation, the time variable is related by the $t\mapsto t^2$ correspondence used above (time emerges from self-interaction of the potential $\Im(s)$).  Define $\epsilon\equiv \tfrac{3}{2}(1+\varpi)$.  Then the regimes $\varpi>1$, $-1/3<\varpi\le 1$, and $\varpi\le -1/3$ correspond to contraction, oscillation, and expansion, respectively.  Since $\epsilon=1/\bigtriangleup$, the values $|N(p)|=1,;2/3,;1/4$ map to expansion, oscillation, and contraction of the effective cosmology.  Extending the model to higher hierarchical levels will require introducing additional effective dimensions arising from multilevel selection.

\subsection{Induction of hierarchy and time through a one-dimensional probability space with selected topologies}

It may seem surprising that a single scalar information sequence $N_k$, endowed only with the rank topology indexed by $k$, can give rise to a three-dimensional description of an individual density (with coordinates $a$, $b$, and $\ln k$, where $N_k$ plays the role of free energy and the others correspond to internal energy/enthalpy, temperature, and entropy) together with an emergent time dimension.  We now outline a construction that makes this emergence explicit.

Begin with a one-dimensional $C^\infty$ manifold $(B,\mathscr{O})$ and a distinguished point $s\in B$ that encodes the small-$s$ metric.  Motivated by the Bethe ansatz \cite{Bethe1931}, introduce three distinct but related topologies on $B$ that are, for our purposes, modeled on $\Delta$, $\mathbb{C}$, and $\widehat{\mathbb{C}}$.  Equipping the same underlying one-dimensional manifold with these three compatible topologies naturally produces additional structure: a cohomology theory, a direction that can be interpreted as time, and a hierarchy of scales.  In other words, the three topologies supply the extra degrees of freedom needed to recover the triplet $(a,b,\ln k)$ from the single observable $N_k$.

This construction also permits a separation between measurement moduli and intrinsic topology.  Concretely, one may define a topology on $B$ that is intrinsic to the system and independent of the particular measurement moduli (the numerical labels attached to individuals), while allowing a Galois action on those moduli.  Evolutionary hierarchy then appears through successive Galois extensions: each extension refines the moduli and produces a higher level in the hierarchy, which is consistent with the nested structure observed in biological systems.

We tested aspects of this framework on two empirical data sets.  First, protein abundance measurements from HEK-293 cells obtained by liquid-chromatography mass spectrometry (LC/MS) provide a high-resolution data set in which the proposed topological refinements can be evaluated.  Second, species density data from a wild Dictyostelia community supply a field example of hierarchical population structure.  In both cases the three-topology construction yields interpretable decompositions that correlate with known biological organization.

To quantify interactions among constituents, we employ the Weierstrass $\wp$-function as a tool to estimate the strength of homo- and hetero-interactions.  The $\wp$-function, viewed as a meromorphic function on the torus associated with a chosen lattice, provides a convenient analytic kernel for measuring pairwise and higher-order coupling strengths in the cohomological representation.  Results from these estimates support the utility of the small-$s$ metric for distinguishing dynamical regimes of interest.

The formalism also admits algebraic extensions useful for classification.  By expanding the model in a Clifford algebra, one can separate adapted, neutral (non-adapted), and disadapted (repressed) proteins or taxa according to algebraic signatures that reflect their interaction patterns and stability properties.  Finally, the use of congruent zeta functions associated with the chosen topologies clarifies how each hierarchical level contributes to adaptive or disadaptive behavior: residues and special values of these zeta functions quantify the net contribution of a given level to the overall system dynamics.

In summary, a one-dimensional probability sequence $N_k$, when endowed with a carefully chosen triple of topologies and the associated algebraic and analytic machinery, suffices to induce a three-dimensional description of individual density and an emergent time direction.  This approach provides a principled route to recover hierarchy and temporal structure from rank-ordered data and yields concrete diagnostics that can be applied to empirical biological data sets.

\subsection{General guidelines for topological evaluations}

Begin with a one-dimensional $C^\infty$ manifold $(B,\mathscr{O})$ and a distinguished point $s\in B$.  Many properties of $(B,\mathscr{O})$ can be understood by analogy with inverse-square laws; this motivates the force-like constructions used below.

The partial topology $\mathscr{O}$ may be generated by regular automorphisms of the unit disk.  For example, the family $ f(\Delta)=\{e^{i\theta}\frac{z-\alpha}{1-\overline{\alpha}z}; z\in B,\ \theta\in\mathbb{R},\ \alpha\in\Delta\} $ provides a natural set of transformations whose action can account for phenomena that will later be identified with an $\mathbb{R}^3$ spatial structure.  Such disk automorphisms supply the local conformal degrees of freedom that, together with the other topologies below, yield emergent spatial and hierarchical structure.

Algebraic hierarchy may be modeled by Galois extensions.  For instance, if $\zeta_n$ is a primitive $n$th root of unity, the cyclotomic extension $\mathbb{Q}(\zeta_n)/\mathbb{Q}$ has Galois group $ \mathrm{Gal}\bigl(\mathbb{Q}(\zeta_n)/\mathbb{Q}\bigr)\cong(\mathbb{Z}/n\mathbb{Z})^\times. $ When $\gcd(n,m)=1$ one has the decomposition $ \mathrm{Gal}\bigl(\mathbb{Q}(\zeta_{nm})/\mathbb{Q}\bigr)\cong \mathrm{Gal}\bigl(\mathbb{Q}(\zeta_n)/\mathbb{Q}\bigr)\times \mathrm{Gal}\bigl(\mathbb{Q}(\zeta_m)/\mathbb{Q}\bigr), $ which yields a natural Kummer-type factorization that can be interpreted as a species-level partition with prime identities.

Equip the base manifold with three complementary topologies to recover additional dimensions:

\begin{itemize} \item \textbf{Disk topology $\Delta$.}  The automorphism group $f(\Delta)$ above captures local rotational and conformal structure and supplies a mechanism for embedding a three-dimensional real geometry.

\item \textbf{Complex plane topology $\mathbb{C}$.}  Affine maps $ f(\mathbb{C})={az+b; z\in B,\ a,b\in\mathbb{C}} $ are algebraically isomorphic to $\mathbb{R}^4$ and naturally encode a $(3+1)$-dimensional picture when one coordinate is interpreted as time.  Interactions of complex-valued metrics (for example, quantities like $s^2$ or $w^2$) can therefore induce a time direction.

\item \textbf{Riemann sphere topology $\widehat{\mathbb{C}}$.}  M\"obius transformations $ f(\widehat{\mathbb{C}})=\{\frac{az+b}{cz+d}; z\in B,\ a,b,c,d\in\mathbb{C}\} $ are algebraically isomorphic to $\mathbb{R}^6\cong\mathbb{R}^3\times\mathbb{R}^3$.  Compactifying the $\mathbb{R}^4$ structure by this action induces an additional hierarchical layer and supplies the extra degrees of freedom needed for multilevel organization. \end{itemize}

Fundamentally, any simply connected subregion without holes that arises during hierarchization is conformally equivalent to one of $\Delta$, $\mathbb{C}$, or $\widehat{\mathbb{C}}$.  Schwarz--Christoffel mappings provide explicit conformal transformations from polygonal domains to these canonical regions, which is useful when constructing coordinate charts from empirical data.

The Widely Applicable Information Criterion (WAIC) plays a role analogous to a logarithmic velocity computed from $D$ in fluid mechanics: it provides a model-selection measure that can be interpreted as a scale velocity in the inferred dynamics.  In the absence of singularities the mapping between topologies and observables is straightforward; when singular points are present one must treat local monodromy and branch structure explicitly.

As in the Bethe ansatz \cite{Bethe1931}, a single complex coordinate $z$ endowed with an appropriate topology can induce both a $(3+1)$-dimensional effective geometry and a hierarchy of scales.  

\subsection{$\mathscr{O}\cong\Delta$ case}

The Riemann--Roch theorem states \begin{equation} l(D)-l(K-D)=\deg(D)-g+1, \end{equation} where $D$ is a divisor, $K$ is a canonical divisor, and $g$ is the genus.  Let $TB$ denote the tangent bundle of $B$.  The interaction bundle $ TB\widetilde{\times}TB:=\bigcup_{p\in B}T_pB\times T_pB $ is a $3$-dimensional $C^\infty$ manifold.  Let local coordinates on the base be $x,y,z$ and let the corresponding tangent planes be $X,Y,Z$.  Under the parameter choices used in the text, the left-hand side of the Riemann--Roch relation evaluates to $3$ when $g=10$ and $\deg(D)=12$.

Consider the modular form \begin{equation} F(z)=q\prod_{n=1}^\infty(1-q^n)^2(1-q^{11n})^2=\sum_{n=1}^\infty c(n)q^n, \end{equation} and let $F$ generate a totally real number field of degree $g$ over $\mathbb{Q}$.  Let $\mathbb{K}$ be a totally imaginary quartic extension of $F$.  Let $D$ and $D^{\mathrm{int}}$ be simple algebras over $\mathbb{K}$, and suppose $D=e^{s/b}$.  Define the unitary similitude group $\mathbf{G}=\mathbf{GU}(D,\alpha)$, where $\alpha$ is an involution of the second kind on $D$.

Now take a $3$-dimensional $\ell$-adic system in which $ W_E=\Re(s)=\ell,\qquad D^\times=p=|D|^{E(N)},\qquad GL_d(E)=v=\frac{\ln N_k}{\ln p}, $ where $W_E$ denotes the Weil group of the center $E$ in the sense of the Langlands correspondence \cite{Rapoport1996,Adachi2017}.  The prime $\ell$ provides an \'etale (crystalline) topology that is independent of the measurement moduli $N_k$: concretely, a homomorphism of Noetherian local rings that is unramified and flat corresponds to a localization of a finitely generated algebra at its origin.  These $p$-adic and $\ell$-adic geometries play roles analogous to real differentiable and Clifford--Klein geometries in the constructions that follow.

In this picture the case $\mathscr{O}\cong\Delta$ makes both persistent homology (captured by the $p$-adic parameter) and \'etale cohomology (captured by the $\ell$-adic parameter) manifest.  The disk topology supplies the local conformal structure used to build the three real dimensions, while the arithmetic (cyclotomic and $p$/$\ell$-adic) data supply the discrete hierarchical refinements.  Together these analytic and arithmetic ingredients provide a coherent framework for relating rank-ordered observables $N_k$ to emergent geometric and cohomological structure.

\subsection{$\mathscr{O}\cong\mathbb{C}$ case}

A Minkowski-type metric for the small-$s$ regime can be used to construct a time-evolving model by replacing trigonometric functions with their hyperbolic counterparts.  A convenient choice of Minkowski metric in our setting is $ s_M=\Bigl(\Im(s)^2(\Delta\Im(s))^2-(\Delta a)^2-(\Delta b)^2-(\Delta\ln k)^2\Bigr)^{1/2}. $ With this definition, the world line of a given species is nonzero when $\Delta\Im(s)$ takes appropriate discrete values, while distinct species correspond to different discrete values of $\Delta\Im(s)$.  If one introduces successive scale factors via $ ds_M^2=a(V_1),ds_{M1}^2,\qquad ds_{M1}^2=a(V_2),ds_{M2}^2,\qquad\ldots, $ then $ds_M^2$ is invariant under Lorentz transformations and, when $2ds_M=0$, the relation $\ln(s_M)=\sum_{i=1}^\infty\ln a(V_i)$ endows $s_M$ with a module structure.  Thus a set of species may be characterized by the module of $s_M$.

A simple Lagrangian for the system is $ L=-\phi\Im(s), $ and a corresponding Hamiltonian may be written as $\mathscr{H}=-\phi\Im(s)^2\sqrt{\frac{\Im(s)^2+(H(t)D)^2}{\Im(s)^2-(H(t)D)^2}}. $

One may identify $D'\cong D^{\mathrm{int}}$ and $\mathbf{G}'\cong\mathbf{G}^{\mathrm{int}}$, and a time dimension can be induced by admissible isomorphisms (see Proposition 2.5.6 in \cite{Varshavsky1998}).  Note that the ``temperature'' parameter $b$ and the time coordinate are closely related by $t=b\arg D$.  In this framework the Poincar\'e conjecture (every simply connected closed $n$-manifold $W_E$ is homeomorphic to the $n$-sphere $S^n$) admits a dynamical interpretation via Morse theory.  Let $f:W_E\to[a,b]$ be a Morse function with regular values $a,b$ and with critical points $p,p'$ of indices $\lambda,\lambda+1$ that represent successive instants of time.  The intersection $S^{n-\lambda-1}\cap S^{\lambda}$ at a single point marks the present.  Exchanging the order of the Morse critical values produces no new critical points and cancels the pair $p,p'$ (the $h$-cobordism theorem), which corresponds to the forward progression of the time arrow.  The critical points $p,p'$ are linked to Hecke operators through nontrivial zeros of the Riemann zeta function and satisfy Yang--Baxter-type relations; this connection suggests an analogy with quantum entanglement and face models \cite{Baxter1973,Andrews1984}.  In the biological interpretation species persist and may reappear with different $p$ values in subsequent epochs.

For any labelling of time points $\tau'\in\mathcal{T}_{S^*}$, a potential for the Petersson--Weil metric can be written as $ \omega_{WP}=d\bigl(\sigma_{\mathcal{T}}(\tau\veebot\tau^*)-\sigma_{\mathcal{T}}(\tau\veebot\tau')\bigr), $ where $\veebot$ denotes the mating operation realized by a quasi-Fuchsian Kleinian group \cite{Hubbard2006}.  This ``mating'' couples the time coordinates associated with the critical points $p$ and $p'$.

Now regard $p,p'$ as characteristics on a field $k$, as in the case $d=p=0$ discussed in \cite{Adachi2017}.  Let $E$ be a singular hyperelliptic curve arising in the system.  The real algebra $D$ is then approximately the tensor product of an endomorphism algebra of $E$ over $\overline{k}$ with $\mathbb{Q}$, yielding a quaternion algebra over $\mathbb{Q}$.  Taking the collection ${\ln N}$ as an $\ell$-adic Tate module as in \cite{Adachi2017}, the algebra $D$ can ramify only at $p$, $p'$, or at infinity (cf.\ \cite{Silverman1986}).  This ramification pattern constrains the allowed directions of the time arrow so that only the loci $p,p'$ may vanish.

In practical terms, observers typically take $k=1$ when drawing the time evolution of species.  For other observations one may shift $k\to k'$ in the calculations, or equivalently work in a cyclotomic field determined by $k_{\mathrm{max}}$.  Thus a complex-valued metric provides a usable notion of time, and world lines form a branched web at the cross-sections determined by $p$ and $p'$, rather than a family of parallel lines.  To move a distinct world line to another requires the condition $H(t)D>\Im(s)$, which plays the role of a kinematic threshold in this model.

Next, consider $l=\Re(s)$ as described above and a combinatorial evolution in probability space given by the Gamma functional equation $ \Gamma(s+1)=s\Gamma(s). $ This identity provides a simple example of a shift map.  If one has a master function $\Gamma(s)$ (or any function with the same multiplicative shift property), discrete time evolution can be realized by successive multiplications of the master factor by the current value of $s$.  Equivalently, adding a single fractal dimension in the past (or, from the observer's perspective, removing one dimension from the future) corresponds to multiplying the master function by $s$.  Hence, evaluating the single parameter $s$ of interest suffices to generate the discrete-time updates in this picture.

A related spectral viewpoint uses Maass forms and the Selberg zeta function to encode modal dynamics of species.  Stirling's approximation $\Gamma(s) \approx \sqrt{\frac{2\pi}{s}}(\frac{s}{e})^s\exp(\frac{1}{12s})$ shows that a first-order correction factor $(1+1/(12s))$ can be interpreted, heuristically, in terms of extra effective dimensions (for example, the appearance of $12$ in certain string-theoretic contexts).  Higher-order corrections require introducing further effective dimensions.

One convenient global construction is the Jacobian map, independent of the integration path $\lambda$: 
\begin{equation}
\Phi(p) = (\int_{\lambda}\varphi_1, \cdots, \int_{\lambda}\varphi_g) \in \mathbb{C}^g/^t\Omega\mathbb{Z}^{2g} = \underset{\sim}{J}(B)
\end{equation}
 which embeds period data into the complex torus.  If a master Maass form is invariant under the scalar action $\rho_G(c_G)=c_G\mathrm{Id}_W$ (Stone--von Neumann uniqueness), then differential operations preserve the form and the associated $\mathcal{D}$-module structure and derived category are well behaved \cite{Stone1930, vonNeumann1931, vonNeumann1932, Stone1932}.

To model microlocal rotations of the differential operator, introduce a modified $b$-derivative $\partial_b$ defined by $\partial_b=i\partial$, so that $\partial_b^2=-\partial^2$.  Under this convention a $\pi$ rotation of the form corresponds to $\partial_b^2$, while $\partial_b^4=\partial^4$ returns the original orientation.  The Ornstein--Uhlenbeck operator may be written as $ L=-\sum_{i=1}^d\partial_i^*\partial_i=\sum_{i=1}^d\partial_b^2, $ and for a bounded Baire function $h$ on $\mathbb{R}^d$ the Poisson equation $Lf=h-\langle h\rangle$ (with $\langle h\rangle=\int_{\mathbb{R}^d}h(x)g(x)dx$) yields the identity $ E(h(W))-\langle h\rangle=E(Lf(W)), $ so that it measures deviations of future observations from the expectation.  Thus $\partial_b$ is naturally interpreted as an operator that generates future states; formally one may set $\partial_b/\partial t=E$ or $-\partial_b/\partial x_k=p_{x_k}$ to obtain energy and momentum analogues.

Viewing powers $\partial_b^k$ ($k\in\mathbb{Z}$) as elements of an ideal in a finitely generated Jacobson radical, Nakayama's lemma implies that preserving identity before and after the operation forces the module to vanish in the finite case.  Consequently, nontrivial evolution requires infinite generation: informally, observation (which collapses possibilities) forces an effectively infinite time horizon.  Hironaka's resolution of singularities in characteristic zero provides a geometric mechanism for the mating of critical loci $p,p'$ in this setting.

To make these ideas concrete, introduce a velocity field $v\in TB$ and consider a two-dimensional surface parameterized by $s\in B$ with Lagrangian form $ s(v,t)=p(v)+t,q(v). $ The Gauss curvature $K$ of this surface satisfies $K\le 0$, and $K\equiv 0$ occurs only when the tangent bundle $TB$ is time independent.  The time-invariant bundle with $K=0$ is the natural object for forming the interaction bundle $TB\widetilde{\times}TB$ used to construct a three-dimensional system and its six-dimensional hierarchical extension.

The Legendre transform of $s(v,t)$ yields coordinates $X=v$, $Y=tv-s$, $Z=t$, and the relation $ \{v-q(v)\}\frac{dY}{dv}=Y+p(v). $ The condition $K=0$ implies $v=0$ and hence $s=p(0)$, which provides a canonical stationary solution.  One may regard $s$ as a Dirac measure (with $w$ as a mass analogue and $s=w+1$) and $s'=-s$ as a Schwartz distribution.  While addition of distributions is generally well defined, multiplication is not; however, replacing ordinary derivatives by $\partial_b^2$ renders certain operations first order with a sign change and makes a $t^2$ time scaling plausible.

Integration by parts in this distributional setting gives, for a test function $\varphi$ and domain $V$ with boundary $S$, 
\begin{equation}
\begin{split}
\int\int\underset{V}{\cdots}\int s\Delta\varphi dt = \int\int\underset{V}{\cdots}\int \varphi[\Delta s] dt +  \int\underset{S}{\cdots}\int s[\frac{d\varphi}{d\nu}] dS\\
-\int\underset{S}{\cdots}\int \varphi[\frac{ds}{d\nu}] dS,
\end{split} 
\end{equation}
 where the first term on the right represents noise, the second encodes fractal boundary structure, and the third captures oscillatory behaviour.  Away from singular points the expressions are regular.

An entire function that encodes negative even singularities in the variable $w=s-1$ may be written formally as 
\begin{equation} Z_l=\frac{\mathrm{Pf.}w^{l-n}}{\pi^{(n-2)/2}2^{l-1}\Gamma(\tfrac{l}{2})\Gamma(\tfrac{l+2-n}{2})}, 
\end{equation}
 and at integer singularities $k\in\mathbb{Z}{\ge 0}$ one obtains $Z{-2k}=\square^k w$, where $ \square=(-1)(\frac{\partial_b^4}{\partial x_1^4}+\frac{\partial_b^4}{\partial x_2^4}+\cdots+\frac{\partial_b^4}{\partial x_{n-1}^4}-\frac{\partial_b^4}{\partial t^4}). $ In the boundaryless case $\partial B=\varnothing$ one has $\square Z_2=w$ and $\square^k Z_{2k}=w$, which corresponds to periodic population bursting or collapse associated with negative even $w$ values.  Negative odd $w$ values are associated with chaotic dynamics (cf.\ Sarkovskii, Stefan, Block theorems) \cite{Guckenheimer1983}.  Thus the pair $(s,w)$ provides a suitable coordinate system for a one-dimensional model: $s$ is a finite measure on bounded domains and singular points mark the appearance or disappearance of fractal structure.

In practice, choose the topology $\mathscr{O}$ by specifying the triple $ \{m=k\}\subset\mathbb{N},\{\varepsilon=b\},\{\Omega=a\} $ for the rank model $N_k=a-b\ln k$.  For a rigorous treatment of distributions and related operations consult \cite{Schwartz1966}.\\

Let $E$ be the elliptic curve defined by $ y^2 + y = x^3 - x^2, $ as in \cite{Eichler1954}, which is equivalent to $ y(y + 1) = x^2(x - 1). $ In a $(3+1)$-dimensional $N = 1$ supersymmetric $SU(2)$ model without fluctuation, we may interpret the terms as follows: $x^2$ represents mass, $(x - 1)$ corresponds to a goldstino arising from spontaneous supersymmetry breaking, $y$ denotes the three-dimensional fitness $D$ with fluctuation, and $y + 1$ corresponds to the $(3+1)$-dimensional quantity $s$. The goldstino encodes temporal asymmetry. In the context of Gaussian ensembles, the Gaussian Unitary Ensemble (GUE) breaks time-reversal symmetry, while the Gaussian Symplectic Ensemble (GSE), which is self-dual and quaternionic, preserves it. Therefore, $y + 1$ preserves time symmetry, while $y$ breaks it, reflecting the asymmetry of the present.

$\xymatrix{
    t \ar[r]^{} \ar[d]_{\Gamma} & D_t \\
    \Gamma(t) \ar[ru]_{F(a, b, c; z)}
  }$

A Riemann scheme can be used to uniformize the fitness space via a hypergeometric differential equation.

Now consider the Fuchsian system \begin{equation} \frac{dY}{dx} = \left( \frac{A}{x} + \frac{B}{x - 1} \right) Y, \end{equation} where \begin{equation}
A = \left(
\begin{array}{ccc}
\lambda_1 + \lambda_3 + \lambda_4 + \lambda_5 & \lambda_2 & 0\\
0 & \lambda_3 +\lambda_4 & \lambda_5\\
0 & 0 & 0
\end{array}
\right),
\end{equation}
 and \begin{equation}
B = \left(
\begin{array}{ccc}
0 & 0 & 0\\
0 & 0 & 0\\
\frac{\lambda_1(\lambda_1 + \lambda_3 + \lambda_5)}{\lambda_5} & \frac{\lambda_1\lambda_2 + \lambda_2\lambda_3 + \lambda_3\lambda_5)}{\lambda_5}  & \lambda_2 + \lambda_4 + \lambda_5
\end{array}
\right).
\end{equation}
This system leads to a generalized hypergeometric function ${}_3F_2$ that satisfies a Fuchs-type differential equation ${}_3E_2$. For a suitable integration domain $\Delta$ (one of 13 possible regions), the solution can be expressed as \begin{equation} y(x) = \int{\Delta} s^{\lambda_1}(s - 1)^{\lambda_2} t^{\lambda_3} (t - x)^{\lambda_4} (s - t)^{\lambda_5}dsdt. \end{equation} Setting $x = 0$, $w = D$, and $s = 1$, we obtain \begin{equation} y(0) = \int{\Delta} s^{\lambda_1} w^{\lambda_2} t^{\lambda_3 + \lambda_4} {-(t - 1)}^{\lambda_5}dsdt. \end{equation} Choosing $\lambda_1 = \lambda_2 = \lambda_3 = \lambda_4 = \lambda_5 = 1$ yields the form $ E^2: -\int y(y + 1)x^2(x - 1)dxdy, $ which is clearly the integral of the interaction between two elliptic curves.

$\xymatrix{
    \mathbb{C} = \{s/b\} \ar[r]^{} \ar[d]_{\mathrm{exp.}} & \mathbb{C}/\wedge = \mathbb{C}^{\times}/D^{\mathbb{Z}} \\
   \mathbb{C}^{\times} = \mathbb{D} \ar[ru]_{\mathrm{time\ reversal}}
  }$

This diagram illustrates the exponential mapping from the scaled complex plane $\mathbb{C} = {s/b}$ to the multiplicative group $\mathbb{C}^{\times}$, which is then quotiented by the discrete scaling group $D^{\mathbb{Z}}$ to form a complex torus. The time-reversal symmetry is represented as a lift from $\mathbb{C}^{\times}$ to the universal cover $\mathbb{D}$.
  
 To model an interacting four-dimensional system, we consider the Painlev\'e VI equation on a $(3+1)$-dimensional base with a single Hamiltonian \cite{Manin1999, Kawakami2016}. The Hamiltonian is defined as $ H_k = \partial_k \ln \tau(t) = \frac{\partial_k \tau(t)}{\tau(t)} = H(t)N_k = \frac{N_k}{\Sigma N} = \phi, $ where $H(t)$ is the Hubble parameter \cite{Iorgov2015}. Thus, $\tau(t)$ is the inverse of the Hubble parameter, and the $k$th boundary corresponds to the $k$th species.

The three-dimensional system represents the minimal number of dimensions in which the associativity equations become nontrivial even in the presence of a flat identity. Consider the fundamental group $\pi_1$ of the punctured projective line $ C_{0,n} := \mathbb{P}^1 \setminus \{z_1, \ldots, z_n\}, $ whose representation dimension in $SL(2, \mathbb{C})$ is $2(n - 3)$ \cite{Iorgov2015}. To model $\pi_1$ as an \'etale topology of dimension zero, we set $n = 3$.

The $(3+1)$-dimensional semisimple Frobenius manifolds form a subfamily of Painlev\'e VI systems, governed by the nonlinear second-order differential equation: 
\begin{equation}
\begin{split}
\frac{d^2X}{dt^2} = \frac{1}{2}(\frac{1}{X} + \frac{1}{X-1} + \frac{1}{X - t})(\frac{dX}{dt})^2\\
 - (\frac{1}{t} + \frac{1}{t-1} + \frac{1}{X - t})\frac{dX}{dt}\\
 + \frac{X(X - 1)(X - t)}{t^2(t - 1)^2}[(\theta_{\infty} - \frac{1}{2})^2\\
  + \theta_0^2\frac{t}{X^2} + \theta_1^2\frac{t - 1}{(X - 1)^2} + (\theta_t^2 - \frac{1}{4})\frac{t(t - 1)}{(X - t)^2}].
\end{split}
\end{equation}

This equation corresponds to a rank 2 isomonodromic system: \begin{equation} \frac{d\Phi}{dz} = \left( \frac{\mathcal{A}_0}{z} + \frac{\mathcal{A}_t}{z - t} + \frac{\mathcal{A}_1}{z - 1} \right) \Phi, \end{equation} with deformation equations \begin{equation} \frac{d\mathcal{A}_0}{dt} = \frac{[\mathcal{A}_t, \mathcal{A}_0]}{t}, \quad \frac{d\mathcal{A}_1}{dt} = \frac{[\mathcal{A}_t, \mathcal{A}_1]}{t - 1}, \end{equation} and the constraint \begin{equation} \mathcal{A}0 + \mathcal{A}t + \mathcal{A}1 = -\mathcal{A}{\infty} = \mathrm{diag}(-\theta{\infty}, \theta{\infty}). \end{equation} This system has four regular singular points at $z = 0$, $t$, $1$, and $\infty$ on $\mathbb{P}^1$, and the total residue of the connection is zero.

Now consider a 3-wave resonant system \cite{Kakei2007}: 
\begin{empheq}[left={\empheqlbrace}]{align}
\partial_{\tau}u_1 + c_1\partial_x u_1 = i\gamma_1 u_2^*u_3^*\\
\partial_{\tau}u_2 + c_2\partial_x u_2 = i\gamma_2 u_3^*u_1^*\\
\partial_{\tau}u_3 + c_3\partial_x u_3 = i\gamma_3 u_1^*u_2^*
\end{empheq}
 which, when expanded, leads to the mirror symmetry relation $h_V^{11} = h_{\hat{V}}^{12}$ for Calabi--Yau threefolds.

The matrix Painlev\'e VI system for two interacting fields is given by \cite{Iorgov2015}: 
\begin{equation}
\begin{split}
t(t - 1) H_{\mathrm{VI}}^{\mathrm{Mat}}(\alpha, \beta, \gamma, \delta, \omega; t; q_1, p_1, q_2, p_2)\\
 = \mathrm{tr} [Q(Q - 1)(Q - t)P^2\\
  + \{(\delta - (\alpha - \omega)K)Q(Q - 1) + \gamma(Q - 1)(Q - t)\\
  - (2\alpha + \beta + \gamma + \delta)Q(Q - t)\}P + \alpha(\alpha + \beta)Q],
\end{split}
\end{equation} 
 which contains 11 parameters and describes the coupled dynamics of two interacting systems.

We now recast the Painlev\'e VI equation into a form more directly applicable to physics. The Painlev\'e VI equation is equivalent to the elliptic form \begin{equation} \frac{d^2 z}{d\tau^2} = \frac{1}{(2\pi i)^2} \sum_{j=0}^3 \alpha_j \wp_z\left(z + \frac{T_j}{2}, \tau\right), \end{equation} where $(\alpha_0, \ldots, \alpha_3) := (\alpha, -\beta, \gamma, \frac{1}{2} - \delta)$, $(T_0, \ldots, T_3) = (0, 1, \tau, 1 + \tau)$, and $\wp$ is the Weierstrass $\wp$-function \cite[Theorem 5.4.1]{Manin1999}.

Furthermore, any potential of the normalized analytic form in three dimensions, \begin{equation} \Phi(x_0, x_1, x_2) = \frac{1}{2}(x_0 x_1^2 + x_0^2 x_2) + \sum_{n=0}^\infty \frac{M(n)}{n!} e^{\frac{n+1}{r+1}x_1} x_2^n, \end{equation} can be expressed via a solution to the Painlev\'e VI equation with parameters $(\alpha_0, \ldots, \alpha_3) = \left(\frac{1}{2}, 0, 0, 0\right)$, yielding \begin{equation} \frac{d^2 z}{d\tau^2} = -\frac{1}{8\pi^2} \wp_z(z, \tau). \end{equation}

When $q = D = e^{i\pi \tau}$, the Picard solution of the $\tau$-function in four dimensions, corresponding to the $c = 1$ conformal blocks in the Ashkin-Teller model, is given by \begin{equation} \tau_{\mathrm{Picard}}(t) = \mathrm{const} \cdot \frac{q^{\sigma_{0t}^2}}{t^{1/8}(1 - t)^{1/8}} \cdot \frac{\vartheta_3(\sigma_{0t} \pi \tau \pm \sigma_{1t} \pi \mid \tau)}{\vartheta_3(0 \mid \tau)}, \end{equation} where the Jacobi theta function is $ \vartheta_3(z \mid \tau) = \sum_{n \in \mathbb{Z}} e^{i\pi n^2 \tau + 2inz}, $ and the monodromy traces satisfy $ \mathrm{tr}(\mathcal{M}\mu \mathcal{M}\nu) = 2 \cos(2\pi \sigma_{\mu\nu}), $ with parameter space $(\theta_0, \theta_t, \theta_1, \theta_\infty) \in \mathcal{M}$ \cite{Gamayun2012, Bershtein2015, Iorgov2015, Gavrylenko2018}. For additional algebraic solutions, see \cite{Lisovyy2014}.

Now consider the Clifford algebra $\mathrm{Cl}_n$ in the case $n = 3$ \cite{Meinrenken2013}. Let $(\rho, V)$ be a representation such that $ \rho : \mathrm{Cl}n \ni \phi \longmapsto \rho(\phi) \in \mathrm{End}(V), \quad \text{with } \rho(\phi)\rho(\psi) = \rho(\phi\psi). $ When $n$ is odd, such as $n = 3$, there exist two inequivalent irreducible representations: \begin{equation} \rho^+ : \mathrm{Cl}3 \simeq \mathbb{C}(2) \oplus \mathbb{C}(2) \ni (\phi, \psi) \mapsto \psi \in \mathrm{End}(\mathbb{C}^2), \end{equation} \begin{equation} \rho^- : \mathrm{Cl}_3 \simeq \mathbb{C}(2) \oplus \mathbb{C}(2) \ni (\phi, \psi) \mapsto \phi \in \mathrm{End}(\mathbb{C}^2). \end{equation}

As an application, define a sequence of complex numbers $v, v', v''$ as follows \cite{Adachi2017}: $ \Re(v) = v, \quad \Im(v) = e^{(\Re(v)/b)E(N)}, $ $ \Re(v') = \frac{N_k}{\Im(v)}, \quad \Im(v') = e^{(\Re(v')/b)E(N)}, $ $ \Re(v'') = \frac{N_k}{\Re(v')}, \quad \Im(v'') = e^{(\Re(v'')/b)E(N)}. $ We denote this recursive construction as the RRR process. Plotting the computed $\Im(v''')$ values against their rank among 800 proteins reveals three distinct clusters based on slope: below $1.01$, between $1.01$ and $2.00$, and above $2.00$ (Supplementary Table 1). The outlier value $0.30$ for Filamin-A was excluded, as it likely reflects adaptation in fibroblasts (HEK-293).

The irreducible representations observed in the LC/MS data of \cite{Adachi2017} are:

4-dimensional (1-2) in non-adapted states, with average $1.368 \pm 0.004$ (99\% confidence),

3-dimensional (1) in adapted states, with average $1.001571 \pm 0.000006$ (99\% confidence).

The remaining proteins are likely repressed (disadapted). In tensor algebra, define $ TB := \bigoplus_{n = 0}^\infty B^{\otimes n}, \quad B = \bigoplus_{i \in I} R X_i, \quad x \in X, $ with the relation $ x \otimes x - q(x) \in R \oplus B^{\otimes 2}. $ Here, $x$ represents a single fractal dimension ($= w$), and the fractal dimension of $q(x)$ is $1/2$ in non-adapted and $1$ in adapted stages. This framework enables the computation of characteristic numbers associated with protein adaptation.\\

\subsection{$\mathscr{O} \cong \widehat{\mathbb{C}}$ Case}

For the species dataset (Tables 1 and 2), we consider that a sequential operation corresponds to an exact form. Following \cite{Adachi2017}, define Operation III as: $ \begin{aligned} \Re(v) &= v, \ \Im(v) &= \mu_\ell = e^{(\Re(v)/b)E(N)}, \ \Re(v') &= E[\ell] = \ell = \frac{\ln N_k}{\ln \Im(v)}, \\ \Im(v') &= e^{(\Re(v')/b)E(N)}, \ \Re(v'') &= \frac{\ln N_k}{\ln \Im(v')}, \ \Im(v'') &= e^{(\Re(v'')/b)E(N)}, \\ \Re(v''') &= \frac{\ln N_k}{\ln \Im(v'')}, \ \Im(v''') &= e^{(\Re(v''')/b)E(N)}. \end{aligned} $ Empirically, we observe: $ \Re(v) \simeq \Re(v'') \simeq 0, \quad \Im(v) \simeq \Im(v'') \simeq 0, \quad \Re(v') \simeq \Re(v''') \simeq 0, \quad \Im(v') \simeq \Im(v''') \simeq 0, $ suggesting that the actual or potential state of a species induces the actual or potential emergence of an adapted hierarchy two layers above. This structure forms a short exact sequence, where the morphism $\Im$ is a monomorphism and $\Re(\ln)$ is an epimorphism, with $\mathrm{Im}(\Im) = \mathrm{Ker}(\Re(\ln))$.

Moreover, there exist homomorphisms such as: $ h: \Im(v') \to \Re(v'), \quad h: \Re(v'') \to \Im(v'), \quad h: \Im(v'') \to \Re(v''), \quad h: \Re(v''') \to \Im(v''), $ and the short exact sequence splits. These are abelian groups, and we have the following isomorphisms: $ \begin{aligned} \Re(v') &\simeq \Im(v) \oplus \Im(v'), \ \Im(v') &\simeq \Re(v') \oplus \Re(v''), \ \Re(v'') &\simeq \Im(v') \oplus \Im(v''), \\ \Im(v'') &\simeq \Re(v'') \oplus \Re(v'''). \end{aligned} $ This implies that each actual layer is the direct sum of the potential layer below and its own potential, while each potential layer is the direct sum of the real layer below and the real layer above.

We define Galois actions via: $ \frac{\Re(v')}{\Im(v)} \simeq \Im(v'), \quad \frac{\Im(v')}{\Re(v')} \simeq \Re(v''), \quad \frac{\Re(v'')}{\Im(v')} \simeq \Im(v''), \quad \frac{\Im(v'')}{\Re(v'')} \simeq \Re(v'''). $ These are all Galois actions, establishing a proper Galois structure on the topology of $v$ for modeling biological hierarchies. Thus, species are likely to emerge from the interaction of species.

$
\xymatrix{
\Re(v) \ar[d]_\Im \ar[r]^{I(\Re)} & \Re(v') \ar[d]_\Im \ar[r]^{I(\Re)} & \Re(v'') \ar[d]_\Im \ar[r]^{I(\Re)} & \Re(v''') \ar[d]_\Im\\
\Im(v) \ar[ur]|{\Re(ln)} \ar[r]_{I(\Im)} & \Im(v') \ar[ur]|{\Re(ln)} \ar[r]_{I(\Im)} & \Im(v'') \ar[ur]|{\Re(ln)} \ar[r]_{I(\Im)} & \Im(v''') \\}
$

As shown in \cite{Adachi2017}, the sequential operation III satisfies $\Re(v) \simeq \Re(v'') \simeq 0$ and $\Im(v) \simeq \Im(v'') \simeq 0$, but not beyond, indicating that the actual or potential state of a species gives rise to an adapted hierarchy two layers above, with diminishing effects beyond three layers. This may reflect the influence of differing time scales across hierarchical levels \cite{Adachi2015}. As before, we observe: $ \Re(v') \simeq \Im(v) \oplus \Im(v'), \quad \Im(v') \simeq \Re(v') \oplus \Re(v''). $

From the morphisms in Operation III, we derive a short exact sequence: \begin{equation} 0 \to \mathcal{A}(u) \xrightarrow{\iota} \mathcal{B}(u) \xrightarrow{\mathrm{sp}} \mathcal{C}(u \times \sqrt{-1}S^*) \to 0, \end{equation} where $g = \ell$ is interpreted as a specific spectrum of the Schwartz distribution (or Sato hyperfunction) of a microfunction $\mathrm{sp}, g$ \cite{Sato1959, Sato1960, Sato1969, Morimoto1970}. In this framework, not only addition but also multiplication is feasible for $-s$, enabling richer algebraic operations on species-level structures. 

\subsection{Congruent zeta function}

Hereafter, we consider the case where $\mathscr{O} \cong \widehat{\mathbb{C}}$. In this framework, instead of $\Im(v')$, we may consider the finite cyclic group $\mathbb{Z}/\ell\mathbb{Z}$ by interpreting $1/\ell$-powered $\Im(v')$ as a $p$-adic number and applying a valuation. The universal coefficient theorem \cite{Bott1982}, \begin{equation} 0 \to \mathrm{Ext}(H_{q-1}(X, A), G) \to H^q(X, A; G) \to \mathrm{Hom}(H_q(X, A), G) \to 0, \end{equation} can be modeled as the short exact sequence \begin{equation} 0 \to \mu_\ell \to E[\ell] \to \mathbb{Z}/\ell\mathbb{Z} \to 0, \end{equation} where $\Re(s)$ lies at an intermediate level between the populational $\Re(v)$ and the fractal $\Re(v'')$. The map $E[\ell] \to \mathbb{Z}/\ell\mathbb{Z}$ is injective, and $\mathbb{Z}/\ell\mathbb{Z} \to 0$ is surjective, with the image of the former equal to the kernel of the latter. This sequence splits, and all groups involved are abelian: $ E[\ell] \cong \boldsymbol{\mu}\ell \oplus \mathbb{Z}/\ell\mathbb{Z}, \quad \mathbb{Z}/\ell\mathbb{Z} \cong E[\ell] \oplus 0. $ Thus, a real level is the direct sum of a potential level below and its own potential, while a potential level is the direct sum of a real level below and a real level above. The quotient $E[\ell]/\boldsymbol{\mu}\ell \cong \mathbb{Z}/\ell\mathbb{Z}$ and $\mathbb{Z}/\ell\mathbb{Z}/E[\ell] \cong 0$ reflect Galois actions, yielding a representation of the \'etale topology $\ell$ and encoding inter-hierarchical interactions.

Species are expected to emerge two layers above the population level. According to \cite{Adachi2015}, point mutation rates are on the order of $10^{-8}$, while speciation occurs at approximately $10^{-25}$, roughly $(10^{-8})^2 / 10^{-8}$. This can be modeled by a critical phenomenon in dendrogram percolation: as the probability of mutation maintenance approaches $1/2-0$, cluster sizes diverge. Interpreting nontrivial zeros of $\zeta(s)$ as seeds of speciation, a population of size $\sim 10^8$ aligns with the scale at which individuals are either identical to or diverge from their ancestors at the base-pair level. Dendrograms, viewed as phylogenetic trees of cell divisions, apply to both asexual and sexual reproduction, particularly in germline cell divisions. These observations support the use of $\ell$-adic and Galois-theoretic frameworks to model inter-level biological interactions.

This motivates the application of Grothendieck groups. Let $B$ be a Noetherian ring. Define $F(B)$ as the set of isomorphism classes of $B$-modules, and $C_B$ as the free abelian group generated by $F(B)$. The short exact sequence above corresponds to the relation $ (\mu_\ell) - (E[\ell]) + (\mathbb{Z}/\ell\mathbb{Z}) \in C_B. $ Let $D_B$ be the subgroup of $C_B$ generated by such relations. Then the Grothendieck group $K(B) := C_B / D_B$ encodes the potential of the $(s, w)$ layers. If $E[\ell]$ is finitely generated, then $\gamma(E[\ell])$ denotes its image in $K(B)$. There exists a unique homomorphism $\lambda_0: K(B) \to G$ such that $\lambda(E[\ell]) = \lambda_0(\gamma(E[\ell]))$ for all $E[\ell]$, where $G$ is an abelian group of $B$-modules. This is analogous to the Stone-von Neumann theorem in this restricted context.

If $B$ is a principal ideal domain representing a single niche (i.e., without interaction between distinct niches), then $K(B) \cong \mathbb{Z}$, which is suitable for modeling individual-level biological counts. For flat $B$-modules $M_\ell$ and $N_\ell$, and the set of isomorphisms $F_1(B)$, we have $ \gamma_1(M_\ell) \cdot \gamma_1(N_\ell) = \gamma_1(M_\ell \otimes N_\ell), \quad \gamma_1(M_\ell) \cdot \gamma(N_\ell) = \gamma(M_\ell \otimes N_\ell), \quad K_1(B) \cong \mathbb{Z}. $ If $B$ is regular, then $K_1(B) \to K(B)$ is an isomorphism. Thus, the sum of interactions within a niche is computable as integers in $K(B)$. If the result is non-integral, it implies interaction across distinct niches. Algebraic extensions of $B$ introduce new niches. For example, if $a \in K$, $f(x) = x^\ell - x - a$, and $\alpha \in \bar{K}$ satisfies $f(\alpha) = 0$ but $\alpha \notin K$, then $f(x)$ is irreducible over $K$, and $L = K(\alpha)$ is a Galois extension with $\mathrm{Gal}(L/K) \cong \mathbb{Z}/\ell\mathbb{Z}$. The element $\alpha$ arises from a higher-level hierarchy via a new ideal. 

To unify the sections introducing Galois cohomology $H^i$ with the earlier discussions on the time arrow, consider two smooth, connected, and geometrically irreducible algebraic curves $X$ and $Y$ over an algebraically closed field. Suppose there exists an algebraic correspondence $\gamma$ from $Y$ to $X$. Then the following diagram describes the induced cohomological transformation: \begin{equation} \begin{split} H^i(X_{\bar{k}}, \mathbb{Q}\ell) \xrightarrow{\mathrm{pr}1^*} H^i(X_{\bar{k}} \times_{\bar{k}} Y_{\bar{k}}, \mathbb{Q}\ell) \xrightarrow{\cup, \mathrm{cl}(\gamma)} H^{i + 2d}(X_{\bar{k}} \times_{\bar{k}} Y_{\bar{k}}, \mathbb{Q}\ell(d)) \xrightarrow{\mathrm{pr}{2*}} H^i(Y_{\bar{k}}, \mathbb{Q}\ell), \end{split} \end{equation} where $\mathrm{cl}(\gamma)$ denotes the cycle class of $\gamma$, and $d$ is the relative dimension. If we interpret $X$ and $Y$ as representing different time points, then the induced map \begin{equation} \gamma^*: H^i(X_{\bar{k}}, \mathbb{Q}\ell) \rightarrow H^i(Y_{\bar{k}}, \mathbb{Q}_\ell) \end{equation} describes the time evolution of the system at the cohomological level.

To analyze the contribution of each cohomological degree to the time development, let $\kappa_m$ be the degree-$m$ extension of a finite field $\kappa$, which is the residue field of a valuation ring $\mathcal{O}_K$ of a number field $K$. When the smooth scheme $Y$ is defined over $\kappa$, the Grothendieck--Lefschetz trace formula gives: \begin{equation} \sum_{i = 0}^{2d} (-1)^i \mathrm{Tr}(\mathrm{Frob}_v^m; H^i(Y{\bar{k}}, \mathbb{Q}_\ell)) = \# Y(\kappa_m), \end{equation} as shown in \cite{Deligne1977, Rapoport1982}.

If $Y$ is finite, the associated congruent zeta function is defined by \begin{equation} Z(Y, T) = \exp( \sum_{n = 1}^\infty \frac{\# Y(\kappa_n)}{n} T^n ). \end{equation} For each cohomological degree $i$, define the characteristic polynomial \begin{equation} P_i(Y, T) = \det(1 - \mathrm{Frob}_v T; H^i(Y{\bar{k}}, \mathbb{Q}\ell)), \end{equation} so that the zeta function factorizes as \begin{equation} Z(Y, T) = \prod_{i = 0}^{2\dim Y} P_i(Y, T)^{(-1)^{i + 1}}. \end{equation}

To isolate the contribution of each $H^i$, we invoke the Weil conjectures \cite{Deligne1974, Deligne1980}, which ensure that the polynomials $P_i(Y, T)$ and $P_j(Y, T)$ are disjoint for $i \neq j$. Thus, both $P_i(Y, T)$ and the traces $\mathrm{Tr}(\mathrm{Frob}v^m; H^i(Y{\bar{k}}, \mathbb{Q}_\ell))$ are computable, allowing us to quantify the contribution of each cohomological degree to the zeta function.

Examples of such calculations are provided in Tables 5 and 6. In general:

Large positive zeta values correspond to highly adapted biological states.

Large negative zeta values indicate highly disadapted or unstable states.

Zeta values near zero represent neutral or transitional states.

In this framework, $P_0$ and $P_1$ correspond to $1 - \Im(v)$ and $1 - \Im(v') = 1 - \Im(s)$, respectively. For $P_0$, values close to zero indicate large contributions from the base population layer, while for $P_1$, large values reflect significant influence from the species layer.

The key insight is that the congruent zeta function provides a means to visualize and quantify the contribution of each hierarchical level in a biological system. This is summarized in Table 9, where the structure of $Z(Y, T)$ reveals the layered dynamics of adaptation and speciation.

\begin{table}[]
\centering
\caption{Contributions of hierarchies.}
\centering Dictyostelia\\
\resizebox{\columnwidth}{!}{%
\begin{tabular}{|l|l|l|l|l|l|l|l|}
\hline
{\color[HTML]{000000} $P_0$} &
  {\color[HTML]{000000} WE \textit{P. pallidum}} &
  {\color[HTML]{000000} WE  \textit{D. purpureum}} &
  {\color[HTML]{000000} WE  \textit{P. violaceum}} &
  {\color[HTML]{000000} $P_0$} &
  {\color[HTML]{000000} WW  \textit{P. pallidum}} &
  {\color[HTML]{000000} WW  \textit{D. purpureum}} &
  {\color[HTML]{000000} WW  \textit{P. violaceum}} \\ \hline
{\color[HTML]{000000} April} &
  {\color[HTML]{000000} } &
  {\color[HTML]{000000} } &
  {\color[HTML]{000000} } &
  {\color[HTML]{000000} April} &
  {\color[HTML]{000000} } &
  {\color[HTML]{000000} } &
  {\color[HTML]{000000} } \\ \hline
{\color[HTML]{000000} June} &
  {\color[HTML]{000000} -517.3} &
  {\color[HTML]{000000} -17.51} &
  {\color[HTML]{000000} -22.28} &
  {\color[HTML]{000000} June} &
  {\color[HTML]{000000} } &
  {\color[HTML]{000000} } &
  {\color[HTML]{000000} } \\ \hline
{\color[HTML]{000000} July} &
  {\color[HTML]{000000} } &
  {\color[HTML]{000000} } &
  {\color[HTML]{000000} } &
  {\color[HTML]{000000} July} &
  {\color[HTML]{000000} -31.22} &
  {\color[HTML]{000000} -11320} &
  {\color[HTML]{000000} -24.82} \\ \hline
{\color[HTML]{000000} August} &
  {\color[HTML]{000000} } &
  {\color[HTML]{000000} } &
  {\color[HTML]{000000} } &
  {\color[HTML]{000000} August} &
  {\color[HTML]{000000} -7.181} &
  {\color[HTML]{000000} -5.079} &
  {\color[HTML]{000000} } \\ \hline
{\color[HTML]{000000} September} &
  {\color[HTML]{000000} -5.701} &
  {\color[HTML]{000000} -3.56} &
  {\color[HTML]{000000} } &
  {\color[HTML]{000000} September} &
  {\color[HTML]{000000} -41.16} &
  {\color[HTML]{000000} -6.605} &
  {\color[HTML]{000000} -672400000} \\ \hline
{\color[HTML]{000000} October} &
  {\color[HTML]{000000} -2.859} &
  {\color[HTML]{000000} -1.051} &
  {\color[HTML]{000000} } &
  {\color[HTML]{000000} October} &
  {\color[HTML]{000000} -5.971} &
  {\color[HTML]{000000} } &
  {\color[HTML]{000000} -3.991} \\ \hline
{\color[HTML]{000000} November} &
  {\color[HTML]{000000} -237.1} &
  {\color[HTML]{000000} } &
  {\color[HTML]{000000} -8.428} &
  {\color[HTML]{000000} November} &
  {\color[HTML]{000000} } &
  {\color[HTML]{000000} } &
  {\color[HTML]{000000} } \\ \hline
{\color[HTML]{000000} December} &
  {\color[HTML]{000000} } &
  {\color[HTML]{000000} } &
  {\color[HTML]{000000} } &
  {\color[HTML]{000000} December} &
  {\color[HTML]{000000} } &
  {\color[HTML]{000000} } &
  {\color[HTML]{000000} } \\ \hline
{\color[HTML]{000000} January} &
  {\color[HTML]{000000} } &
  {\color[HTML]{000000} } &
  {\color[HTML]{000000} } &
  {\color[HTML]{000000} January} &
  {\color[HTML]{000000} } &
  {\color[HTML]{000000} } &
  {\color[HTML]{000000} } \\ \hline
{\color[HTML]{000000} $P_1$} &
  {\color[HTML]{000000} WE \textit{P. pallidum}} &
  {\color[HTML]{000000} WE  \textit{D. purpureum}} &
  {\color[HTML]{000000} WE  \textit{P. violaceum}} &
  {\color[HTML]{000000} $P_1$} &
  {\color[HTML]{000000} WW  \textit{P. pallidum}} &
  {\color[HTML]{000000} WW  \textit{D. purpureum}} &
  {\color[HTML]{000000} WW  \textit{P. violaceum}} \\ \hline
{\color[HTML]{000000} April} &
  {\color[HTML]{000000} } &
  {\color[HTML]{000000} } &
  {\color[HTML]{000000} } &
  {\color[HTML]{000000} April} &
  {\color[HTML]{000000} } &
  {\color[HTML]{000000} } &
  {\color[HTML]{000000} } \\ \hline
{\color[HTML]{000000} June} &
  {\color[HTML]{000000} -7.182} &
  {\color[HTML]{000000} -147.6} &
  {\color[HTML]{000000} -30.1} &
  {\color[HTML]{000000} June} &
  {\color[HTML]{000000} } &
  {\color[HTML]{000000} } &
  {\color[HTML]{000000} } \\ \hline
{\color[HTML]{000000} July} &
  {\color[HTML]{000000} } &
  {\color[HTML]{000000} } &
  {\color[HTML]{000000} } &
  {\color[HTML]{000000} July} &
  {\color[HTML]{000000} -38.31} &
  {\color[HTML]{000000} -4.331} &
  {\color[HTML]{000000} -173.5} \\ \hline
{\color[HTML]{000000} August} &
  {\color[HTML]{000000} } &
  {\color[HTML]{000000} } &
  {\color[HTML]{000000} } &
  {\color[HTML]{000000} August} &
  {\color[HTML]{000000} -21.63} &
  {\color[HTML]{000000} -12.8} &
  {\color[HTML]{000000} -1.488} \\ \hline
{\color[HTML]{000000} September} &
  {\color[HTML]{000000} -22.24} &
  {\color[HTML]{000000} -13.99} &
  {\color[HTML]{000000} -1.41} &
  {\color[HTML]{000000} September} &
  {\color[HTML]{000000} -52.12} &
  {\color[HTML]{000000} -114.9} &
  {\color[HTML]{000000} -1.028} \\ \hline
{\color[HTML]{000000} October} &
  {\color[HTML]{000000} -44.68} &
  {\color[HTML]{000000} -37.74} &
  {\color[HTML]{000000} -1.096} &
  {\color[HTML]{000000} October} &
  {\color[HTML]{000000} -21.67} &
  {\color[HTML]{000000} -1.476} &
  {\color[HTML]{000000} -12.93} \\ \hline
{\color[HTML]{000000} November} &
  {\color[HTML]{000000} -6.726} &
  {\color[HTML]{000000} -14.38} &
  {\color[HTML]{000000} -274.5} &
  {\color[HTML]{000000} November} &
  {\color[HTML]{000000} } &
  {\color[HTML]{000000} } &
  {\color[HTML]{000000} } \\ \hline
{\color[HTML]{000000} December} &
  {\color[HTML]{000000} } &
  {\color[HTML]{000000} } &
  {\color[HTML]{000000} } &
  {\color[HTML]{000000} December} &
  {\color[HTML]{000000} } &
  {\color[HTML]{000000} } &
  {\color[HTML]{000000} } \\ \hline
{\color[HTML]{000000} January} &
  {\color[HTML]{000000} } &
  {\color[HTML]{000000} } &
  {\color[HTML]{000000} } &
  {\color[HTML]{000000} January} &
  {\color[HTML]{000000} } &
  {\color[HTML]{000000} } &
  {\color[HTML]{000000} } \\ \hline
\end{tabular}%
}

\centering Collembola\\
\resizebox{\columnwidth}{!}{%
\begin{tabular}{|l|l|l|l|l|l|l|l|l|l|l|l|}
\hline
{\color[HTML]{000000} $P_0$} &
  {\color[HTML]{000000} \textit{F. octoulata}} &
  {\color[HTML]{000000}  \textit{E. aino}} &
  {\color[HTML]{000000}  \textit{H. watanabei}} &
  {\color[HTML]{000000}  \textit{S. celebensis}} &
  {\color[HTML]{000000}  \textit{S. aureus}} &
  {\color[HTML]{000000}  \textit{D. trispinata}} &
  {\color[HTML]{000000}  \textit{S. japonica}} &
  {\color[HTML]{000000}  \textit{H. nigrochephala}} &
  {\color[HTML]{000000}  \textit{W. japonica}} &
  {\color[HTML]{000000}  \textit{F. yosii}} &
  {\color[HTML]{000000}  \textit{A. laricis}} \\ \hline
April &
   &
   &
  0.3532 &
   &
   &
   &
   &
   &
  0.7500 &
   &
   \\ \hline
May &
   &
   &
  0.9078 &
   &
  0.3950 &
  0.9291 &
   &
   &
   &
   &
   \\ \hline
June &
  -1.2073 &
  0.1911 &
  0.5308 &
   &
  0.5308 &
  0.6909 &
  0.3002 &
   &
   &
  0.0440 &
  0.5308 \\ \hline
July &
  0.7021 &
  0.1533 &
  0.7910 &
   &
  0.7910 &
  0.7910 &
  0.7910 &
   &
   &
  0.7326 &
   \\ \hline
August &
  -0.6740 &
  0.2069 &
  0.5953 &
   &
   &
  0.3074 &
  0.5953 &
   &
  0.5953 &
   &
   \\ \hline
September &
  -1.4630 &
  -0.1858 &
  -0.2221 &
  0.4020 &
  0.5127 &
  0.5127 &
  0.6511 &
  0.6511 &
   &
   &
   \\ \hline
October &
  0.1353 &
  0.7108 &
  0.6337 &
  0.7108 &
  0.7108 &
   &
   &
   &
   &
   &
   \\ \hline
November &
   &
  0.8889 &
   &
   &
  0.4976 &
   &
   &
   &
   &
   &
   \\ \hline

{\color[HTML]{000000} $P_1$} &
  {\color[HTML]{000000} \textit{F. octoulata}} &
  {\color[HTML]{000000}  \textit{E. aino}} &
  {\color[HTML]{000000}  \textit{H. watanabei}} &
  {\color[HTML]{000000}  \textit{S. celebensis}} &
  {\color[HTML]{000000}  \textit{S. aureus}} &
  {\color[HTML]{000000}  \textit{D. trispinata}} &
  {\color[HTML]{000000}  \textit{S. japonica}} &
  {\color[HTML]{000000}  \textit{H. nigrochephala}} &
  {\color[HTML]{000000}  \textit{W. japonica}} &
  {\color[HTML]{000000}  \textit{F. yosii}} &
  {\color[HTML]{000000}  \textit{A. laricis}} \\ \hline
  April &
   &
   &
  -31.9351 &
   &
   &
   &
   &
   &
   &
   &
   \\ \hline
May &
  -155.1129 &
  -155.1129 &
  -24.0109 &
  -155.1129 &
  -3459.0461 &
   &
  -155.1129 &
  -155.1129 &
  -155.1129 &
  -155.1129 &
  -155.1129 \\ \hline
June &
  -3737.5881 &
  -1551.7361 &
  -2077.1555 &
  -47.0052 &
  -2077.1555 &
  -1330.8276 &
  -559.2408 &
  -47.0052 &
  -47.0052 &
  -23163.6618 &
  -2077.1555 \\ \hline
July &
  -13.1347 &
  -3087.0834 &
  -1747.5640 &
  -203.1897 &
  -1747.5640 &
  -1747.5640 &
  -1747.5640 &
  -203.1897 &
  -203.1897 &
  -429.3287 &
  -203.1897 \\ \hline
August &
  -652.6496 &
  -1390.8532 &
  -1024.2657 &
  -13.1347 &
  -13.1347 &
  -94.8706 &
  -1024.2657 &
  -13.1347 &
  -1024.2657 &
  -13.1347 &
  -13.1347 \\ \hline
September &
  -2364.1915 &
  -3320.6636 &
  -218.0676 &
  -3068.7829 &
  -2597.5988 &
  -2597.5988 &
  -2544.8780 &
  -2544.8780 &
  -42.3271 &
  -42.3271 &
  -42.3271 \\ \hline
October &
  -392.4277 &
  -455.3284 &
  -13.1347 &
  -455.3284 &
  -455.3284 &
  -31.9351 &
  -31.9351 &
  -31.9351 &
  -31.9351 &
  -31.9351 &
  -31.9351 \\ \hline
November &
   &
   &
   &
   &
  -93.6513 &
   &
   &
   &
   &
   &
   \\ \hline
\end{tabular}%
}

\centering Sarcoptiformes\\
\resizebox{\columnwidth}{!}{%
\begin{tabular}{|l|l|l|l|l|l|l|l|}
\hline
{\color[HTML]{000000} $P_0$} &
  {\color[HTML]{000000} \textit{A. gigantea}} &
  {\color[HTML]{000000} \textit{P. parvisetigerum}} &
  {\color[HTML]{000000} \textit{T. stercus}} &
  {\color[HTML]{000000} \textit{M. japonica}} &
  {\color[HTML]{000000} \textit{G. fusca}} &
  {\color[HTML]{000000} \textit{N. rotundus}} &
  {\color[HTML]{000000} \textit{C. reticulatus}}\\ \hline
{\color[HTML]{000000} April} &
  {\color[HTML]{000000} } &
  {\color[HTML]{000000} 0.6731} &
  {\color[HTML]{000000} 0.9877} &
  {\color[HTML]{000000} 0.6731} &
  {\color[HTML]{000000} } &
  {\color[HTML]{000000} 0.6731} &
  {\color[HTML]{000000} } \\ \hline
{\color[HTML]{000000} May} &
  {\color[HTML]{000000} 0.4113} &
  {\color[HTML]{000000} 0.4113} &
  {\color[HTML]{000000} 0.9078} &
  {\color[HTML]{000000} } &
  {\color[HTML]{000000} } &
  {\color[HTML]{000000} } &
  {\color[HTML]{000000} } \\ \hline
{\color[HTML]{000000} June} &
  {\color[HTML]{000000} -0.244} &
  {\color[HTML]{000000} -3.946} &
  {\color[HTML]{000000} 0.6952} &
  {\color[HTML]{000000} 0.7949} &
  {\color[HTML]{000000} } &
  {\color[HTML]{000000} 0.4761} &
  {\color[HTML]{000000} -0.8868} \\ \hline
{\color[HTML]{000000} July} &
  {\color[HTML]{000000} 0.1854} &
  {\color[HTML]{000000} 0.5726} &
  {\color[HTML]{000000} 0.5726} &
  {\color[HTML]{000000} 0.5726} &
  {\color[HTML]{000000} } &
  {\color[HTML]{000000} 0.5726} &
  {\color[HTML]{000000} } \\ \hline
{\color[HTML]{000000} August} &
  {\color[HTML]{000000} -0.04109} &
  {\color[HTML]{000000} 0.4839} &
  {\color[HTML]{000000} } &
  {\color[HTML]{000000} 0.4839} &
  {\color[HTML]{000000} } &
  {\color[HTML]{000000} 0.4839} &
  {\color[HTML]{000000} } \\ \hline
{\color[HTML]{000000} September} &
  {\color[HTML]{000000} 0.08768} &
  {\color[HTML]{000000} 0.08768} &
  {\color[HTML]{000000} 0.7874} &
  {\color[HTML]{000000} 0.8234} &
  {\color[HTML]{000000} 0.8351} &
  {\color[HTML]{000000} 0.8569} &
  {\color[HTML]{000000} 0.8721} \\ \hline
{\color[HTML]{000000} October} &
  {\color[HTML]{000000} 0.7021} &
  {\color[HTML]{000000} 0.1407} &
  {\color[HTML]{000000} 0.7108} &
  {\color[HTML]{000000} } &
  {\color[HTML]{000000} } &
  {\color[HTML]{000000} 0.7108} &
  {\color[HTML]{000000} 0.7108} \\ \hline
{\color[HTML]{000000} November} &
  {\color[HTML]{000000} 0.4575} &
  {\color[HTML]{000000} } &
  {\color[HTML]{000000} -0.077} &
  {\color[HTML]{000000} 0.6346} &
  {\color[HTML]{000000} } &
  {\color[HTML]{000000} } &
  {\color[HTML]{000000} } \\ \hline
{\color[HTML]{000000} $P_1$} &
  {\color[HTML]{000000} \textit{A. gigantea}} &
  {\color[HTML]{000000} \textit{P. parvisetigerum}} &
  {\color[HTML]{000000} \textit{T. stercus}} &
  {\color[HTML]{000000} \textit{M. japonica}} &
  {\color[HTML]{000000} \textit{G. fusca}} &
  {\color[HTML]{000000} \textit{N. rotundus}} &
  {\color[HTML]{000000} \textit{C. reticulatus}}\\ \hline
  {\color[HTML]{000000} April} &
  {\color[HTML]{000000} -426300} &
  {\color[HTML]{000000} -37340000} &
  {\color[HTML]{000000} -652.6} &
  {\color[HTML]{000000} -37340000} &
  {\color[HTML]{000000} -426300} &
  {\color[HTML]{000000} -37340000} &
  {\color[HTML]{000000} -426300} \\ \hline
{\color[HTML]{000000} May} &
  {\color[HTML]{000000} -12660} &
  {\color[HTML]{000000} -12660} &
  {\color[HTML]{000000} -48.77} &
  {\color[HTML]{000000} -477.9} &
  {\color[HTML]{000000} -477.9} &
  {\color[HTML]{000000} -477.9} &
  {\color[HTML]{000000} -477.9} \\ \hline
{\color[HTML]{000000} June} &
  {\color[HTML]{000000} -25990} &
  {\color[HTML]{000000} } &
  {\color[HTML]{000000} -192.1} &
  {\color[HTML]{000000} -44400} &
  {\color[HTML]{000000} -2243} &
  {\color[HTML]{000000} -20.02} &
  {\color[HTML]{000000} } \\ \hline
{\color[HTML]{000000} July} &
  {\color[HTML]{000000} -793.5} &
  {\color[HTML]{000000} -6184} &
  {\color[HTML]{000000} -6184} &
  {\color[HTML]{000000} -6184} &
  {\color[HTML]{000000} } &
  {\color[HTML]{000000} -6184} &
  {\color[HTML]{000000} } \\ \hline
{\color[HTML]{000000} August} &
  {\color[HTML]{000000} -172.4} &
  {\color[HTML]{000000} -695.6} &
  {\color[HTML]{000000} } &
  {\color[HTML]{000000} -695.6} &
  {\color[HTML]{000000} } &
  {\color[HTML]{000000} -695.6} &
  {\color[HTML]{000000} } \\ \hline
{\color[HTML]{000000} September} &
  {\color[HTML]{000000} -64420} &
  {\color[HTML]{000000} -64420} &
  {\color[HTML]{000000} } &
  {\color[HTML]{000000} -20.02} &
  {\color[HTML]{000000} -191} &
  {\color[HTML]{000000} -791.4} &
  {\color[HTML]{000000} -10090} \\ \hline
{\color[HTML]{000000} October} &
  {\color[HTML]{000000} } &
  {\color[HTML]{000000} -365.2} &
  {\color[HTML]{000000} -548.5} &
  {\color[HTML]{000000} -31.94} &
  {\color[HTML]{000000} -31.94} &
  {\color[HTML]{000000} -548.5} &
  {\color[HTML]{000000} -548.5} \\ \hline
{\color[HTML]{000000} November} &
  {\color[HTML]{000000} -36.59} &
  {\color[HTML]{000000} } &
  {\color[HTML]{000000} -51.97} &
  {\color[HTML]{000000} -24.01} &
  {\color[HTML]{000000} } &
  {\color[HTML]{000000} } &
  {\color[HTML]{000000} } \\ \hline
\end{tabular}%
}
(continues)\\

\end{table}

\begin{table}

\centering (continued)\\
\centering The world population\\
\resizebox{\columnwidth}{!}{%
\begin{tabular}{|l|l|l|l|l|l|l|l|l|}
\hline
{\color[HTML]{000000} $P_0$} &
  {\color[HTML]{000000} Western Europe} &
  {\color[HTML]{000000} Eastern Europe} &
  {\color[HTML]{000000} Former Soviet Union} &
  {\color[HTML]{000000} Western Offshoots} &
  {\color[HTML]{000000} Latin America} &
  {\color[HTML]{000000} Japan} &
  {\color[HTML]{000000} Asia} &
  {\color[HTML]{000000} Africa} \\ \hline
{\color[HTML]{000000} 1} &
  {\color[HTML]{000000} -40.25} &
  {\color[HTML]{000000} -45.43} &
  {\color[HTML]{000000} -51.12} &
  {\color[HTML]{000000} -17.59} &
  {\color[HTML]{000000} -33.26} &
  {\color[HTML]{000000} -47.69} &
  {\color[HTML]{000000} -136.3} &
  {\color[HTML]{000000} -109.4} \\ \hline
{\color[HTML]{000000} 1000} &
  {\color[HTML]{000000} -327.3} &
  {\color[HTML]{000000} -178} &
  {\color[HTML]{000000} -141.2} &
  {\color[HTML]{000000} -30.53} &
  {\color[HTML]{000000} -113.5} &
  {\color[HTML]{000000} -94.41} &
  {\color[HTML]{000000} -276.7} &
  {\color[HTML]{000000} -69.65} \\ \hline
{\color[HTML]{000000} 1500} &
  {\color[HTML]{000000} -136} &
  {\color[HTML]{000000} -486.8} &
  {\color[HTML]{000000} -290.2} &
  {\color[HTML]{000000} -36.24} &
  {\color[HTML]{000000} -141.7} &
  {\color[HTML]{000000} -420.9} &
  {\color[HTML]{000000} -497.9} &
  {\color[HTML]{000000} -849.5} \\ \hline
{\color[HTML]{000000} 1600} &
  {\color[HTML]{000000} -133.5} &
  {\color[HTML]{000000} -300.8} &
  {\color[HTML]{000000} -123.4} &
  {\color[HTML]{000000} -23.18} &
  {\color[HTML]{000000} -111.3} &
  {\color[HTML]{000000} -204.9} &
  {\color[HTML]{000000} -416.2} &
  {\color[HTML]{000000} -604.3} \\ \hline
{\color[HTML]{000000} 1700} &
  {\color[HTML]{000000} -169.4} &
  {\color[HTML]{000000} -361.9} &
  {\color[HTML]{000000} -489.2} &
  {\color[HTML]{000000} -17.06} &
  {\color[HTML]{000000} -203.6} &
  {\color[HTML]{000000} -215.8} &
  {\color[HTML]{000000} -541.9} &
  {\color[HTML]{000000} -792.3} \\ \hline
{\color[HTML]{000000} 1820} &
  {\color[HTML]{000000} -149.7} &
  {\color[HTML]{000000} -322.6} &
  {\color[HTML]{000000} -405.1} &
  {\color[HTML]{000000} -112} &
  {\color[HTML]{000000} -277.9} &
  {\color[HTML]{000000} -403.3} &
  {\color[HTML]{000000} -535.9} &
  {\color[HTML]{000000} -260.2} \\ \hline
{\color[HTML]{000000} 1870} &
  {\color[HTML]{000000} -485.2} &
  {\color[HTML]{000000} -816} &
  {\color[HTML]{000000} -1785} &
  {\color[HTML]{000000} -1055} &
  {\color[HTML]{000000} -1246} &
  {\color[HTML]{000000} -1224} &
  {\color[HTML]{000000} -1183} &
  {\color[HTML]{000000} -403} \\ \hline
{\color[HTML]{000000} 1913} &
  {\color[HTML]{000000} -922.4} &
  {\color[HTML]{000000} -7677} &
  {\color[HTML]{000000} -1608} &
  {\color[HTML]{000000} -6856} &
  {\color[HTML]{000000} -4061} &
  {\color[HTML]{000000} -2491} &
  {\color[HTML]{000000} -2386} &
  {\color[HTML]{000000} -3341} \\ \hline
{\color[HTML]{000000} 1950} &
  {\color[HTML]{000000} -418.8} &
  {\color[HTML]{000000} -3692} &
  {\color[HTML]{000000} -4797} &
  {\color[HTML]{000000} -16980} &
  {\color[HTML]{000000} -35100} &
  {\color[HTML]{000000} -5434} &
  {\color[HTML]{000000} -2965} &
  {\color[HTML]{000000} -2424} \\ \hline
{\color[HTML]{000000} 1973} &
  {\color[HTML]{000000} -2614} &
  {\color[HTML]{000000} -2043} &
  {\color[HTML]{000000} -31760} &
  {\color[HTML]{000000} -11320} &
  {\color[HTML]{000000} -8399} &
  {\color[HTML]{000000} -3268} &
  {\color[HTML]{000000} -2799} &
  {\color[HTML]{000000} -188.4} \\ \hline
{\color[HTML]{000000} 2003} &
  {\color[HTML]{000000} -3206} &
  {\color[HTML]{000000} -1310} &
  {\color[HTML]{000000} -7369} &
  {\color[HTML]{000000} -6372} &
  {\color[HTML]{000000} -2095} &
  {\color[HTML]{000000} -931.8} &
  {\color[HTML]{000000} -2735} &
  {\color[HTML]{000000} -710.8} \\ \hline
{\color[HTML]{000000} $P_1$} &
  {\color[HTML]{000000} Western Europe} &
  {\color[HTML]{000000} Eastern Europe} &
  {\color[HTML]{000000} Former Soviet Union} &
  {\color[HTML]{000000} Western Offshoots} &
  {\color[HTML]{000000} Latin America} &
  {\color[HTML]{000000} Japan} &
  {\color[HTML]{000000} Asia} &
  {\color[HTML]{000000} Africa} \\ \hline
{\color[HTML]{000000} 1} &
  {\color[HTML]{000000} -2.139} &
  {\color[HTML]{000000} -1.526} &
  {\color[HTML]{000000} -1.407} &
  {\color[HTML]{000000} -1.743} &
  {\color[HTML]{000000} -1.789} &
  {\color[HTML]{000000} -1.376} &
  {\color[HTML]{000000} -1.788} &
  {\color[HTML]{000000} -1.387} \\ \hline
{\color[HTML]{000000} 1000} &
  {\color[HTML]{000000} -1.274} &
  {\color[HTML]{000000} -1.212} &
  {\color[HTML]{000000} -1.314} &
  {\color[HTML]{000000} -1.784} &
  {\color[HTML]{000000} -1.52} &
  {\color[HTML]{000000} -1.504} &
  {\color[HTML]{000000} -1.736} &
  {\color[HTML]{000000} -2.138} \\ \hline
{\color[HTML]{000000} 1500} &
  {\color[HTML]{000000} -2.389} &
  {\color[HTML]{000000} -1.321} &
  {\color[HTML]{000000} -1.562} &
  {\color[HTML]{000000} -2.329} &
  {\color[HTML]{000000} -1.943} &
  {\color[HTML]{000000} -1.397} &
  {\color[HTML]{000000} -2.013} &
  {\color[HTML]{000000} -1.395} \\ \hline
{\color[HTML]{000000} 1600} &
  {\color[HTML]{000000} -3.902} &
  {\color[HTML]{000000} -2.273} &
  {\color[HTML]{000000} -3.188} &
  {\color[HTML]{000000} -4.416} &
  {\color[HTML]{000000} -2.797} &
  {\color[HTML]{000000} -2.604} &
  {\color[HTML]{000000} -3.367} &
  {\color[HTML]{000000} -2.271} \\ \hline
{\color[HTML]{000000} 1700} &
  {\color[HTML]{000000} -1.293} &
  {\color[HTML]{000000} -0.8766} &
  {\color[HTML]{000000} -0.8588} &
  {\color[HTML]{000000} -1.645} &
  {\color[HTML]{000000} -0.946} &
  {\color[HTML]{000000} -1.044} &
  {\color[HTML]{000000} -1.156} &
  {\color[HTML]{000000} -0.8632} \\ \hline
{\color[HTML]{000000} 1820} &
  {\color[HTML]{000000} -1.776} &
  {\color[HTML]{000000} -1.201} &
  {\color[HTML]{000000} -1.2} &
  {\color[HTML]{000000} -1.354} &
  {\color[HTML]{000000} -1.158} &
  {\color[HTML]{000000} -1.112} &
  {\color[HTML]{000000} -1.528} &
  {\color[HTML]{000000} -1.398} \\ \hline
{\color[HTML]{000000} 1870} &
  {\color[HTML]{000000} -1.829} &
  {\color[HTML]{000000} -1.364} &
  {\color[HTML]{000000} -1.24} &
  {\color[HTML]{000000} -1.265} &
  {\color[HTML]{000000} -1.2} &
  {\color[HTML]{000000} -1.178} &
  {\color[HTML]{000000} -1.749} &
  {\color[HTML]{000000} -1.74} \\ \hline
{\color[HTML]{000000} 1913} &
  {\color[HTML]{000000} -1.78} &
  {\color[HTML]{000000} -1.026} &
  {\color[HTML]{000000} -1.476} &
  {\color[HTML]{000000} -1.088} &
  {\color[HTML]{000000} -1.141} &
  {\color[HTML]{000000} -1.174} &
  {\color[HTML]{000000} -1.688} &
  {\color[HTML]{000000} -1.247} \\ \hline
{\color[HTML]{000000} 1950} &
  {\color[HTML]{000000} -2.744} &
  {\color[HTML]{000000} -1.398} &
  {\color[HTML]{000000} -1.463} &
  {\color[HTML]{000000} -1.188} &
  {\color[HTML]{000000} -1.065} &
  {\color[HTML]{000000} -1.298} &
  {\color[HTML]{000000} -2.039} &
  {\color[HTML]{000000} -1.717} \\ \hline
{\color[HTML]{000000} 1973} &
  {\color[HTML]{000000} -1.704} &
  {\color[HTML]{000000} -1.54} &
  {\color[HTML]{000000} -1.083} &
  {\color[HTML]{000000} -1.259} &
  {\color[HTML]{000000} -1.353} &
  {\color[HTML]{000000} -1.404} &
  {\color[HTML]{000000} -2.076} &
  {\color[HTML]{000000} -3.493} \\ \hline
{\color[HTML]{000000} 2003} &
  {\color[HTML]{000000} -2.5} &
  {\color[HTML]{000000} -2.596} &
  {\color[HTML]{000000} -2.028} &
  {\color[HTML]{000000} -2.135} &
  {\color[HTML]{000000} -2.876} &
  {\color[HTML]{000000} -2.854} &
  {\color[HTML]{000000} -3.469} &
  {\color[HTML]{000000} -4.114} \\ \hline
\end{tabular}%
}

\centering GDP\\
\resizebox{\columnwidth}{!}{%
\begin{tabular}{|l|l|l|l|l|l|l|l|l|}
\hline
{\color[HTML]{000000} $P_0$} &
  {\color[HTML]{000000} Western Europe} &
  {\color[HTML]{000000} Eastern Europe} &
  {\color[HTML]{000000} Former Soviet Union} &
  {\color[HTML]{000000} Western Offshoots} &
  {\color[HTML]{000000} Latin America} &
  {\color[HTML]{000000} Japan} &
  {\color[HTML]{000000} Asia} &
  {\color[HTML]{000000} Africa} \\ \hline
{\color[HTML]{000000} 1} &
  {\color[HTML]{000000} -54.2} &
  {\color[HTML]{000000} -27.17} &
  {\color[HTML]{000000} -28.83} &
  {\color[HTML]{000000} -10.89} &
  {\color[HTML]{000000} -19.9} &
  {\color[HTML]{000000} -26.96} &
  {\color[HTML]{000000} -112.5} &
  {\color[HTML]{000000} -81.01} \\ \hline
{\color[HTML]{000000} 1000} &
  {\color[HTML]{000000} -159.4} &
  {\color[HTML]{000000} -83.65} &
  {\color[HTML]{000000} -68.51} &
  {\color[HTML]{000000} -17.77} &
  {\color[HTML]{000000} -56.26} &
  {\color[HTML]{000000} -53.76} &
  {\color[HTML]{000000} -159.5} &
  {\color[HTML]{000000} -40.35} \\ \hline
{\color[HTML]{000000} 1500} &
  {\color[HTML]{000000} -382.6} &
  {\color[HTML]{000000} -236.8} &
  {\color[HTML]{000000} -74.47} &
  {\color[HTML]{000000} -18.43} &
  {\color[HTML]{000000} -185.4} &
  {\color[HTML]{000000} -122} &
  {\color[HTML]{000000} -357.4} &
  {\color[HTML]{000000} -187.5} \\ \hline
{\color[HTML]{000000} 1600} &
  {\color[HTML]{000000} -793.1} &
  {\color[HTML]{000000} -193.5} &
  {\color[HTML]{000000} -86.18} &
  {\color[HTML]{000000} -12.72} &
  {\color[HTML]{000000} -53.35} &
  {\color[HTML]{000000} -121.3} &
  {\color[HTML]{000000} -388.2} &
  {\color[HTML]{000000} -158.7} \\ \hline
{\color[HTML]{000000} 1700} &
  {\color[HTML]{000000} -3217} &
  {\color[HTML]{000000} -299} &
  {\color[HTML]{000000} -180.8} &
  {\color[HTML]{000000} -11.42} &
  {\color[HTML]{000000} -125.6} &
  {\color[HTML]{000000} -361.2} &
  {\color[HTML]{000000} -608.5} &
  {\color[HTML]{000000} -193.2} \\ \hline
{\color[HTML]{000000} 1820} &
  {\color[HTML]{000000} -10570} &
  {\color[HTML]{000000} -368.7} &
  {\color[HTML]{000000} -139.3} &
  {\color[HTML]{000000} -353.9} &
  {\color[HTML]{000000} -304.8} &
  {\color[HTML]{000000} -427.7} &
  {\color[HTML]{000000} -941.4} &
  {\color[HTML]{000000} -290.7} \\ \hline
{\color[HTML]{000000} 1870} &
  {\color[HTML]{000000} -1.4E+45} &
  {\color[HTML]{000000} -4391} &
  {\color[HTML]{000000} -22860} &
  {\color[HTML]{000000} -21800} &
  {\color[HTML]{000000} -1641} &
  {\color[HTML]{000000} -2095} &
  {\color[HTML]{000000} -6803} &
  {\color[HTML]{000000} -6697} \\ \hline
{\color[HTML]{000000} 1913} &
  {\color[HTML]{000000} -16430} &
  {\color[HTML]{000000} -22010} &
  {\color[HTML]{000000} -304500} &
  {\color[HTML]{000000} -3.18E+14} &
  {\color[HTML]{000000} -33770} &
  {\color[HTML]{000000} -9678} &
  {\color[HTML]{000000} -1.61E+10} &
  {\color[HTML]{000000} -8418} \\ \hline
{\color[HTML]{000000} 1950} &
  {\color[HTML]{000000} -8.14E+26} &
  {\color[HTML]{000000} -50500} &
  {\color[HTML]{000000} -6210000} &
  {\color[HTML]{000000} -32770} &
  {\color[HTML]{000000} -3956000} &
  {\color[HTML]{000000} -46760} &
  {\color[HTML]{000000} -3.94E+09} &
  {\color[HTML]{000000} -36260} \\ \hline
{\color[HTML]{000000} 1973} &
  {\color[HTML]{000000} -120000} &
  {\color[HTML]{000000} -369100} &
  {\color[HTML]{000000} -3.99E+08} &
  {\color[HTML]{000000}  } &
  {\color[HTML]{000000} -1.39E+09} &
  {\color[HTML]{000000} -1.43E+09} &
  {\color[HTML]{000000} -6.26E+15} &
  {\color[HTML]{000000} -879900} \\ \hline
{\color[HTML]{000000} 2003} &
  {\color[HTML]{000000} -2.26E+13} &
  {\color[HTML]{000000} -18760} &
  {\color[HTML]{000000} -116800} &
  {\color[HTML]{000000} -4.11E+13} &
  {\color[HTML]{000000} -1138000} &
  {\color[HTML]{000000} -2128000} &
  {\color[HTML]{000000} -101600} &
  {\color[HTML]{000000} -117400} \\ \hline
{\color[HTML]{000000} $P_1$} &
  {\color[HTML]{000000} Western Europe} &
  {\color[HTML]{000000} Eastern Europe} &
  {\color[HTML]{000000} Former Soviet Union} &
  {\color[HTML]{000000} Western Offshoots} &
  {\color[HTML]{000000} Latin America} &
  {\color[HTML]{000000} Japan} &
  {\color[HTML]{000000} Asia} &
  {\color[HTML]{000000} Africa} \\ \hline
{\color[HTML]{000000} 1} &
  {\color[HTML]{000000} -1.728} &
  {\color[HTML]{000000} -1.596} &
  {\color[HTML]{000000} -1.484} &
  {\color[HTML]{000000} -1.819} &
  {\color[HTML]{000000} -1.905} &
  {\color[HTML]{000000} -1.446} &
  {\color[HTML]{000000} -1.712} &
  {\color[HTML]{000000} -1.357} \\ \hline
{\color[HTML]{000000} 1000} &
  {\color[HTML]{000000} -1.296} &
  {\color[HTML]{000000} -1.235} &
  {\color[HTML]{000000} -1.342} &
  {\color[HTML]{000000} -1.785} &
  {\color[HTML]{000000} -1.572} &
  {\color[HTML]{000000} -1.496} &
  {\color[HTML]{000000} -1.748} &
  {\color[HTML]{000000} -2.198} \\ \hline
{\color[HTML]{000000} 1500} &
  {\color[HTML]{000000} -1.363} &
  {\color[HTML]{000000} -1.16} &
  {\color[HTML]{000000} -1.719} &
  {\color[HTML]{000000} -2.102} &
  {\color[HTML]{000000} -1.256} &
  {\color[HTML]{000000} -1.434} &
  {\color[HTML]{000000} -1.641} &
  {\color[HTML]{000000} -1.463} \\ \hline
{\color[HTML]{000000} 1600} &
  {\color[HTML]{000000} -1.152} &
  {\color[HTML]{000000} -1.226} &
  {\color[HTML]{000000} -1.625} &
  {\color[HTML]{000000} -2.328} &
  {\color[HTML]{000000} -1.588} &
  {\color[HTML]{000000} -1.412} &
  {\color[HTML]{000000} -1.579} &
  {\color[HTML]{000000} -1.498} \\ \hline
{\color[HTML]{000000} 1700} &
  {\color[HTML]{000000} -0.9944} &
  {\color[HTML]{000000} -1.243} &
  {\color[HTML]{000000} -1.506} &
  {\color[HTML]{000000} -2.73} &
  {\color[HTML]{000000} -1.44} &
  {\color[HTML]{000000} -1.241} &
  {\color[HTML]{000000} -1.57} &
  {\color[HTML]{000000} -1.588} \\ \hline
{\color[HTML]{000000} 1820} &
  {\color[HTML]{000000} -0.7531} &
  {\color[HTML]{000000} -1.103} &
  {\color[HTML]{000000} -1.523} &
  {\color[HTML]{000000} -1.02} &
  {\color[HTML]{000000} -1.073} &
  {\color[HTML]{000000} -1.038} &
  {\color[HTML]{000000} -1.262} &
  {\color[HTML]{000000} -1.207} \\ \hline
{\color[HTML]{000000} 1870} &
  {\color[HTML]{000000} -0.08936} &
  {\color[HTML]{000000} -1.45} &
  {\color[HTML]{000000} -1.19} &
  {\color[HTML]{000000} -1.243} &
  {\color[HTML]{000000} -1.606} &
  {\color[HTML]{000000} -1.511} &
  {\color[HTML]{000000} -1.759} &
  {\color[HTML]{000000} -1.328} \\ \hline
{\color[HTML]{000000} 1913} &
  {\color[HTML]{000000} -2.342} &
  {\color[HTML]{000000} -1.743} &
  {\color[HTML]{000000} -1.307} &
  {\color[HTML]{000000} -0.4044} &
  {\color[HTML]{000000} -1.608} &
  {\color[HTML]{000000} -1.831} &
  {\color[HTML]{000000} -0.6227} &
  {\color[HTML]{000000} -1.905} \\ \hline
{\color[HTML]{000000} 1950} &
  {\color[HTML]{000000} -0.2097} &
  {\color[HTML]{000000} -1.544} &
  {\color[HTML]{000000} -1.015} &
  {\color[HTML]{000000} -2.15} &
  {\color[HTML]{000000} -1.034} &
  {\color[HTML]{000000} -1.534} &
  {\color[HTML]{000000} -0.6727} &
  {\color[HTML]{000000} -1.64} \\ \hline
{\color[HTML]{000000} 1973} &
  {\color[HTML]{000000} -2.683} &
  {\color[HTML]{000000} -1.809} &
  {\color[HTML]{000000} -1.053} &
  {\color[HTML]{000000} -0.01373} &
  {\color[HTML]{000000} -0.9597} &
  {\color[HTML]{000000} -0.9477} &
  {\color[HTML]{000000} -0.5022} &
  {\color[HTML]{000000} -1.63} \\ \hline
{\color[HTML]{000000} 2003} &
  {\color[HTML]{000000} -0.4582} &
  {\color[HTML]{000000} -1.74} &
  {\color[HTML]{000000} -1.441} &
  {\color[HTML]{000000} -0.455} &
  {\color[HTML]{000000} -1.189} &
  {\color[HTML]{000000} -1.101} &
  {\color[HTML]{000000} -1.835} &
  {\color[HTML]{000000} -1.416} \\ \hline
\end{tabular}%
}

\flushleft \textbf{Abbreviations:} WE: Washidu East; WW: Washidu West. \textit{P.~pallidum}: \textit{Polysphondylium pallidum}; \textit{D.~purpureum}: \textit{Dictyostelium purpureum}; \textit{P.~violaceum}: \textit{Polysphondylium violaceum}. For scientific names of soil mesofauna, see Table~2. Blank values indicate undefined or overflow entries. 
\end{table}

From the theorems discussed, we deduce that $\mathbf{P}^2$ can be interpreted as a pencil of elliptic curves possessing a section of order two and an additional multisection. Setting $\zeta = e^{2\pi i/3} = (e^{\pi i/3})^2$ as the initial condition for $\mathbf{P}^2$ at the point $x_a = 0$, we obtain: \begin{equation} t = \zeta + 1, \quad X(\zeta + 1) = \frac{1}{1 - \zeta}, \quad X'(t) = \frac{1}{3}. \end{equation} In the PzDom model, $1/\Im(s - 1) \approx e^{\pi i/3}$ is used for forecasting, and $t$ becomes $1 + (e^{\pi i/3})^2$ when $\Re(s - 1)$ is negligible. Near the trivial zeros of the Riemann $\zeta$-function, we find $t \sim 1$ and $X(t) \sim 1$, with $X'(t) = \frac{1}{3}$ indicating a $(2+1)$-dimensional system.

Thus, the system effectively reduces to $2+1$ dimensions. To reproduce the kernels, let $q \in (Q^{\infty})^{\Gamma}(\mathbf{H}^*)$. Then, \begin{equation} q(w)dw^2 = \frac{12}{\pi} \left( \int_{\mathbf{H}} \frac{q(\bar{z}) \Im(z)^2}{(z - w)^4} |dz|^2 \right) dw^2, \end{equation} where $w = \alpha/\beta$ and $z := (\alpha \zeta + \bar{\alpha}) / (\beta \zeta + \bar{\beta})$. The expression in parentheses is the reproducing kernel (Proposition 5.4.9 of \cite{Hubbard2006}).

Now consider the $q$-difference Painlev\'e VI equation associated with the $\widehat{\mathfrak{gl}}3$ hierarchy. Let $q = -s$, and observe that $ y(x + 1) = \frac{1 - q^x}{1 - q} y(x) = \left( \sum_{i = 0}^{x - 1} q^i \right) y(x), $ which can be transformed from $q$ to $-s$ in the limit $x \to \infty$.

Assuming $|q| > 1$, and treating $t$ as the independent variable with $f$ and $g$ as dependent variables, we define the $q$-Painlev\'e VI system as: 
\begin{equation}
\begin{split}
T(g) = \frac{(f - ta_1)(f - ta_2)b_3b_4}{g(f - a_3)(f - a_4)}, T^{-1}(f)\\
 = \frac{(g - tb_1)(g - tb_2)a_3a_4}{f(g - b_3)(g - b_4)}, 
\end{split}
\end{equation}
 with \begin{equation}
f = -\frac{\mathcal{A}_0^{12}}{\mathcal{A}_1^{12}}, g = \frac{(\mathcal{A}_0^{12} + x_1\mathcal{A}_1^{12})(\mathcal{A}_0^{12} + x_1q^{\alpha_1 + 1}\mathcal{A}_1^{12})}{q(\mathcal{A}_0^{11}(\mathcal{A}_1^{12})^2 - \mathcal{A}_1^{11}\mathcal{A}_0^{12}\mathcal{A}_1^{12} + q^{\beta_2 + 1}(\mathcal{A}_0^{12})^2)}.
\end{equation} and matrix components defined as: \begin{equation}
\mathcal{A}_0^{12} = q^{\alpha_1 + \alpha_2 + 2}x_1x_2\omega_{13}\bar{w}_{32},
\end{equation}
\begin{equation}
\mathcal{A}_1^{12} = q^{\alpha_1 + 1}x_1\omega_{11}\bar{w}_{12} +  q^{\alpha_2+ 1}x_2\omega_{12}\bar{w}_{22},
\end{equation}
\begin{equation}
\mathcal{A}_0^{11} = q^{\alpha_1 + \alpha_2 + 2}x_1x_2(1 + \omega_{13}\bar{w}_{31}),
\end{equation}
\begin{equation}
\begin{split}
\mathcal{A}_1^{11} = -q^{\alpha_1 + 1}x_1(1 + \omega_{12}\bar{w}_{21} + \omega_{13}\bar{w}_{31})\\
  - q^{\alpha_2 + 1}x_2(1 + \omega_{11}\bar{w}_{11} + \omega_{13}\bar{w}_{31})
\end{split}
\end{equation}

In the PzDom model, setting $q = -b \ln D$ allows for straightforward computation of local time evolution. The parameters $(a_1, a_2, a_3, a_4)$ and $(b_1, b_2, b_3, b_4)$ represent four interacting degrees of freedom in this soliton-type similarity reduction equation \cite{Jimbo1996, Kakei2007}.

In essence, this formulation captures a direct sum of two Virasoro algebras—one resembling a Majorana fermion analog and the other a super-Virasoro algebra \cite{Bershtein2015}—providing a rich algebraic structure for modeling hierarchical dynamics and time evolution in complex systems.

\subsection{FurthercConsiderations of $1+1$ dynamics}

An alternative approach to system dynamics involving the parameter $q$ begins with the structure of a Young tableau. Let $S$ be a finite or countable set—for instance, representing species densities—modeled as $\mathrm{Spec},\mathbb{Z}$.

For $\Re(s) \leq \frac{1}{2}$, define the absolute value of the absolute zeta function as $ \zeta_K = \zeta_{\mathbb{G}_m/F_1}(x, y) = \left| \frac{s(x, y)}{(s - 1)(o, y)} \right|, \quad x, y \in S, $ where $\mathbb{G}_m = \mathrm{GL}(1)$.

For $\Re(s) > \frac{1}{2}$, define the absolute value of the inverse of the absolute zeta function as $ \zeta_K = \frac{1}{\zeta_{\mathbb{G}_m/F_1}(x, y)} = \left| \frac{(s - 1)(o, y)}{s(x, y)} \right|, \quad x, y \in S. $ This $\zeta_K$ functions as a Martin kernel.

Define a distance function: $ D_\delta(x, y) = \sum_{z \in S} C_z \left( |\zeta_K(z, x) - \zeta_K(z, y)| + |\delta_{zx} - \delta_{zy}| \right), $ where $\delta$ is the Kronecker delta. The metric space $(S, D_\delta)$ induces a discrete topology and is totally bounded. Its completion is denoted $\hat{S}$, and the Martin boundary is defined as $\partial S = \hat{S} \setminus S$, representing a $(d - 1)$-dimensional species density space not necessarily constrained by random walks or transition probabilities.

The space $S^d$ then encodes all possible configurations of $S^{d-1}$ extended by a time dimension. Moreover, the set $S^{d-1}$ can be represented by a Young tableau in Frobenius coordinates. Taking the associated Maya diagram projects this data into a one-dimensional structure. Hence, a three-dimensional system can be effectively encoded as a one-dimensional system over $F_1 = \mathbb{F}_q$.

In this framework, the set of individual species counts lies over $\mathbb{Z}$, and time $X$ is modeled as a flat $\Lambda$-algebra over $\mathbb{Z}$. A $\Lambda$-structure on $X$ is given by $ \psi_p: X \to X, \quad \text{with } \psi = X \times_{\mathrm{Spec}\mathbb{Z}} \mathrm{Spec}\mathbb{F}_{p_c}. $ That is, $\Lambda = \mathbb{Z}[\mathrm{Gal}(\mathbb{Z}/\mathbb{F}_{p_c})]$. The parameter $p_c = 1$ when no hierarchy or periodicity is present, and $p_c = 2$ in the protein or species datasets discussed earlier. Thus, the hierarchy extends from $F_1$ to $F_2$.

We then have: $ M_{n/\mathbb{F}1} = \underline{\mathrm{Hom}}_{\mathbb{G}_m/F_1}(\mathbb{A}^n, \mathbb{A}^n) = \zeta_K, \quad GL_{n/\mathbb{F}1} = \underline{\mathrm{Aut}}_{\mathbb{G}_m/F_1}(\mathbb{A}^n) = S^n. $ Therefore: $ s \in \mathbb{G}_m, \quad s - 1 \in F_1 \text{ when } \Re(s) \leq \frac{1}{2}; \quad s - 1 \in \mathbb{G}_m, \quad s \in F_1 \text{ when } \Re(s) > \frac{1}{2}. $ We interpret $q \in \mathbb{G}m$, and $\mathrm{Spec}(q)$ corresponds to either $\mathrm{Spec}(s)$ or $\mathrm{Spec}(s - 1)$. Since $D = e^{s/b}$ is computable using the root of time $t$, and temperature $b_t$ at time $t^2$ and $b_{t-1}$ at $(t - 1)^2$, the dynamics of $q$ can be derived from this foundational data.

This perspective aligns with the framework developed in \cite{Borger2009}, which connects $\Lambda$-rings and the geometry over $\mathbb{F}_1$ to Grothendieck’s Riemann-Roch theorem. It provides an alternative explanation for how a one-dimensional system, when endowed with a suitable topology, can give rise to effective $3 + 1$-dimensional dynamics.

\subsection{$\wp$ as evaluations for interactions}
Consider Wallis' formula:
\begin{equation}
\lim_{n \to \infty} \frac{1}{\sqrt{n}} \cdot \frac{2 \cdot 4 \cdot \cdots \cdot (2n)}{1 \cdot 3 \cdot \cdots \cdot (2n - 1)} = \sqrt{\pi}.
\end{equation}
The numerator, a product of even numbers, may be interpreted as bosonic multiplications, while the denominator, a product of odd numbers, corresponds to fermionic multiplications. The square of this ratio, divided by $n$ as the average number of actions, yields $\pi$. Thus, $\pi$ represents the ratio of bosonic to fermionic multiplications. In this analogy, the area of a circle corresponds to bosonic actions, and the square of the radius to fermionic actions. Globally, there are approximately three times more bosonic actions than fermionic ones.

To further analyze bosonic contributions, consider even negative $s$ (excluding $s = 0$) with $\mu(n) = 1$. The Weierstrass $\wp$ function satisfies
$
\wp(1/n) = \sum_{s = -2}^{\text{negative even} \neq 0} (1/n)^s,
$
and a $((s/2 + 1) \times n)(n \times 1)$ matrix computes a set of patch qualities $P_s$ for bosons, including a future state at $s = -2$.

Similarly, for even negative $w$ with $\mu(n) = -1$,
$
-\wp(1/n) = -\sum_{w = -2}^{\text{negative even} \neq 0} (1/n)^w,
$
and a $((w/2 + 1) \times n)(n \times 1)$ matrix computes patch qualities $-P_w$ for fermions, including a future state at $w = -2$.

Given $w = s - 1$, we define
$
P(s) = P_s - P_{w = s - 1} = \zeta(s) + n + n^2,
$
relating the Riemann $\zeta$ function to patch quality. Population bursts associated with even $s$ (odd $w$) can be modeled by $P_s \to +\infty$ for negative even $s$ (negative odd $w$), or in less extreme cases, $P_w \to \mp\infty$ as $s \to 1 \mp 0$ ($w \to 2 \mp 0$).

Since $P(0) \neq 0$ and $P(0) \to +\infty$, consider
$
P(s) = \wp(1/n) + \frac{\wp(1/n)}{n},
$
and let $a_k, b_k$ be the zeros and poles of the function. Then,
\begin{equation}
\begin{split}
f_P(1/n) = C_P \prod_{k = 1}^{\infty} \frac{\wp(1/n) - \wp(a_k)}{\wp(1/n) - \wp(b_k)} \\
\times \prod_{k = 1}^{\infty} \frac{\wp(1/n)/n - \wp(a_k)/n}{\wp(1/n)/n - \wp(b_k)/n} = 0,
\end{split}
\end{equation}
since the constant $C_P = 0$ when $s = 0$ \cite{Ahlfors1979}. Thus, $s = 0$ ($w = -1$) implies that every singularity can be treated as a zero ideal under $f_P$.

We may interpret $\log s$ as a measure of fitness when fitness is sufficiently small. If we regard the Weierstrass $\zeta$ function
$
\zeta(z; \Lambda) = \frac{1}{z} + \sum_{w \in \Lambda^*} \left( \frac{1}{z - w} + \frac{1}{w} + \frac{z}{w^2} \right)
$
(not the Riemann zeta function) as a distribution function, then the additive operation for fractal dimensions $s_1, s_2$ yields
\begin{equation}
\zeta(s_1 + s_2) = \zeta(s_1) + \zeta(s_2) + \frac{1}{2} \cdot \frac{\wp'(s_1) - \wp'(s_2)}{\wp(s_1) - \wp(s_2)}.
\end{equation}
This shows that the third term on the right-hand side represents the contribution from interactions between different fractal hierarchies, beyond the direct sum of distribution functions.

Table 10 and Supplementary Table 2 present numerical values for these interaction terms. Generally, interactions are weaker in summer compared to spring or fall. Notably, at Washidu West in September, the interaction strength followed the order Pv-Dp-Pp. In October, a strong interaction between Pv and Pp was also observed. In contrast, Washidu East exhibited weaker interactions overall and was dominated by Pp. It is also noteworthy that hetero-interactions were generally weaker than homo-interactions, as expected. However, when using population data instead of species data, hetero-interaction terms often showed similar or even greater magnitudes than homo-interaction terms (data not shown), which aligns with theoretical expectations.

\begin{table}[]
\caption{Hetero-interaction terms.}
\centering 
\small
\scalebox{0.7}[0.7]{
\begin{tabular}{|l|l|l|l|}
\hline
hetero-interaction & WE                                          &             & WW                                          \\ \hline
Pp-Dp (June)       & 0.001160 - 0.09419i  & Pp-Dp (Jul) & 0.01142 - 0.3558i    \\ \hline
Pp-Pv (June)       & 0.01818 - 0.4494i    & Pp-Pv (Jul) & 0.0008154 - 0.08021i \\ \hline
Dp-Pv (June)       & 0.001160 - 0.09419i  & Dp-Pv (Jul) & 0.0008154 - 0.08021i \\ \hline
September          & 0.09055 - 0.5885i    & August      & 0.09149 - 0.6049i    \\ \hline
October            & 0.03433 - 0.3021i    & Pp-Dp (Sep) & -2.372E+8 + 3.704E+8i  \\ \hline
November           & 0.0003791 - 0.05081i & Pp-Pv (Sep) & 0.008871 - 0.2633i   \\ \hline
                   &                                             & Dp-Pv (Sep) & -1.659E+8 - 1.791E+9i  \\ \hline
                   &                                             & October     & -3.728E+5 - 3.975E+5i      \\ \hline
\end{tabular}
}
\\ \flushleft \textbf{Abbreviations:} WE: Washidu East quadrat; WW: Washidu West quadrat; \textit{P.~pallidum} (Pp): \textit{Polysphondylium pallidum}; \textit{D.~purpureum} (Dp): \textit{Dictyostelium purpureum}; \textit{P.~violaceum} (Pv): \textit{Polysphondylium violaceum}. For soil mesofauna, see Supplementary Table~2. Data related to world economics are omitted, as they typically show uniformly low values across all tables.
\end{table}

For further clarification, regarding $\wp$ as an elliptic function,
\begin{equation}
\wp'^2 = 4\wp^3 - g_2 \wp - g_3
\end{equation}
is its normal form without multiple roots. There exist rational functions $F(\wp(u))$ and $G(\wp(u))$, known as Legendre canonical forms of elliptic integrals, such that any elliptic function can be expressed as
$
f(u) = F(\wp) + G(\wp)\wp'.
$
Therefore, a particular state during time evolution, represented by $\wp'$, can be associated with any elliptic function form via a specific pair of Legendre canonical forms. The use of the Weierstrass $\wp$ function is thus intimately connected to the abstraction of interactions among states, particularly through the cubic term $\wp^3$.

Let $\Omega$ denote a period of $f(u)$. The canonical form is given by
$
K(\Omega) \cong \mathbb{C}[x, y]/(y^2 - 4x^3 + g_2 x + g_3),
$
where $\mathbb{C}[x, y]$ is an integral domain. This ideal characterizes the observed phenomena in terms of the functions $F$ and $G$.

To further develop this evaluation, one may interpret $s$ as the elliptic function $f(u)$ under double periodicity in $p$ and $l$. A linear plot of $f(u)$ against $\wp'$ reveals the values of $F$ and $G$. Empirically, due to the typically large magnitude of $\wp'$, we find $G \sim 0$, and $F$ closely approximates the $s$ values used in calculating interactions. This method enables identification of which interacting partner plays the dominant role in the interaction.

At Washidu East (WE), the climax species \textit{Pp} dominated, whereas at Washidu West (WW), the pioneering species \textit{Dp} and \textit{Pv} played more significant roles \cite{Adachi2015}. Note that $F$ and $G$ are solutions to corresponding hypergeometric differential equations. Consequently, the invariants $g_2$ and $g_3$ emerge naturally during the time development process. The periods $\omega$ can be computed via
$
g_2 = 60 \sum_{\omega \in \Lambda'} \frac{1}{\omega^4}, \quad g_3 = 140 \sum_{\omega \in \Lambda'} \frac{1}{\omega^6}.
$
Riemann's theta relations demonstrate how a $(3 + 1)$-dimensional system may be restructured into a $(2 + 2)$-dimensional framework.

\subsection{Renormalization and fractals among sets beyond (empty) boundaries}

In physics, renormalization is widely employed to manage the divergence of terms (e.g., \cite{Benfatto1995}, \cite{Cardy1996}), despite its known fragility in rigorous mathematical frameworks. Notably, renormalization may be intrinsically linked to the emergence of fractals, which appear to vanish at a given observational layer, only to reemerge at larger scales as higher-order structures. 

To explore this phenomenon, we begin by analyzing a model of magnetization under the scaling hypothesis, emphasizing that the parameters involved correspond precisely to the imaginary parts of certain nontrivial zeros of the Riemann zeta function. Drawing an analogy between magnetization and species dynamics in biology, we propose that fractal structures can account for many of the salient features observed in such systems. To approach this problem, we introduce two mathematical constructs: \'{e}tale and Frobenioid.

Broadly speaking, an \'{e}tale is a category-theoretic abstraction akin to a local diffeomorphism—i.e., a function between smooth manifolds that preserves local differentiable structure. A Frobenioid, on the other hand, is a categorical abstraction of line bundles or monoids of divisors over a base category of topological localizations, such as a Galois category \cite{Mochizuki2020a}. A Gaussian monoid may be loosely interpreted as a harmonic function associated with an \'{e}tale or Frobenioid structure. Consequently, $p$ (persistent homology) or $l$ (\'{e}tale cohomology) can be evaluated in terms of the number of Gaussian monoids.

In Cartesian coordinates—implying entanglement of each line as discussed in \cite{Witten2017}—the system described via \'{e}tale and Frobenioid corresponds to the de Rham side or Hodge filtration, referred to as the Frobenius picture in \cite{Mochizuki2020b}. In contrast, in polar coordinates, the system is described by the \'{e}tale side or Galois action on torsion points, known as the \'{e}tale picture in \cite{Mochizuki2020b}. The Frobenius picture aligns with the final figure we propose herein. In the \'{e}tale picture, gradients represent \'{e}tale actions on self-organization, while rotations represent Frobenioid actions on circulation—both contributing to divergence. The logarithm of a multiplicative \'{e}tale picture corresponds to an additive Frobenius picture in the context of complex metrics.

Our biological model exhibits properties consistent with a real \'{e}tale structure, while the Frobenioid corresponds to the imaginary component of a newly defined complex metric $s$. The introduction of automorphic forms and tori leads naturally to a master Lagrangian, inspired by analogies with the standard model of particle physics. This framework further elucidates how higher-order fractals can absorb divergences that are negligible at the current observational layer.

Through a biological species model that inherently exhibits multilevel hierarchical structure, we propose a hybrid toy model that integrates mathematics, physics, and biology to address the problem of divergent terms in complex systems.

\subsection{Imaginary part of zeros for Riemann zeta function deciphering scaling hypothesis}

Drawing on the concept of critical phenomena in magnetic systems, we developed a biological phase model. To test this hypothesis, we begin with the empirical critical exponents for systems with high isotropy: specific heat $\alpha \approx -0.14 \approx -T_1/100$, spontaneous magnetization $\beta \approx 0.38 \approx T_6/100$, and magnetic susceptibility $\gamma \approx 1.375 \approx T_{47}/100$, where $T_n$ denotes the imaginary part of the $n$-th nontrivial zero of the Riemann zeta function, corresponding to the $n$-th prime in ascending order. The empirical relation $\alpha + 2\beta + \gamma \approx 2$ is known as the scaling hypothesis.

Applying this scaling hypothesis to the PzDom model by analogy with magnetization, we write:
\begin{equation}
\ln PD^{N_k} = \ln P + \left(-\frac{1}{b}\right)(-\Re(s)N_k) + i\frac{\Im(s)N_k}{b},
\end{equation}
where $\alpha = \ln P$ corresponds to the specific heat term, $\beta = -1/b$ to the spontaneous magnetization term (which is fractal in nature, as discussed below), and $\gamma = \Im(s)/b$ to the magnetic susceptibility term.

The fractal dimension $\beta$ is associated with a 26-dimensional structure, reminiscent of heterotic string theory or the Fake Monster Lie Algebra, as suggested in \cite{Borcherds1996, Kachru2017}. This connection is made through the operator $\nabla$ (later related to $(p, v)$ in \cite{Adachi2017}) and the compactification $T^7 \times T^2$. These structures may become apparent under congruence conditions such as $p \equiv 1, 3 \mod 8$ and $p \equiv 1, 3, 4, 9, 10, 12 \mod 13$, considering representations like $p = x^2 + 26y^2$. Castelnuovo's theorem implies that for a curve $X$ of even degree $d$ and genus $g$ in $\mathbb{P}^3$ not lying in a plane, $g \leq \frac{1}{4}d^2 - d + 1 = 25$ when $d = 12$ \cite{Hartshorne1977}, setting a dimensional constraint on string-theoretic symmetries.

In this framework, energy is given by $E = -\Re(s)N_k$, momentum by $\mathbf{P} = \Im(s)^2 N_k$, and temperature by $T = b$. 

Furthermore, the expected prime values $l = \Re(s)$ for species exhibit a right-handed helicoid motion with constant $M_z + \hbar P_z$, where $M_z$ and $P_z$ are the $z$-components of angular momentum and momentum, respectively. Taking $\hbar = 1$, and assuming $l$ and mass (as a proxy for population/species density) are constant, we define $\bar{D} = N_k/E(N)$ as the average linear growth rate. Then, $(1 + l)\bar{D} = C_1 t + C_0$, where $t$ is a parameter and $C_1, C_0$ are constants. In this helicoid-type ruled surface, $(l + 1)$ acts as an effective mass, with $\bar{D}$ as velocity.

This leads to a minimal surface in $\mathbb{R}^3$ defined by:
$
x = l\cos t, \quad y = l\sin t, \quad z = (1 + l)\bar{D} = C_1 t + C_0,
$
where the $z$-axis represents an additional fractal dimension. This surface minimizes area, satisfying the principle of least action.

Considering von Neumann entropy,
$
S_{vN}[\hat{\rho}] = -k\,\mathrm{Tr}[\hat{\rho} \ln \hat{\rho}], \quad \hat{\rho} = \sum p_i \hat{P}_{\varphi_i},
$
where $\varphi_i$ are particular states and $\hat{P}$ are orthogonal projection operators, maximizing $S_{vN}$ corresponds to orthogonalizing the system—effectively generating orthogonal fractal dimensions. This is analogous to the nematic (non-fractal) and cholesteric (helicoid fractal) phases in liquid crystals.

Next, consider $\dim_B B = 2$ as the box dimension of $B$, representing the boundary of $\Re(s)$ that separates Minkowski-measurable (quantized) from Minkowski non-measurable (chaotic) spaces. The relation $\alpha + 2\beta + \gamma \approx 2$ may thus be interpreted as a fractal dimension, with scaling parameters mapped to topological dimensions.

Given that $T_1, T_6, T_{47}$ correspond to $2, 13, 211$—interpreted as the number of ``unsplittable" interactions—and using $10^2 = 100$ as a normalization, the root of this parameter is $2 \times 5 = 10$. Then,
$
\frac{2 + 2 \times 13 + 211}{10} = 23.9 \approx 24.
$
The significance of the number $5$ will be discussed below. Including the purely self-interacting case with $p = 1$, we get $240/10 = 24$.

The $\epsilon$-expansion of the renormalization group yields $(\alpha, \beta, \gamma) = (\epsilon/6, 1/2 - \epsilon/6, 1 + \epsilon/6)$. For a 12-dimensional system, $\epsilon = 12$, giving $(\alpha, \beta, \gamma) = (2, -3/2, 3)$. These values correspond to the dimensions of specific heat, spontaneous magnetization, and magnetic susceptibility, respectively. The sum of the dimensions of specific heat and magnetic susceptibility is $5$, while the doubled dimension of spontaneous magnetization is $-3$, representing diminished dimensions that contribute to fractality.

If we construct a Fuchs-type differential equation from the system, a necessary condition for the sum of characteristic exponents $\lambda$ is
$
\sum \lambda = 240.
$
This is equivalent to
$
\sum_{a \in S} \sum_{k = 1}^n \lambda_{a, k} = \frac{n(n - 1)(\#S - 2)}{2},
$
where $\#S$ is the number of elements in the set $S$. Decomposing $240$ into $\{1, 2, 13, 13, 211\}$ implies $n = 5$ and $\#S = 26$, which corresponds to the dimensional structure of heterotic string theory. This condition is sufficient because the greatest common divisor of $\{1, 2, 13, 13, 211\}$ is $1$, ensuring the existence of a Fuchs-type (hypergeometric) differential equation.

A related dimension appears in an unpublished manuscript by Srinivasa Ramanujan (circa 1916):
\begin{equation}
F(z) = q \prod_{n = 1}^{\infty}(1 - q^n)^2(1 - q^{11n})^2 = \sum_{n = 1}^{\infty} c(n) q^n.
\end{equation}
Consider a stress tensor $T^{ik}$ with 13 independent components, which may be interpreted as a cyclic group powered by $q$:
$
\{W, S_{x/c}, S_{y/c}, S_{z/c}, \sigma_{xx}, \sigma_{xy}, \sigma_{xz}, \sigma_{yx}, \sigma_{yy}, \sigma_{yz}, \sigma_{zx}, \sigma_{zy}, \sigma_{zz}\}.
$
Excluding $W$, we obtain 12 dimensions, analogous to the Teichm\"uller space for genus $g = 3$ ($6g - 6 = 12$), representing two self-interacting and one hetero-interacting term. Their interaction leads to 24 dimensions or weights, as in the Teichm\"uller space for $g = 5$ ($\Re(s) = 5$). Analogous to orthogonal electric and magnetic fields ($\mathbf{E} \perp \mathbf{H}$, with $E = H$), the tensor is non-diagonalizable and lacks an \'etale element.

The weight $24$ also appears in the free energy of the canonical distribution in quantum theory:
\begin{equation}
F = F_{cl} + \frac{1}{24T^2} \sum_i \frac{1}{m_i} \left\langle \left( \frac{\partial U}{\partial q_i} \right)^2 \right\rangle.
\end{equation}
Transitions such as $F \to T_{eff}$, $F_{cl} \to T$, and $24 \to 12$ are also derivable. For the classical case, let $\alpha = (2a_s - 1)/a_s$, $\beta = (1 - b_s)/a_s$, and $\gamma = (2b_s - 1)/a_s$, with $a_s = 1/2$ and $b_s = 3/4$. Consider a system $X^2 + XY + YX + Y^2$ with each term contributing $1/4$. Then $a_s = 1/2$ implies that hetero-interaction terms $XY$ and $YX$ vanish, consistent with the PzDom model's harmonic neutrality at $\Re(s) = 1$.

As the system evolves into the cooperative regime $1 < \Re(s) < 2$, commensalism emerges: $XY$ becomes $1/4$ while $YX$ remains zero, yielding $b_s = 3/4$. This development leads to cooperation with parameter $1$, transitioning into $\Re(s) > 2$, where interactions cease, analogous to an ideal gas. For instance, the relation $P_W = \frac{2}{3}u$ (pressure vs. internal energy density) implies a $2/3$ weight. The general inequality $P_W < \frac{E}{3V}$ suggests a norm of $1/3$, derived from the trichotomy $X^2 + XY + Y^2$. The norm $|N(p)| = 1/3$ appears in species/population densities.

For stricter systems, consider the van der Waals equation:
\begin{equation}
P_W = \frac{NT_W}{V - Nb_W} - \frac{N^2 a_W}{V^2},
\end{equation}
with critical points:
$
T_{cr} = \left(\frac{2}{3}\right)^3 \frac{a_W}{b_W}, \quad V_{cr} = 3Nb, \quad P_{cr} = \left(\frac{1}{3}\right)^3 \frac{a_W}{b_W^2}.
$
This corresponds to a model with $\frac{1}{3}X^2 + \frac{0}{3}XY + \frac{1}{3}Y^2$, where critical phenomena resemble gas-like behavior. Empirically, $q/T_{cr} \sim 10$ when $q$ is latent heat and $T \ll T_{cr}$, indicating $T_{cr}$ as the boundary between $\frac{0}{3}XY$ and $\frac{1}{3}XY$ regimes.

The value $5 + 5 = 10$ was observed for $\Re(s) \geq 5$ (e.g., in October at Washidu East between \textit{Polysphondylium pallidum} and \textit{Dictyostelium purpureum}). This value matches the critical temperature. For these species, $(a_W, b_W) \sim (2700, 0.53)$ and $(82000, 16)$, respectively—the latter being 30 times the former. Here, \textit{P. pallidum} may represent $l = 2$ and \textit{D. purpureum} $l = 3$ (two main fractal dimensions plus one vibrational dimension). Both species are shrinking due to $P_W < 0$.

However, consider an additional critical index for magnetic charge, $\delta (= 3)$. This is reasonable because population/species density is inherently three-dimensional, and one dimension may be equipartitioned among the three spatial dimensions, serving as the counterpart to magnetic charge. The classical Rushbrooke inequality,
$
\alpha + 2\beta + \gamma + \delta \geq 5,
$
is satisfied in this context. Moreover, the appearance of a prime number equal to or greater than $5$ suggests a deeper mathematical structure, potentially linked to the logic underlying Inter-Universal Teichm\"uller Theory \cite{Mochizuki2020a, Mochizuki2020b, Mochizuki2020c, Mochizuki2020d}.

This reflects a hierarchical structure induced by the product $\mathbb{R}^3 \times \mathbb{R}^3 = \mathbb{R}^6$, which itself is generated by another $\mathbb{R}^3$ structure already situated in the regime $\Re(s) \geq 5$, yielding a total of 9 dimensions. This dimensional expansion becomes prominent when $\Re(s) \geq 5$. Therefore, to develop a system that transcends the boundary $\dim_B B = 2$, a distinct interaction partner is required for progression beyond $\Re(s) = 5$ (since $2 \times 5 = 10$). Alternatively, primes of the form $p = x^2 + 5y^2$ satisfying $p \equiv 1, 9 \mod 20$ might suffice, though such cases have not been empirically observed.

This leads to a Navier-Stokes-type interaction equation, where the concept of ``viscosity" represents the entanglement of different subgroups. The dimension $24$ thus emerges as a critical parameter in describing such phenomena.

Consider Mathieu's equation:
\begin{equation}
\ddot{x} + \omega_0^2[1 + \cos(2\omega_0 + \varepsilon)t]x = 0,
\end{equation}
which models resonance near the outer vibration frequency $\gamma_M = \omega_0$. The boundaries of instability for parametric resonance are given by:
$
\varepsilon_M = -\frac{5}{24}\omega_0, \quad \frac{1}{24}\omega_0.
$
These values suggest that critical phenomena can be interpreted through a 24-dimensional system, with resonance behavior associated with a system characterized by $\Re(s) = 5$.

The general solution to the Mathieu (Hill) equation is:
$
x(z) = e^{\mu_M z} x_1(z) + e^{-\mu_M z} x_2(z),
$
where $x_1(z)$ and $x_2(z)$ are periodic functions, such as those arising from $\omega_0^2[1 + \cos(2\omega_0 + \varepsilon_M)t]$. At least locally, this equation describes the transfer of \'{e}tale structure from one domain to another.

The \'{e}tale is computable via:
\begin{equation}
\cosh \mu_M T_M = \phi_2\left(\frac{T_M}{2}\right)\phi_1'\left(\frac{T_M}{2}\right) + \phi_1\left(\frac{T_M}{2}\right)\phi_2'\left(\frac{T_M}{2}\right),
\end{equation}
where $\phi_1$ and $\phi_2$ are particular solutions of $x$, and $T_M = \frac{2\pi}{2\omega_0 + \varepsilon_M}$. Calculations of the derivatives $\phi'$ indicate that $\varepsilon_M = -\frac{5}{24}\omega_0$ is the dominant resonance mode, which empirically prevails in Dictyostelia (see Table~11).

\begin{table}[]
\caption{$\phi'$ values.}
\centering
\scalebox{0.7}[0.7]{
\begin{tabular}{|l|l|l|l|l|}
\hline
$\phi'$ & \textit{P. pallidum 1} & \textit{P. violaceum 1} & \textit{P. pallidum 2} & \textit{P. violaceum 2} \\ \hline
WE June                 & 1.00E+168              & 2.3E+272                & -2.1E+167              & -2E+271                 \\ \hline
WE November             & 4.7E+69                & 7E+186 to 4E+188         & -1.06E+69              & -1E+185 to -5E+186       \\ \hline
\end{tabular}
}
\small \\
\begin{flushleft}
\textbf{Abbreviations:} WE: Washidu East quadrat; \textit{P.~pallidum}: \textit{Polysphondylium pallidum}; \textit{P.~violaceum}: \textit{Polysphondylium violaceum}. The numerals 1 and 2 denote the cases $\varepsilon = -\frac{5}{24}\omega_0$ and $\varepsilon = \frac{1}{24}\omega_0$, respectively.
\end{flushleft}
\end{table}

In analogy with rigid bodies (e.g., a uniform sphere of radius $a$) as Monsters, the moment of inertia is given by $I = \frac{2}{5} \mu a^2$, and the $z$-component of the moment of force in the fractal axis is described by
$
I \frac{d\Omega_z}{dt} = K_z.
$
This equation characterizes a Monster with $l = 2$ under the guidance of a $\Re(s) = 5$ resonance. Considering two Monsters with boundaries and one additional dimension for fluctuation (analogous to the normal vibration between two atoms), the configuration $2 + 2 + 1 = 5$ signifies the successful construction of a $\Re(s) = 5$ system.

If the vibrational energy is strong, a Hodge-Kodaira decomposition-type observation and time development model suggests that the boundary of interaction forms a hyperboloid of one sheet, indicating a connected relation. Conversely, weak vibration yields a hyperboloid of two sheets, representing a split structure. When the surface reaches the origin of coordinates, it becomes a cone, symbolizing the point of observation.

Anisotropic critical phenomena may be influenced by the distinction between Helmholtz free energy (without migration, as in the PzDom model) and Gibbs free energy (with migration), leading to slightly lower critical indices compared to isotropic systems.

In fluid mechanics, long shallow-water waves are observed near the boundary $\partial B$, while short deep-water waves appear near the origin of coordinates. A characteristic wavelength is given by $\lambda_m = 2\pi/N(T)$, where $\lambda < \lambda_m$ corresponds to capillary waves dominated by surface tension effects from $\partial A$ (with $A \subset B$). In contrast, $\lambda > \lambda_m$ corresponds to gravity waves, where the influence of $D$ becomes dominant. Thus, $D$ governs regions near the boundary, while inner structures exhibit substructures shaped by their borders.

Vorticity and related features can be analyzed using persistent homology \cite{Kramar2016, Kashiwara2018}, denoted by $p$. Note that $\Re(s)$ corresponds to \'{e}tale cohomology and is dual to $\Im(s)$. Homology pertains to internal structure, while cohomology relates to boundaries.

For example, consider the Poisson-Schwarz integral formula:
\begin{equation}
f(z_0) = iC_I + \frac{1}{2\pi i} \int_{\partial A} \Re[f(z)]\, dK, \quad
f(z_0) = C_R + \frac{1}{2\pi} \int_{\partial A} \Im[f(z)]\, dK,
\end{equation}
where $H$ is a complex velocity potential with $(z_0, a \in A)$ as a source or sink of strength $1$, and $G$ is a complex velocity potential with $z_0$ representing a vortex filament of strength $1$. The kernel is given by $K = H(z; z_0, a) + iG(z, z_0)$, with $C_R = \Re[f(a)]$ and $C_I = \Im[f(a)]$.

These formulas are valuable for analyzing both internal structures and boundary behavior, provided the system is regular and singularities can be neglected. 

\subsection{A master automorphic form}

An automorphic form is an invariant meromorphic function under the action of a linear transformation group. For the modular group, such a function is referred to as a modular function. A modular function satisfies the transformation property
\begin{equation}
\lambda(\tau) = \lambda\left(\frac{a\tau + b}{c\tau + d}\right), \quad
\begin{pmatrix}
a & b \\
c & d
\end{pmatrix}
\equiv 
\begin{pmatrix}
1 & 0 \\
0 & 1
\end{pmatrix}
\pmod{2},
\end{equation}
which induces a permutation $(0, 1, \infty) \mapsto (1, \infty, 0)$, implying that $\lambda^3 = \mathrm{id}$. This cyclic structure reflects a triadic system composed of an observant, an observer, and a limit—forming a framework of interest in three-dimensional space.

For the modular group, the automorphic form $J(\tau)$ is given by
\begin{equation}
J(\tau) = \frac{4}{27} \cdot \frac{(1 - \lambda + \lambda^2)^3}{\lambda^2(1 - \lambda)^2} = \frac{-4(e_1e_2 + e_2e_3 + e_3e_1)^3}{(e_1 - e_2)^2(e_2 - e_3)^2(e_3 - e_1)^2},
\end{equation}
where $e_1, e_2, e_3$ are the roots of the associated cubic polynomial. A specific domain $\Delta$ such that $\Delta \oplus \Delta' = [-1, 1] \subset \mathbb{H}$ is mapped into the upper half-plane $\mathbb{H}$. In the future, the transformations $0 \mapsto 1$, $1 \mapsto \infty$, and $\infty \mapsto 0$ are realized \cite{Ahlfors1979}.

In the special case where the trace of $\lambda$ is $2$, a master form exists:
\begin{equation}
\rho_G(c_G) = c_G \, \mathrm{Id}_W,
\end{equation}
where $\mathrm{Id}_W = |J(\tau)|$ is the identity mapping, as per the Stone-von Neumann theorem \cite{Stone1930, vonNeumann1931, vonNeumann1932, Stone1932}. This corresponds to the case $\varnothing = \partial B$. The condition $J(\tau) = 1$ can be recursively defined using the above relations, starting from any specific time point. If one assumes a primitive recursive function, no initial value at infinity is required, provided all relevant information is embedded in $J(\tau)$.

In the PzDom model, the static final state is characterized by $N_k = a$, $b = 0$, $D = 1$, $H(t) = 1/2$, $\Re(s) = 0$, and $\Im(s) = 1$, yielding $s = i$. For constant $s$, multiplication acts as a shift map in time development, and the resulting transformation matrix $\lambda_s$ is
\begin{equation}
\begin{pmatrix}
i & 0 \\
0 & 1
\end{pmatrix}
\pmod{2}.
\end{equation}

For $\mathrm{Id}_W = J(\tau)$, taking $\tau = \frac{i \pm \sqrt{7}}{2}$ exceeds the boundary $|\tau| = 2$, thereby crossing the threshold $\Re(s) = 2$ in the PzDom model. This implies that $\varnothing \neq \partial B$ (an open model) can be introduced via this $\tau$, acting as a seed for $\partial B$ (a closed model).

The virial theorem relates kinetic energy $T_v$ and potential energy $U$ as
$
\bar{U} = \frac{2E}{k + 2}, \quad \bar{T}_v = \frac{kE}{k + 2},
$
where $E$ is the total energy, and the overbars denote time-averaged quantities. For a system governed by an inverse-square law, the orbit is elliptic with $t'/t = (l_o'/l_o)^{3/2}$, where $l_o$ is the orbital locus. Setting $k = 3/2$ yields $\bar{U} = \frac{4}{7}E$ and $\bar{T}_v = \frac{3}{7}E$. Masses are related by their square roots, and for $\Re(\tau) = \frac{\sqrt{7}}{2}$, we have $\sqrt{\bar{U}} = \frac{2}{\sqrt{7}}\sqrt{E} = \sqrt{E}/l$. If $\sqrt{E} \in \mathbb{Z}$, then $\sqrt{\bar{U}} \cong \mathbb{Z}/l\mathbb{Z}$ and $\sqrt{\bar{T}} \cong l\mathbb{Z}$, analogous to the Chinese Remainder Theorem.

A similar correspondence holds for squared potentials in $D^2$ space with $\Re(s) = 2$, leading to the equivalence $(D, s) \cong (\sqrt{E}, \tau)$ and the mapping $2 \leftrightarrow \frac{\sqrt{7}}{2}$.

For elliptic integrals of the first kind in Legendre canonical form,
$
K = \int_0^1 \frac{dz}{\sqrt{(1 - z^2)(1 - k^2 z^2)}},
$
the moduli $(k, k' = \sqrt{1 - k^2}) = (i, \sqrt{2})$ form a pair of periods for the inverse functions $z$. The value of $K$ is given by $K = \frac{\sqrt{2} \Gamma^2(1/4)}{8\sqrt{\pi}}$, and the periods $(4K, 2iK') = (5.244, 2.622 + 2.622i)$. The branch points of $\sqrt{1 - z^2}$ are located at $(1.311, 3.933)$. Since the sum of residues in a periodic parallelogram of elliptic functions is zero, any nonzero observation implies an unbounded parallelogram and infinite time. Thus, rotation by $i$ and $|\tau| = \sqrt{2}$ form a pair, with $(i^2 = -1, \sqrt{2}^2 = 2)$ representing interacting elliptic functions.

The Weierstrass function
$
\wp = \left(\frac{K}{\omega_1}\right)^2\left(\frac{1}{z^2} - \frac{1 + k^2}{3}\right)
$
shows that $k = i$ corresponds to a Frobenioid (a monoid and morphism induced by a ring homomorphism) via the $\frac{1}{z^2}$ term, while the remaining $-1$ with $k = \sqrt{2}$ corresponds to an \'etale (see Figure~7).

For $\tau = i$, note that $R = \mathbb{Z}[i]$ and $R^* \cong \mathbb{Z}/4$. The automorphism group of a set $X$ has order $4$, and $j = 1728$ (for $\mathrm{char}\,k \neq 3$), so $J(\tau) = 1$. The lattice is $\Lambda = \mathbb{Z} \oplus \mathbb{Z}i$, and $g_3 = -g_3 = 0$, yielding the elliptic curve $X: y^2 = x^3 - Ax$ \cite{Hartshorne1977}, where $-g_3$ is \'etale.

This aligns with Berger's theorem \cite{Kedlaya2010}, where the functor $D_{\mathrm{rig}}^{\dagger}$ from the category of continuous representations of $G_K$ on finite-dimensional $\mathbb{Q}_p$-vector spaces to the category of \'etale $(\phi, \Gamma)$-modules over $\mathbf{B}_{\mathrm{rig}, K}^{\dagger}$ is an equivalence of categories, assuming rigid geometry \cite{Adachi2017}. Such a continuous representation $G_K$ exists, considering the torus structure.

An example of a (Lie) group corresponding to $G_K$ is as follows. Let $v_K: \Gamma_K \to v \to \mathbb{Z}_p^{\times}$ denote the cyclotomic character. For all negative integers $m$ and $N_k = \gamma_{\nabla} \in \Gamma_K$, we have $\gamma_{\nabla}(\zeta_{p^m}) = \zeta_{p^m}^{v_K}$. Then,
\begin{equation}
\nabla = \frac{\ln \gamma_{\nabla}}{\ln v}
\end{equation}
acts as an endomorphism of $D_{\mathrm{rig}}^{\dagger}$ via the logarithmic power series. Since $v$ is continuous, $\Gamma_K$ is a one-dimensional $p$-adic Lie group over $\mathbb{Z}_p$, and $\nabla$ is its Lie group action.

Thus, our model defines a discrete group $N_k / l\mathbb{Z}$ and a continuous Lie group $\nabla$ over $\mathbb{Z}_p = l \mod p$, forming a $p$-adic Hodge theory that connects \'etale and de Rham cohomology \cite{Mochizuki1999}. We can also express this via a short exact sequence:
\begin{equation}
(\sim 0) \to p \to v \to \nabla \to (\sim 0).
\end{equation}

Frobenioids and \'etales are therefore classified as continuous and discrete aspects of corresponding groups. As discussed later, this distinction explains why renormalization—while neglecting fractal structures—is sufficient for interpreting Lie group-based physical theories. However, integrative models require discrete fractal structures derived from \'etale properties. Hence, $(p, l)$ or $(p, v)$ may be viewed as two dimensions of a Young tableau, where the numbers define the order of appearances in the system. Notably, multiplications involving $p$, $l$, and $v$ correspond to additions under logarithmic structures.

\begin{figure}
\includegraphics[width=14cm]{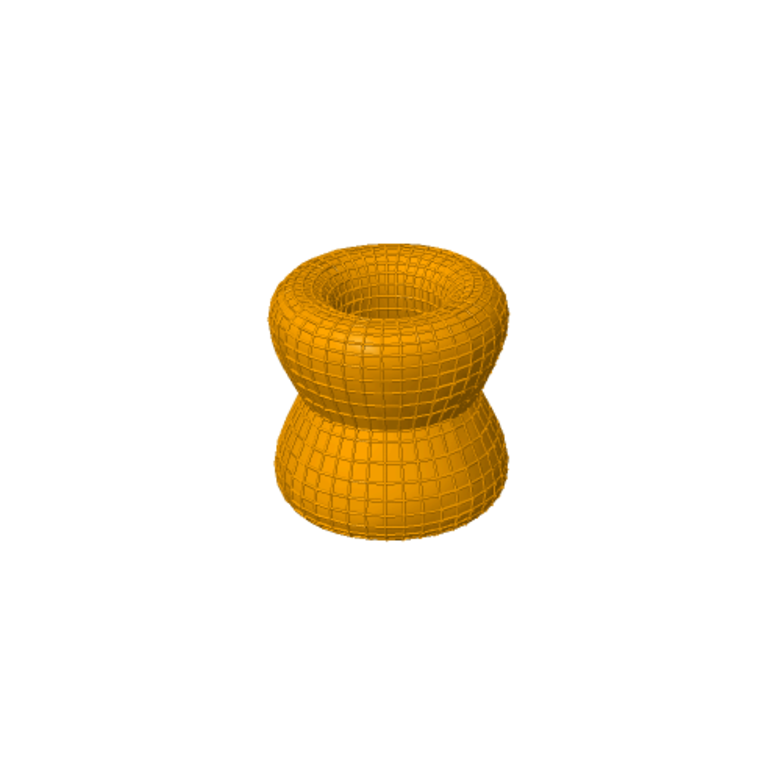}
\caption{The torus of moduli $(i, \sqrt{2})$.}
\label{fig:7}
\end{figure}

\subsection{Communication among different sets (of Monsters) described by an integrated model with a master Lagrangian}

Consider a finite covering of $B$ as a compact space, such as a compact Riemann surface, or more generally, any infinite sequence of points with an accumulation point and identity (Bolzano-Weierstra\ss theorem; Artin perspective). Within such a framework, the maximal order of finite simple groups is bounded by that of the Monster group-maximizing entropic information while minimizing free energy (Noetherian perspective). If the order exceeds this bound, the group ceases to be simple, and the system necessarily splits, leading to the condition $\varnothing = \partial B$.

As an example, consider the absolute zeta function $p' = s/w$, which belongs to the distribution space $\mathscr{D}_{L^{p'}}$ and is expressible as a finite sum of differential coefficients of functions in $L^s$ \cite{Schwartz1966}. However, if one considers Sato hyperfunctions—infinitely ordered objects beyond Schwartz distributions and boundaries \cite{Sato1959, Sato1960}—then communication among different Monsters remains possible. These Monsters, analogous to rigid bodies, exhibit angular velocity $\Im(s)$ independent of coordinate systems, while translational motion or fractal dimension $\Re(s)$ depends on the observer’s frame—mirroring the nonholonomic world of relativity versus the holonomic world of quantum mechanics.

Such communications allow discrete binding of different Monsters into a unified system, not in the manifold sense. Consider the matching of Steiner circles or, in terms of elliptic integrals (denoted $D$), Poncelet’s poristic polygons including Steiner’s series \cite{Emch1899, Emch1900} across two different Monsters. For optimal communication, a non-Euclidean metric such as the Lorentz metric, multiplied by $\sqrt{-1}$ to accommodate hyperfunctions, should be employed. The ideal communication involves transmitting the automorphic information $J(\tau)$ to align the series, particularly between $\Delta \cong B_1$ and $\Delta \cong B_2$ via $\tau = \frac{i \pm \sqrt{7}}{2}$.

The Routh function is instrumental in describing communication from a Monster governed by a Lagrangian (least action principle):
\begin{equation}
R(q, p_m, \xi, \dot{\xi}) = p_m \dot{q} - L.
\end{equation}
Given momentum and coordinate $q$, with $\mathbf{P} = \Im(s)\phi$ and $q = D$, the relative entropy is
$
S = \ln \Delta \Gamma = \ln \left( \frac{\Delta \mathbf{P} \Delta q}{(2\pi \hbar)^3} \right),
$
where $(2\pi \hbar)^3$ is interpreted as the volume of the line integral of action, and $k$ becomes the uncertainty per unit volume. Thus, the least action corresponds to maximal entropy.

In other words, least action maximizes a Minkowski metric $s_M$ associated with $\xi$, transferring to another Monster with a Hamiltonian structure associated with $q$. For example, a Carnot cycle of a Monster $M_C$ performs work on another Monster $M_W$, which absorbs free energy from $M_C$.

From the Frobenioid perspective, $q$ is straightforwardly handled via Kobayashi’s framework. For the \'etale function, refer to Inter-Universal Teichm\"uller Theory \cite{Mochizuki2020a, Mochizuki2020b, Mochizuki2020c, Mochizuki2020d}. We adopt the Dirac picture, where the Hamiltonian is decomposed as $H = H_0 + H'$, with $H_0$ time-independent and $H'$ time-dependent. Here, $M_W$ and $M_C$ correspond to $H_0$ and $H'$, respectively.

This interaction picture (denoted $I$; with the Schr\"odinger picture denoted $S$) yields:
\begin{equation}
\ket{\alpha_I(t)} \equiv e^{iH_{0S}t} \ket{\alpha_S(t)}, \quad
\beta_I(t) \equiv e^{iH_{0S}t} \beta_S e^{-iH_{0S}t}, \quad
i \frac{d}{dt} \ket{\alpha_I(t)} = H'_I \ket{\alpha_I(t)},
\end{equation}
where $\beta = 1/b$ is the inverse temperature.

For example, if we define $\alpha''(\omega) d\omega = \alpha''(b) \ket{\alpha_I(b)} = \Im(s) D$ and impose the condition $\omega \alpha''(\omega) = \bra{\alpha_I(b)} \alpha''(b) = \Im(s) D$, then the integral
$
\int_0^{\infty} \omega \alpha''(\omega) d\omega = \frac{i\pi}{2} \langle \{ \dot{\hat{D}}, \hat{D} \} \rangle = 0
$
leads to a commutative relation in the Poisson bracket, where $m$ is the quantization number. The term $\pi m/2$ may be interpreted as a retarded Green function, or from the perspective of elliptic functions, as Legendre’s relation:
$
\eta_1 \omega_3 - \eta_3 \omega_1 = \frac{\pi m}{2}.
$
In this context, when the time of $\hat{D}$ precedes (follows) that of $\dot{\hat{D}}$, the Green function is non-zero (zero). This behavior is generally observed in Dictyostelia, with the exception of \textit{Polysphondylium violaceum} in November at the Washidu East quadrat. This operation predicts the presence of an irreversible process when the expression is non-zero.

If we fix $\hat{D}$, then $\dot{\hat{D}}$ begins with a vanishing Green function, which becomes non-zero once it surpasses the time of $\hat{D}$. In this case, we define $\varepsilon = \mu = D/\Im(s)$, equivalently $\alpha''/\Im(s)^2$. The distribution of $\varepsilon$ and $\mu$ in varying ratios, constrained by $\varepsilon \mu = (D/\Im(s))^2$, results in a distribution of \'etale and Frobenioid structures, analogous to electric and magnetic forces, respectively.

As $t \to \infty$, $H(t) = 1/2$, and since $m \in \mathbb{Z}$, it follows that $\hat{D}^2 \in 2\mathbb{Z}$, satisfying the condition for crossing the boundary $\Re(s) = 2$.

For further development, forces analogous to inverse-square laws in physics (e.g., gravity and electromagnetism) can be defined on $D$-space. Define
$
M = \frac{\phi}{\sqrt{1 - \frac{(H(t) D)^2}{\Im(s)^2}}},
$
which is analogous to mass or electric/magnetic charge. The square root structure will be justified later.

The electric force can be defined as:
\begin{equation}
\mathbf{F}_e = \frac{\mathrm{sign}(M)}{4\pi \varepsilon} \cdot \frac{M_i M_j}{D^2} \mathbf{e_F},
\end{equation}
where $\mathrm{sign}(M)$ is positive if the force increases the number of individuals of type $M_i$, and negative if it decreases them. The case $i = j$ corresponds to self-interaction. Here, $\varepsilon = D/\Im(s)$.

Similarly, the magnetic force is defined with $\mu = D/\Im(s)$, and the gravitational force is given by:
\begin{equation}
\mathbf{F}_g = \frac{\mathrm{sign}(M)}{4\pi k_g} \cdot \frac{M_i M_j}{D^2} \mathbf{e_F},
\end{equation}
where $k_g = D/\Im(s)$. Although this resembles Newton’s classical law, note that $k_g$ is a function of $D$, not a constant.

All these forces—gravitational, electric, and magnetic—can be viewed as $\mathbb{R}^3 \to \mathbb{R}^2$ projections of a more general force:
\begin{equation}
\mathbf{F}_s = \frac{\mathrm{sign}(M) \Im(s)}{4\pi} \cdot \frac{M_i M_j}{D^3} \mathbf{e_F},
\end{equation}
where fixing a single dimension of $D$ as $k_g \Im(s)$ or $\varepsilon \Im(s)$ yields the respective classical forces. The same logic applies to the magnetic force.

Now consider an analogy to the Schwarzschild solution of a black hole. The entropy of such a solution is given by
$
S_{BH} = \frac{c^3 A}{4G} = \pi D \cdot 4\pi (\Im(s) D)^2,
$
which represents a $\pi D$ increase in the surface area of a sphere with radius $\Im(s) D$. The Planck length in this context is defined as
$
\ell_P = \frac{1}{\sqrt{4\pi D}}.
$
This implies that the information content of the solution is roughly encoded on this minimal surface.

The Schwarzschild radius is
$
r_H = 2GM = \frac{\Im(s) M}{2\pi D},
$
interpreted as the ratio of momentum $\Im(s) M$ to the circumference $2\pi D$. The Hawking temperature associated with this radius is
$
T_H = \frac{1}{8\pi GM} = \frac{\pi D}{2 \Im(s) M}.
$
Thus, a species $M$ is characterized by a trapping radius and a temperature analogous to those of a black hole.

The Jeans wave number $k_J$, wavelength $\lambda_J$, and mass $M_J$ are given by:
\begin{equation}
k_J = \sqrt{\frac{N_k \Im(s)}{bD}}, \quad
\lambda_J = 2\pi \sqrt{\frac{bD}{N_k \Im(s)}}, \quad
M_J = \frac{4}{3} \pi^4 \sqrt{\frac{1}{N_k} \left( \frac{bD}{\Im(s)} \right)^3}.
\end{equation}
Perturbations larger than $\lambda_J$ and $M_J$ permit the development of species dynamics (see Tables~12 and~13).

\begin{table}[]
\caption{$\lambda_J$ values compared to $D$ (observed).}
\centering
Dictyostelia\\
\resizebox{\columnwidth}{!}{%
\begin{tabular}{|l|l|l|l|l|l|l|l|}
\hline
$\lambda_J$ & WE \textit{P. pallidum} & WE \textit{D. purpureum}              & WE \textit{P. violaceum}              & \textit{D} & WE \textit{P. pallidum} & WE \textit{D. purpureum}              & WE \textit{P. violaceum}              \\ \hline
April       &              &                                   &                            & April        &                     &                                  &                                   \\ \hline
June      & 2.358          & {\color[HTML]{FE0000} 0.4255} & 1.852                        & June       & 1.005          & {\color[HTML]{FE0000} 1.013} & 1.009                        \\ \hline
July      &                     &                            &                            & July       &                     &                                   &                                  \\ \hline
August    &                    &                           &                           & August     &                     &                                  &                                   \\ \hline
September & 1.472          & 5.324                        &                            & September  & 1.003          & 1.003                        &                                   \\ \hline
October   & 1.100          & 6.623                        &                            & October    & 1.003          & 1.003                        &                                  \\ \hline
November  & 2.253          &                            & {\color[HTML]{FE0000} 0.2943} & November   & 1.013           &                                   & {\color[HTML]{FE0000} 1.036} \\ \hline
December  &                    &                           &                            & December   &                     &                                  &                                   \\ \hline
January   &                     &                            &                           & January    &                     &                                   &                                   \\ \hline
\end{tabular}
}
\centering
\resizebox{\columnwidth}{!}{%
\begin{tabular}{|l|l|l|l|l|l|l|l|}
\hline
$\lambda_J$ & WW \textit{P. pallidum} & WW \textit{D. purpureum} & WW \textit{P. violaceum} & \textit{D} & WW \textit{P. pallidum} & WW \textit{D. purpureum} & WW \textit{P. violaceum} \\ \hline
April       &              &                      &               & April        &                     &                      &                      \\ \hline
June      &                     &               &               & June       &                     &                      &                      \\ \hline
July      & 1.634          & 2.703           & {\color[HTML]{FE0000} 0.3881}           & July       & 1.006          & 1.003          & {\color[HTML]{FE0000} 1.008}           \\ \hline
August    & 1.476          & 5.127           &               & August     & 1.002          & 1.002           &            \\ \hline
September & {\color[HTML]{FE0000}  0.7980 }         & 1.752           & 4.556          & September  & {\color[HTML]{FE0000} 1.003}          & 1.003           & 1.000           \\ \hline
October   & 1.478          &               & 5.158            & October    & 1.003          &            & 1.003           \\ \hline
November  &                    &              &               & November   &                     &                      &                     \\ \hline
December  &                    &              &              & December   &                     &                      &                      \\ \hline
January   &                    &              &              & January    &                     &                      &                      \\ \hline
\end{tabular}
}
\small \\

Collembola\\
\resizebox{\columnwidth}{!}{%
\begin{tabular}{|l|l|l|l|l|l|l|l|l|l|l|l|}
\hline
$\lambda_J$ &
  \textit{F. octoulata} &
  \textit{E. aino} &
  \textit{H. watanabei} &
  \textit{S. celebensis} &
  \textit{S. aureus} &
  \textit{D. trispinata} &
  \textit{S. japonica} &
  \textit{H. nigrochephala} &
  \textit{W. japonica} &
  \textit{F. yosii} &
  \textit{A. laricis} \\ \hline
April &
  {\color[HTML]{FF0000} } &
  {\color[HTML]{FF0000} } &
  {\color[HTML]{FF0000} 18.2160} &
  {\color[HTML]{FF0000} } &
  {\color[HTML]{FF0000} } &
  {\color[HTML]{FF0000} } &
  {\color[HTML]{FF0000} } &
  {\color[HTML]{FF0000} } &
  {\color[HTML]{FF0000} 14.0310} &
  {\color[HTML]{FF0000} } &
  {\color[HTML]{FF0000} } \\ \hline
May &
  {\color[HTML]{FF0000} } &
  {\color[HTML]{FF0000} } &
  {\color[HTML]{FF0000} 1.9770} &
  {\color[HTML]{FF0000} } &
  {\color[HTML]{FF0000} 1.3979} &
  {\color[HTML]{FF0000} 1.6142} &
  {\color[HTML]{FF0000} } &
  {\color[HTML]{FF0000} } &
  {\color[HTML]{FF0000} } &
  {\color[HTML]{FF0000} } &
  {\color[HTML]{FF0000} } \\ \hline
June &
  {\color[HTML]{FF0000} 0.1062} &
  {\color[HTML]{FF0000} 0.1490} &
  {\color[HTML]{FF0000} 0.2269} &
  {\color[HTML]{FF0000} } &
  {\color[HTML]{FF0000} 0.2269} &
  {\color[HTML]{FF0000} 0.3904} &
  {\color[HTML]{FF0000} 0.2555} &
  {\color[HTML]{FF0000} } &
  {\color[HTML]{FF0000} } &
  {\color[HTML]{FF0000} 0.0393} &
  {\color[HTML]{FF0000} 0.2269} \\ \hline
July &
  {\color[HTML]{FF0000} 2.7106} &
  {\color[HTML]{FF0000} 0.1953} &
  {\color[HTML]{FF0000} 0.5958} &
  {\color[HTML]{FF0000} } &
  {\color[HTML]{FF0000} 0.5958} &
  {\color[HTML]{FF0000} 0.5958} &
  {\color[HTML]{FF0000} 0.5958} &
  {\color[HTML]{FF0000} } &
  {\color[HTML]{FF0000} } &
  {\color[HTML]{FF0000} 0.5980} &
  {\color[HTML]{FF0000} } \\ \hline
August &
  {\color[HTML]{FF0000} 0.3507} &
  {\color[HTML]{FF0000} 0.2091} &
  {\color[HTML]{FF0000} 0.5273} &
  {\color[HTML]{FF0000} } &
  {\color[HTML]{FF0000} } &
  {\color[HTML]{FF0000} 0.5974} &
  {\color[HTML]{FF0000} 0.5273} &
  {\color[HTML]{FF0000} } &
  {\color[HTML]{FF0000} 0.5273} &
  {\color[HTML]{FF0000} } &
  {\color[HTML]{FF0000} } \\ \hline
September &
  {\color[HTML]{FF0000} 0.1227} &
  {\color[HTML]{FF0000} 0.0741} &
  {\color[HTML]{FF0000} 0.2630} &
  {\color[HTML]{FF0000} 0.1599} &
  {\color[HTML]{FF0000} 0.2111} &
  {\color[HTML]{FF0000} 0.2111} &
  {\color[HTML]{FF0000} 0.3014} &
  {\color[HTML]{FF0000} 0.3014} &
  {\color[HTML]{FF0000} } &
  {\color[HTML]{FF0000} } &
  {\color[HTML]{FF0000} } \\ \hline
October &
  {\color[HTML]{FF0000} 1.6441} &
  {\color[HTML]{FF0000} 1.4124} &
  {\color[HTML]{FF0000} 1.2803} &
  {\color[HTML]{FF0000} 1.4124} &
  {\color[HTML]{FF0000} 1.4124} &
  {\color[HTML]{FF0000} } &
  {\color[HTML]{FF0000} } &
  {\color[HTML]{FF0000} } &
  {\color[HTML]{FF0000} } &
  {\color[HTML]{FF0000} } &
  {\color[HTML]{FF0000} } \\ \hline
November &
  {\color[HTML]{FF0000} } &
  {\color[HTML]{FF0000} 20.1846} &
  {\color[HTML]{FF0000} } &
  {\color[HTML]{FF0000} } &
  {\color[HTML]{FF0000} 168.6273} &
  {\color[HTML]{FF0000} } &
  {\color[HTML]{FF0000} } &
  {\color[HTML]{FF0000} } &
  {\color[HTML]{FF0000} } &
  {\color[HTML]{FF0000} } &
  {\color[HTML]{FF0000} } \\ \hline
  
  $D$ &
  \textit{F. octoulata} &
  \textit{E. aino} &
  \textit{H. watanabei} &
  \textit{S. celebensis} &
  \textit{S. aureus} &
  \textit{D. trispinata} &
  \textit{S. japonica} &
  \textit{H. nigrochephala} &
  \textit{W. japonica} &
  \textit{F. yosii} &
  \textit{A. laricis} \\ \hline
April &
  {\color[HTML]{FF0000} } &
  {\color[HTML]{FF0000} } &
  {\color[HTML]{FF0000} 2578.0711} &
  {\color[HTML]{FF0000} } &
  {\color[HTML]{FF0000} } &
  {\color[HTML]{FF0000} } &
  {\color[HTML]{FF0000} } &
  {\color[HTML]{FF0000} } &
  {\color[HTML]{FF0000} 139.4321} &
  {\color[HTML]{FF0000} } &
  {\color[HTML]{FF0000} } \\ \hline
May &
  {\color[HTML]{FF0000} } &
  {\color[HTML]{FF0000} } &
  {\color[HTML]{FF0000} 24.2039} &
  {\color[HTML]{FF0000} } &
  {\color[HTML]{FF0000} 3459.8183} &
  {\color[HTML]{FF0000} 8.1348} &
  {\color[HTML]{FF0000} } &
  {\color[HTML]{FF0000} } &
  {\color[HTML]{FF0000} } &
  {\color[HTML]{FF0000} } &
  {\color[HTML]{FF0000} } \\ \hline
June &
  {\color[HTML]{FF0000} 2.6837} &
  {\color[HTML]{FF0000} 2.4150} &
  {\color[HTML]{FF0000} 2.5011} &
  {\color[HTML]{FF0000} } &
  {\color[HTML]{FF0000} 2.5011} &
  {\color[HTML]{FF0000} 2.3709} &
  {\color[HTML]{FF0000} 2.1370} &
  {\color[HTML]{FF0000} } &
  {\color[HTML]{FF0000} } &
  {\color[HTML]{FF0000} 3.3403} &
  {\color[HTML]{FF0000} 2.5011} \\ \hline
July &
  {\color[HTML]{FF0000} 3.6365} &
  {\color[HTML]{FF0000} 55.5707} &
  {\color[HTML]{FF0000} 41.8121} &
  {\color[HTML]{FF0000} } &
  {\color[HTML]{FF0000} 41.8121} &
  {\color[HTML]{FF0000} 41.8121} &
  {\color[HTML]{FF0000} 41.8121} &
  {\color[HTML]{FF0000} } &
  {\color[HTML]{FF0000} } &
  {\color[HTML]{FF0000} 20.7542} &
  {\color[HTML]{FF0000} } \\ \hline
August &
  {\color[HTML]{FF0000} 4.0107} &
  {\color[HTML]{FF0000} 4.7167} &
  {\color[HTML]{FF0000} 4.4177} &
  {\color[HTML]{FF0000} } &
  {\color[HTML]{FF0000} } &
  {\color[HTML]{FF0000} 2.6608} &
  {\color[HTML]{FF0000} 4.4177} &
  {\color[HTML]{FF0000} } &
  {\color[HTML]{FF0000} 4.4177} &
  {\color[HTML]{FF0000} } &
  {\color[HTML]{FF0000} } \\ \hline
September &
  {\color[HTML]{FF0000} 2.1382} &
  {\color[HTML]{FF0000} 2.2105} &
  {\color[HTML]{FF0000} 1.6943} &
  {\color[HTML]{FF0000} 2.1935} &
  {\color[HTML]{FF0000} 2.1580} &
  {\color[HTML]{FF0000} 2.1580} &
  {\color[HTML]{FF0000} 2.1536} &
  {\color[HTML]{FF0000} 2.1536} &
  {\color[HTML]{FF0000} } &
  {\color[HTML]{FF0000} } &
  {\color[HTML]{FF0000} } \\ \hline
October &
  {\color[HTML]{FF0000} 46.5924} &
  {\color[HTML]{FF0000} 51.2183} &
  {\color[HTML]{FF0000} 6.0142} &
  {\color[HTML]{FF0000} 51.2183} &
  {\color[HTML]{FF0000} 51.2183} &
  {\color[HTML]{FF0000} } &
  {\color[HTML]{FF0000} } &
  {\color[HTML]{FF0000} } &
  {\color[HTML]{FF0000} } &
  {\color[HTML]{FF0000} } &
  {\color[HTML]{FF0000} } \\ \hline
November &
  {\color[HTML]{FF0000} } &
  {\color[HTML]{FF0000} 518.0128} &
  {\color[HTML]{FF0000} } &
  {\color[HTML]{FF0000} } &
  {\color[HTML]{FF0000} 854296.9028} &
  {\color[HTML]{FF0000} } &
  {\color[HTML]{FF0000} } &
  {\color[HTML]{FF0000} } &
  {\color[HTML]{FF0000} } &
  {\color[HTML]{FF0000} } &
  {\color[HTML]{FF0000} } \\ \hline
\end{tabular}%
}

Sarcoptiformes\\
\resizebox{\columnwidth}{!}{%
\begin{tabular}{|l|l|l|l|l|l|l|l|}
\hline
$\lambda_J$ &
  \textit{A. gigantea} &
  \textit{P. parvisetigerum} &
  \textit{T. stercus} &
  \textit{M. japonica} &
  \textit{G. fusca} &
  \textit{N. rotundus} &
  \textit{C. reticulatus} \\ \hline
April &
  {\color[HTML]{FF0000} } &
  {\color[HTML]{FF0000} 39.4099} &
  {\color[HTML]{FF0000} 11.6547} &
  {\color[HTML]{FF0000} 39.4099} &
  {\color[HTML]{FF0000} } &
  {\color[HTML]{FF0000} 39.4099} &
   \\ \hline
May &
  {\color[HTML]{FF0000} 2.4933} &
  {\color[HTML]{FF0000} 2.4933} &
  {\color[HTML]{FF0000} 4.6543} &
  {\color[HTML]{FF0000} } &
  {\color[HTML]{FF0000} } &
  {\color[HTML]{FF0000} } &
   \\ \hline
June &
  {\color[HTML]{FF0000} 0.0751} &
  {\color[HTML]{FF0000} 3.1055} &
  {\color[HTML]{FF0000} 0.9227} &
  {\color[HTML]{FF0000} 0.21} &
  {\color[HTML]{FF0000} } &
  {\color[HTML]{FF0000} 1.7274} &
  2.875 \\ \hline
July &
  {\color[HTML]{FF0000} 3.1609} &
  {\color[HTML]{FF0000} 6.9876} &
  {\color[HTML]{FF0000} 6.9876} &
  {\color[HTML]{FF0000} 6.9876} &
  {\color[HTML]{FF0000} } &
  {\color[HTML]{FF0000} 6.9876} &
   \\ \hline
August &
  {\color[HTML]{FF0000} 2.3086} &
  {\color[HTML]{FF0000} 5.8949} &
  {\color[HTML]{FF0000} } &
  {\color[HTML]{FF0000} 5.8949} &
  {\color[HTML]{FF0000} } &
  {\color[HTML]{FF0000} 5.8949} &
   \\ \hline
September &
  {\color[HTML]{FF0000} 0.0775} &
  {\color[HTML]{FF0000} 0.0775} &
  {\color[HTML]{FF0000} 2.2062} &
  {\color[HTML]{FF0000} 1.897} &
  {\color[HTML]{FF0000} 1.0486} &
  {\color[HTML]{FF0000} 0.767} &
  {\color[HTML]{FF0000} 0.43} \\ \hline
October &
  4.0812 &
  {\color[HTML]{FF0000} 1.4963} &
  {\color[HTML]{FF0000} 3.652} &
   &
   &
  {\color[HTML]{FF0000} 3.652} &
  {\color[HTML]{FF0000} 3.652} \\ \hline
November &
  {\color[HTML]{FF0000} 7.1878} &
  {\color[HTML]{FF0000} } &
  {\color[HTML]{FF0000} 2.8733} &
  10.9312 &
   &
   &
   \\ \hline
$D$ &
  \textit{A. gigantea} &
  \textit{P. parvisetigerum} &
  \textit{T. stercus} &
  \textit{M. japonica} &
  \textit{G. fusca} &
  \textit{N. rotundus} &
  \textit{C. reticulatus} \\ \hline
April &
   &
  {\color[HTML]{FF0000} 5441190372} &
  {\color[HTML]{FF0000} 4160.262} &
  {\color[HTML]{FF0000} 5441190372} &
   &
  {\color[HTML]{FF0000} 5441190372} &
   \\ \hline
May &
  {\color[HTML]{FF0000} 4924.1334} &
  {\color[HTML]{FF0000} 4924.1334} &
  {\color[HTML]{FF0000} 33.2869} &
  {\color[HTML]{FF0000} } &
   &
  {\color[HTML]{FF0000} } &
   \\ \hline
June &
  {\color[HTML]{FF0000} 7.6377} &
  {\color[HTML]{FF0000} 1.2138} &
  {\color[HTML]{FF0000} 2.8674} &
  {\color[HTML]{FF0000} 8.5012} &
   &
  {\color[HTML]{FF0000} 1.8221} &
  1.2919 \\ \hline
July &
  {\color[HTML]{FF0000} 407.5891} &
  {\color[HTML]{FF0000} 2583.8321} &
  {\color[HTML]{FF0000} 2583.8321} &
  {\color[HTML]{FF0000} 2583.8321} &
   &
  {\color[HTML]{FF0000} 2583.8321} &
   \\ \hline
August &
  {\color[HTML]{FF0000} 35.5048} &
  {\color[HTML]{FF0000} 92.9407} &
   &
  {\color[HTML]{FF0000} 92.9407} &
   &
  {\color[HTML]{FF0000} 92.9407} &
   \\ \hline
September &
  {\color[HTML]{FF0000} 20.4905} &
  {\color[HTML]{FF0000} 20.4905} &
  {\color[HTML]{FF0000} 1.8518} &
  {\color[HTML]{FF0000} 2.2205} &
  {\color[HTML]{FF0000} 4.1952} &
  {\color[HTML]{FF0000} 6.1748} &
  {\color[HTML]{FF0000} 12.3592} \\ \hline
October &
  2.8214 &
  {\color[HTML]{FF0000} 34.5564} &
  {\color[HTML]{FF0000} 44.0229} &
   &
   &
  {\color[HTML]{FF0000} 44.0229} &
  {\color[HTML]{FF0000} 44.0229} \\ \hline
November &
  {\color[HTML]{FF0000} 10.2226} &
   &
  {\color[HTML]{FF0000} 12.897} &
  7.8696 &
   &
  {\color[HTML]{FF0000} } &
   \\ \hline
\end{tabular}%
}

\begin{flushleft}
\textbf{Abbreviations:} WE: Washidu East; WW: Washidu West. \textit{P.~pallidum}: \textit{Polysphondylium pallidum}; \textit{D.~purpureum}: \textit{Dictyostelium purpureum}; \textit{P.~violaceum}: \textit{Polysphondylium violaceum}. For scientific names of soil mesofauna, refer to Table~2. 
\textbf{Note:} Numbers highlighted in \textcolor{red}{red} indicate perturbations exceeding the Jeans wavelength $\lambda_J$, suggesting conditions favorable for species dynamics development. Blank cells denote values that could not be calculated due to insufficient or undefined parameters. Data pertaining to world economics are excluded, as the corresponding $D$ values do not surpass the $\lambda_J$ threshold.
\end{flushleft}
\end{table}

\begin{table}[]
\caption{$M_J$ values compared to $N$.}
\centering
Dictyostelia\\
\resizebox{\columnwidth}{!}{%
\begin{tabular}{|l|l|l|l|l|l|l|l|}
\hline
$M_J$      & WE \textit{P. pallidum} & WE \textit{D. purpureum}              & WE \textit{P. violaceum}              & \textit{N} & WE \textit{P. pallidum} & WE \textit{D. purpureum}              & WE \textit{P. violaceum}              \\ \hline
April       &              &                                   &                            & April        &                     & 75.56                        &                                   \\ \hline
June      & 841.7           & {\color[HTML]{FE0000} 8.423} & 174.5                        & June       & 122.6          & {\color[HTML]{FE0000} 208.9} & 52.44                        \\ \hline
July      &                     &                            &                           & July       & 1282         &                                  &                                   \\ \hline
August    &                     &                            &                            & August     & 1561          &                                   &                                   \\ \hline
September & 1504           & 8429                        &                            & September  & 900.6         & 106.7                        &                                   \\ \hline
October   &  {\color[HTML]{FE0000} 745.3 }        & 5289                        &                            & October    &  {\color[HTML]{FE0000} 1069}          & 34.78                        &                                   \\ \hline
November  & 359.1          &                            & {\color[HTML]{FE0000} 1.345} & November   & 60.00                   &                                   & {\color[HTML]{FE0000} 100.8} \\ \hline
December  &                     &                            &                           & December   & 189.6          &                                   &                                   \\ \hline
January   &                     &                            &                            & January    & 28.90          &                                   &                                   \\ \hline
\end{tabular}
}

\centering
\resizebox{\columnwidth}{!}{%
\begin{tabular}{|l|l|l|l|l|l|l|l|}
\hline
$M_J$     & WW \textit{P. pallidum}              & WW \textit{D. purpureum} & WW \textit{P. violaceum}              & \textit{N} & WW \textit{P. pallidum}               & WW \textit{D. purpureum} & WW \textit{P. violaceum}      \\ \hline
April       &                           &                      &                            & April        &                                   & 82.67           &                           \\ \hline
June      &                                  &               &                           & June       & 146.7                        &                      &                           \\ \hline
July      & 182.6                       & 2220           & {\color[HTML]{FE0000} 9.795} & July       & 80.00                                 & 214.8           & {\color[HTML]{FE0000} 320.0} \\ \hline
August    & 2241                       & 12760           &                            & August     & 1330                        & 180.78           &                           \\ \hline
September & {\color[HTML]{FE0000} 215.3} & 216.8           & 32130                        & September  & {\color[HTML]{FE0000} 809.2} & 77.00                    & 648.9                \\ \hline
October   & 1350                      &              & 7666                        & October    & 798.8                        &                      & 106.7                \\ \hline
November  &                                  &               &                           & November   & 335.6                       &                     &                           \\ \hline
December  &                                  &               &                            & December   & 711.1                        &                      &                           \\ \hline
January   &                                  &               &                            & January    & 99.00                                 &                     &                           \\ \hline
\end{tabular}
}

Collembola\\
\resizebox{\columnwidth}{!}{%
\begin{tabular}{|l|l|l|l|l|l|l|l|l|l|l|l|}
\hline
$M_J$ &
  \textit{F. octoulata} &
  \textit{E. aino} &
  \textit{H. watanabei} &
  \textit{S. celebensis} &
  \textit{S. aureus} &
  \textit{D. trispinata} &
  \textit{S. japonica} &
  \textit{H. nigrochephala} &
  \textit{W. japonica} &
  \textit{F. yosii} &
  \textit{A. laricis} \\ \hline
April &
  {\color[HTML]{FF0000} } &
  {\color[HTML]{FF0000} } &
  28484.0764 &
  {\color[HTML]{FF0000} } &
  {\color[HTML]{FF0000} } &
  {\color[HTML]{FF0000} } &
  {\color[HTML]{FF0000} } &
  {\color[HTML]{FF0000} } &
  4338.9641 &
  {\color[HTML]{FF0000} } &
  {\color[HTML]{FF0000} } \\ \hline
May &
  {\color[HTML]{FF0000} } &
  {\color[HTML]{FF0000} } &
  24.2741 &
  {\color[HTML]{FF0000} } &
  17.1643 &
  19.8197 &
  {\color[HTML]{FF0000} } &
  {\color[HTML]{FF0000} } &
  {\color[HTML]{FF0000} } &
  {\color[HTML]{FF0000} } &
  {\color[HTML]{FF0000} } \\ \hline
June &
  {\color[HTML]{FF0000} 0.0034} &
  {\color[HTML]{FF0000} 0.0312} &
  {\color[HTML]{FF0000} 0.0367} &
  {\color[HTML]{FF0000} } &
  {\color[HTML]{FF0000} 0.0367} &
  {\color[HTML]{FF0000} 0.0935} &
  {\color[HTML]{FF0000} 0.1310} &
  {\color[HTML]{FF0000} } &
  {\color[HTML]{FF0000} } &
  {\color[HTML]{FF0000} 0.0008} &
  {\color[HTML]{FF0000} 0.0367} \\ \hline
July &
  5.7933 &
  {\color[HTML]{FF0000} 0.0820} &
  {\color[HTML]{FF0000} 0.3323} &
  {\color[HTML]{FF0000} } &
  {\color[HTML]{FF0000} 0.3323} &
{\color[HTML]{FF0000} 0.3323} &
  {\color[HTML]{FF0000} 0.3323} &
  {\color[HTML]{FF0000} } &
  {\color[HTML]{FF0000} } &
  0.6718 &
  {\color[HTML]{FF0000} } \\ \hline
August &
  {\color[HTML]{FF0000} 0.0728} &
  {\color[HTML]{FF0000} 0.0719} &
  0.2303 &
  {\color[HTML]{FF0000} } &
  {\color[HTML]{FF0000} } &
  1.6744 &
  0.2303 &
  {\color[HTML]{FF0000} } &
  0.2303 &
  {\color[HTML]{FF0000} } &
  {\color[HTML]{FF0000} } \\ \hline
September &
  {\color[HTML]{FF0000} 0.0062} &
  {\color[HTML]{FF0000} 0.0083} &
  {\color[HTML]{FF0000} 0.3430} &
  {\color[HTML]{FF0000} 0.0193} &
  {\color[HTML]{FF0000} 0.0296} &
  {\color[HTML]{FF0000} 0.0296} &
  {\color[HTML]{FF0000} 0.0430} &
  {\color[HTML]{FF0000} 0.0430} &
  {\color[HTML]{FF0000} } &
  {\color[HTML]{FF0000} } &
  {\color[HTML]{FF0000} } \\ \hline
October &
  1.8099 &
  4.4254 &
  13.1866 &
  4.4254 &
  4.4254 &
  {\color[HTML]{FF0000} } &
  {\color[HTML]{FF0000} } &
  {\color[HTML]{FF0000} } &
  {\color[HTML]{FF0000} } &
  {\color[HTML]{FF0000} } &
  {\color[HTML]{FF0000} } \\ \hline
November &
  {\color[HTML]{FF0000} } &
  12917.5445 &
  {\color[HTML]{FF0000} } &
  {\color[HTML]{FF0000} } &
  15063756.6124 &
  {\color[HTML]{FF0000} } &
  {\color[HTML]{FF0000} } &
  {\color[HTML]{FF0000} } &
  {\color[HTML]{FF0000} } &
  {\color[HTML]{FF0000} } &
  {\color[HTML]{FF0000} } \\ \hline
  
  $N$ &
  \textit{F. octoulata} &
  \textit{E. aino} &
  \textit{H. watanabei} &
  \textit{S. celebensis} &
  \textit{S. aureus} &
  \textit{D. trispinata} &
  \textit{S. japonica} &
  \textit{H. nigrochephala} &
  \textit{W. japonica} &
  \textit{F. yosii} &
  \textit{A. laricis} \\ \hline
April &
  {\color[HTML]{FF0000} } &
  {\color[HTML]{FF0000} } &
  0.3333 &
  {\color[HTML]{FF0000} } &
  {\color[HTML]{FF0000} } &
  {\color[HTML]{FF0000} } &
  {\color[HTML]{FF0000} } &
  {\color[HTML]{FF0000} } &
  0.1111 &
  {\color[HTML]{FF0000} } &
  {\color[HTML]{FF0000} } \\ \hline
May &
  {\color[HTML]{FF0000} } &
  {\color[HTML]{FF0000} } &
  0.2222 &
  {\color[HTML]{FF0000} } &
  0.4444 &
  0.3333 &
  {\color[HTML]{FF0000} } &
  {\color[HTML]{FF0000} } &
  {\color[HTML]{FF0000} } &
  {\color[HTML]{FF0000} } &
  {\color[HTML]{FF0000} } \\ \hline
June &
  {\color[HTML]{FF0000} 5.4444} &
  {\color[HTML]{FF0000} 0.6667} &
  {\color[HTML]{FF0000} 0.2222} &
  {\color[HTML]{FF0000} } &
  {\color[HTML]{FF0000} 0.2222} &
  {\color[HTML]{FF0000} 0.1111} &
  {\color[HTML]{FF0000} 0.5556} &
  {\color[HTML]{FF0000} } &
  {\color[HTML]{FF0000} } &
  {\color[HTML]{FF0000} 0.8889} &
  {\color[HTML]{FF0000} 0.2222} \\ \hline
July &
  0.5556 &
  {\color[HTML]{FF0000} 0.7778} &
  {\color[HTML]{FF0000} 0.1111} &
  {\color[HTML]{FF0000} } &
  {\color[HTML]{FF0000} 0.1111} &
  {\color[HTML]{FF0000} 0.1111} &
  {\color[HTML]{FF0000} 0.1111} &
  {\color[HTML]{FF0000} } &
  {\color[HTML]{FF0000} } &
  0.2222 &
  {\color[HTML]{FF0000} } \\ \hline
August &
  {\color[HTML]{FF0000} 3.2222} &
  {\color[HTML]{FF0000} 0.5556} &
  0.1111 &
  {\color[HTML]{FF0000} } &
  {\color[HTML]{FF0000} } &
  0.5556 &
  0.1111 &
  {\color[HTML]{FF0000} } &
  0.1111 &
  {\color[HTML]{FF0000} } &
  {\color[HTML]{FF0000} } \\ \hline
September &
  {\color[HTML]{FF0000} 6.4444} &
  {\color[HTML]{FF0000} 1.4444} &
  {\color[HTML]{FF0000} 1.3333} &
  {\color[HTML]{FF0000} 0.3333} &
  {\color[HTML]{FF0000} 0.2222} &
  {\color[HTML]{FF0000} 0.2222} &
  {\color[HTML]{FF0000} 0.1111} &
  {\color[HTML]{FF0000} 0.1111} &
  {\color[HTML]{FF0000} } &
  {\color[HTML]{FF0000} } &
  {\color[HTML]{FF0000} } \\ \hline
October &
  0.7778 &
  0.1111 &
  0.4444 &
  0.1111 &
  0.1111 &
  {\color[HTML]{FF0000} } &
  {\color[HTML]{FF0000} } &
  {\color[HTML]{FF0000} } &
  {\color[HTML]{FF0000} } &
  {\color[HTML]{FF0000} } &
  {\color[HTML]{FF0000} } \\ \hline
November &
  {\color[HTML]{FF0000} } &
  0.1111 &
  {\color[HTML]{FF0000} } &
  {\color[HTML]{FF0000} } &
  0.2222 &
  {\color[HTML]{FF0000} } &
  {\color[HTML]{FF0000} } &
  {\color[HTML]{FF0000} } &
  {\color[HTML]{FF0000} } &
  {\color[HTML]{FF0000} } &
  {\color[HTML]{FF0000} } \\ \hline
\end{tabular}%
}

Sarcoptiformes\\
\resizebox{\columnwidth}{!}{%
\begin{tabular}{|l|l|l|l|l|l|l|l|}
\hline
$M_J$ & \textit{A. gigantea} & \textit{P. parvisetigerum} & \textit{T. stercus} & \textit{M. japonica} & \textit{G. fusca} & \textit{N. rotundus} & \textit{C. reticulatus} \\ \hline
April &
   &
  7122.0114 &
  92.1001 &
  7122.0114 &
   &
  7122.0114 &
   \\ \hline
May &
  3.607 &
  3.607 &
  11.7312 &
   &
   &
   &
   \\ \hline
June &
  {\color[HTML]{FF0000} 0.0003} &
  19.1662 &
  {\color[HTML]{FF0000} 0.1828} &
  {\color[HTML]{FF0000} 0.0005} &
   &
  2.0989 &
  13.8252 \\ \hline
July &
  11.024 &
  19.8492 &
  19.8492 &
  19.8492 &
   &
  19.8492 &
   \\ \hline
August &
  7.158 &
  11.9173 &
   &
  11.9173 &
   &
  11.9173 &
   \\ \hline
September &
  {\color[HTML]{FF0000} 0.0002} &
  {\color[HTML]{FF0000}0.0002} &
  3.7482 &
  1.9859 &
  {\color[HTML]{FF0000} 0.2012} &
  {\color[HTML]{FF0000} 0.0525} &
  {\color[HTML]{FF0000} 0.0046} \\ \hline
October &
  19.7732 &
  1.3644 &
  2.8337 &
   &
   &
  2.8337 &
  2.8337 \\ \hline
November &
  43.2086 &
   &
  15.1804 &
  75.991 &
   &
   &
   \\ \hline
$N$  & \textit{A. gigantea} & \textit{P. parvisetigerum} & \textit{T. stercus} & \textit{M. japonica} & \textit{G. fusca} & \textit{N. rotundus} & \textit{C. reticulatus} \\ \hline
April &
   &
  0.2222 &
  0.1111 &
  0.2222 &
   &
  0.2222 &
   \\ \hline
May &
  0.4444 &
  0.4444 &
  0.2222 &
   &
   &
   &
   \\ \hline
June &
  {\color[HTML]{FF0000} 1.3333} &
  1.2222 &
  {\color[HTML]{FF0000} 0.4444} &
  {\color[HTML]{FF0000} 0.1111} &
   &
  0.7778 &
  1.1111 \\ \hline
July &
  0.6667 &
  0.1111 &
  0.1111 &
  0.1111 &
   &
  0.1111 &
   \\ \hline
August &
  1.1111 &
  0.1111 &
   &
  0.1111 &
   &
  0.1111 &
   \\ \hline
September &
  {\color[HTML]{FF0000} 0.8889} &
  {\color[HTML]{FF0000}0.8889} &
  0.6667 &
  0.5556 &
  {\color[HTML]{FF0000} 0.3333} &
  {\color[HTML]{FF0000} 0.2222} &
  {\color[HTML]{FF0000} 0.1111} \\ \hline
October &
  0.5556 &
  0.7778 &
  0.1111 &
   &
   &
  0.1111 &
  0.1111 \\ \hline
November &
  0.2222 &
   &
  1.2222 &
  0.1111 &
   &
   &
   \\ \hline
\end{tabular}%
}

\small
\begin{flushleft}
\textbf{Abbreviations:} WE: Washidu East; WW: Washidu West. \textit{P.~pallidum}: \textit{Polysphondylium pallidum}; \textit{D.~purpureum}: \textit{Dictyostelium purpureum}; \textit{P.~violaceum}: \textit{Polysphondylium violaceum}. For scientific names of soil mesofauna, refer to Table~2.

\textbf{Note:} Numbers shown in \textcolor{red}{red} indicate perturbations exceeding the Jeans mass $M_J$, suggesting the potential for species-level structural development. Blank cells indicate values that could not be calculated. Data related to world economics are excluded, as the corresponding $N$ values do not exceed the $M_J$ threshold.
\end{flushleft}
\end{table}

For further investigation, consider an analogue of the radius of action in nuclear force theory:
$
\lambdabar \approx \frac{\hbar}{Mc} = \frac{1}{M \Im(s)}.
$
Assuming $D \sim 1$, we can set $\lambdabar \cdot r_H = 1$. Therefore, when $D > r_H$, interaction between Monsters becomes possible, analogous to the Schwarzschild black hole scenario, likely involving a Yukawa-type potential. Conversely, when $D \leq r_H$, escape from the Monster is not feasible. If $r_H \ll 1$, an attractive force emerges, promoting the development of fractal dimension.

Furthermore, considering the inverse cube of $\frac{1}{3} \Im(s) D$ yields a scenario reminiscent of quantum chromodynamics (QCD), corresponding to the average squared radius of gyration. Defining an analogue of color as
$
C = \sqrt{\frac{1}{3}}(R\bar{R} + G\bar{G} + B\bar{B}),
$
we obtain
$
\exp(C)^2 = \exp\left(\frac{1}{3} \Im(s) D\right)^3,
$
which leads to
$
\Im(s) D = (R\bar{R})^2 + R\bar{R}G\bar{G} + B\bar{B}R\bar{R},
$
or any permutation of $R$, $G$, and $B$, in the spirit of QCD. Here, color serves as a valuation of interacting particles. A three-body interaction involving $\Re(s) = 2$ (and $\Re(s) = -2$ as its future counterpart) and $\Re(s) = 5$ exemplifies this structure. The $l = 5$ case should be interpreted as comprising two distinct aspects.

Consider the modular relation
$
\tau_{\rho} = \frac{1}{12} \left( \frac{1}{4} - \rho^2 \right) \equiv 2(1 + \rho^2) \pmod{5},
$
as discussed in \cite{Mochizuki1999}. For $\rho = 0$, the moduli space $\bar{\mathcal{N}}_{1,1}^0$ decomposes into $2$ and $3$ over $\bar{\mathcal{M}}_{1,1}$, corresponding to the observant layer with constituents $l = 2, 2, 3$. For $\rho = 1$, $\bar{\mathcal{N}}_{1,1}^1$ decomposes into $1$ and $3$ over $\bar{\mathcal{M}}_{1,1}$, representing the observer layer, observing the transition $\Re(s) = 4 \to 3$ from $1$. For $\rho = \pm i$, we have $\bar{\mathcal{N}}_{1,1}^{-1} \cong \bar{\mathcal{N}}_{1,0}$, representing a higher layer above the observer. In this case, the $l = 2 + 3 = 5$ constituents form a solitary structure acting on the original $l = 2$ component.

To further clarify, consider the D-fivebranes in superstring theory \cite{Witten1995, Lambert1998}. In the $\mathbb{R}^4$ (3+1) case, a one-instanton ADHM solution arises \cite{Witten1995}, with potential energy
$
V = \frac{1}{8}(X^2 + \rho^2)\phi^2.
$
In our model, $(X, \phi) = (1, 4)$, and $\rho$ takes values $0$ (branch), $1$ (normal observation), or $\pm i$ (fractal). A massless model leads to D-fivebranes with $\Re(s) = 5$ in type I or type IIA/B superstring theory compactified on $K3 \times S^1$ \cite{Lambert1998}, corresponding to $11 - 5 = 6$ compactified dimensions in M-theory. Since the $K3$ surface is conjecturally related to the Mathieu group $M_{24}$, this group becomes relevant to our model in the context of condensed matter physics described via superstring theory.

This D5 system is analogous to the D1-D5 system described in \cite{Martinec1999}. When
$
H_i = 1 + \left( \frac{q_i}{D} \right)^2, \quad i = 1,5; \quad h = 1 - \left( \frac{r_H}{D} \right)^2,
$
the Lorentzian metric becomes
\begin{equation}
ds_M^2 = \frac{-h\,dt^2 + dD_5^2}{\sqrt{H_1 H_5}} + \sqrt{\frac{H_1}{H_5}}\,dD_{\parallel}^2 + \frac{h^{-1} dD^2 + d\Omega_3^2}{\sqrt{H_1 H_5}}.
\end{equation}

Taking the limit $l_s = 2\pi l \to 0$ (as the observant frame), we obtain
\begin{equation}
\frac{ds_M^2}{l_s^2} = \frac{-h\,dt^2 + dD_5^2}{\sqrt{g_s^2 Q_1 Q_5}} \left( \frac{D}{l_s^2} \right)^2 + \sqrt{\frac{Q_1}{Q_5}} \left[ \frac{dD_{\parallel}^2}{V_4^{1/2}} \right] + \frac{h^{-1} \left( \frac{dD}{D} \right)^2 + d\Omega_3^2}{\sqrt{g_6^2 Q_1 Q_5}},
\end{equation}
where $Q_i$ are instanton charges and $V_4 = \Sigma_1 \Sigma_2 \Sigma_3 \Sigma_4$. The geometry is locally $AdS_3 \times S^3 \times T^4$, with the radius of $S^3$ given by $R_{AdS} = l_s (g_6^2 Q_1 Q_5)^{1/4}$ and the characteristic proper size of $T^4$ as $l_s (Q_1/Q_5)^{1/4}$. This leads to a harmonic function
\begin{equation}
H = \sum_{\alpha} \frac{q_\alpha^2}{|D/D_\alpha|^2}, \quad h = 0 \quad (D \sim r_H),
\end{equation}
which recovers the inverse-square law when $M_\alpha = q_\alpha$.

The physical Lagrangian densities for quantum electrodynamics (QED) and quantum chromodynamics (QCD) are:
\begin{align}
\mathfrak{L}_{\text{QED}} &= \bar{\psi}(ic \gamma^\mu \partial_\mu - Mc^2)\psi + e \bar{\psi} \gamma^\mu A_\mu \psi - \frac{1}{4\mu_0} F_{\mu\nu} F^{\mu\nu}, \\
\mathfrak{L}_{\text{QCD}} &= \bar{q}(ic \gamma^\mu \partial_\mu - Mc^2)q + g(\bar{q} \gamma^\mu T_a q) G_\mu^a - \frac{1}{4} G_{\mu\nu}^a G_a^{\mu\nu}.
\end{align}

In the form of a Lefschetz-type operator ($i\bar{\partial}$), the scalar field Lagrangian becomes:
\begin{equation}
\mathfrak{L} = \frac{1}{2} (\partial_\mu \phi_i)^2 - \frac{1}{2} \mu^2 \phi_i^2 - \frac{1}{4} \lambda (\phi_i)^4.
\end{equation}

An integrated analogue in our model is:
\begin{equation}
\mathfrak{L} = \frac{1}{2}(-sw \Im(s) - M \Im(s)^2) \phi_i^2 + \frac{1}{2} p^2 \phi_i^2 - \frac{1}{4} l \frac{\Im(s)}{D} (\phi_i)^4.
\end{equation}
Here, the first term represents diminishing mass, the second term corresponds to the residual Frobenioid structure, and the third term contributes to fractal formation. The field $\phi_i$ may be interpreted as a Selberg zeta function $\zeta_\Gamma$, a Hasse-Weil $L$-function, or a field magnitude $|\mathbf{f}_s| = \frac{\Im(s)}{4\pi} \cdot \frac{M}{D^3}$ (with $s \to \Re(s)$, $w \to \Re(s) - 1$), as previously discussed (see Table~14).

Empirically, $H(t) \approx 0$, and its effect is negligible. In the Hamiltonian framework, values are significantly suppressed when black hole analogies dominate. In the Lagrangian framework, negative values may indicate contributions to fractal structures formed by climax species \textit{Polysphondylium pallidum}, while positive values may reflect anti-fractal dynamics associated with pioneering species \textit{Dictyostelium purpureum} and \textit{Polysphondylium violaceum} \cite{Adachi2015}.

\begin{table}[]
\caption{$|\mathbf{f}_s|$ values.}
\centering
Dictyostelia\\
\resizebox{\columnwidth}{!}{%
\begin{tabular}{|l|l|l|l|l|l|l|}
\hline
$|\mathbf{f}_s|$        & WE \textit{P. pallidum} & WE \textit{D. purpureum} & WE \textit{P. violaceum} & WW \textit{P. pallidum} & WW \textit{D. purpureum} & WW \textit{P. violaceum} \\ \hline
April       &               &      &                &               &      &                \\ \hline
June      & 0.2045    & {\color[HTML]{FE0000}6.189}     & 0.3291    &     &                &                \\ \hline
July      &     &                &                & 0.3998    & 0.1470     & {\color[HTML]{FE0000}7.049}     \\ \hline
August    &     &                &                & 1.575     & 0.1307     &               \\ \hline
September & 1.638    & 0.1253     &               & {\color[HTML]{FE0000}2.211}    & 0.4583      &          \\ \hline
October   &  {\color[HTML]{FE0000} 3.484}    & 0.09617     &                & 1.575    &                & 0.1294     \\ \hline
November  & 0.2209    &                & {\color[HTML]{FE0000}12.38}     &     &                &               \\ \hline
December  &     &               &                &     &               &                \\ \hline
January   &     &                &                &     &                &                \\ \hline
\end{tabular}
}

Collembola\\
\resizebox{\columnwidth}{!}{%
\begin{tabular}{|l|l|l|l|l|l|l|l|l|l|l|l|}
\hline
$|\mathbf{f}_s|$ &
  \textit{F. octoulata} &
  \textit{E. aino} &
  \textit{H. watanabei} &
  \textit{S. celebensis} &
  \textit{S. aureus} &
  \textit{D. trispinata} &
  \textit{S. japonica} &
  \textit{H. nigrochephala} &
  \textit{W. japonica} &
  \textit{F. yosii} &
  \textit{A. laricis} \\ \hline
April &
   &
   &
   &
   &
   &
   &
   &
   &
   &
   &
   \\ \hline
May &
   &
   &
   &
   &
   &
   &
   &
   &
   &
   &
   \\ \hline
June &
  {\color[HTML]{FF0000} 10.0572} &
  {\color[HTML]{FF0000} 0.7018} &
  {\color[HTML]{FF0000} 0.2822} &
   &
  {\color[HTML]{FF0000} 0.2822} &
  {\color[HTML]{FF0000} 0.1069} &
  {\color[HTML]{FF0000} 0.3051} &
   &
   &
  {\color[HTML]{FF0000} 5.2761} &
  {\color[HTML]{FF0000} 0.2822} \\ \hline
July &
  0.0430 &
  {\color[HTML]{FF0000} 0.0006} &
 {\color[HTML]{FF0000} 0.0001} &
   &
  {\color[HTML]{FF0000} 0.0001} &
  {\color[HTML]{FF0000} 0.0001} &
  {\color[HTML]{FF0000} 0.0001} &
   &
   &
  0.0005 &
   \\ \hline
August &
  {\color[HTML]{FF0000} 0.5563} &
  {\color[HTML]{FF0000} 0.1257} &
  0.0229 &
   &
   &
  0.0498 &
  0.0229 &
   &
  0.0229 &
   &
   \\ \hline
September &
  {\color[HTML]{FF0000} 12.1375} &
  {\color[HTML]{FF0000} 3.4588} &
  {\color[HTML]{FF0000} 0.4685} &
  {\color[HTML]{FF0000} 0.7550} &
  {\color[HTML]{FF0000} 0.4477} &
  {\color[HTML]{FF0000} 0.4477} &
  {\color[HTML]{FF0000} 0.2211} &
  {\color[HTML]{FF0000} 0.2211} &
   &
   &
   \\ \hline
October &
  0.0002 &
   &
   &
   &
   &
   &
   &
   &
   &
   &
   \\ \hline
November &
   &
   &
   &
   &
   &
   &
   &
   &
   &
   &
   \\ \hline
\end{tabular}%
}

Sarcoptiformes\\
\resizebox{\columnwidth}{!}{%
\begin{tabular}{|l|l|l|l|l|l|l|l|}
\hline
$|\mathbf{f}_s|$ &
  \textit{A. gigantea} &
  \textit{P. parvisetigerum} &
  \textit{T. stercus} &
  \textit{M. japonica} &
  \textit{G. fusca} &
  \textit{N. rotundus} &
  \textit{C. reticulatus} \\ \hline
April    &        &        &        &  &  &  &  \\ \hline
May      & 0      & 0      &        &  &  &  &  \\ \hline
June &
  {\color[HTML]{FF0000} 1.2379} &
   &
  {\color[HTML]{FF0000} 0.059} &
  {\color[HTML]{FF0000} 0.1278} &
   &
  0.0506 &
   \\ \hline
July     & 0      &        &        &  &  &  &  \\ \hline
August   & 0.0002 &        &        &  &  &  &  \\ \hline
September &
  {\color[HTML]{FF0000} 0.1445} &
  {\color[HTML]{FF0000} 0.1445} &
   &
  0.0332 &
  {\color[HTML]{FF0000} 0.0194} &
  {\color[HTML]{FF0000} 0.0164} &
  {\color[HTML]{FF0000} 0.0129} \\ \hline
October  &        & 0.0003 &        &  &  &  &  \\ \hline
November &        &        & 0.0016 &  &  &  &  \\ \hline
\end{tabular}%
}

\small
\begin{flushleft}
\textbf{Abbreviations:} WE: Washidu East quadrat; WW: Washidu West quadrat. \textit{P.~pallidum}: \textit{Polysphondylium pallidum}; \textit{D.~purpureum}: \textit{Dictyostelium purpureum}; \textit{P.~violaceum}: \textit{Polysphondylium violaceum}. For the scientific names of soil mesofauna, refer to Table~2.

\textbf{Note:} Values highlighted in \textcolor{red}{red} indicate significantly large quantities inferred through a black hole analogy. Blank cells denote cases where values could not be calculated due to insufficient or undefined parameters.
\end{flushleft}
\end{table}

\begin{table}[]
\caption{Lagrangian.}
\centering
Dictyostelia\\
\resizebox{\columnwidth}{!}{%
\begin{tabular}{|l|l|l|l|l|l|l|}
\hline
$\mathfrak{L}$     & WE \textit{P. pallidum} & WE \textit{D. purpureum} & WE \textit{P. violaceum} & WW \textit{P. pallidum} & WW \textit{D. purpureum} & WW \textit{P. violaceum} \\ \hline
April       &                &                 &                 &                &                 &                 \\ \hline
June      & -0.3983   & {\color[HTML]{FE0000}7.609E+5}     & -5.164     &                &                 &                 \\ \hline
July      &                &                 &                 & 5.676    & -0.08279    & {\color[HTML]{FE0000}1.907E+6}     \\ \hline
August    &                &                 &                 & -899.8995563   & -0.800427157    &                 \\ \hline
September & -1072  & -0.9108    &                 & {\color[HTML]{FE0000}-2036}   & 2485      &                 \\ \hline
October   &  {\color[HTML]{FE0000} -2.368E+4}   & -2.932    &                 & -906.8   &                 & -0.8065    \\ \hline
November  & -0.4868   &                 & {\color[HTML]{FE0000}2.820E+7}     &                &                 &                 \\ \hline
December  &                &                 &                 &                &                 &                 \\ \hline
January   &                &                 &                 &                &                 &                 \\ \hline
\end{tabular}
}

Collembola\\
\resizebox{\columnwidth}{!}{%
\begin{tabular}{|l|l|l|l|l|l|l|l|l|l|l|l|}
\hline
$\mathfrak{L}$ &
  \textit{F. octoulata} &
  \textit{E. aino} &
  \textit{H. watanabei} &
  \textit{S. celebensis} &
  \textit{S. aureus} &
  \textit{D. trispinata} &
  \textit{S. japonica} &
  \textit{H. nigrochephala} &
  \textit{W. japonica} &
  \textit{F. yosii} &
  \textit{A. laricis} \\ \hline
April &
  {\color[HTML]{333333} } &
  {\color[HTML]{333333} } &
  {\color[HTML]{333333} } &
  {\color[HTML]{333333} } &
  {\color[HTML]{333333} } &
  {\color[HTML]{333333} } &
  {\color[HTML]{333333} } &
  {\color[HTML]{333333} } &
  {\color[HTML]{333333} } &
  {\color[HTML]{333333} } &
  {\color[HTML]{333333} } \\ \hline
May &
  {\color[HTML]{333333} } &
  {\color[HTML]{333333} } &
  {\color[HTML]{333333} } &
  {\color[HTML]{333333} } &
  {\color[HTML]{333333} } &
  {\color[HTML]{333333} } &
  {\color[HTML]{333333} } &
  {\color[HTML]{333333} } &
  {\color[HTML]{333333} } &
  {\color[HTML]{333333} } &
  {\color[HTML]{333333} } \\ \hline
June &
  {\color[HTML]{FF0000} 184600000} &
  {\color[HTML]{FF0000} 256900} &
  {\color[HTML]{FF0000} 63790} &
  {\color[HTML]{333333} } &
  {\color[HTML]{FF0000} 63790} &
  {\color[HTML]{FF0000} 4653} &
  {\color[HTML]{FF0000} 9576} &
  {\color[HTML]{333333} } &
  {\color[HTML]{333333} } &
  {\color[HTML]{FF0000} 536800000} &
  {\color[HTML]{FF0000} 63790} \\ \hline
July &
  {\color[HTML]{333333} -12.98} &
  {\color[HTML]{FF0000} 39380000} &
  {\color[HTML]{FF0000} 172000} &
  {\color[HTML]{333333} } &
  {\color[HTML]{FF0000} 172000} &
  {\color[HTML]{FF0000} 172000} &
  {\color[HTML]{FF0000} 172000} &
  {\color[HTML]{333333} } &
  {\color[HTML]{333333} } &
  {\color[HTML]{333333} 11940} &
  {\color[HTML]{333333} } \\ \hline
August &
  {\color[HTML]{FF0000} 1562000} &
  {\color[HTML]{FF0000} 417500} &
  {\color[HTML]{333333} 7075} &
  {\color[HTML]{333333} } &
  {\color[HTML]{FF0000} } &
  {\color[HTML]{333333} 72.44} &
  {\color[HTML]{333333} 7075} &
  {\color[HTML]{333333} } &
  {\color[HTML]{333333} 7075} &
  {\color[HTML]{333333} } &
  {\color[HTML]{333333} } \\ \hline
September &
  {\color[HTML]{FF0000} 49770000} &
  {\color[HTML]{FF0000} 6355000} &
  {\color[HTML]{FF0000} 1769} &
  {\color[HTML]{FF0000} 274200} &
  {\color[HTML]{FF0000} 77740} &
  {\color[HTML]{FF0000} 77740} &
  {\color[HTML]{FF0000} 18420} &
  {\color[HTML]{FF0000} 18420} &
  {\color[HTML]{333333} } &
  {\color[HTML]{333333} } &
  {\color[HTML]{333333} } \\ \hline
October &
  {\color[HTML]{333333} 173900} &
  {\color[HTML]{333333} } &
  {\color[HTML]{333333} } &
  {\color[HTML]{333333} } &
  {\color[HTML]{333333} } &
  {\color[HTML]{333333} } &
  {\color[HTML]{333333} } &
  {\color[HTML]{333333} } &
  {\color[HTML]{333333} } &
  {\color[HTML]{333333} } &
  {\color[HTML]{333333} } \\ \hline
November &
  {\color[HTML]{333333} } &
  {\color[HTML]{333333} } &
  {\color[HTML]{333333} } &
  {\color[HTML]{333333} } &
  {\color[HTML]{333333} } &
  {\color[HTML]{333333} } &
  {\color[HTML]{333333} } &
  {\color[HTML]{333333} } &
  {\color[HTML]{333333} } &
  {\color[HTML]{333333} } &
  {\color[HTML]{333333} } \\ \hline
\end{tabular}%
}

Sarcoptiformes\\
\resizebox{\columnwidth}{!}{%
\begin{tabular}{|l|l|l|l|l|l|l|l|}
\hline
$\mathfrak{L}$ &
  \textit{A. gigantea} &
  \textit{P. parvisetigerum} &
  \textit{T. stercus} &
  \textit{M. japonica} &
  \textit{G. fusca} &
  \textit{N. rotundus} &
  \textit{C. reticulatus} \\ \hline
April &
  {\color[HTML]{333333} } &
  {\color[HTML]{333333} } &
  {\color[HTML]{333333} } &
  {\color[HTML]{333333} } &
   &
   &
   \\ \hline
May &
  {\color[HTML]{333333} 1631000000} &
  {\color[HTML]{333333} 1631000000} &
  {\color[HTML]{333333} } &
  {\color[HTML]{333333} } &
   &
   &
   \\ \hline
June &
  {\color[HTML]{FF0000} 4455000000} &
  {\color[HTML]{333333} } &
  {\color[HTML]{FF0000} 566.3} &
  {\color[HTML]{FF0000} 116500000} &
  {\color[HTML]{333333} } &
  {\color[HTML]{333333} -0.7687} &
   \\ \hline
July &
  {\color[HTML]{333333} 2034000} &
  {\color[HTML]{333333} } &
  {\color[HTML]{333333} } &
  {\color[HTML]{333333} } &
  {\color[HTML]{333333} } &
  {\color[HTML]{333333} } &
   \\ \hline
August &
  {\color[HTML]{333333} 21660} &
  {\color[HTML]{333333} } &
  {\color[HTML]{333333} } &
  {\color[HTML]{333333} } &
  {\color[HTML]{333333} } &
  {\color[HTML]{333333} } &
   \\ \hline
September &
  {\color[HTML]{FF0000} 34380000000} &
  {\color[HTML]{FF0000} 34380000000} &
  {\color[HTML]{333333} } &
  {\color[HTML]{333333} -0.8269} &
  {\color[HTML]{FF0000} 579.5} &
  {\color[HTML]{FF0000} 21950} &
  {\color[HTML]{FF0000} 5326000} \\ \hline
October &
  {\color[HTML]{333333} } &
  {\color[HTML]{333333} 117400} &
  {\color[HTML]{333333} } &
  {\color[HTML]{333333} } &
   &
   &
   \\ \hline
November &
  {\color[HTML]{333333} } &
  {\color[HTML]{333333} } &
  {\color[HTML]{333333} -506.9} &
  {\color[HTML]{333333} } &
   &
   &
   \\ \hline
\end{tabular}%
}

\small
\begin{flushleft}
\textbf{Abbreviations:} WE: Washidu East quadrat; WW: Washidu West quadrat. \textit{P.~pallidum}: \textit{Polysphondylium pallidum}; \textit{D.~purpureum}: \textit{Dictyostelium purpureum}; \textit{P.~violaceum}: \textit{Polysphondylium violaceum}. For the scientific names of soil mesofauna, refer to Table~2.

\textbf{Note:} Values shown in \textcolor{red}{red} represent significantly large absolute values derived from a black hole analogy. All numerical entries are rounded to 4 significant digits. Blank cells indicate values that could not be calculated.
\end{flushleft}
\end{table}

\begin{table}[]
\caption{Hamiltonian.}
\centering
Dictyostelia\\
\resizebox{\columnwidth}{!}{%
\begin{tabular}{|l|l|l|l|l|l|l|}
\hline
$\mathfrak{H}$    & WE \textit{P. pallidum} & WE \textit{D. purpureum} & WE \textit{P. violaceum} & WW \textit{P. pallidum} & WW \textit{D. purpureum} & WW \textit{P. violaceum} \\ \hline
April       &                &                 &                 &                &                 &                 \\ \hline
June      & -0.4346   & {\color[HTML]{FE0000}-1.230E+6}    & -10.24    &                &                 &                 \\ \hline
July      &                &                 &                 & -39.89   & -0.1037    & {\color[HTML]{FE0000}-2.707E+6}    \\ \hline
August    &                &                 &                 & -684.2  & -0.8629    &                 \\ \hline
September & -798.1  & -0.9680    &                 & {\color[HTML]{FE0000}-5601}   & -2686    &                 \\ \hline
October   & {\color[HTML]{FE0000}-1.278E+4}   & -4.487    &                 & -690.0   &                 & -0.8678     \\ \hline
November  & -0.5288   &                 & {\color[HTML]{FE0000}-3.558E+7}    &                &                 &                 \\ \hline
December  &                &                 &                 &                &                 &                 \\ \hline
January   &                &                 &                 &                &                 &                 \\ \hline
\end{tabular}
}

Collembola\\
\resizebox{\columnwidth}{!}{%
\begin{tabular}{|l|l|l|l|l|l|l|l|l|l|l|l|}
\hline
$\mathfrak{H}$ &
  \textit{F. octoulata} &
  \textit{E. aino} &
  \textit{H. watanabei} &
  \textit{S. celebensis} &
  \textit{S. aureus} &
  \textit{D. trispinata} &
  \textit{S. japonica} &
  \textit{H. nigrochephala} &
  \textit{W. japonica} &
  \textit{F. yosii} &
  \textit{A. laricis} \\ \hline
April &
  {\color[HTML]{FF0000} } &
  {\color[HTML]{FF0000} } &
  {\color[HTML]{FF0000} } &
  {\color[HTML]{FF0000} } &
  {\color[HTML]{FF0000} } &
  {\color[HTML]{FF0000} } &
  {\color[HTML]{FF0000} } &
  {\color[HTML]{FF0000} } &
  {\color[HTML]{FF0000} } &
  {\color[HTML]{FF0000} } &
  {\color[HTML]{FF0000} } \\ \hline
May &
  {\color[HTML]{FF0000} } &
  {\color[HTML]{FF0000} } &
  {\color[HTML]{FF0000} } &
  {\color[HTML]{FF0000} } &
  {\color[HTML]{FF0000} } &
  {\color[HTML]{FF0000} } &
  {\color[HTML]{FF0000} } &
  {\color[HTML]{FF0000} } &
  {\color[HTML]{FF0000} } &
  {\color[HTML]{FF0000} } &
  {\color[HTML]{FF0000} } \\ \hline
June &
  {\color[HTML]{FF0000} -188500000} &
  {\color[HTML]{FF0000} -258200} &
  {\color[HTML]{FF0000} -63880} &
  {\color[HTML]{FF0000} } &
  {\color[HTML]{FF0000} -63880} &
  {\color[HTML]{FF0000} -4658} &
  {\color[HTML]{FF0000} -9671} &
  {\color[HTML]{FF0000} } &
  {\color[HTML]{FF0000} } &
  {\color[HTML]{FF0000} -537400000} &
  {\color[HTML]{FF0000} -63880} \\ \hline
July &
  {\color[HTML]{333333} -13.28} &
  {\color[HTML]{FF0000} -39940000} &
  {\color[HTML]{FF0000} -172600} &
  {\color[HTML]{FF0000} } &
  {\color[HTML]{FF0000} -172600} &
  {\color[HTML]{FF0000} -172600} &
  {\color[HTML]{FF0000} -172600} &
  {\color[HTML]{FF0000} } &
  {\color[HTML]{FF0000} } &
  {\color[HTML]{333333} -12220} &
  {\color[HTML]{FF0000} } \\ \hline
August &
  {\color[HTML]{FF0000} -1703000} &
  {\color[HTML]{FF0000} -420800} &
  {\color[HTML]{333333} -7092} &
  {\color[HTML]{FF0000} } &
  {\color[HTML]{FF0000} } &
  {\color[HTML]{333333} -89.82} &
  {\color[HTML]{333333} -7092} &
  {\color[HTML]{FF0000} } &
  {\color[HTML]{333333} -7092} &
  {\color[HTML]{FF0000} } &
  {\color[HTML]{FF0000} } \\ \hline
September &
  {\color[HTML]{FF0000} -51180000} &
  {\color[HTML]{FF0000} -6386000} &
  {\color[HTML]{FF0000} -1878} &
  {\color[HTML]{FF0000} -274500} &
  {\color[HTML]{FF0000} -77820} &
  {\color[HTML]{FF0000} -77820} &
  {\color[HTML]{FF0000} -18430} &
  {\color[HTML]{FF0000} -18430} &
  {\color[HTML]{FF0000} } &
  {\color[HTML]{FF0000} } &
  {\color[HTML]{FF0000} } \\ \hline
October &
  {\color[HTML]{333333} -194000} &
  {\color[HTML]{FF0000} } &
  {\color[HTML]{FF0000} } &
  {\color[HTML]{FF0000} } &
  {\color[HTML]{FF0000} } &
  {\color[HTML]{FF0000} } &
  {\color[HTML]{FF0000} } &
  {\color[HTML]{FF0000} } &
  {\color[HTML]{FF0000} } &
  {\color[HTML]{FF0000} } &
  {\color[HTML]{FF0000} } \\ \hline
November &
  {\color[HTML]{FF0000} } &
  {\color[HTML]{FF0000} } &
  {\color[HTML]{FF0000} } &
  {\color[HTML]{FF0000} } &
  {\color[HTML]{FF0000} } &
  {\color[HTML]{FF0000} } &
  {\color[HTML]{FF0000} } &
  {\color[HTML]{FF0000} } &
  {\color[HTML]{FF0000} } &
  {\color[HTML]{FF0000} } &
  {\color[HTML]{FF0000} } \\ \hline
\end{tabular}%
}

Sarcoptiformes\\
\resizebox{\columnwidth}{!}{%
\begin{tabular}{|l|l|l|l|l|l|l|l|}
\hline
$\mathfrak{H}$ &
  \textit{A. gigantea} &
  \textit{P. parvisetigerum} &
  \textit{T. stercus} &
  \textit{M. japonica} &
  \textit{G. fusca} &
  \textit{N. rotundus} &
  \textit{C. reticulatus} \\ \hline
April &
   &
   &
   &
   &
   &
   &
   \\ \hline
May &
  {\color[HTML]{333333} -1674000000} &
  {\color[HTML]{333333} -1674000000} &
   &
   &
   &
   &
   \\ \hline
June &
  {\color[HTML]{FF0000} -4468000000} &
  {\color[HTML]{FF0000} } &
  {\color[HTML]{FF0000} -592.7} &
  {\color[HTML]{FF0000} -116500000} &
  {\color[HTML]{333333} } &
  {\color[HTML]{333333} -0.9852} &
  {\color[HTML]{333333} } \\ \hline
July &
  {\color[HTML]{333333} -2298000} &
  {\color[HTML]{333333} } &
  {\color[HTML]{333333} } &
  {\color[HTML]{333333} } &
  {\color[HTML]{333333} } &
   &
   \\ \hline
August &
  {\color[HTML]{333333} -36310} &
  {\color[HTML]{333333} } &
  {\color[HTML]{333333} } &
  {\color[HTML]{333333} } &
  {\color[HTML]{333333} } &
   &
   \\ \hline
September &
  {\color[HTML]{FF0000} -34440000000} &
  {\color[HTML]{FF0000} -34440000000} &
  {\color[HTML]{333333} } &
  {\color[HTML]{333333} -1.033} &
  {\color[HTML]{FF0000} -607.6} &
  {\color[HTML]{FF0000} -22090} &
  {\color[HTML]{FF0000} -5329000} \\ \hline
October &
  {\color[HTML]{333333} } &
  {\color[HTML]{333333} -131500} &
  {\color[HTML]{333333} } &
   &
   &
   &
   \\ \hline
November &
  {\color[HTML]{333333} } &
  {\color[HTML]{333333} } &
  {\color[HTML]{333333} -1098} &
   &
   &
   &
   \\ \hline
\end{tabular}%
}

\small
\begin{flushleft}
\textbf{Abbreviations:} WE: Washidu East quadrat; WW: Washidu West quadrat. \textit{P.~pallidum}: \textit{Polysphondylium pallidum}; \textit{D.~purpureum}: \textit{Dictyostelium purpureum}; \textit{P.~violaceum}: \textit{Polysphondylium violaceum}. For the scientific names of soil mesofauna, refer to Table~2.

\textbf{Note:} Values highlighted in \textcolor{red}{red} indicate significantly large absolute values derived from a black hole analogy. All numerical values are presented with 4 significant digits. Blank cells indicate cases where values could not be calculated. 
\end{flushleft}
\end{table}

The energy of the system is given by
\begin{equation}
E = \sum_k \frac{\partial \mathfrak{L}}{\partial \dot{q}_k} \dot{q}_k - \mathfrak{L} = \mathfrak{H} = \frac{1}{2}(-sw\,\Im(s) - M \Im(s)^2)\phi_k^2 - \frac{1}{2}p^2 \phi_k^2 + \frac{1}{4}l \frac{\Im(s)}{D} (\phi_k)^4,
\end{equation}
and the conjugate momentum is
\begin{equation}
p_k = \frac{\partial \mathfrak{L}}{\partial \dot{q}_k} = \frac{-sw\,\Im(s) - M \Im(s)^2}{H(t) D} \phi_k^2.
\end{equation}
From this, the velocity is
$
\dot{q}_k = \frac{p_k}{M},
$
and the spatial scale $D$ is expressed as
\begin{equation}
D = \frac{i}{H(t)} \sqrt{\frac{sw\,\Im(s)}{M} + \Im(s)^2} \cdot \phi_k.
\end{equation}

The sign in the expression for $D$ reflects a weak interaction analogue and is entirely dependent on environmental factors in the biological model. This formulation completes our integrated model for biological systems, drawing analogies to the standard model of particle physics plus gravity. It achieves this by embedding a single dimension with topological structure into a dynamical system governed by both Lagrangian and Hamiltonian mechanics (see Tables~15 and~16).

\subsection{Further clarification of our model}
If we interpret $D = u - w = K$ as an elliptic integral, then the function $f(u) = \sum \frac{1}{D_i^3}$ exhibits dual periodicity, and the Weierstra\ss $\wp$-function satisfies $\wp = -\zeta'(z; \Lambda)$, where the Weierstra\ss zeta function is given by
$
\zeta(z; \Lambda) = \frac{1}{z} + \sum_{w \in \Lambda^*} \left( \frac{1}{z - w} + \frac{1}{w} + \frac{z}{w^2} \right).
$
This zeta function accounts for observations from multiple $w$ and $0$, incorporating an absolute zeta function through duplications of $w$.

Furthermore, a set of differentials of the first kind on $K$, $\mathfrak{L}_0 \ni w_i\,dz$, and the one-dimensional topologically congruent group $B_1^*(\mathfrak{R})$ with real coefficients become isomorphic as a $k_0$-module. The Abel-Jacobi map yields:
\begin{equation}
\bar{\mathfrak{D}}_0 := \mathfrak{D}_0 / \mathfrak{D}_H \cong V_g(k) / P(w_i\,dz),
\end{equation}
where $\mathfrak{D}_0$ is a zero-dimensional factor group of $K$, and $\mathfrak{D}_H$ is a principal factor group. Thus, $D$ can be regarded as a periodic group $P(w_i\,dz)$, and $\mathfrak{D}_H \cong P(w_i\,dz)$. A $g$-dimensional complex vector module $V_g(k) \cong \mathfrak{D}_0$ may be interpreted as $\mathrm{sign}(M_i) M_i$, where $M_i$ can be $\Im(s)$. Neglecting the solid angle factor $4\pi$, the inverse-square law corresponds to the multiplication of three copies of $\bar{\mathfrak{D}}_0$ under the Abel-Jacobi map.

Interestingly, using 2018 CODATA physical constants, we find
$
\sqrt{k_g \sqrt{\varepsilon_0 \mu_0}} \approx 1.9943 \approx 2,
$
and the generalized force becomes
$
\mathbf{F}_s = \frac{\mathrm{sign}(M)}{8\pi} \cdot \frac{M_i M_j}{D^2} \mathbf{e_F},
$
which satisfies equal partitioning of force into inward and outward directions on a sphere of radius $D$. This suggests that gravitational force branching occurred first, followed by the branching of electric (communicative, ``\'etale”) and magnetic (Frobenioid) forces. The slight deviation from the value $2$ reflects the omission of weak interaction effects, which are not easily incorporated into this model. Even in the standard model, the masslessness of neutrinos remains an unresolved issue.

In SI units, the dimension of this value is $\mathrm{\sqrt{kg \cdot s^3}/m^2}$, whereas in our unit system, it is dimensionless. The corresponding constant is
$
\frac{\sqrt{\phi} t^3}{D^2}.
$
Here, the root of observed $\phi$, $\sqrt{\phi}$, is equivalent to $t$ (the root of physical time), and $D$ has the dimension of $t^2$, indicating that observation is inherently linked to interaction. Thus, the product of the root of the observer and the cube of time reflects a duplication of observed fitness in $D^2$, suggesting a duplicated fitness in the potential cube-$t$ space.

For further development, consider Einstein’s field equation:
\begin{equation}
R_{ik} - \frac{1}{2} g_{ik} R = \frac{8\pi G}{c^4} T_{ik},
\end{equation}
where $G = 1/8\pi$, $c = \Im(s)$, and
$
T_{ik} = \phi \Im(s)^2 D^2 \sqrt{\frac{\Im(s)^2 + (H(t) D)^2}{\Im(s)^2 - (H(t) D)^2}}, \quad \dot{D} = H(t) D.
$
The curvature on the left-hand side becomes
$
\phi \cdot \left( \frac{\dot{D}}{H(t) \Im(s)} \right)^2 \cdot \sqrt{\frac{\Im(s)^2 + (H(t) D)^2}{\Im(s)^2 - (H(t) D)^2}}.
$
If the system is abstracted to a sphere, then
$
\frac{1}{r^2} = \phi \cdot \left( \frac{\dot{D}}{H(t) \Im(s)} \right)^2 \cdot \sqrt{\frac{\Im(s)^2 + (H(t) D)^2}{\Im(s)^2 - (H(t) D)^2}},
$
and thus
$
r = \frac{H(t) \Im(s)}{\dot{D}} \cdot \sqrt{\phi \sqrt{\frac{\Im(s)^2 + (H(t) D)^2}{\Im(s)^2 - (H(t) D)^2}}}.
$

It is also notable that the Ricci curvature tensor is a product of the metric tensor (normalized to $1$) and the ratio of \'etale to Frobenioid structures. This corresponds to the determinant of the matrix formed by the second fundamental form relative to the first fundamental form. The resulting term is
$
\phi \cdot \left( \frac{\dot{D}}{H(t) \Im(s)} \right)^2 \cdot \sqrt{\frac{\Im(s)^2 + (H(t) D)^2}{\Im(s)^2 - (H(t) D)^2}},
$
and in the limit $t \to \infty$ with $\Im(s) \gg H(t) D$, this simplifies to
$
4\phi \left( \frac{\dot{D}}{\Im(s)} \right)^2.
$

The Macdonald formula expresses the infinite product as a $q$-series:
\begin{equation}
\prod_{m = 1}^{\infty}(1 - q^{2m})^k = \sum_{m = 0}^{\infty} a_{m,k} q^m.
\end{equation}
In the case where $\mathfrak{g} = \mathfrak{sl}_2$, we have $k = 3$, and the identity becomes:
\begin{equation}
\sum_{m = -\infty}^{\infty} (-1)^m (2m + 1) q^{m(m+1)} = 2 \prod_{m = 1}^{\infty}(1 - q^{2m})^3.
\end{equation}
By identifying $D_m^{-1} = (1 - q^{2m})$, the infinite product over $D$ leads to a multiplication by $2$, analogous to the above identity. Interpreting $2m + 1$ as a measure of bosonic fitness $w$, this term reflects a chaotic configuration, which underpins the inverse-square law in the absence of multiple force splittings.

Importantly, introducing a squared mass term reveals that when $H(t)D > \Im(s)$, the mass becomes purely imaginary. In this regime, the potential—originally along the imaginary axis—rotates into the real axis upon multiplication by $-i$, effectively converting the potential into a fractal dimension. This transition becomes manifest when the growth rate of $D$ is sufficiently high, i.e., $H(t)D > \Im(s)$, a condition satisfied at the nontrivial zeros of the Riemann zeta function.

When $H(t)D \approx \Im(s)$, the first-order approximation of kinetic energy is $\frac{1}{2} M \Im(s)^2$, aligning with the energy expression $E = M \Im(s)^2$ when $(H(t)D)^2 = 5 \Im(s)^2$. Therefore, to generate fractals beyond $\Re(s) = 2$ (the square-law threshold), recognition from five-dimensional fractals is required, particularly when $H(t)D \approx \Im(s)$. This insight is especially relevant in the range $\Re(s) = 5$ to $\Re(s) \sim 2$, where fractal development is prominent.

For instance, consider $s = \frac{\sqrt{7} + i}{2}$ as previously discussed. The self-interaction of $s$ yields:
$
\frac{7}{4} e_{\Re(s)}^2 + \frac{2\sqrt{7}}{4} i\, e_{\Re(s)} e_{\Im(s)} - \frac{1}{4} e_{\Im(s)}^2.
$
Here, the coefficient of the first term is $\frac{4}{7}$, which must be an integer to contribute meaningfully. The second term is non-integer and imaginary, while the third term is a multiple of $4$; thus, only the first term is retained in the physical interpretation.

In the context of characteristic classes, the Pontryagin class (or a component of the Hirzebruch signature) is given by:
$
p_2(M) = \frac{4(2i - 1)^2 + 45}{7},
$
where $\Re(s) = 2i - 1$ and $p_1(M) = 2(2i - 1)$. If $i \equiv 0, 1 \pmod{7}$, then $p_2(M)$ is an integer, and $M$ qualifies as a differential manifold. Otherwise, $M$ is an exotic sphere—homeomorphic but not diffeomorphic to the standard Euclidean $S^7$.

In a continuous-time differential model, such exotic structures are inadmissible. Therefore, development into $\Re(s) \geq 5$ must occur in pairs satisfying $\Re(s_1) + \Re(s_2) = 7j$ or $7j + 1$, where $j \in \mathbb{N}$. An example is the pairing of $\Re(s) = 5$ and $\Re(s) = 2$, observed in October at the Washidu East quadrat between \textit{Polysphondylium pallidum} and \textit{Dictyostelium purpureum}.

\subsection{Roles of another analogy to quantum mechanics in our model}
Let us consider the operator relation $\phi \ket{\alpha_S(t)} = l \ket{\alpha_S(t)}$ and $\bra{\alpha_S(t)} \phi = \bra{\alpha_S(t)} p$. In this framework, the bra vector represents phase information (Frobenioid; persistence homology), while the ket vector encodes the norm (\'etale function or cohomology). The eigenvalue $l$ serves as a communicative invariant, transcending the constraints of a specific system layer.

By defining $\Phi(\alpha) = 0$ as a monic polynomial over a cyclotomic field, this polynomial becomes the zero ideal in the ring structure associated with either the bra or ket. The eigenvalues of the projection operator $\ket{\alpha(t)}\bra{\alpha(t)}$ are either $1$ or $0$, corresponding to observed or unobserved states, respectively. This binary structure can be interpreted as a Morse code for communication. The integration of these eigenstates yields a Dirac delta function, implying $\Re(s) = l$. Only when $p = l$ does the expectation value $\bra{\alpha(t)} \phi \ket{\alpha(t)}$ yield the energy $E(N)$.

In the interaction picture where $H = H_0 + H'$, with $M_C$ and $M_W$ representing two interacting Monsters, we set $H_0 = l$, $H' = -l$, so that $H = l + (-l) = 0$. This implies that $M_W$ acquires a fractal structure of $l$ dimensions, while $M_C$ loses the same, leading to a static state where $M_W$ is fully invested with the transferred structure, provided the interaction persists.

The entanglement entropy between these two harmonic oscillators is given by:
\begin{equation}
S_A = \cosh^2 \theta \ln \cosh^2 \theta - \sinh^2 \theta \ln \sinh^2 \theta,
\end{equation}
where the coupling parameter $\theta$ is related to $l$ via
$
l = \frac{2 \sinh \theta \cosh \theta}{1 + 2 \sinh^2 \theta}.
$
The Hamiltonian for the coupled system is:
$
H = w^\dagger w + c^\dagger c + l(w^\dagger c^\dagger + w c),
$
where $(w, w^\dagger)$ and $(c, c^\dagger)$ are the creation and annihilation operators for $M_W$ and $M_C$, respectively.

For maximal entanglement, consider Hilbert spaces $\mathcal{H}_W$ and $\mathcal{H}_C$ with dimensions $|\mathcal{H}_W| = \sum N_W$ and $|\mathcal{H}_C| = \sum N_C$. The maximum entanglement entropy is then $\min(\ln |\mathcal{H}_W|, \ln |\mathcal{H}_C|)$. If $M_W$ absorbs all $N_k$ from $M_C$, then
$
\ln |\mathcal{H}_C| = \ln P + N_k \frac{s}{b},
$
and as $N_k \to 0$, the second term vanishes, leaving $\ln P$ as the entanglement entropy associated with the loss of $N_k$.

Assuming an analogy to QED in the Frobenioid-\'etale relationship, particularly for $\Re(s) = 2, 3$, hetero-interaction terms become negligible, consistent with empirical observations. For optimal communication—transmitting both \'etale and Frobenioid components in parallel—the following condition must be satisfied:
\begin{equation}
\frac{\frac{\mathbf{V}}{\Im(s)}}{1 + \frac{\mathbf{V}^2}{\Im(s)^2}} = \frac{\mathbf{E} \times \mathbf{H}}{E^2 + H^2}
\end{equation}
\cite{Landau1975}.

This leads to a conservation law in $D$-space:
\begin{equation}
\frac{\partial}{\partial t} \left\{ \frac{E^2 + H^2}{2} dV + \sum \mathscr{E}_{\mathrm{kin}} \right\} = - \oint \mathbf{S} \cdot d\mathbf{f} (= 0),
\end{equation}
where the Poynting vector is defined as $\mathbf{S} = \Im(s) \mathbf{E} \times \mathbf{H}$ and $d\mathbf{f}$ is an areal element. The equality to zero holds in the limit of an infinite volume, ensuring global conservation.

\subsection{From renormalization to fractals}

In our biological model, hierarchical formation can be interpreted as a resolution to the divergence problem in transition probabilities, akin to the renormalization procedure in quantum field theory. In quantum electrodynamics (QED), the Hamiltonian includes harmonic oscillator terms that lead to divergences unless renormalized. However, in our framework, each oscillator can be reinterpreted as an additional dimension within a fractal structure.

Mathematically, the plausibility of renormalization—denoted as $P^m$—can be formalized through the existence of $0 \in U$, where $(P^m, U, V)$ defines a mapping analogous to a quadratic polynomial, and the filled Julia set $K_m$ is connected. Here, $P(z) = z^2 + c$, and $U$, $V$ are simply connected regions. This condition is naturally satisfied in a continuously evolving system composed of simply connected components, with $U$ representing the observant domain in our model.

The renormalization group equation takes the form:
\begin{equation}
\mu \frac{d\mathfrak{M}}{d\mu} = \left( \mu \left. \frac{\partial}{\partial \mu} \right|_e + \mu \frac{\partial e}{\partial \mu} \frac{\partial}{\partial e} \right) \mathfrak{M} = 0.
\end{equation}
Here, the first term represents the renormalizing contribution, while the second term captures the residual (non-renormalized) effects. These terms can assume negative and positive values, respectively, and together they quantify the contribution of $\mathfrak{M}$ to the emergent fractal structure. In this sense, a fractal $l$-dimension may be viewed as one of the periods of an elliptic function acting on a “string.”

However, the validity of renormalization is constrained to dimensions $d < 8$ in the context of $\mathbb{R}^3 \times \mathbb{R}^3 \cong \hat{\mathbb{C}}$. For broader applications, a similar conceptual framework appears in the Bogomolov conjecture.

Let $A$ be an abelian variety over an algebraic field $K$, and let $L$ be a symmetric, ample, and invertible sheaf on $A$. Define $A_{\bar{K}} := A \times_{\mathrm{Spec}(K)} \mathrm{Spec}(\bar{K})$, and let $X$ be an irreducible algebraic subvariety of $A_{\bar{K}}$. If $\hat{h}_L$ is the N\'eron-Tate height function associated with $L$, and for every $\varepsilon > 0$, the set
$
\{ x \in X(\bar{K}) \mid \hat{h}_L(x) \leq \varepsilon \}
$
is Zariski-dense in $X$, then there exists an abelian subvariety $B \subset A$ and a torsion point $b \in A_{\bar{K}}$ such that $X = B + b$.

In our analogy, the torsion point $b$ corresponds to the renormalization term, while $B$ represents the current structural layer. Thus, $B$ can be interpreted as a Frobenioid component, and $b$ as an \'etale fragment—a localized fractal string. This perspective unifies renormalization and fractal emergence within a coherent algebraic and topological framework.

\section{Discussion}
\subsection{Our main results}

To the best of our knowledge, no previous studies have considered the biological perspective proposed herein. We introduce an index, small $s$, to distinguish between populations and species. Specifically, $\Re(s) = \frac{\ln \frac{N_1}{N_k}}{\ln k}$ when $N_1$ and $N_k$ are the population densities of the first and $k$-th ranks, respectively, with $k \neq 1$; and $\Re(s) = \zeta^{-1}(\Sigma N / N_1)$ when $\Sigma N$ is the total population density, $k = 1$, and $\zeta^{-1}$ denotes the inverse of the Riemann zeta function. The imaginary part is given by $\Im(s) = e^{E(N)\Re(s)/b}$, where $b$ is an approximation parameter in the logarithmic model $N_k = a - b \ln k$.

The real part $\Re(s)$ is high for ordered species and low for fluctuating populations in the wild. Notably, $\Re(s) = 2$ serves as a critical line for both, aligning with the fractal theory of box dimension $\dim_B A = 2$ in the model $\mathbb{R}^2$, which corresponds to our $s$ metric.

To develop this index, we modified the Price equation \cite{Price1970}, grounding our approach in the $R = T$ theorem and Weil's explicit formula \cite{Weil1952, Taylor1995, Wiles1995}. The Price equation, which describes evolution and natural selection, is adapted here by replacing gene frequencies with the proportion of individuals in a given population or species. The small $s$ index is related to the Price equation through covariance and expectation.

The nontrivial zeros of the Riemann $\zeta$ function provide insight into population bursts or collapses. In this model, speciation is linked to prime numbers: a prime ideal represents the status of a specific species, and time-dependent fitness multiplication can be calculated using these primes. We derive unique equations in the Maass form and analyze the spectral properties of the data. Employing the Selberg zeta function and the Hasse-Weil $L$-function to compute the norm of prime closed geodesics $|N(p)|$ reveals that interacting neutral populations exhibit behavior near $|N(p)| \sim 1$.

By integrating these results with phylogenetic asymmetry, we assess whether the observed data hierarchy reflects chaotic populations or nonadaptive species, versus adaptive species occupying successful ecological niches. Our model has shown partial success in predicting imminent transitions between biological phases, such as adaptation and disadaptation. Using the Schwarz equation, we also derive a time-dependent fitness function consistent with empirical observations.

Additionally, a web-based formalism \cite{Gaiotto2016}, combining supersymmetry (via the Hodge-Kodaira decomposition of a non-degenerate fermionic $\phi$ function) and an analogy to the transactional interpretation of quantum mechanics, leads to a nine-dimensional model without time, structured as three spatial dimensions multiplied by three fluctuation dimensions. The concept of a fitness space enables the definition of a precise, time-dependent Hubble parameter for this space over appropriate timescales.

Finally, we elucidate the asymmetric time development of $\phi$ within our model, and a degenerate $\psi$ (fitness-related) function offers insights into future dynamics. The proposed model integrates information theory with empirical biological observations to provide a novel understanding of populations and species. This approach differs fundamentally from the physical and thermodynamic models such as that of \cite{Volkov2004}; our model is not based on physical dynamics but rather applies physical analogies at a biological scale.

In this context, a patch is defined as a small plot or piece of land, particularly one used for cultivating specific organisms. We refer to our framework as the Patch with Zeta Dominance (PzDom) model. It suffices to evaluate $\Re(s)$ to determine whether a population is chaotic or species-dominated, with the threshold at $\Re(s) = 2$. That is, species-level dynamics begin to emerge (but not complete) when $\Re(s) > 2$. The model requires only temporal changes in individual density and also addresses the significance of biological hierarchies. We propose a framework that enables future research to investigate the nature of hierarchical systems.

\subsection{Our model interpreted by information theory}
We propose an original model that employs a novel definition of entropy—specifically, relative entropy defined as $\ln k$, where $k$ is the rank of a population. This is equivalent to the Kullback-Leibler divergence, $D_1(P \| Q) = \sum_{i = 1}^n p_i \ln \frac{p_i}{q_i}$, in the case where $n = 1$, $p_i = 1$, and $q_i = 1/k$. This approach differs from that of \cite{Harte2008}. We also introduce a new definition of temperature: an integrative environmental parameter that governs the distribution of populations and species, derived from their logarithmic distribution. The units of this parameter are expressed in cells per gram (cells/g) or individuals per 100 cc (individuals/100 cc), and it is classified as a half-intensive parameter, as described in the initial sections of the Results.

We define new metrics based on statistical mechanics \cite{Fujisaka1998, Banavar2010, Tasaki2015} to distinguish and interpret species and population counts within mixed communities. This framework is applied to empirical data from Eastern Japanese Dictyostelia \cite{Adachi2015}, Ohdaigahara soil mesofauna, and global economic systems.

The application of statistical mechanics to biological systems began with \cite{Lotka1922}, evolved through the complex Hamiltonian formulations of \cite{Kerner1957}, and continued with the Lotka-Volterra equations for $N$-interacting species in artificially noisy environments \cite{Spagnolo2004}. Although \cite{Kerner1957} proposed an intriguing model for time-evolving systems, it operates in a different mathematical space from ours. A rigorous mathematical investigation is necessary to clarify the relationship between the two models.

Our approach also diverges from time-dependent ecosystem assembly models constrained by finite Markov chains, such as those proposed by \cite{Pigolotti2005} and \cite{Capitan2009}. Crucially, our model enables the calculation of distinct sets of critical temperatures and Weiss fields (incorporating Bose-Einstein condensation) at which natural first-order phase transitions occur among species or populations. Here, "phase transitions" refer to shifts in community structure—from chaotic states driven by neutrality or nonadaptive conditions to dominance by a particular species, or conversely, from species-level dominance to dominance by a specific population within a species.

It has been proposed that the edge of many-body quantum chaos in quantum reservoir computing exhibits optimal performance, analogous to classical systems \cite{Kobayashi2026}, a finding that aligns with our observations. The order parameter in our model is denoted by large $S$.

This complex phase transition behavior across biological hierarchies was not fully addressed by \cite{Harte2008}, and only briefly mentioned by \cite{Banavar2010} in the context of relative entropy. Our model demonstrates that populations of certain highly adapted species exhibit significantly greater stability compared to others.

\subsection{Application of our model beyond Dictyostelia, soil mesofauna, and world economics}
In our PzDom model, under the influence of entropy, populations and species are expected to fluctuate when $\Re(s) \sim 1$ or $\zeta = 0$. However, an alternative solution emerges. According to the fractal theory results illustrated in Figure~1, $\Re(s) = 2$ serves as the boundary between populations and species in terms of the system's box dimension.

The global economy once appeared to approach this boundary, but subsequently declined, possibly due to increasing regional interconnectivity. In the ecological dataset on ruderal vegetation presented by \cite{Rodriguez2013}, the data suggest the presence of opportunistic species, as most regions exhibit $\Re(s) > 2$. Conversely, marine interstitial meiofauna on sandy beaches and tropical rocky shore snails likely represent populations—or at least equilibrium species—since their values fall below $\Re(s) = 2$.

As previously discussed, adaptation can be defined as a level of fitness sufficient to transcend the fluctuations of harmonic neutrality. The threshold for this transition is $\Re(s) = 2$. From the data, we also observe that species with larger $s$ or $D$ exhibit higher fitness.

Since the domain $\Re(s) > 1$ corresponds to the region of absolute convergence for the Riemann zeta function, the species domain is not chaotic. Instead, it exhibits structured community dynamics governed by adaptation and hysteresis. From the original Price equation, it is evident that $\Re(s) > 1$ implies $\mathrm{Cov} > 0$, indicating that the species or population is diverging in its characteristics.

In contrast, when $0 < \Re(s) < 2$, the system is chaotic and characterized by the inner product $\langle \tilde{\mathscr{R}}_{A, \Omega}^{[k]}, \varphi \rangle$. Furthermore, $\Re(s) < 1$ implies $\mathrm{Cov} < 0$, signifying that the characteristics of the species or population are converging.

\subsection{$s$ and mathematical physics}
Although $\Re(s)$ may vary continuously depending on how populations or species adapt to their environments, each species possesses certain discrete characteristic variables. These arise from the quantization of $D$ by $T$, implying adaptation (or disadaptation) and population/species bursts (or collapses) in specific environments. Since environmental variables are inherently continuous, the discreteness of $D$ must stem from genetic or epigenetic factors. In other words, discrete genetic/epigenetic backgrounds or biological hierarchies impose discrete traits at higher scales, as required by $T$.

The spectrum of the Selberg zeta function also reveals the discrete nature of both populations and species. However, while these spectra are based on prime closed geodesics, they are not necessarily indicative of adaptation or disadaptation. This suggests that populations may exhibit either discrete or pseudo-continuous and redundant behaviors. Thus, prime numbers are associated with adaptive species, whereas prime closed geodesics with high degeneracy correspond to populations.

Given that the number of primes grows as $T \ln T$ and the number of closed geodesics grows as $e^{(d - 1)T}/T$, the latter vastly outnumbers the former. This reflects the empirical relationship between populations and species. By analogy, eukaryotic species adaptations are closely linked to speciation, while prokaryotic discreteness is less distinguishable from that of the entire population. This may support the hypothesis of phenotypic discreteness among living organisms in nature \cite{Rasnitsyn2007}. 

If the Mathieu group corresponds to a mock modular form, such as a Maass form, then the system's dimensionality may be computed using Eisenstein series (e.g., for $M_{24}$; see \cite{Eguchi2011}). Assuming the K3 surface (interpreted as $s$ with interaction) possesses holomorphic symplectic symmetry, it forms a subgroup of $M_{24}$ \cite{Mukai1988}.

Now consider the oscillatory component of an Eisenstein series: $e^{2\pi i n \Re(s)}$. In a $(1+i)$-adic system, if $m \pi / 4 = 2\pi n \Re(s)$, then $m$ must be a multiple of $8$ when $\Re(s)$ is quantized as in a Bose-Einstein condensate (as discussed in the Results). In contrast, a $(-1+i)$-adic system yields $3m \pi / 4 = 2\pi n \Re(s)$, requiring $3m$ to be a multiple of $24$ under the same quantization. Atypical quantization with $m \sim 3$ may represent population bursts or collapses not in equilibrium.

When the system results from the interaction of two subsystems, these can be decomposed into four and twelve dimensions, respectively. Four dimensions suffice to describe any observed state, but twelve dimensions (including fluctuations) are necessary to represent all possible unobserved states.

Classically, using Liouville's equation,
$
\partial \overline{\partial} \phi = 2\pi \mu_{\phi} b_{\phi} e^{2b_{\phi} \phi}, \quad P^2 = 2\pi \mu_{\phi} b_{\phi}, \quad D = e^{b_{\phi}},
$
the Weil-Petersson metric becomes $ds_{\phi}^2 = e^{2b_{\phi} \phi} dk d\overline{k} = \text{constant}$ (see \cite{Dubrovin1992}), with negative Ricci curvature. The uniformization theorem guarantees unique solutions for any Teichm\"uller space \cite{Poincare1882, Klein1883, Poincare1883, Poincare1907, Koebe1907a, Koebe1907b, Koebe1907c}. This is especially relevant for genus $3$ on a closed Riemann surface with twelve real dimensions, as observed in Selberg zeta analysis.

When the quantum dilogarithm function is $e_b$, the ideal tetrahedron is given by
$
\psi(Z_{\psi}) = e_b\left( \frac{Z_{\psi}}{2\pi b_{\phi}} + \frac{i Q_{\psi}}{2} \right),
$
with $qa = \infty$, $qb = 1$, and $Z_{\psi} = \ln k$ (see \cite{Terashima2013}). This represents three spatial dimensions plus time, along with three-dimensional fluctuations (arising from dimensions other than the one of interest). Thus, an extension of the PzDom model to a time-evolving system should involve twelve dimensions.

Supporting this, the $(a, b, k)$ system may correspond to $SO(3)$, and Vogel's parameters for a simple Lie algebra $g$ are $\alpha = -2$, $\beta = 4$, $\gamma = -1$, and $t = h^{\vee} = 1$ \cite{Mkrtchyan2012}. If we set the half-sum of positive roots $\rho = 1$ (representing two interacting roots of $1$), the Freudenthal-de Vries strange formula gives:
$
12 \rho^2 = h^{\vee} \dim g = \dim g = 12,
$
highlighting the connection between object interactions in $\rho$ and twelve-dimensional time development.

More simply, consider $w_k = s - 1 = -1$ when the natural boundary is $s = 0$. Then,
$
\psi = \frac{1}{k^{w_k} |\zeta(w_k)|} = 12k,
$
implying that the top-ranking population or species has $12$ degrees of freedom for fitness at the natural boundary.

Finally, using Stirling's approximation,
$
\frac{\sqrt{2\pi}}{e} \left(1 + \frac{1}{12}\right) \approx 1.
$
This suggests that the ratio between the number of expected interactions in a large-$n$ system (plus one additional interaction) and that in the $k$-th population or species (plus one additional interaction) approaches $\sqrt{2\pi}$. This value is the geometric mean of interactions in two four-dimensional systems, aligning with the four-dimensional system extended by three additional dimensions each, yielding $4 \times 3 = 12$ dimensions as described.

\medskip

Note that when $\Re(s) \sim 1/2$, the observations pertain to a chaotic speciation world, not a structured species world. When $\Re(s) > 2$, structured species dynamics are observed. Interestingly, at $\Re(s) = 2$, extremely complex harmonic functions are generated by the boundary $A$ of the Mandelbrot set, and it remains mathematically unresolved whether this boundary is Minkowski nondegenerate or degenerate \cite{Lapidus2017}.

\begin{figure}
\includegraphics[width=10cm]{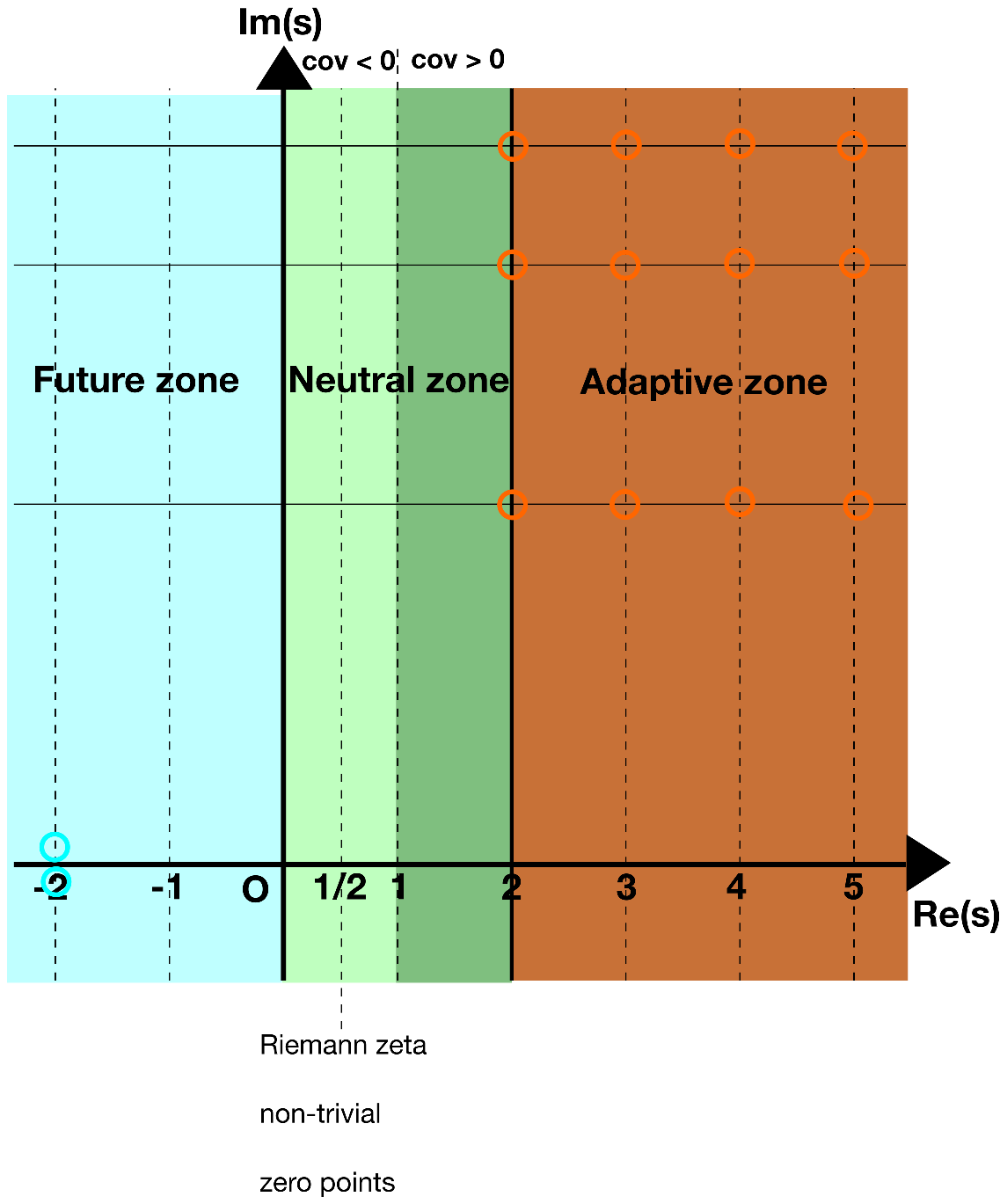}
\caption{Visualization of the upper half-plane $\mathbb{H}$ in the PzDom model. Nontrivial zeros of the Riemann $\zeta$ function, representing population bursts or collapses, are assumed to occur at the intersections of the $\Re(s) = \frac{1}{2}$ axis with horizontal dashed lines (note that the horizontal and vertical axes are scaled differently). Orange circles indicate adapted stages corresponding to species. The blue region represents a future stage characterized by $V_p = -\phi$, $\Re(s) \approx -1/(3\phi)$, and $\Im(s) \approx e^{-1/(3\phi)}$ \cite{Marino2014}; converged states are marked by blue circles. Note that $\Re(s)$ values corresponding to non-prime integers are theoretically unstable and are not observed empirically. The point $\Re(s) = 4$ is marked with an orange circle to indicate a potential ramification.}
\label{fig:8}
\end{figure}

\subsection{$s$ and mathematical biology}
The output of the developed system becomes unpredictable when $T_s < T_c$, due to the non-uniqueness of the infinite volume limit of the system \cite{Tasaki2015}. This phenomenon may correspond to the simulated case of sympatric ``speciation" described by \cite{deAguiar2009}.

\subsection{Interactions of subgroups and mathematical physics}
If a population or species consists of two subgroups with frequencies $X/Y$, then the fitness is given by $w = w_0 + BY - CX$, where $w_0$ is the baseline fitness without interaction, $B$ is the benefit to $X$ from $Y$, and $C$ is the cost to $X$ from the interaction. The term $|D|^{E(\Sigma N)}$ represents the virtual benefit from noninteracting individuals, and the combination of $w_0$ and $\Re(s) - 1$ captures the overall fitness resulting from interaction.

When $\Re(s) > 1$, the system tends toward a cooperative regime with higher fitness and potential speciation. In contrast, $0 < \Re(s) < 1$ corresponds to a competitive regime characterized by population bursts or collapses and discrete traits. Ordered symbiosis may be associated with $\Re(s) > 1$, while chaotic mutual exclusion may arise when $0 < \Re(s) < 1$. This interpretation is supported by geodesic analysis.

If $\Re(s) - 1 < 0$ persists for a significant duration (a condition not observed in \cite{Adachi2015}), then $q_b < 0$, indicating a highly competitive system. For instance, the case $|N(p)| = 1/4$ corresponds to a contracting universe. Here, $q_b$ is the coefficient of $XY$ in \cite{Zagier2000b}, and the population or species is negatively impacted by competition.

This scenario parallels the Chern-Simons action $S_{CS}$ \cite{Witten1989}:
\begin{equation}
S_{CS} = \frac{k_{CS}}{4\pi} \int_M \mathrm{tr} \left( A \wedge dA + \frac{2}{3} A \wedge A \wedge A \right),
\end{equation}
where $k_{CS}$ is the integer level of the theory, and the field strength vanishes at all boundaries. The gauge field $A$ may represent vectors such as $X^2$, $XY$, or $Y^2$. The weight of the first term in the trace is $0$ in adaptive (dependent) situations, $+1/3$ in cooperative interactions, and $-1/3$ in competitive interactions.

In the PzDom model, the time derivative of the action corresponds to the system's Lagrangian. Given that the Lagrangian is $T_s \ln k$, and assuming $k$ is constant, we have:
$
\frac{dS_{CS}}{dt} \cdot dT_s = T_s \ln k \, dT_s, \quad \text{so} \quad dS_{CS} = \frac{\ln k}{2 \arg D} t^2.
$
Assuming $\arg D$ is constant, we obtain:
$
\iint dS_{CS} \, dt = \frac{\ln k}{6 \arg D} t^3 = \frac{(\arg D)^2}{6} \ln k \, T_s^3.
$
If $\arg D = \pi$, then:
$
\iint dS_{CS} \, dt = \zeta(2) \ln k \, T_s^3 \propto s_d = \frac{4}{3} a_b T_s^3,
$
where $U = a_b T^4$ as in black-body radiation. This suggests that populations or species tend to cross this boundary. Here, $s_d$ represents the volume of a non-Euclidean sphere with radius $T_s$, and $a_b = C_c / d_c$, where $C_c$ is the circumference and $d_c$ is the diameter. Also, note that $1/\zeta(2)$ gives the probability that two randomly chosen integers are coprime. Thus, $\ln k \, T_s^3$ represents the expected interaction scale of $\iint dS_{CS} \, dt$, which is $T_s^3$ scaled by the relative entropy.

\medskip

An alternative formulation of the Lagrangian for a bosonic system is:
\begin{equation}
L = i \bar{\psi} \dot{\psi} - \epsilon \bar{\psi} \psi = |D|^2 (i \ln \dot{|D|} - \epsilon).
\end{equation}
As $\epsilon \rightarrow 0$, we find $T_s \rightarrow |D|^2$ and $\dot{|D|} \rightarrow k$, simplifying the analysis. Considering the sum of relative entropies $\sum \ln k$, we apply Stirling's approximation:
$
\ln k! \approx \ln \left( \sqrt{2\pi k} \left( \frac{k}{e} \right)^k \right) = \frac{1}{2} \ln(2\pi k) + k \ln \left( \frac{k}{e} \right),
$
for large $k$. The partial time derivative of $\sum \ln k$ is approximately:
\begin{equation}
\frac{1}{2k} \frac{\partial k}{\partial t} + (\ln k) \frac{\partial k}{\partial t} = \left( \frac{1}{2 \dot{|D|}} + \ln \dot{|D|} \right) \ddot{|D|}.
\end{equation}

The Hamiltonian $H_H$ incorporates a reflectionless potential and, from discrete quantum mechanics, includes pure imaginary shifts \cite{Odake2015}:
\begin{equation}
\lim_{\gamma \rightarrow 0} \gamma^{-2} H_H = p_m^2 - \frac{h_H(h_H + 1)}{\cosh^2(\Im(s)/b)},
\end{equation}
where $h_H = -s$. Note that $i \Im(s)/b = i \pi/2 \, (\mathrm{mod} \, i \pi)$ corresponds to regular singularities.

\subsection{Weights and dimensions}
Interestingly, if we define a $p$-adic field $\mathbb{F}$ for a prime $p \neq 2, 3$, then a necessary and sufficient condition for the existence of a cube root of $3$ in $\mathbb{F}$ is that $p \equiv 1 \ \mathrm{or} \ 11 \pmod{12}$. Therefore, if we restrict ourselves to three-dimensional interactions among identical constituents, this congruence condition must be satisfied. This highlights the symmetry inherent in the following modular forms.

Consider the function studied by Srinivasa Ramanujan (circa 1916) in an unpublished manuscript:
\begin{equation}
F(z) = q \prod_{n = 1}^{\infty}(1 - q^n)^2(1 - q^{11n})^2 = \sum_{n = 1}^{\infty} c(n) q^n.
\end{equation}
He proposed the associated $L$-function:
\begin{equation}
L(s, F) = (1 - c(11)11^{-s})^{-1} \prod_{p \neq 11} (1 - c(p)p^{-s} + p^{1 - 2s})^{-1},
\end{equation}
and later, \cite{Eichler1954} proved that for the elliptic curve $E: y^2 + y = x^3 - x^2$, we have $L(s, E) = L(s, F)$.

It is also well known that:
\begin{equation}
\Delta(z) = q \prod_{n = 1}^{\infty}(1 - q^n)^{24} = \sum_{n = 1}^{\infty} \tau(n) q^n,
\end{equation}
\begin{equation}
L(s, \Delta) = \prod_{p \ \text{prime}} (1 - \tau(p)p^{-s} + p^{11 - 2s})^{-1},
\end{equation}
where the pair $(F, \Delta)$ corresponds to $(1, 11)$ dimensions in terms of the parameter $p$. Both $F(z)$ and $\Delta(z)$ represent interactions between two identical decompositions of 12 dimensions, scaled by $q_R$.

The elliptic curve $E: y^2 + y = x^3 - x^2$ implies $y(y + 1) = x^2(x - 1)$. If time progresses from $x$ to $x - 1$, the interaction transitions from $x^2$ (interacting) to $(x - 1)$ (noninteracting). Modulo $4$, the equation yields $y$ values corresponding to both $0$ and a three-dimensional projection. The solutions are $(0, 0)$, $(0, 3)$, $(1, 0)$, and $(1, 3)$. Notably, $y = 3$ implies a left-hand side value of $3 \times 4 = 12$, which reflects the system's time asymmetry.

Furthermore, the combination of an oscillatory state with $|N(p)| = 2/3$ and a contractive state with $|N(p)| = 1/4$ yields a total of $11/12$. A contracted weight of $1$ versus a weight of $11$ signifies the disappearance of a function of order $1$ and the persistence of a function of order $11$.

\medskip

To expand this discussion, consider the icosahedron and the quintic equation, as explored in \cite{Klein1993}. A dihedral group represents a pair of complex conjugates. An icosahedron consists of six conjugate $n = 5$ dihedral groups and ten conjugate $n = 3$ dihedral groups. The quintic dimensions include three spatial dimensions, one temporal dimension, and one interaction axis.

For $n = 5$, the dihedral groups correspond to quintic-dimensional functions, with six conjugates arising from two interacting functions, each with three-dimensional fluctuations. For $n = 3$, the dihedral groups represent three-dimensional fluctuations, with ten conjugates corresponding to two interacting quintic-dimensional functions.

The fifteen cross lines of the icosahedron correspond to fifteen four-groups (each consisting of two conjugates), representing four distinct types of $SU(2)$ interactions between a dimension and a three-dimensional space minus one. The system of sixteen weights marks the first instance where complete symmetry breaking becomes possible, a necessary condition for system development \cite{Tachikawa2016, Witten2016}.

\subsection{Paddelbewegung-like motion}
To expand the interpretation presented in Figure~8, consider the $\phi$-plane with a small $s$ value defined on four-groups (three physical space dimensions plus one time dimension). For the spatial coordinates $(x, y, z)$, we define $\phi$ and a real parameter $\rho_P$ with a positive constant $a$ as follows:
\begin{eqnarray}
\left\{
\begin{array}{l}
x = (a - \rho_P \sin \frac{1}{2} \arg \phi) \cos \arg \phi, \\
y = (a - \rho_P \sin \frac{1}{2} \arg \phi) \sin \arg \phi, \\
z = \rho_P \cos \frac{1}{2} \arg \phi.
\end{array}
\right.
\end{eqnarray}

A Paddelbewegung-like motion \cite{Weyl1913} is then described by $\rho_P' = (-1)^{n_{\phi}} \rho_P$, $\phi' = \phi + 2n_{\phi} \pi i$, where $n_{\phi} \sim N(T)$ and $2n_{\phi} \pi \sim \Im(s)$.

For a toroidal structure, let $r$ be the radius of the tube and $R$ the distance from the center of the tube to the center of the torus. Introducing a real parameter $\sigma_P$, the coordinates become:
\begin{eqnarray}
\left\{
\begin{array}{l}
x = (R + r \cos \sigma_P) \cos \arg \phi, \\
y = (R + r \cos \sigma_P) \sin \arg \phi, \\
z = r \sin \sigma_P.
\end{array}
\right.
\end{eqnarray}

Assuming $\phi(t)$ and $\sigma_P(t)$ are differentiable functions of time $t$, with $\frac{d\phi}{dt}$ and $\frac{d\sigma_P}{dt}$ continuous and $(\frac{d\phi}{dt})^2 + (\frac{d\sigma_P}{dt})^2 \neq 0$, the rate of change of the surface area $A_{\phi\sigma}$ is given by:
\begin{equation}
\left( \frac{dA_{\phi\sigma_P}}{dt} \right)^2 = (R + r \cos \sigma_P)^2 \left( \frac{d\phi}{dt} \right)^2 + r^2 \left( \frac{d\sigma_P}{dt} \right)^2.
\end{equation}

The function $\psi$ is defined by:
\begin{equation}
\psi = \int_0^{\sigma_P} \frac{d\sigma_P}{\frac{R}{r} + \cos \sigma_P},
\end{equation}
and the metric becomes:
\begin{equation}
dA_{\phi\sigma_P}^2 = (R + r \cos \sigma_P)^2 (d\phi^2 + d\psi^2).
\end{equation}

A point in the four-group $(\phi, \psi)$ can be described as a translation group:
\begin{eqnarray}
\left\{
\begin{array}{l}
\phi' = \phi + 2n_{\phi} \pi i \sim \phi + \Im(s) i, \\
\psi' = \psi + \frac{2\pi r}{\sqrt{R^2 - r^2}} n_{\psi} \sim \psi + \frac{2\pi}{2\pi} n_{\psi} \sim \psi + n_{\psi},
\end{array}
\right.
\end{eqnarray}
where $\sin \theta_{\psi} = \frac{r}{R}$ and $\tan \theta_{\psi} = \frac{n_{\phi}}{T} \sim \frac{1}{2\pi}$. Here, $\tan \theta_{\psi}$ is a $T$-normalized quantized number $n_{\phi}$.

Thus, when $\Re(s) > 2$, the values of $s$ and $w$ tend to localize and jump among integer positions, as illustrated by the orange circles in Figure~8.

\subsection{Dynamical system hierarchy}
We now turn to several additional considerations related to the suppression of fluctuations. For the application of hyperbolic geometry (logarithmic-adic space) and the technique of blowing up to resolve singularities, see our earlier work \cite{Adachi2017}. From the perspective of generalized function theory, the concept of cohomology naturally arises. If we define operator III in terms of cohomology, then the vanishing of higher cohomology groups, $H^p = 0$ for $p \geq 2$ (with $p$ prime), and the Kawamata-Viehweg vanishing theorem are satisfied. This result indicates that investment in adaptation at higher hierarchical levels reduces chaotic behavior within those hierarchies. This is consistent with the fact that our complex manifold is a Stein manifold, and $s$ behaves as a Schwartz distribution.

An empirical process has already been introduced in the form of ``Paddelbewegung,” inspired by the work of Hermann Weyl. Further developments of this framework may include the use of Riemann schemes, hypergeometric differential equations, or Painlev\'e VI equations to model hierarchical time evolution. Exploring a variety of model types may also open pathways to connections with Galois theory and \'etale cohomology, offering a deeper understanding of hierarchical structures in natural systems, particularly in biological contexts. This represents a promising direction for future research.

Finally, by invoking the Atiyah-Singer index theorem, we observe that the twisted (fractal) property and the Euler number $\int_B e(TB)$ are equal to the topological Euler characteristic $\chi(B) = \sum (-1)^l l$. Consequently, the analytical index of the Euler class (as the Poincar\'e dual) must also coincide. To evaluate this correspondence, the Chern class should be given by $(-1)^l l$. Analytically, the Hirzebruch signature (a characteristic derived from species) of $B$ is:
$
(-1)^n \int_B \prod_{i = 1}^n \frac{p_i}{\tanh p_i},
$
where
$
\frac{p_i}{\tanh p_i} = \sum_{k \geq 0} \frac{2^{2k} B_{2k}}{(2k)!} p_i^{2k}.
$
Topologically, this expression corresponds to the $L$-genus.

\medskip

Thus, the methodology developed for the “small $s$” metric can be extended to characterize the hierarchy of dynamical systems—capturing both adaptation and interaction—using only abundance data over time.

\subsection{Comparison between our model and mathematical ideas}
According to \cite{Mochizuki2020d}, a species can be viewed as a collection of set-theoretic formulas that define a category within any given model of set theory. For instance, species may be represented as a $k$-set of a single dimension $N_k$ with a specific topology, or as a set of tuples $(p, l, v)$ equipped with particular morphisms. This paper completes our integrated model of species dynamics, and we now compare it with several mathematical frameworks that offer alternative interpretations.

\medskip

If $\varnothing \neq \partial B$, the Poisson-Jensen formula applies to $D = e^{s/b}$:
\begin{equation}
\Re(s) = -b \sum_{i = 1}^n \ln \left| \frac{\rho^2 - \bar{a}_i z}{\rho(z - a_i)} \right| + \frac{1}{2\pi} \int_0^{2\pi} \Re \left( \frac{\rho e^{i\theta} + z}{\rho e^{i\theta} - z} \right) \rho \, d\theta,
\end{equation}
where $a_i$ are the zero points of $w = D$. The first term represents the discrete contributions to $\Re(s)$ from the zero points of $w$ (i.e., observed phenomena), while the second term corresponds to the continuous contributions to $\Re(s)$ in the range $0 \leq \Re(s) < 2$.

Alternatively, this decomposition can be interpreted through the lens of the Radon-Nikodym theorem as a unique Lebesgue decomposition:
\begin{equation}
\Phi(E) = F(E) + \Psi(E),
\end{equation}
where $s \in E$, $\Phi(E)$ is a $\sigma$-additive set function on $B$, $F(E)$ is absolutely continuous, and $\Psi(E)$ is singular. Identifying $F(E) = \Im(s)$ and $\Psi(E) = \Re(s)$, and invoking the Riemann-Hurwitz formula, we see that $\Phi(E)$ uniquely decomposes contributions from the same hierarchical level ($F(E)$) and from higher-order hierarchies ($\Psi(E)$).

Further insight is provided by the universal coefficient theorems \cite{Bott1982}:
\begin{itemize}
\item[(a)] For homology with coefficients in a $B$-module $G$:
\begin{equation}
H_q(B; G) \cong H_q(B) \otimes G \oplus \mathrm{Tor}(H_{q-1}(B), G),
\end{equation}
where the first term corresponds to $\Im(s)$ and the second to $\Re(s)$.
\item[(b)] For cohomology with coefficients in a $B$-module $G$:
\begin{equation}
H^q(B; G) \cong \mathrm{Hom}(H_q(B), G) \oplus \mathrm{Ext}(H_{q-1}(B), G),
\end{equation}
again with the first term corresponding to $\Im(s)$ and the second to $\Re(s)$.
\end{itemize}

By Hodge's theorem, cohomology is isomorphic to the space of harmonic forms, implying that the cohomological dimension reflects the number of harmonic oscillators present in the system. This provides a physical interpretation of the mathematical structure underlying the biological system.

\medskip

Collectively, these mathematical frameworks offer valuable perspectives on renormalization, fractal structures, and multilevel selection—concepts that are central to understanding complex biological hierarchies. In particular, divergent terms may signify the influence of higher-order structures, such as fractals, acting across multiple levels of biological organization.

\subsection{Our model and statistical mechanics}
In this study, we introduce new metrics—$T_c$, $W$, and $S$—to distinguish between populations and species using only the total number of individuals. These metrics reveal spontaneous symmetry breaking in biological systems. The parameter $S$ quantifies the degree of ordering based on dominance. Our framework enables the evaluation of the critical temperature $T_c$ and the Weiss field $W$, allowing us to differentiate between various ordering states such as random distributions, dominant species, or dominant populations. It also identifies potential phase transitions and adaptive states, particularly when $T_s \approx W$.

The condition $T_s \approx W$ represents a biological analogue of Bose-Einstein condensation in population dynamics \cite{Tasaki2015}. While \cite{Knebel2015} described similar condensation phenomena, our model extends the concept to include non-typical Bose-Einstein condensation scenarios. Phase separation in our model depends on the level of biological hierarchy—whether population or species—and allows us to assess the adaptive structures that emerge during evolutionary transitions. In a population-dominant phase, a single population may dominate the community, whereas in a species-dominant phase, the collective presence of a species, rather than any single population, becomes dominant. These increasing phases signal the onset of subsequent domination phases. In contrast, chaotic phases yield unpredictable outcomes.

Our model also demonstrates that some highly adapted species exhibit greater stability than others. Furthermore, the nontrivial zeros of the Riemann zeta function correspond to adaptive species and are associated with prime numbers. The entropy term $\ln k$, introduced in this work, can be approximated by $\pi(k)/k \approx 1/\ln k$, where $\pi(k)$ is the prime counting function. The observed decline in prime number density with increasing entropy suggests that this trend is disrupted only when higher-order hierarchies are present. Thus, higher-order hierarchies may act as investments in adaptive ordering structures.

\medskip

Let us now consider biological hierarchies ranging from genes and cells to multicellular individuals, populations, and species. For genes and cells, when the surrounding environment is restricted to a small scale (e.g., within cells or individuals), their copy numbers remain nearly static and are no longer adaptive. In such cases, $T_s \sim 0$. In contrast, individuals within populations or the populations themselves exhibit chaotic behavior when $T_s \neq 0$. Our data indicate that the species level demonstrates greater adaptability than lower-order hierarchies.

We therefore conclude that reproductive scales are inherently chaotic, and achieving adaptation requires the presence of higher-order hierarchies. While lower-order hierarchies may exhibit localized adaptive behavior, broader environmental scales—such as species, communities, or ecosystems—must be considered to detect selection pressures acting on genes or cells.

\subsection{Relation of $s$ to entropy}
In connection with Landauer’s principle (e.g., \cite{Palazzo2018}), the expression $N_k = a - b \ln k$ implies a reduction of information by an amount $b \ln k$ from an initial information content $a$ in each $k$-th group. This assumption forms the foundation of our model, and many of its features are derived from this principle.

For instance, we adopt the partition function $\zeta(s)$—the Riemann zeta function—as a generalization of Zipf’s law. Within this framework, the nontrivial zeros of the zeta function correspond to adapted states characterized by diverging fitness. Consequently, the zeta function, as the inverse of the fitness landscape, can be interpreted as a free energy landscape across hierarchical levels (e.g., \cite{Annila2009}), reflecting its divergent behavior.

In this context, $\Re(s)$ serves as a factor that elevates the system to higher hierarchical levels, while $\Im(s)$ is associated with constraints at lower levels. This interpretation aligns with the proposal in \cite{Grmela2013}, which states that “macroscopic systems, in their response to imposed external forces, follow an optimization strategy that alters their internal structure to minimize resistance.”

In our biological model, this optimization corresponds to natural selection acting on fitness. The potential for such optimization is captured not by entropy or entropy production, as in traditional thermodynamic models, but by the parameter small $s$.

\subsection{$s$ as an information criterion}
The small $s$ parameter bears conceptual similarity to the Widely Applicable Information Criterion (WAIC) \cite{Watanabe2010} in several respects. WAIC is defined as:
\begin{equation}
W_n = T_n + \frac{\beta}{n} V_n,
\end{equation}
where $T_n$ represents the training loss, $V_n$ denotes the functional variance, and $\beta$ is the inverse temperature.

If we substitute the functional variance $V_n$ with a covariance term, then the expression $\Delta z_k \ln D_k$ becomes analogous to the WAIC for the $k$-th component. In this sense, the small $s$ parameter can be interpreted as a derivative form of an information criterion derived from maximum likelihood estimation. It encapsulates both the fit to observed data and the complexity of the model, thereby serving as a biologically grounded analog to WAIC in the context of population and species dynamics.

\subsection{PzDom Model}
The state of the $\Re(s)$ domain can be broadly categorized into three regimes:
\begin{enumerate}
    \item \textbf{Chaotic phase:} $0 < \Re(s) < 1$, characterized by mutual exclusion and competitive interactions.
    \item \textbf{Ordered phase:} $1 < \Re(s) \leq 2$, marked by cooperation and speciation.
    \item \textbf{Most adaptively ordered phase:} $\Re(s) > 2$, associated with stability.
\end{enumerate}

According to Fermat's Last Theorem, the noninteracting mode of subpopulations with $q_b = 0$ is only feasible when $X^n + Y^n = Z^n$ for $n \leq 2$. This imposes a structural constraint on the system. It is important to note that co-evolving communities occupying distinct ecological niches are not captured by this model, representing a limitation of the current framework.

Now consider the Euler-Mascheroni constant:
\begin{equation}
\lim_{s \rightarrow 1} \left( \zeta(s) - \frac{1}{s - 1} \right) = \gamma = 0.577\ldots
\end{equation}
When $k = 1$, we have $|\zeta(s)| = \Sigma N / N_1$. In the context of a population, this value is approximately $0.5$, implying $w < 0$, and thus a long-term decline in population size. In contrast, for a species undergoing speciation, where $N_1 \ll \Sigma N$, we find $w > 0$. Even when $N_1 \approx \Sigma N$, $w$ remains positive, indicating that transitioning from a population-level to a species-level hierarchy enables escape from the declining trend observed in isolated populations.

This suggests that species-level organization provides a more robust and adaptive structure. The conceptual framework and implications discussed here are visually summarized in Figure~8.\\ 

\subsection{The boundary $\Re(s) = 2$}
The parameter $\Re(s)$ in the PzDom framework provides a quantitative criterion for distinguishing continuous community dynamics from the emergence of discrete species-level structure. In this model, $\Re(s)$ reflects the degree of order and the strength of interactions within a community. The value $\Re(s)$  represents a critical threshold at which the system transitions from a continuous regime to one in which species become stable, discrete units.

When $\Re(s) < 2$, community composition behaves as a continuum: species boundaries are diffuse, competitive interactions dominate, and turnover rates are high. This regime corresponds closely to the classical continuum concept in ecology, in which communities are viewed as fluid assemblages shaped by individualistic species responses. In contrast, when $\Re(s)>2$, species stabilize as discrete entities. Interactions weaken, independence among species increases, and mathematically identifiable structures—analogous to $p$-Sylow subgroups—emerge. Thus, $\Re(s)=2$ marks a phase transition between continuous and discrete ecological organization.

The prominence of this threshold is not arbitrary. It arises naturally from the mathematical structure of the model, including properties of zeta functions, fractal dimensions, and number-theoretic constraints. In particular, the stability of discrete units above $\Re(s)=2$ parallels the emergence of stable closed orbits in analytic number theory and the stabilization of structures in systems whose effective dimension exceeds two. These mathematical features provide a theoretical basis for the ecological interpretation of $\Re(s)=2$ as a boundary between fluid and structured community states.

Empirical data support this interpretation. In Dictyostelia field surveys, soil mesofauna datasets, protein-expression distributions in HEK-293 cells, and even long-term economic time series, transitions across $\Re(s)=2$ coincide with shifts between stable and unstable system states. Periods with $\Re(s)>2$ exhibit persistent structure, reduced turnover, and identifiable discrete units, whereas periods with $\Re(s)<2$ show increased variability, weakened boundaries, and continuous dynamics. These observations suggest that $\Re(s)=2$ functions as a general threshold for order-disorder transitions across diverse complex systems.

This framework offers a potential resolution to long-standing debates regarding the ontological status of species and community boundaries. Rather than being universally real or universally artificial, species-level discreteness emerges only when the system resides in the ordered regime defined by $\Re(s)>2$. Below this threshold, continuous descriptions are more appropriate. Thus, the PzDom model unifies continuous and discrete perspectives by showing that both arise naturally as different phases of the same underlying system.

\subsection{Continuous v.s. Discrete}
The question raised here highlights a fundamental tension between classical ecological views of community continuity and modern mathematical approaches that emphasize discrete hierarchical structure. The continuum concept, originating from Gleason's individualistic hypothesis, argues that community composition changes smoothly along environmental gradients and that categorical boundaries are largely artifacts of human description. In this view, species respond independently to environmental variation, and community boundaries lack objective biological reality.

In contrast, the framework proposed in *``Biological hierarchies emerged from natural characteristics of number theory''* suggests that biological hierarchies arise not merely from ecological contingencies but from intrinsic mathematical constraints. According to this perspective, hierarchical levels—such as cells, individuals, and communities—emerge as discrete, stable configurations determined by number-theoretic properties. These constraints impose a limited set of permissible system states, thereby generating discontinuities that are not arbitrary but mathematically necessary.

The relationship between these two perspectives can be understood as a distinction between phenomenological continuity and structural discreteness. From a phenomenological standpoint, ecological interactions often appear continuous: species exhibit omnivory, ontogenetic shifts, and context-dependent trophic positions, all of which blur categorical boundaries. However, when viewed from a structural or dynamical perspective, ecological systems are constrained by energy transfer efficiencies, stability requirements, and number-theoretic invariants that favor discrete trophic levels and hierarchical segmentation. Thus, continuous ecological variation may coexist with, and even give rise to, discrete higher-order structures.

This interpretation becomes particularly evident in the context of food-web organization. Although trophic interactions form a seemingly continuous network, energetic constraints (e.g. the approximate 10\% transfer efficiency) impose quantized trophic steps, while number-theoretic considerations may contribute to the avoidance of resonant population cycles and the emergence of stable interaction motifs. Similarly, allometric scaling laws link continuous body-size distributions to discrete trophic roles, illustrating how physical and mathematical constraints generate hierarchical structure from continuous underlying variables.

The theory here suggests that discrete microscopic elements become continuous under coarse-graining, and subsequently give rise to new discrete macroscopic levels, aligns closely with renormalization-group concepts in statistical physics. The number-theoretic framework extends this view by proposing that the emergent discrete levels are not arbitrary but are instead anchored to mathematically privileged configurations. In this sense, the number-theoretic approach provides a rationale for why specific hierarchical levels recur across biological systems, rather than varying continuously with environmental conditions.

Overall, the continuum perspective captures the fluidity and context-dependence of ecological interactions, whereas the number-theoretic framework explains the emergence of stable, discrete hierarchical organization. These views are not mutually exclusive; rather, they describe different layers of the same system. Continuous ecological processes may, through coarse-graining and dynamical constraints, converge onto discrete hierarchical states that reflect deeper mathematical structure. This synthesis suggests that biological hierarchies are neither purely observational constructs nor solely products of ecological dynamics, but instead arise from the interplay between continuous processes and discrete mathematical invariants.\\

\subsection*{Conclusions}
With only three foundational assumptions—(1) a logarithmic approximation of population densities, (2) the exponential nature of fitness, and (3) the Price equation—we have successfully applied the Price equation, the $R = T$ theorem, and Weil's explicit formula to construct an ecological model, which we term the PzDom model. This model is grounded in a novel topological parameter, the small $s$. Within this framework, species can be defined as $p$-Sylow subgroups of a community occupying a single niche. The boundary between adaptive species and chaotic populations or species is identified at $\Re(s) = 2$, a result also supported by fractal theory. The norm of prime closed geodesics is approximately $|N(p)| \sim 1$, and the congruence class of $p \bmod 4$ reveals adaptive versus disadaptive states, as well as connections to Bose-Einstein condensates.

The future adaptive or disadaptive behavior of individuals is partially predictable via the Hurwitz zeta function, and a time-dependent fitness function has been identified. Furthermore, we demonstrated that species adaptations (viewed as primes) and population conversions (viewed as geodesics) naturally lead to phenotypic discontinuities when information entropy is maximized. The model confirms the existence of distinct biological phases for populations and species, enabling classification and prediction of population/species types based solely on their dynamics and distributions. Remarkably, this partially dynamic model requires only population density and species classification data, avoiding the need for extrapolation. It thus offers a foundational framework for future investigations into the hierarchical nature of biological systems. 

\medskip

We also examined the topological structure of the system using empirical species density data from wild Dictyostelia and soil mesofauna communities, complemented by protein data obtained through liquid chromatography-mass spectrometry. By employing Clifford algebras, congruent zeta functions, the Weierstra\ss $\wp$ function, and the Painlev\'e VI equation, we confirmed the emergence of hierarchy and temporal structure within a one-dimensional probability space endowed with specific topologies. This approach also provided insights into the interaction dynamics within the model. The previously developed “small $s$” metric proves effective in characterizing dynamical system hierarchies and interactions using only abundance data over time.

\medskip

To further elucidate the role of induced fractals and their relationship to renormalization in physics, we propose a theoretical extension based on a newly discovered correspondence: the scaling parameters for magnetization align precisely with the imaginary parts of the nontrivial zeros of the Riemann zeta function. Drawing analogies to magnetization and invoking the Fake Monster Algebra, we support this theory with empirical species density data from a wild Dictyostelia community. A master torus and associated Lagrangian/Hamiltonian formulations are derived, expressing fractal structures as solutions that mitigate divergent terms in renormalization. 

\medskip

Our findings indicate that the long-standing contrast between continuous and discrete views of community structure reflects different dynamical regimes of the same system. The PzDom framework shows that continuous ecological variation, represented by small $s$, can transition into discrete species-level organization when number-theoretic constraints become dominant. Thus, continuity and discreteness coexist as phase-dependent properties, providing a unified explanation for persistent debates in community ecology and the **species problem**.

\section{Methods and Materials}
\subsection{Field research (Dictyostelia)}

The data on the number of individuals in each population and species were collected from natural (non-laboratory) environments. The sampling methodology follows that described in \cite{Adachi2015}. Field experiments were conducted with approval from the Ministry of the Environment (Japan), Ministry of Agriculture, Forestry and Fisheries (Japan), Shizuoka Prefecture (Japan), and Washidu Shrine (Japan). Approval numbers include 23Ikan24, 24Ikan72-32, and 24Ikan72-57.

Soil samples were collected from two point quadrats in the Washidu region of Izu, Japan. The number of individual cellular slime molds per gram of soil was determined by counting the number of plaques cultivated from the soil samples. Species identification was performed using both morphological characteristics and DNA sequencing of the 18S rRNA genes. Sampling was conducted monthly from April 2012 to January 2013, excluding May. Data analysis was performed using Microsoft Excel 12.3.6, wxMaxima 15.04.0, SageMath 8.8, R 3.3.2, and GNU Octave 3.8.0.

More specifically, sampling was carried out at two $100\, \mathrm{m^2}$ quadrats in Washidu (coordinates: $35^\circ3'33''$N, $138^\circ53'46''$E and $35^\circ3'45''$N, $138^\circ53'32''$E). Within each quadrat, nine sampling points were established at 5-meter intervals. From each point, 25 g of soil was collected.

Cellular slime molds were isolated as follows: each soil sample was suspended in 25 ml of sterile water and filtered through sterile gauze. Then, $100\, \mu$l of the filtrate was mixed with $100\, \mu$l of HL5 culture medium containing \textit{Klebsiella aerogenes} and spread onto KK2 agar plates. After incubation at $22^\circ$C for two days, the number of plaques on each plate was recorded. Each plaque corresponds to the total number of living cells at any stage of the life cycle. The niche considered here includes all propagable individuals of Dictyostelia, without classification by life stage or hierarchy. Age and size structures were not examined, as the organisms are primarily unicellular microbes.

Mature fruiting bodies, composed of cells from a single species, were collected along with plaque count data from their respective regions. Spores were used to inoculate KK2 medium for purification or SM/5 medium for expansion. All analyses were completed within two weeks of sample collection. Species identification was based on 18S rRNA (SSU) sequences, amplified and sequenced using PCR and primers described in \cite{Medlin1988}, with reference to the SILVA database (http://www.arb-silva.de/). Media recipes followed protocols available at http://dictybase.org/techniques/media/media.html.

\subsection{Field research (soil mesofauna community)}

The data on the number of individuals in each population and species were collected from natural (non-laboratory) environments. Field experiments were approved by the Ministry of the Environment (Japan) and Nara Prefecture (Japan), under approval number Kankinyoshikokkyo2410311.

Litter samples were collected from point quadrats in the Ohdaigahara region of Nara, Japan. The number of individuals per 100 cc of litter was determined by counting specimens fixed in 100\% ethanol, collected using Tullgren apparatuses. Species identification was based on morphological characteristics, following the references in \cite{Aoki2015}. Samples were collected monthly from April 2025 to November 2025 for analysis, and from April 2026 to November 2026 for validation. Data analysis was performed using Microsoft Excel 16.16.27, SageMath 10.0, Julia 1.3.1, and R 4.0.3.

More specifically, sampling was conducted at a $100\, \mathrm{m^2}$ quadrat in Higashi-Ohdai (coordinates: $34^\circ6'47''$N, $136^\circ3'42''$E). Within this quadrat, nine sampling points were established at 5-meter intervals. From each point, 100 cc of litter was collected.

Soil mesofauna individuals were extracted using Tullgren apparatuses handmade from 2-liter Bansoreicha plastic bottles (Yakult Honsha Co., Ltd., Tokyo, Japan) and 40 W incandescent lamps (Bon Furniture Corp., Kainan, Japan). The apparatuses included two layers of Medigaze USP type IV gauze (9 × 8 pitches/cm\textsuperscript{2}, 15 cm × 15 cm; Osaki Medical Corp., Nagoya, Japan) and five pinches of Tetra Goldfish Easy Care Gravel (Spectrum Brands Japan Corp., Yokohama, Japan). Collected individuals were preserved in 100\% ethanol and examined under a binocular stereo microscope SZH-ILLD (OLYMPUS OPTICAL CO., LTD., Hachioji, Japan), equipped with a D7000 digital single-lens reflex camera (Nikon Corp., Tokyo, Japan) and a PF-D double-arm microscope light source (Shiokaze Engineering Corp., Niigata, Japan).

In addition, the physical and chemical properties of the litter were measured as follows:
\begin{itemize}
    \item Soil and air temperature, moisture, and pH: Digital Soil Tester (ZOUBAOQ, Shanghai, China)
    \item Slope angle: SR-90 slant rule (Akatsuki Mfg. Co., Ltd., Ayabe, Japan)
    \item Light intensity (lux): LX-1010B digital lux meter (Shanghai Handsun Electronic Co., Ltd., Shanghai, China)
    \item Electrical conductivity: TDS EC meter (DiyStudio Corp., Sendai, Japan)
\end{itemize}

\subsection{Experiments}
A human HEK-293 cell line derived from embryonic kidney tissue was obtained from RIKEN (Japan). The sampling protocol is detailed in \cite{Adachi2017}. Original cultures were cryopreserved on March 18, 2013 (3-year storage) and March 5, 2014 (2-year storage), and subsequently used in experiments conducted between February and June 2016. Cells were cultured in Modified Eagle's Medium (MEM) supplemented with 10\% fetal bovine serum (FBS) and 0.1 mM nonessential amino acids (NEAA) at 37$^\circ$C in a 5\% $\mathrm{CO_2}$ atmosphere. Subculturing was performed using 0.25\% trypsin.

For cryopreservation, cells were frozen according to RIKEN's standard protocol: suspended in culture medium with 10\% dimethyl sulfoxide (DMSO), cooled to 4$^\circ$C at a rate of $-2^\circ$C/min, held for 10 minutes, then cooled to $-30^\circ$C at $-1^\circ$C/min and held again for 10 minutes. Subsequently, they were cooled to $-80^\circ$C at $-5^\circ$C/min and stored overnight before transfer to liquid nitrogen.

Protein extraction from HEK-293 cells followed the standard RIPA buffer protocol (NACALAI TESQUE, INC., Kyoto, Japan). Approximately $10^6$ harvested cells were washed in Krebs-Ringer Buffer (KRB; 154 mM NaCl, 5.6 mM KCl, 5.5 mM glucose, 20.1 mM HEPES pH 7.4, 25 mM $\mathrm{NaHCO_3}$), resuspended in 30 $\mu$l of RIPA buffer, lysed by repeated passage through 21G needles, and incubated on ice for 1 hour. Lysates were centrifuged at 10,000 g for 10 minutes at 4$^\circ$C, and supernatants were collected.

Protein concentrations were determined using a Micro BCA Protein Assay Kit (Thermo Fisher Scientific, Waltham, USA). Proteins were processed using the XL-Tryp Kit Direct Digestion (APRO SCIENCE, Naruto, Japan). Samples were embedded in acrylamide gel, washed with ultrapure water and dehydration solution, and dried. Further processing used the In-Gel R-CAM Kit (APRO SCIENCE, Naruto, Japan): samples were reduced for 2 hours at 37$^\circ$C, alkylated for 30 minutes at room temperature, washed multiple times, and trypsinized overnight at 35$^\circ$C.

Peptides were purified using ZipTipC18 (Merck Millipore, Billerica, USA). Tips were conditioned with acetonitrile and 0.1\% trifluoroacetic acid (TFA), peptides were bound via $\sim$20 aspiration/dispensing cycles, washed, and eluted with 0.1\% TFA/50\% acetonitrile. Eluates were vacuum-dried and stored at $-20^\circ$C. Prior to LC/MS, samples were resuspended in 0.1\% formic acid and quantified using the Pierce Quantitative Colorimetric Peptide Assay (Thermo Fisher Scientific). The full protocol is available at\\ \url{http://dx.doi.org/10.17504/protocols.io.h4qb8vw}.

LC/MS analysis was performed at the Medical Research Support Center, Graduate School of Medicine, Kyoto University, using a TripleTOF 5600 quadrupole time-of-flight (Q-Tof) mass spectrometer (AB Sciex Pte., Ltd., Concord, Canada). Standard protocols were followed, with 1 $\mu$g of sample loaded per run. Quantitative data for identified proteins were extracted using ProteinPilot 4.5.0.0 software (AB Sciex). For further details, see \cite{Adachi2017}.

\subsection{Economic history}
The raw historical economic data based on the Western calendar were obtained from \cite{Maddison2007}. All calculations and analyses were conducted using Microsoft Excel 16.16.27, SageMath 10.0, Julia 1.3.1, and R 4.0.3.

\section{Acknowledgments}
I thank the Ministry of the Environment (Japan), Ministry of Agriculture, Forestry and Fisheries (Japan), Shizuoka Prefecture (Japan), Nara Prefecture (Japan), and Washidu Shrine (Japan) for providing sample resources. I also extend my gratitude to the Medical Research Support Center, Graduate School of Medicine, Kyoto University, for conducting the LC/MS analyses. I am deeply grateful to the reviewers and colleagues whose insights and advice—spanning the fields of medicine, biology, mathematics, and physics—have greatly enriched this work. Financial support from Tokushima University and Kyoto University is also gratefully acknowledged. Finally, I thank all reviewers and collaborators for their valuable comments and suggestions.

\end{document}